\documentclass[traditabstract]{aa}  
\usepackage{geometry}
\usepackage{xcolor}
\usepackage{colortbl}
\usepackage{tabularx}
\usepackage{graphicx} 
\usepackage{multirow}
\usepackage{placeins}
\definecolor{lightgray}{rgb}{0.83, 0.83, 0.83}
\definecolor{lightblue}{rgb}{0.67, 0.84, 0.90}
\definecolor{lightgreen}{rgb}{0.56, 0.93, 0.56}

\usepackage{natbib}
\bibpunct{(}{)}{;}{a}{}{,} 

\usepackage{ctable}
\usepackage{graphics}   
\usepackage{graphicx}   
\usepackage{epstopdf}
\usepackage{longtable}   
\usepackage{url}      
\usepackage{bm}        
\usepackage{soul}

\usepackage{caption}
\usepackage{subcaption}

\usepackage{tablefootnote}
\usepackage[varg]{txfonts}
\usepackage{pdflscape}
\usepackage{supertabular}
\usepackage{hyperref}
\usepackage{xcolor}
\usepackage[normalem]{ulem}
\usepackage{natbib}
\usepackage{amsmath}
\definecolor{green}{rgb}{0.3,0.7,0.}
\definecolor{purple}{rgb}{0.77, 0.29, 0.55}

\hypersetup{
    bookmarks=true,         
    unicode=false,          
    colorlinks=true,       
    linkcolor=blue,          
    citecolor=blue,        
    filecolor=blue,      
    urlcolor=blue           
}

\usepackage{color}

\begin{document}

\title{BASS. XLIII: Optical, UV, and X-ray emission properties of unobscured \textit{Swift}/BAT active galactic nuclei\thanks{Full versions of Tables \ref{tab:src_list} to \ref{tab:kbol} are only available in electronic form at the CDS via anonymous ftp to \url{https://cdsarc.u-strasbg.fr} (130.79.128.5) or via \url{http://cdsweb.u-strasbg.fr/cgi-bin/qcat?J/A+A/}.}}  

\titlerunning{Optical, UV, and X-ray emission properties of unobscured \textit{Swift}/BAT AGN}
\author{Kriti K. Gupta\inst{1,2,3}\thanks{E-mail: kkgupta@uliege.be}, Claudio Ricci\inst{1,4}, Matthew J. Temple\inst{1}, Alessia Tortosa\inst{5}, Michael J. Koss\inst{6,7}, Roberto J. Assef\inst{1}, Franz E. Bauer\inst{8,9}, Richard Mushotzy\inst{10}, Federica Ricci\inst{5,11}, Yoshihiro Ueda\inst{12}, Alejandra F. Rojas\inst{1,13}, Benny Trakhtenbrot\inst{14}, Chin-Shin Chang\inst{15}, Kyuseok Oh\inst{16}, Ruancun Li\inst{4,17}, Taiki Kawamuro\inst{18}, Yaherlyn Diaz\inst{1}, Meredith C. Powell\inst{19,20}, Daniel Stern\inst{21}, C. Megan Urry\inst{22}, Fiona Harrison\inst{23}, and Brad Cenko\inst{24}
}
\authorrunning{Gupta et al.}

\institute{Instituto de Estudios Astrofísicos, Facultad de Ingenier\'ia y Ciencias, Universidad Diego Portales, Av. Ej\'ercito Libertador 441, Santiago, Chile \and
STAR Institute, Li\`ege Universit\'e, Quartier Agora - All\'ee du six Ao\^ut, 19c B-4000 Li\`ege, Belgium \and
Sterrenkundig Observatorium, Universiteit Gent, Krijgslaan 281 S9, B-9000 Gent, Belgium \and
Kavli Institute for Astronomy and Astrophysics, Peking University, Beijing 100871, People's Republic of China \and
INAF\textemdash Osservatorio Astronomico di Roma, via di Frascati 33, I-00078 Monte Porzio Catone, Italy \and
Eureka Scientific, 2452 Delmer Street Suite 100, Oakland, CA 94602-3017, USA \and
Space Science Institute, 4750 Walnut Street, Suite 205, Boulder, CO 80301, USA \and
Instituto de Astrof\'isica and Centro de Astroingenier\'ia, Facultad de F\'isica, Pontificia Universidad Cat\'olica de Chile, Campus San Joaquin, Av. Vicu\~na Mackenna 4860, Macul Santiago 7820436, Chile \and
Millennium Institute of Astrophysics, Nuncio Monse\~nor S\'otero Sanz 100, Of 104, Providencia, Santiago, Chile \and
Department of Astronomy and Joint Space-Science Institute, University of Maryland, College Park, MD 20742, USA \and
Dipartimento di Matematica e Fisica, Universita Roma Tre, via della Vasca Navale 84, I-00146, Roma, Italy \and
Department of Astronomy, Kyoto University, Kitashirakawa-Oiwake-cho, Sakyo-ku, Kyoto 606-8502, Japan \and
Centro de Astronomía (CITEVA), Universidad de Antofagasta, Avenida Angamos 601, Antofagasta, Chile \and
School of Physics and Astronomy, Tel Aviv University, Tel Aviv 69978, Israel \and
Joint ALMA Observatory, Avenida Alonso de Cordova 3107, Vitacura 7630355, Santiago, Chile \and
Korea Astronomy \& Space Science institute, 776, Daedeokdae-ro, Yuseong-gu, Daejeon 34055, Republic of Korea \and
Department of Earth and Space Science, Graduate School of Science, Osaka University, 1-1 Machikaneyama, Toyonaka, Osaka 560-0043, Japan \and
6 RIKEN Cluster for Pioneering Research, 2-1 Hirosawa, Wako, Saitama 351-0198, Japan \and
Kavli Institute for Particle Astrophysics and Cosmology, Stanford University, 452 Lomita Mall, Stanford, CA 94305, USA \and
Department of Physics, Stanford University, 382 Via Pueblo Mall, Stanford, CA 94305, USA \and
Jet Propulsion Laboratory, California Institute of Technology, 4800 Oak Grove Drive, MS 169-224, Pasadena, CA 91109, USA \and
Yale Center for Astronomy \& Astrophysics, Physics Department, PO Box 208120, New Haven, CT 06520-8120, USA \and
Cahill Center for Astronomy and Astrophysics, California Institute of Technology, Pasadena, CA 91125, USA \and
Astrophysics Science Division, NASA Goddard Space Flight Center, Mail Code 661, Greenbelt, MD 20771, USA
}

\date{}


\abstract{ 

 We present one of the largest multiwavelength studies of simultaneous optical-to-X-ray spectral energy distributions (SEDs) of unobscured ($N_{\rm H} < 10^{22}\,{\rm cm^{-2}}$) active galactic nuclei (AGN) in the local Universe. Using a representative sample of hard-X-ray-selected AGN from the 70-month \textit{Swift}/BAT catalog, with optical/UV photometric data from \textit{Swift}/UVOT and X-ray spectral data from \textit{Swift}/XRT, we constructed broadband SEDs of 236 nearby AGN ($0.001 < z < 0.3$). We employed \textsc{GALFIT} to estimate host galaxy contamination in the optical/UV and determine the intrinsic AGN fluxes. We used an absorbed power law with a reflection component to model the X-ray spectra and a dust-reddened multi-temperature blackbody to fit the optical/UV SED. We calculated intrinsic luminosities at multiple wavelengths, total bolometric luminosities ($L_{\rm bol}$), optical-to-X-ray spectral indices ($\alpha_{\rm ox}$), and multiple bolometric corrections ($\kappa_{\lambda}$) in the optical, UV, and X-rays. We used black hole masses obtained by reverberation mapping and the virial method to estimate Eddington ratios ($\lambda_{\rm Edd}$) for all our AGN. We confirm the tight correlation (scatter = 0.45 dex) between UV ($2500\,\rm\AA$) and X-ray (2\,keV) luminosity for our sample. We observe a significant decrease in $\alpha_{\rm ox}$ with $L_{\rm bol}$ and $\lambda_{\rm Edd}$, suggesting that brighter sources emit more UV photons per X-rays. We report a second-order regression relation (scatter = 0.15 dex) between the 2--10\,keV bolometric correction ($\kappa_{2-10}$) and $\alpha_{\rm ox}$, which is useful to compute $L_{\rm bol}$ in the absence of multiband SEDs. We also investigate the dependence of optical/UV bolometric corrections on the physical properties of AGN and obtain a significant increase in the UV bolometric corrections ($\kappa_{\rm W2}$ and $\kappa_{\rm M2}$) with $L_{\rm bol}$ and $\lambda_{\rm Edd}$, unlike those in the optical ($\kappa_{\rm V}$ and $\kappa_{\rm B}$), which are constant across five orders of $L_{\rm bol}$ and $\lambda_{\rm Edd}$. We obtain significant dispersions ($\sim$ 0.1--1\,dex) in all bolometric corrections, and hence recommend using appropriate relations with observed quantities while including the reported scatter, instead of their median values.
}


\maketitle
\section{Introduction}\label{sect:intro}

Supermassive black holes (SMBHs) are present at the center of almost all massive galaxies (e.g., \citealp{doi:10.1146/annurev.aa.33.090195.003053}). However, only a fraction of these galaxies are observed as active galactic nuclei (AGN), when the central SMBH is in an actively accreting phase. While most of the light coming from a galaxy can be typically attributed to its stellar population that mainly emits radiations from the ultraviolet (e.g., \citealp{1993ApJS...86....5K}) to the near-infrared bands, AGN are characterized by their multiwavelength emission spanning the entire electromagnetic spectrum (e.g., \citealp{2017A&ARv..25....2P}). This multiwavelength emission is generated in different regions of the AGN via various physical processes (e.g., \citealp{2013peag.book.....N}). The AGN is fueled by the accretion of the surrounding material onto the central SMBH in the form of a disk emitting optical/UV photons. A fraction of these photons can be either absorbed by the dusty torus surrounding the AGN (e.g., \citealp{1993ARA&A..31..473A}) and re-emitted in the infrared (IR) or Comptonized by a corona of hot electrons located close to the SMBH, giving rise to high-energy X-ray radiation (e.g., \citealp{1991ApJ...380L..51H}).

A detailed analysis of the broadband spectral energy distributions (SEDs) of a representative sample of AGN can therefore provide useful insights into the various physical processes at play in the inner regions of AGN, which are otherwise difficult to image with current observational facilities. The AGN SEDs can further be used to investigate relations between emissions at different wavelengths originating in different regions of the SMBH. Additionally, a complete and consistent study of broadband AGN SEDs can be used to constrain AGN accretion models and compute several important quantities such as luminosities in various energy bands ($L_\lambda$), optical to X-ray spectral indices ($\alpha_{\rm ox}$), bolometric luminosities ($L_{\rm bol}$), mass accretion rates ($\dot {M}_{\rm acc}$), and Eddington ratios ($\lambda_{\rm Edd}$) (e.g., \citealp{2009MNRAS.392.1124V}; \citealp{2010MNRAS.402.1081V}; \citealp{2012MNRAS.425..623L}; \citealp{2022MNRAS.515.5617M}; \citealp{2023MNRAS.524.1796M}). Moreover, one can also estimate what fraction of the total luminosity of the AGN is emitted at different wavelengths, in terms of bolometric corrections ($\kappa_{\lambda} = L_{\rm bol}/L_\lambda$; e.g., \citealp{2004IAUS..222...49M}; \citealp{2010A&A...512A..34L}; \citealp{2020A&A...636A..73D}; \citealp{2023A&A...671A..34S}; \citealp{2023arXiv231203552S}). These bolometric correction factors can be extremely useful when we do not have access to multiwavelength data, as they estimate the total luminosity emitted by an AGN and their Eddington ratios ($\lambda_{\rm Edd} = L_{\rm bol}/L_{\rm Edd}$; $L_{\rm Edd} \propto M_{\rm BH}$). They are widely employed by the scientific community to constrain the accretion history of SMBHs from AGN luminosity functions (e.g., \citealp{2003ApJ...598..886U}; \citealp{2020ApJ...903...85A}). However, the construction and fitting of AGN SEDs is not trivial and must consider several key points, including:

Effects of obscuration: AGN are majorly classified into two types: type\,I and type\,II, based on the presence and absence of broad Balmer emission lines in their optical spectra. According to the unified AGN model, these two classes of AGN are fundamentally similar and their observed optical properties simply reflect their different viewing angles (e.g., \citealp{1995PASP..107..803U}). Type\,I AGN are viewed face-on, and hence are less affected by obscuration in the line of sight of the observer, whereas type\,II AGN are viewed edge-on, through the dusty toroidal structure, due to which the broad optical emission lines are missing in their spectra. Consequently, the SEDs of these two classes of AGN have different shapes, with the type\,IIs showing higher levels of absorption, which needs to be considered when doing a comprehensive study of AGN SEDs (e.g., \citealp{2012MNRAS.425..623L}; \citealp{2020A&A...636A..73D}).

Host galaxy contamination: While analyzing multiband AGN SEDs, one needs to include the contribution from the host galaxy, especially in the optical and UV energy bands, where one can expect mild to severe contamination from the host galaxy light. This contamination can be particularly important in low-luminosity AGN (e.g., \citealp{2008ARA&A..46..475H}). Therefore, to get reliable measurements of the total AGN luminosity, one must separate the AGN light from the host galaxy light. Over the years, different works have employed different approaches to handle host galaxy contamination in their SED fitting, based on their specific science goals. Some of these techniques remove the host galaxy light before the SED fitting, either by using an image decomposition software like \textsc{GALFIT} (e.g., \citealp{2009MNRAS.392.1124V}; \citealp{2011ApJS..196....2S}), or by using empirical correction factors based on quantities like the 4000${\rm \AA}$ break (e.g., \citealp{2012A&A...539A..48M}) or the 5100$\rm \AA$ luminosity (e.g., \citealp{2023A&A...671A..34S}). Others follow a different approach, fitting the entire SED with galaxy $+$ AGN templates to also extract the host galaxy properties (e.g., \citealp{2010ApJ...713..970A}; \citealp{2012MNRAS.425..623L}; \citealp{2020A&A...636A..73D}; \citealp{2022MNRAS.515.5617M}; \citealp{2023arXiv231203552S}).

Galactic dust extinction: While studying the SEDs of distant objects, one also needs to take into account Galactic extinction. This is usually handled by including different dust extinction curves during the fitting process like the \cite{1989ApJ...345..245C} extinction curve (e.g., \citealp{2011ApJS..196....2S}), and the \cite{2000ApJ...533..682C} reddening curve (e.g., \citealp{2020A&A...636A..73D}). Additionally, intrinsic extinction at the redshift of the source, due to the host galaxy, can also affect the final estimates of the AGN luminosity and must be included accordingly (e.g., \citealp{2021MNRAS.508..737T}).

Variability: AGN show variability across all wavelengths, with timescales that can range from hours to days to years (e.g., \citealp{1997ARA&A..35..445U}). Since the multiwavelength emission in AGN originates in different regions of the AGN located at different physical scales (the optical/UV arising from the sub-parsec accretion disk and the X-rays originating much closer to the central SMBH), they may show uncorrelated variability with distinct intensities and time. These flux variations can have large amplitudes and be associated with strong spectral variability (e.g., \citealp{2023NatAs...7.1282R} and references therein). Hence, to get a consistent picture of the total emission from the AGN at a certain time, simultaneous observations in different bands are required to minimize the effect of variability and to construct well-defined broadband SEDs of AGN. However, due to the difficulties associated with obtaining simultaneous observations, most of the previous studies of AGN SEDs have either included an error term to account for the effects of variability in the SED (e.g., \citealp{2012A&A...539A..48M}) or used quasi-simultaneous observations (e.g., \citealp{2011ApJS..196....2S}; \citealp{2012MNRAS.422..478R}) to limit their impact.

The first large, high-quality atlas of quasar (high-luminosity AGN) SEDs over the whole accessible range of the electromagnetic spectrum, from radio to X-rays, was reported by \cite{1994ApJS...95....1E}. They calculated the bolometric luminosity for their sample by integrating over the full SED, and derived mean bolometric corrections in different bands. Following this milestone work, many significant contributions were made to the investigation of multiwavelength SEDs of AGN by treating the selection effects of \cite{1994ApJS...95....1E} in the X-rays and removing the contribution of the reprocessed IR emission in the total accretion luminosity of AGN (e.g., \citealp{2003ApJ...590..128K}; \citealp{2004ASSL..308..187R}; \citealp{2004IAUS..222...49M}; \citealp{2006ApJS..166..470R}). \cite{2009MNRAS.392.1124V} and \cite{2009MNRAS.399.1553V} used simultaneous data to study optical to X-ray SEDs of ($\sim30$) nearby type\,I AGN. They successfully estimated the 2--10\,keV bolometric corrections ($\kappa_{2-10}$) for their sample and found a strong dependence of $\kappa_{2-10}$ on the Eddington ratio. A tight correlation between $\alpha_{\rm ox}$ and $\kappa_{2-10}$ was reported by \cite{2010A&A...512A..34L}, which can be employed to estimate bolometric corrections using only optical and X-ray data. While most of the SED studies in the early 2000s (e.g., \citealp{2012A&A...539A..48M}; \citealp{2012MNRAS.422..478R}; \citealp{2012MNRAS.426.2677R}; \citealp{2013ApJS..206....4K}) primarily focused on calculating bolometric corrections for type\,I AGN, \cite{2012MNRAS.425..623L} analyzed a sample of both type\,I and type\,II AGN and derived their bolometric luminosities and corrections, confirming the dependence of $\kappa_{\rm x}$ on $\lambda_{\rm Edd}$ for both classes of AGN. Additionally, \citet{2017ApJ...844...10B} presented X-ray bolometric corrections for Compton-thick AGN ($N_{\rm H} > 10^{24}\,{\rm cm}^{-2}$). More recently, \cite{2020A&A...636A..73D} analyzed $\sim$ 1000 AGN SEDs, including both type\,Is and type\,IIs, and found that while the X-ray bolometric correction is correlated with $L_{\rm bol}$, $\lambda_{\rm Edd}$ and $M_{\rm BH}$, the optical bolometric correction tends to remain constant with all these parameters.


\begin{figure}
  \begin{subfigure}[t]{0.5\textwidth}
    \centering
    \includegraphics[width=\textwidth]{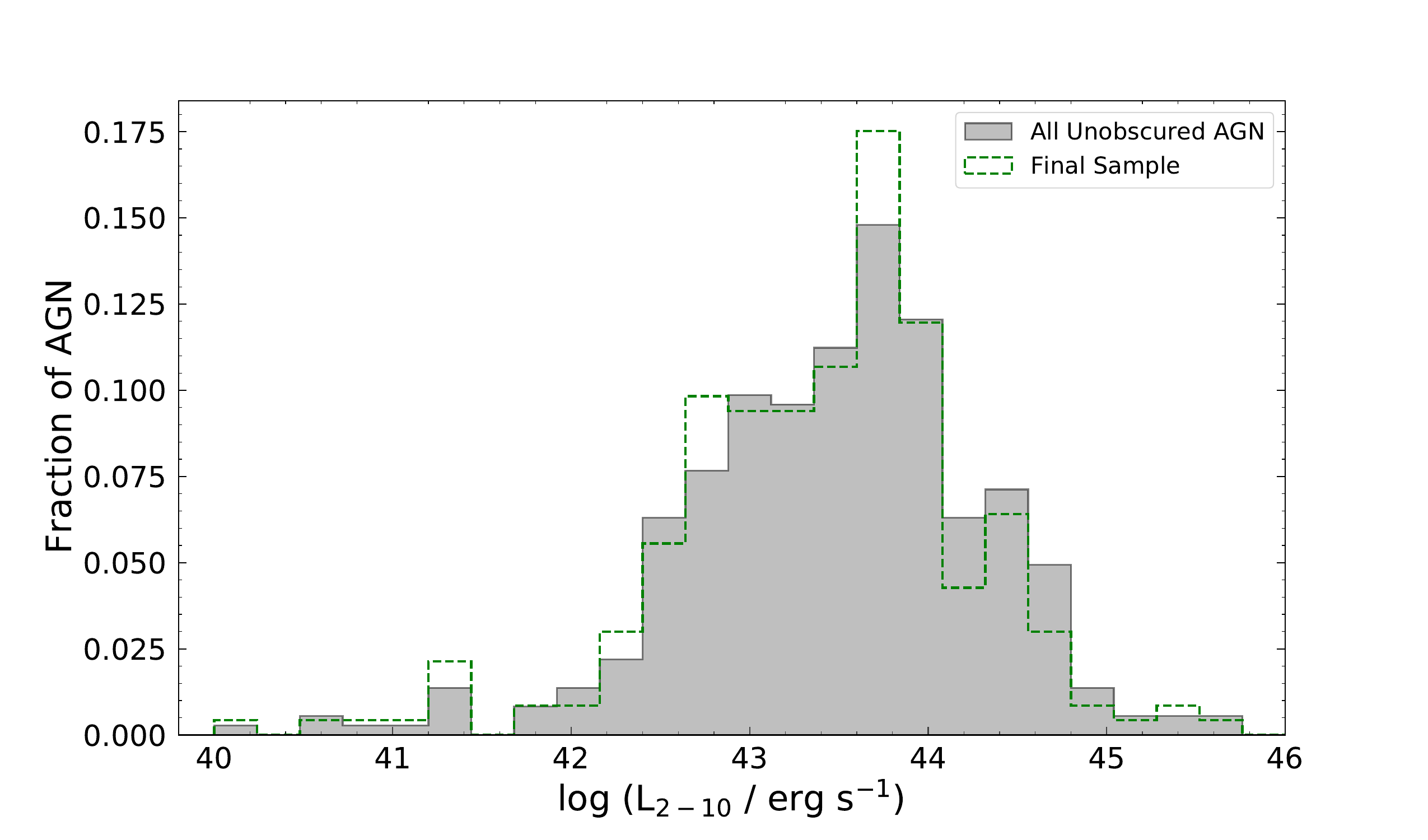}
    \vspace{-0.2cm}
    \caption{}
    \label{fig:lx}
  \end{subfigure}
  
  \begin{subfigure}[t]{0.5\textwidth}
    \centering
    \includegraphics[width=\textwidth]{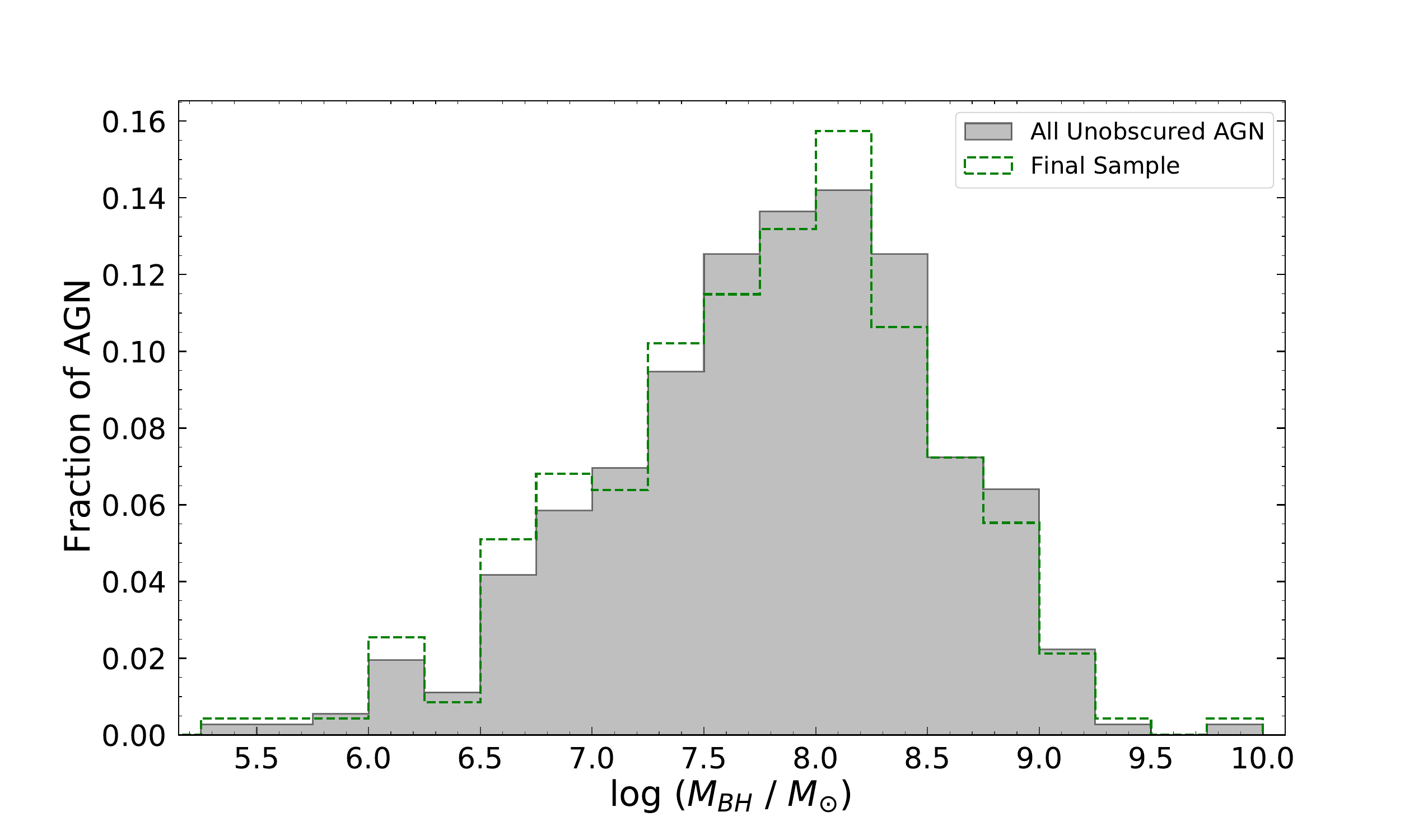}
    \vspace{-0.2cm}
    \caption{}
    \label{fig:M_BH}
  \end{subfigure}
  
  \begin{subfigure}[t]{0.5\textwidth}
    \centering
    \includegraphics[width=\textwidth]{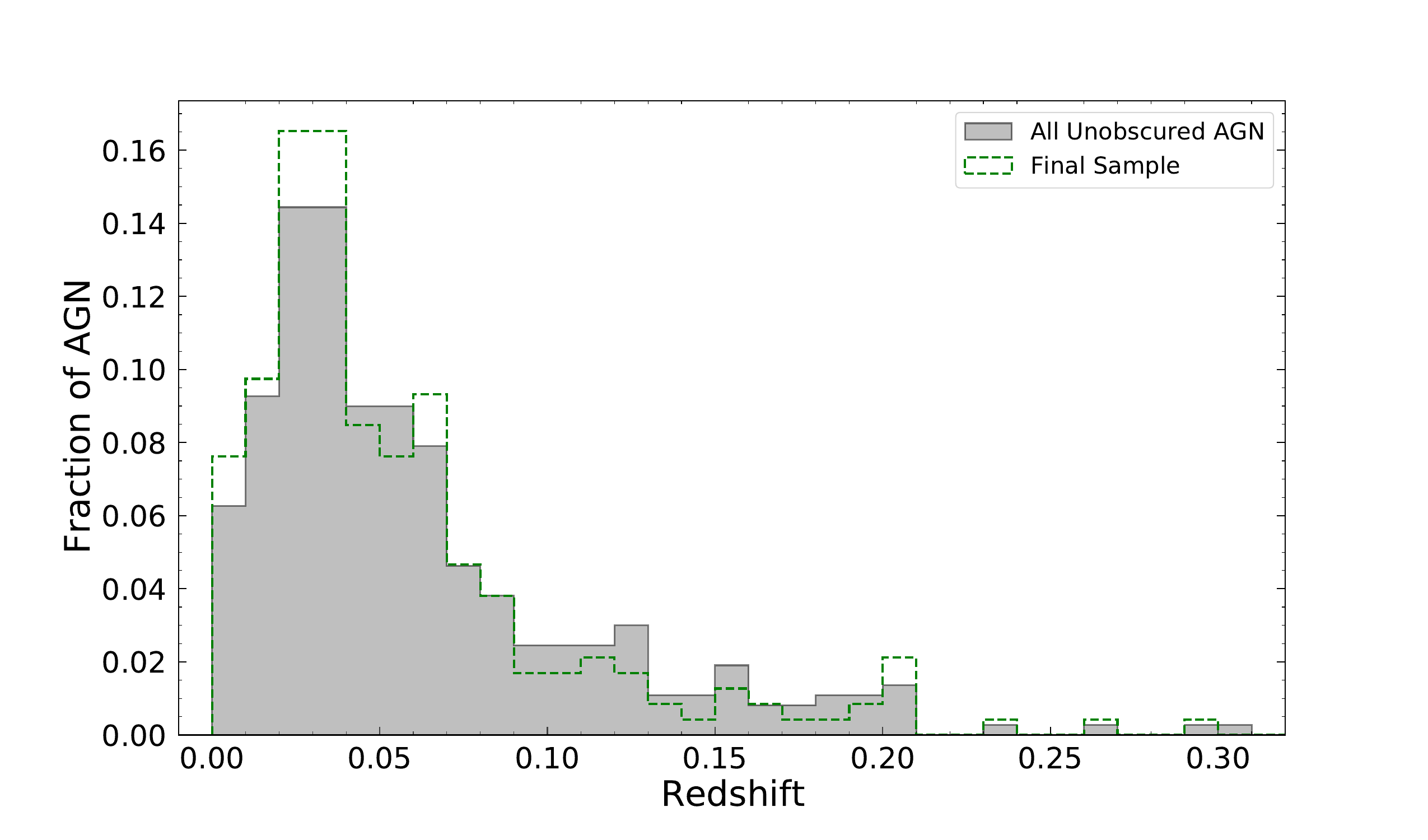}
    \vspace{-0.2cm}
    \caption{}
    \label{fig:z}
  \end{subfigure}
\vspace{-0.2cm}
\caption{Distribution of the (a) black hole mass, (b) 2--10\,keV intrinsic X-ray luminosity, and (c) redshift of the total sample of unobscured \textit{Swift}/BAT AGN (in gray) and our final sample of 236 unobscured AGN (dashed green histogram). The figures show that our sample covers the parameter space probed by the full unobscured \textit{Swift}/BAT AGN sample, in terms of redshift, black hole mass, and luminosity.}
\label{fig:prop}
\end{figure}


In this work, we perform a comprehensive multiwavelength study of a large sample of nearby ($z<0.1$) unobscured ($N_{\rm H} < 10^{22}\,{\rm cm^{-2}}$), hard-X-ray-selected type\,I AGN, with the goal of shedding light on AGN accretion. This was done by creating and modeling optical-to-X-ray SEDs of a sample of 236 \textit {Swift}/BAT-detected AGN using simultaneous optical, UV, and X-ray data. A similar effort was made by \citet{2009MNRAS.399.1553V}, who fit the optical-to-X-ray SEDs of 26 hard-X-ray-selected AGN and reported their X-ray bolometric corrections. More than ten years later, we have successfully improved the sample size by 1 dex. 
To date, this is one of the largest samples of hard-X-ray-selected AGN for which such an analysis has been carried out in a consistent way. We have also refined the \textsc{GALFIT}-based AGN-galaxy decomposition using simulations and upgraded the SED fitting procedure. Based on our analysis, in this paper, we present multiwavelength fluxes and luminosities, along with estimates of $L_{\rm bol}$, $\lambda_{\rm Edd}$, $\alpha_{\rm ox}$, and $\kappa_{\lambda}$ at multiple energies for our AGN sample. We also discuss some important correlations between these quantities that help us better understand the accretion physics of unobscured AGN in the local Universe.

The structure of the paper is as follows. In Sect. \ref{sect:sample}, we describe the sample used in this work, while in Sect. \ref{sect:dr} we illustrate the data reduction process for the optical/UV (Sect. \ref{sect:uvot}) and X-ray data (Sect. \ref{sect:xrt}). In Sects. \ref{sect:xray_spec} and \ref{sect:galfit}, we explain the fitting analysis of the X-ray spectra and the image decomposition of the optical/UV images to correct for host galaxy contamination, respectively. Sect. \ref{sect:sed} describes in detail the broadband SED fitting procedure. Finally, in Sects. \ref{sect:discussion} and \ref{sect:summary}, we present and summarize our main results, including the optical-to-X-ray spectral indices and the X-ray, optical, and UV bolometric corrections over a large range of Eddington ratios [$-5.0 < {\rm log}\,(\lambda_{\rm Edd}) < 0.0$]. We also report the main correlations found for these quantities with various physical properties of AGN, such as their black hole mass, bolometric luminosity, and Eddington ratio. Throughout the paper, we assume a cosmological model with $H_{\rm 0}=70\,\rm km\,s^{-1}\,Mpc^{-1}$, $\Omega_{\rm M}=0.3$, and $\Omega_{\Lambda}=0.7$. Images are shown adopting the standard astronomical orientation; that is, north is up and east is left. All magnitudes quoted throughout the text are AB magnitudes. All errors and uncertainties listed in the tables or shown as error bars or shaded regions in the figures are one sigma, unless otherwise stated. All the correlations were obtained using various functions from the \texttt{statistics}\footnote{\url{https://docs.scipy.org/doc/scipy/reference/stats.html}} module of the Python library \texttt{scipy} (\citealp{2020SciPy-NMeth}) and the Python package \texttt{linmix}\footnote{\url{https://linmix.readthedocs.io/en/latest/}} (\citealp{2007ApJ...665.1489K}).


\section{Sample}\label{sect:sample}

Unlike the optical/UV and soft X-ray radiation ($E<$ 10\,keV), which are highly sensitive to extinction by the circumnuclear and interstellar dust, the hard X-rays ($E>$ 10\,keV) are much less affected by obscuration (up to $N_{\rm H} \sim 10^{24}\,{\rm cm}^{-2}$; see Fig. 1 of \citealp{Ricci_2015}) due to their high penetration power. They are also less contaminated by stellar emission from the host galaxy and can therefore be used to obtain a more complete and unbiased sample of AGN including the most obscured sources, which could be missed by optical/UV surveys. In this work, we use a sample of hard-X-ray-selected AGN detected in the 14--195\,keV band by the Burst Alert Telescope (BAT: \citealp{2005SSRv..120..143B}; \citealp{2013ApJS..209...14K}) on board NASA's \textit{Neil Gehrels Swift} Observatory. \textit{Swift} is a multiwavelength space observatory with two additional instruments: the X-ray Telescope (XRT; \citealp{2005SSRv..120..165B}) operates in the 0.3--10\,keV energy range, while the UV/Optical Telescope (UVOT; \citealp{2008MNRAS.383..627P}; \citealp{2010MNRAS.406.1687B}) observes the sky in six filters spanning $170$ to $650\,{\rm nm}$. Most of the sources detected by BAT in the hard X-rays have eventually been simultaneously observed (over $\sim 2$ days) in the soft X-rays by \textit{Swift}/XRT and in the optical/UV band by \textit{Swift}/UVOT. A systematic multiwavelength study of all the \textit{Swift}/BAT AGN has been carried out by the BAT AGN Spectroscopic Survey (BASS\footnote{\url{www.bass-survey.com}}). BASS recently published their second data release (DR2; \citealp{2022ApJS..261....1K}) with a complete, optical spectroscopic analysis of all the 858 \textit{Swift}/BAT AGN in the 70-month catalog (\citealp{2022ApJS..261....2K}), thanks to a dedicated campaign. BASS also provides a comprehensive study of \textit{Swift}/BAT AGN in the X-rays \citep{2017ApJS..233...17R}, near-IR \citep{2017MNRAS.467..540L}, mid-IR \citep{2019ApJ...870...31I}, millimeter (\citealp{2022ApJ...938...87K,2023ApJ...952L..28R}), and radio \citep{2019MNRAS.488.4317B}. Other data products from this release include measurements of the broad emission lines \citep{2022ApJS..261....5M}, narrow emission lines \citep{2022ApJS..261....4O}, velocity dispersions \citep{2022ApJS..261....6K}, and NIR spectral properties (\citealp{2022ApJS..261....8R,2022ApJS..261....7D}).

\defcitealias{2017ApJS..233...17R}{R17}


\begin{table*}
\centering
\caption{Characteristics of \textit{Swift}/UVOT filters.}
\begin{tabular}{ccccc} 
\hline
\hline
\specialrule{0.1em}{0em}{0.5em}
\vspace{1mm}
Filter & Central Wavelength  & FWHM & AB Zero-point & Counts to Flux ($F_\lambda$) Ratio\\
 & ($\rm \AA$) & ($\rm \AA$) & (mag) & ($10^{-16}\,{\rm erg\,\,count^{-1}\,cm^{-2}\,\AA^{-1}}$)\\
\hline\\
\vspace{1mm}
UVW2 & 1928 & 657 & 19.11 $\pm$ 0.03 & 6.225 $\pm$ 0.1300\\
\vspace{1mm}
UVM2 & 2246 & 498 & 18.54 $\pm$ 0.03 & 8.489 $\pm$ 0.0530\\
\vspace{1mm}
UVW1 & 2600 & 693 & 18.95 $\pm$ 0.03 & 4.623 $\pm$ 0.1400\\
\vspace{1mm}
U & 3465 & 785 & 19.36 $\pm$ 0.02 & 1.663 $\pm$ 0.0250\\
\vspace{1mm}
B & 4392 & 975 & 18.98 $\pm$ 0.02 & 1.480 $\pm$ 0.0056\\
\vspace{1mm}
V & 5468 & 769 & 17.88 $\pm$ 0.01 & 2.611 $\pm$ 0.0087\\
\hline
\end{tabular}
\tablefoot{The central wavelength corresponds to the mid-point between the wavelengths at the FWHM. The zeropoint magnitude represents the magnitude of a source producing one count per second.\footnote{\url{https://heasarc.gsfc.nasa.gov/docs/heasarc/caldb/swift/docs/uvot/uvot\_caldb\_AB\_10wa.pdf}}}
\label{tab:uvot}
\end{table*}


In this study, we use the 70-month \textit{Swift}/BAT catalog \citep{2013ApJS..207...19B}, consisting of 858 hard-X-ray-selected local (median redshift, $z=0.037$) AGN. We focus on unobscured AGN to mitigate host galaxy contamination and carefully study the very central optical and UV emission associated with the AGN. We start with an X-ray-selected sample of unobscured AGN to have minimum obscuration due to the material in the line of sight, and to do so we use the X-ray column densities ($N_{\rm H}$). As part of the first data release (DR1) of BASS, \citeauthor{2017ApJS..233...17R} (\citeyear{2017ApJS..233...17R}; hereafter \citetalias{2017ApJS..233...17R}) performed a detailed X-ray spectral analysis of all BAT AGN and reported the $N_{\rm H}$ values for 836/858 sources. For our primary sample, we shortlist all AGN with $N_{\rm H} < 10^{22}\,{\rm cm}^{-2}$, which gives us a set of 367 unobscured AGN. This sample excludes the sources that are flagged as blazars following the classification presented in \citet{2022ApJS..261....2K} based on the Roma Blazar Catalog (BZCAT; \citealp{2009A&A...495..691M}) and the study done by \citet{2019ApJ...881..154P}. We also have black hole mass estimates for most of these sources from the BASS DR2 \citep{2022ApJS..261....5M}. 


\begin{table*}
\setlength{\tabcolsep}{0.7\tabcolsep}
\centering
\caption{Sources in our sample with simultaneous optical/UV and X-ray observations.}
\begin{tabular}{cccccccc}
\hline
\hline
\specialrule{0.1em}{0em}{0.5em}
\vspace{1mm}
BAT ID & Swift ID & Counterpart Name & R.A. & Dec. & Redshift & Distance & ${\rm log}\,(N_{\rm H}/{\rm cm^{-2}})$\\
\hline\\
\vspace{1mm}
2 & SWIFTJ0001.6-7701 & 2MASXJ00014596-7657144 & \,\,\,0.4420 & $-76.9540$ & 0.0584 & 261.7 & $\le20.00$\\
\vspace{1mm}
3 & SWIFTJ0002.5+0323 & NGC7811 & \,\,\,0.6103 & \,\,\,\,\,\,\,3.3519 & 0.0255 & 111.2 & $\le20.00$\\
\vspace{1mm}
6 & SWIFTJ0006.2+2012 & Mrk335 & \,\,\,1.5814 & \,\,\,\,20.2030 & 0.0258 & 113.2 & \,\,\,\,\,20.48\\
\vspace{1mm}
10 & SWIFTJ0021.2-1909 & LEDA1348 & \,\,\,5.2814 & $-19.1682$ & 0.0956 & 439.9 & \,\,\,\,\,21.98\\
\vspace{1mm}
14 & SWIFTJ0026.5-5308 & LEDA433346 & \,\,\,6.6695 & $-53.1633$ & 0.0629 & 283.7 & $\le20.00$\\
\vspace{1mm}
16 & SWIFTJ0029.2+1319 & PG0026+129 & \,\,\,7.3071 & \,\,\,\,13.2678 & 0.1420 & 671.7 & $\le20.00$\\
\vspace{1mm}
19 & SWIFTJ0034.5-7904 & RHS3 & \,\,\,8.5697 & $-79.0890$ & 0.0740 & 335.9 & $\le20.00$\\
\vspace{1mm}
34 & SWIFTJ0051.6+2928 & UGC524 & 12.8959 & \,\,\,\,29.4013 & 0.0360 & 157.9 & $\le20.00$\\
\vspace{1mm}
36 & SWIFTJ0051.9+1724 & Mrk1148 & 12.9782 & \,\,\,\,17.4329 & 0.0640 & 287.5 & \,\,\,\,\,20.30\\
\vspace{1mm}
39 & SWIFTJ0054.9+2524 & PG0052+251 & 13.7172 & \,\,\,\,25.4275 & 0.1550 & 739.5 & $\le20.00$\\
\hline
\end{tabular}
\tablefoot{We have listed the BAT and Swift IDs, counterpart names, coordinates (in degrees, J2000), redshifts, luminosity distances in Mpc \citep{2022ApJS..261....2K}, and X-ray column density values from \citetalias{2017ApJS..233...17R} for all the sources. The table in its entirety is available at the CDS.}
\label{tab:src_list}
\end{table*}


\begin{table*}
\centering
\caption{Observation IDs of the observations used for each source.}
\begin{tabular}{cccccc}
\hline
\hline
\specialrule{0.1em}{0em}{0.5em}
\vspace{1mm}
BAT ID & Swift ID & Obs. ID & UVOT Exposure (s) & XRT Exposure (s) & UVOT Filters\\
\hline\\
\vspace{1mm}
2 & SWIFTJ0001.6-7701 & 00080854001 & 4419 & 4485 & Only UV\\
\vspace{1mm}
3 & SWIFTJ0002.5+0323 & 00047107014 & 1600 & 1625 & All\\
\vspace{1mm}
6 & SWIFTJ0006.2+2012 & 00090006004 & 7917 & 8136 & All\\
\vspace{1mm}
10 & SWIFTJ0021.2-1909 & 00040691001 & 6485 & 6493 & U, UVW1\\
\vspace{1mm}
14 & SWIFTJ0026.5-5308 & 00041141001 & 2669 & 2670 & UVW1, UVM2\\
\vspace{1mm}
16 & SWIFTJ0029.2+1319 & 00080859001 & 5976 & 6000 & All\\
\vspace{1mm}
19 & SWIFTJ0034.5-7904 & 00080861001 & 4782 & 4806 & All\\
\vspace{1mm}
34 & SWIFTJ0051.6+2928 & 00080867001 & 6222 & 6297 & All\\
\vspace{1mm}
36 & SWIFTJ0051.9+1724 & 00080868001 & 6696 & 6800 & All\\
\vspace{1mm}
39 & SWIFTJ0054.9+2524 & 00080869001 & 6117 & 6146 & All\\
\hline
\end{tabular}
\tablefoot{For each observation, the total exposure times with \textit{Swift}/UVOT and \textit{Swift}/XRT, and the available \textit{Swift}/UVOT filters are also reported. The table in its entirety is available at the CDS.}
\label{tab:obs_list}
\end{table*}


We used the HEASARC archive\footnote{\url{https://heasarc.gsfc.nasa.gov/cgi-bin/W3Browse/w3browse.pl}} to compile the available \textit{Swift} observations for our AGN sample. To obtain the best possible estimates of the optical/UV fluxes, we used the observation with maximum exposure and coverage in all the six \textit{Swift}/UVOT filters (V, B, U, UVW1, UVM2, and UVW2; see Table \ref{tab:uvot}). A total of 226/367 ($\sim$ 62\%) sources had \textit{Swift} observations in all UVOT filters plus simultaneous \textit{Swift}/XRT observations. For the remaining 141 ($\sim$ 38\%) sources, we checked for \textit{Swift} observations with maximum wavelength coverage. 67/141 ($\sim$ 47\%) sources had \textit{Swift} observations with at least one optical and one UV filter, and simultaneous X-ray observations. In total, we recovered 293/367 ($\sim$ 80\%) sources (see Table \ref{tab:src_list}) with simultaneous optical/UV and X-ray observations (Table \ref{tab:obs_list}) that can be used to extract multiwavelength fluxes and construct optical-to-X-ray SEDs, with $\sim$ 76\% of these sources having observations in all \textit{Swift}/UVOT filters. We performed the broadband SED fitting for a final sample of 236/293 ($\sim$ 80\%) sources, after removing (i) two sources with strong contamination from a nearby source affecting the optical/UV or X-ray data extraction (discussed in Sect. \ref{sect:dr}), (ii) one source with a strongly variable X-ray spectrum (see Sect. \ref{sect:xray_spec}), and (iii) 54 sources with fewer than four points in the optical/UV wavebands making them unsuitable for the SED fitting (explained in detail in Sect. \ref{sect:sed}). We show the range of redshift, black hole mass, and X-ray luminosity covered by our final sample, as well as by the total unobscured, \textit{Swift}/BAT AGN sample in Fig. \ref{fig:prop}. Our sample is representative of the \textit{Swift}/BAT parent sample in terms of black hole mass, luminosity and redshift, as it covers the same ranges for these parameters as the original \textit{Swift}/BAT sample.


\section{Data reduction}\label{sect:dr}


\subsection{Optical/UV photometric data from Swift/UVOT} \label{sect:uvot}

\begin{figure*}
  \begin{subfigure}[t]{1.05\textwidth}
    \centering
    \includegraphics[width=\textwidth]{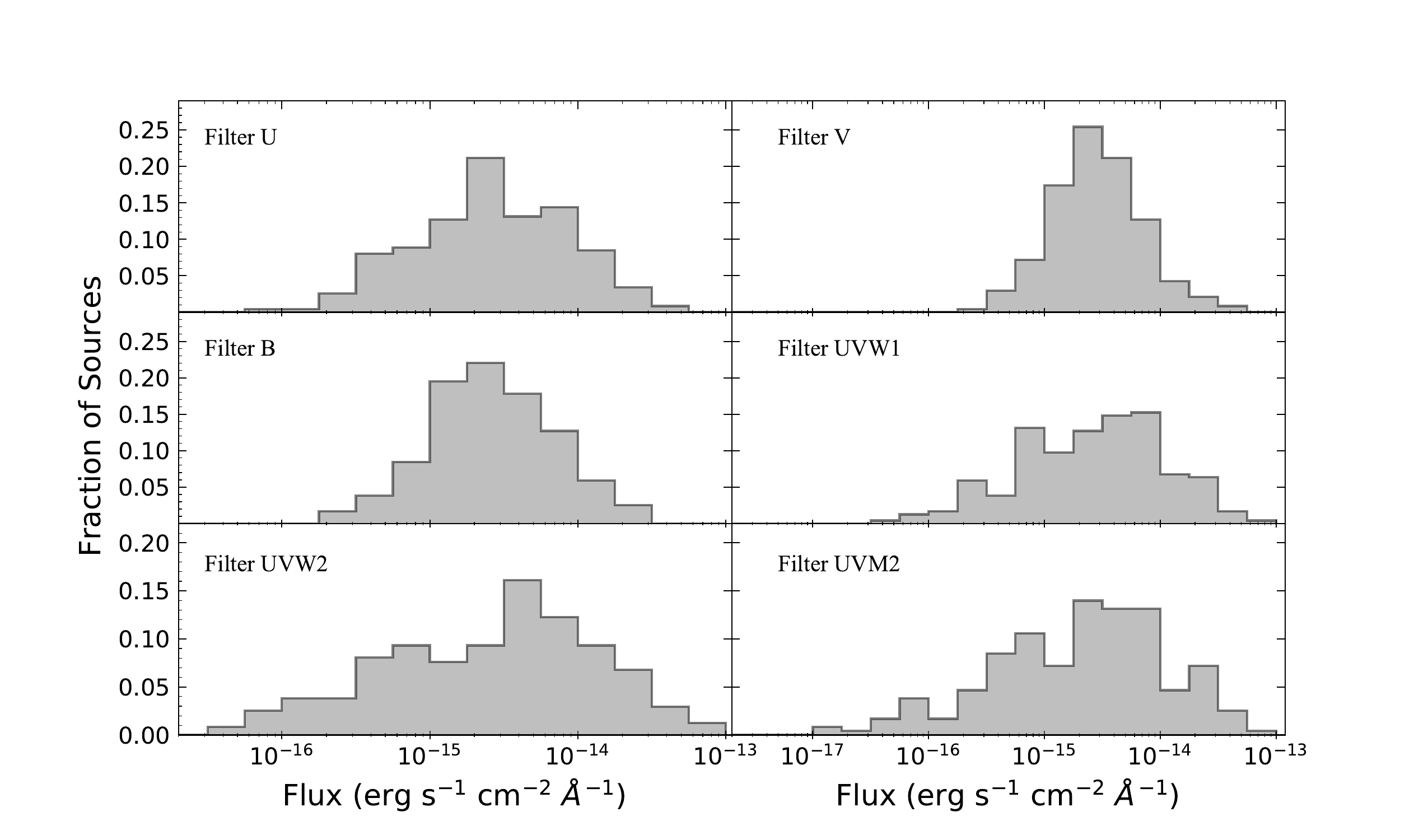}
    \vspace{-1cm}
    \caption{}
    \label{fig:uvot_flux}
  \end{subfigure}

\vspace{-0.5cm}
  
  \begin{subfigure}[t]{1.05\textwidth}
    \centering
    \includegraphics[width=\textwidth]{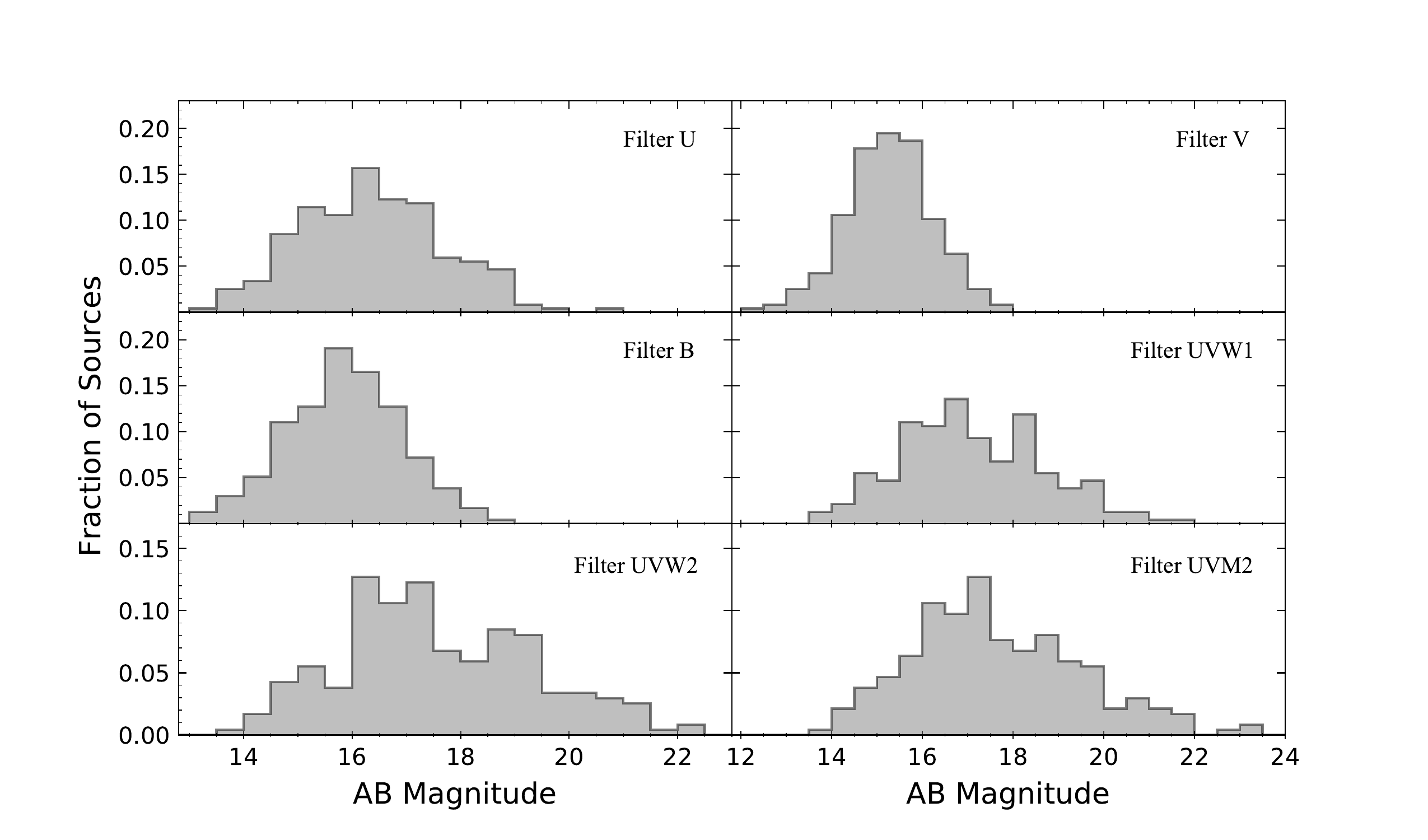}
    \vspace{-1cm}
    \caption{}
    \label{fig:uvot_mag}
  \end{subfigure}
\vspace{-0.2cm}
\caption{Distribution of the source (a) fluxes and (b) AB magnitudes in the six \textit{Swift}/UVOT filters estimated using aperture photometry for our sample of 292 AGN (Sect. \ref{sect:uvot}).}
\label{fig:uvot}
\end{figure*}

The reduction and analysis of the \textit{Swift}/UVOT data for our sample of unobscured AGN was done following the steps recommended by the \textit{Swift}/UVOT Software Guide Version 2.2\footnote{\url{https://swift.gsfc.nasa.gov/analysis/UVOT\_swguide\_v2\_2.pdf}} (released in March 2008). Although the calibrated and pipeline processed images are available in the archive for all our sources, to take into account the possible changes in the data analysis due to the 2020 update in the \textit{Swift} calibration database\footnote{\url{https://heasarc.gsfc.nasa.gov/docs/heasarc/caldb/swift/}} (CALDB), we recreated the fully calibrated images from the raw images for all the sources in our sample. The final sky images were eventually used to extract the flux and magnitudes of the sources. We only used image or event files in the pointing mode. We started with the raw image HDUs in each filter provided in the HEASARC archive and created the corresponding bad pixel maps using the latest UVOT CALDB (\texttt{uvotbadpix}). Next, flat field images were created for each filter using the attitude file and CALDB (\texttt{uvotflatfield}). The bad pixel maps and the flat field images were both used to create the sky images in each filter (\texttt{swiftxform}). One must note that some raw images might consist of multiple image HDUs corresponding to multiple exposures in the same filter that need to be summed to get the final, deeper image. In the case of single HDUs, the resultant sky image was used to extract the source and sky region with radii of $\sim$ 5 and $\sim$ 20 arcseconds, respectively, to obtain the source and background fluxes, magnitudes, and errors (\texttt{uvotsource}). The size of the regions is based on the average curves of growth\footnote{\url{https://heasarc.gsfc.nasa.gov/docs/heasarc/caldb/swift/docs/uvot/uvot_caldb_psf_02.pdf}} of the UVOT point spread function (PSF) for the six UVOT filters. For source images with crowded fields, smaller background regions ($\sim10"$) were selected. In the case of multiple image HDUs corresponding to multiple exposures in the same filter, all the image HDUs were first aspect-corrected (\texttt{uvotskycorr}) and then summed using the \texttt{uvotimsum} tool to create the final, deeper image. These images were all visually inspected to make sure that the source of interest was perfectly aligned. The summed images were then used to extract the source and background regions, as is illustrated above, in order to get the source and background fluxes and magnitudes along with statistical errors. The \texttt{uvotsource} command includes corrections for coincidence loss, which can become significant for sources with fluxes $>$ 20 counts s$^{-1}$. All the source fluxes and magnitudes thus measured are listed in Tables \ref{tab:UVOT_flux} and \ref{tab:UVOT_mag} and their distributions are shown in Fig. \ref{fig:uvot_flux}. The systematic uncertainties in the flux values are $\sim1.5\%$ and always smaller than the reported statistical errors.

One source in our sample, SWIFTJ1940.4-3015 (BAT ID = 1045), has a $V=11.3$ magnitude star (CD-30 17265) close to it (distance between centroids $= 11.2"$). Due to strong contamination from the star within the 5$"$ region of our source of interest, we are unable to extract reliable fluxes for this source in any of the \textit{Swift}/UVOT filters. Moving forward, we have removed this source from our sample, which now consists of 292 sources.


\begin{table*}
\centering
\setlength{\tabcolsep}{0.8\tabcolsep}
\caption{Source fluxes as calculated by the \textit{Swift}/UVOT command \texttt{uvotsource} via 5$"$ aperture photometry.}
\begin{tabular}{cccccccc}
\hline
\hline
\specialrule{0.1em}{0em}{0.5em}
\vspace{1mm}
BAT ID & Swift ID & $F_{\rm U}$ & $F_{\rm V}$ & $F_{\rm B}$ & $F_{\rm W1}$ & $F_{\rm W2}$ & $F_{\rm M2}$\\
\hline\\
\vspace{1mm}
2 & SWIFTJ0001.6-7701 & \,\,\,$-$ & \,\,\,$-$ & \,\,\,$-$ & \,\,\,$2.40\pm0.21$ & \,\,\,$3.30\pm0.2$ & \,\,\,$2.49\pm0.15$\\
\vspace{1mm}
3 & SWIFTJ0002.5+0323 & \,\,\,$3.18\pm0.12$ & \,\,\,$3.66\pm0.15$ & \,\,\,$3.29\pm0.11$ & \,\,\,$4.03\pm0.13$ & \,\,\,$4.73\pm0.14$ & \,\,\,$4.11\pm0.12$\\
\vspace{1mm}
6 & SWIFTJ0006.2+2012 & $16.16\pm0.36$ & \,\,\,$7.43\pm0.17$ & \,\,\,$9.86\pm0.19$ & $22.65\pm0.45$ & $28.35\pm0.56$ & $23.56\pm0.51$\\
\vspace{1mm}
10 & SWIFTJ0021.2-1909 & \,\,\,$0.19\pm0.02$ & \,\,\,$-$ & \,\,\,$-$ & \,\,\,$0.06\pm0.01$ & \,\,\,$-$ & \,\,\,$-$\\
\vspace{1mm}
14 & SWIFTJ0026.5-5308 & \,\,\,$-$ & \,\,\,$-$ & \,\,\,$-$ & \,\,\,$1.33\pm0.17$ & \,\,\,$-$ & \,\,\,$1.06\pm0.07$\\
\vspace{1mm}
16 & SWIFTJ0029.2+1319 & \,\,\,$4.93\pm0.11$ & \,\,\,$2.56\pm0.10$ & \,\,\,$3.26\pm0.09$ & \,\,\,$5.97\pm0.16$ & \,\,\,$8.16\pm0.20$ & \,\,\,$5.93\pm0.18$\\
\vspace{1mm}
19 & SWIFTJ0034.5-7904 & \,\,\,$9.65\pm0.25$ & \,\,\,$4.64\pm0.15$ & \,\,\,$5.79\pm0.15$ & $12.81\pm0.25$ & $17.38\pm0.38$ & $13.74\pm0.35$\\
 \vspace{1mm}
34 & SWIFTJ0051.6+2928 & \,\,\,$1.74\pm0.07$ & \,\,\,$3.41\pm0.12$ & \,\,\,$2.75\pm0.08$ & \,\,\,$1.64\pm0.07$ & \,\,\,$1.61\pm0.04$ & \,\,\,$1.36\pm0.07$\\
 \vspace{1mm}
36 & SWIFTJ0051.9+1724 & \,\,\,$6.08\pm0.13$ & \,\,\,$2.46\pm0.11$ & \,\,\,$3.24\pm0.11$ & \,\,\,$7.83\pm0.22$ & $10.08\pm0.21$ & \,\,\,$7.39\pm0.23$\\
 \vspace{1mm}
39 & SWIFTJ0054.9+2524 & \,\,\,$5.25\pm0.15$ & \,\,\,$2.12\pm0.09$ & \,\,\,$3.05\pm0.09$ & \,\,\,$6.53\pm0.18$ & $10.49\pm0.25$ & \,\,\,$6.63\pm0.14$\\
\hline
\end{tabular}
\tablefoot{These flux values include potential contribution from the host galaxy to the AGN (within 5$"$ region; see Sect. \ref{sect:uvot}). Flux units: $10^{-15}\,{\rm erg\,\,cm^{-2}\,s^{-1}\,\AA^{-1}}$. The table in its entirety is available at the CDS.}
\label{tab:UVOT_flux}
\end{table*}


\subsection{X-ray spectroscopic data from Swift/XRT}\label{sect:xrt}

We downloaded the level 2 \textit{Swift}/XRT event files in photon counting mode from the HEASARC archive to extract the X-ray spectra for our sources. Following the standard procedures reported in the \textit{Swift}/XRT Data Reduction Guide Version 1.2\footnote{\url{https://swift.gsfc.nasa.gov/analysis/xrt\_swguide\_v1\_2.pdf}} (April 2005), we used the \texttt{xselect} package to select, from the event files, circular source and annular background regions with radii of $\sim$ 50 and $\sim$ 150 arcseconds, respectively. For context, 90\% of the XRT PSF at 1.5 keV is enclosed by a circular region of $\sim47"$. These regions were then used to extract the source and background spectra. The \texttt{xrtmkarf} tool was used to create the ARF (ancillary response file) and get the RMF (response matrix file). The source spectra were binned, using \texttt{grppha}, with a minimum of one count/bin. All the source spectra were visually examined to check that the background counts were overall lower than the source counts.

One AGN in our sample, SWIFTJ1043.4+1105 (BAT ID = 512), has strong contamination from an unidentified nearby X-ray source (within 18$"$). Therefore, we decided to remove this source from our sample, giving us an updated sample size of 291 sources.


\section{X-ray spectral fitting}\label{sect:xray_spec}

In this section, we describe in detail the fitting procedure for the 0.3--10\,keV X-ray spectra of the sources in our sample. A detailed broadband X-ray (0.3--150\,keV) spectral analysis of 836 AGN in the 70-month \textit{Swift}/BAT catalog was carried out using \textsc{XSPEC} (\citealp{1996ASPC..101...17A}) as part of the BASS DR1 and is presented in \citetalias{2017ApJS..233...17R}. Following their analysis, we employed the models they used to fit the unobscured AGN in their catalog to fit the AGN in our sample. We used models A1 and A2, which were used by \citetalias{2017ApJS..233...17R} to fit most ($\sim 77\%$) of the unobscured AGN in their sample. These two models are defined as follows:

\begin{itemize}
    \item Model A1 (\texttt{TBABS}\textsubscript{\texttt{Gal}} $\times$ \texttt{ZPHABS} $\times$ \texttt{CABS} $\times$ \texttt{PEXRAV}): This model has a component that accounts for the primary X-ray emission along with reflection by optically thick, neutral circumnuclear material (\texttt{PEXRAV}; \citealp{1995MNRAS.273..837M}).  It also includes Galactic absorption (\texttt{TBABS$_{\rm Gal}$}), and absorption by neutral material via photoelectric absorption (\texttt{ZPHABS}) and Compton scattering (\texttt{CABS}).
    \item Model A2 [\texttt{TBABS}\textsubscript{\texttt{Gal}} $\times$ \texttt{ZPHABS} $\times$ \texttt{CABS} $\times$ (\texttt{PEXRAV} $+$ \texttt{BB})]: This model is the same as model A1 with an additional blackbody component (\texttt{BB}) to account for the soft excess.
\end{itemize}

We began the fitting for the 291 sources in our sample by using the best-fit values estimated by \citetalias{2017ApJS..233...17R} as starting values for most of the model parameters. These include $N_{\rm H}$ for \texttt{ZPHABS} and \texttt{CABS}, photon index ($\Gamma$), high-energy cutoff ($E_{\rm c}$), and a reflection parameter ($R$) for \texttt{PEXRAV}, and the temperature ($kT_{\rm bb}$) of the \texttt{BB} component. In the cases where values for $R$, $E_{\rm c}$, or $kT_{\rm bb}$ were not reported by \citetalias{2017ApJS..233...17R}, either because they were not constrained or because a different model was used to fit the source, we used as starting values $R = 0.0$, $E_{\rm c}=200$\,keV, and $kT_{\rm bb}=0.1$\,keV, respectively. The redshift parameter for both \texttt{ZPHABS} and \texttt{PEXRAV} was fixed using the values reported in \citealp{2022ApJS..261....2K}. The inclination was fixed to $30^\circ$, the metallicity ($Z$) was fixed to solar, and the iron abundance was fixed at its default value, as was done in \citetalias{2017ApJS..233...17R}. The $E_{\rm c}$ and $R$ parameters were also frozen to the values reported by \citetalias{2017ApJS..233...17R}. Hence, only the column density (which was tied to have the same value for \texttt{ZPHABS} and \texttt{CABS}), the photon index, and the normalization ($K_{\rm x}$: for \texttt{PEXRAV}) were free parameters in the fitting process. The $N_{\rm H}$ for Galactic absorption was also fixed at the value obtained at the coordinate of the source from the HI 4-PI Survey (HI4PI; \citealp{2016A&A...594A.116H}), a 21-cm all-sky survey of neutral atomic Hydrogen. In the case of model A2, the temperature ($kT_{\rm bb}$) and normalization ($K_{\rm bb}$) of the blackbody component were allowed to vary during the fitting.


\begin{figure*}
  \begin{subfigure}[t]{0.54\textwidth}
    \centering
    \includegraphics[width=\textwidth]{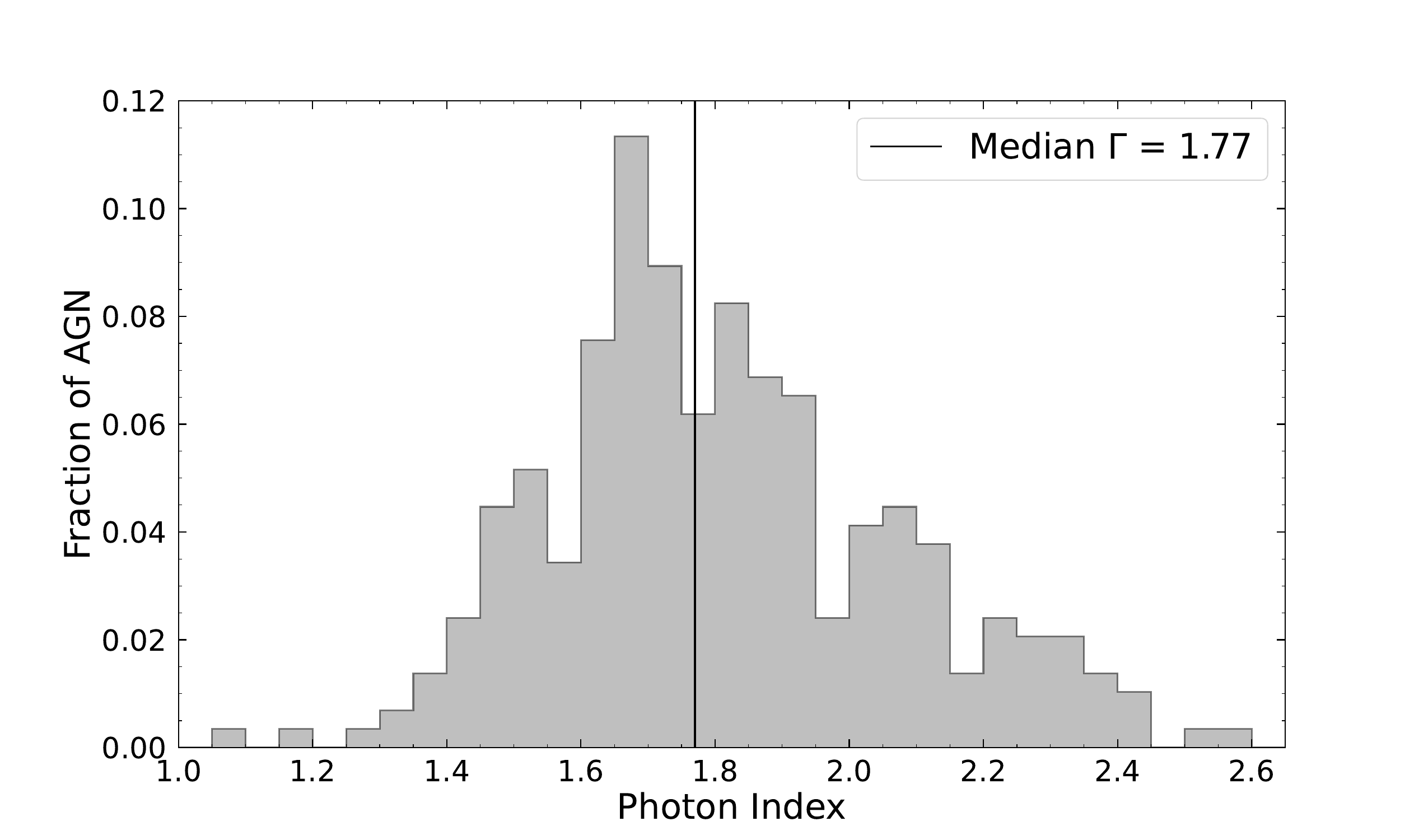} 
    \caption{}        
    \label{fig:gamma}
  \end{subfigure}
  \hspace{-9mm}
  \begin{subfigure}[t]{0.54\textwidth}
    \centering
    \includegraphics[width=\textwidth]{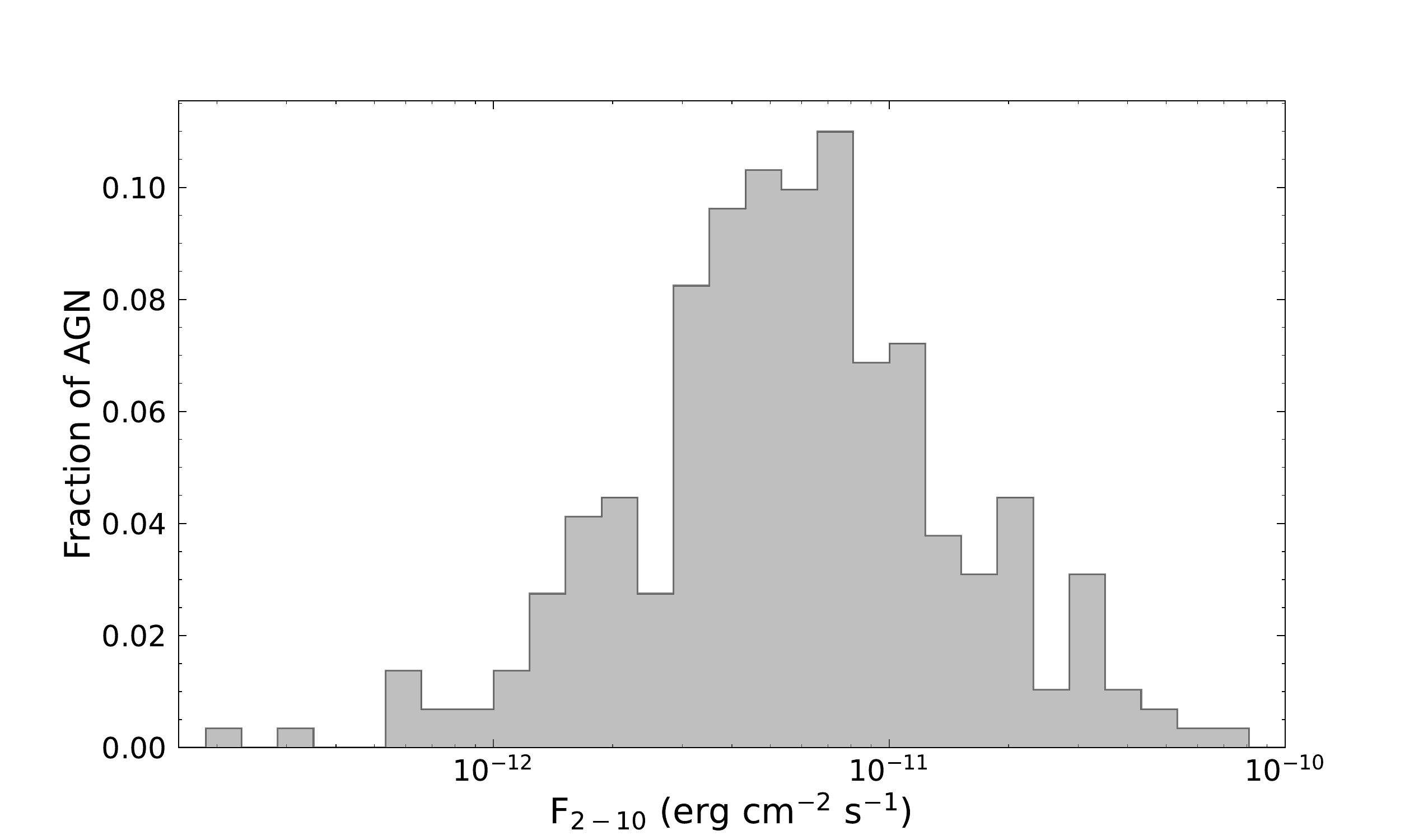}
    \caption{}
    \label{fig:xray_flux}
  \end{subfigure}
\caption{Outputs of the 0.3--10\,keV X-ray spectral fitting (Sect. \ref{sect:xray_spec}). (a) Distribution of the best-fit values of photon index ($\Gamma$). The vertical line marks the median value of $\Gamma = 1.77$. (b) Histogram of the intrinsic 2--10\,keV flux for our AGN sample.}
\label{fig:xray}
\end{figure*}


\begin{figure}
\centering
\includegraphics[width=0.54\textwidth]{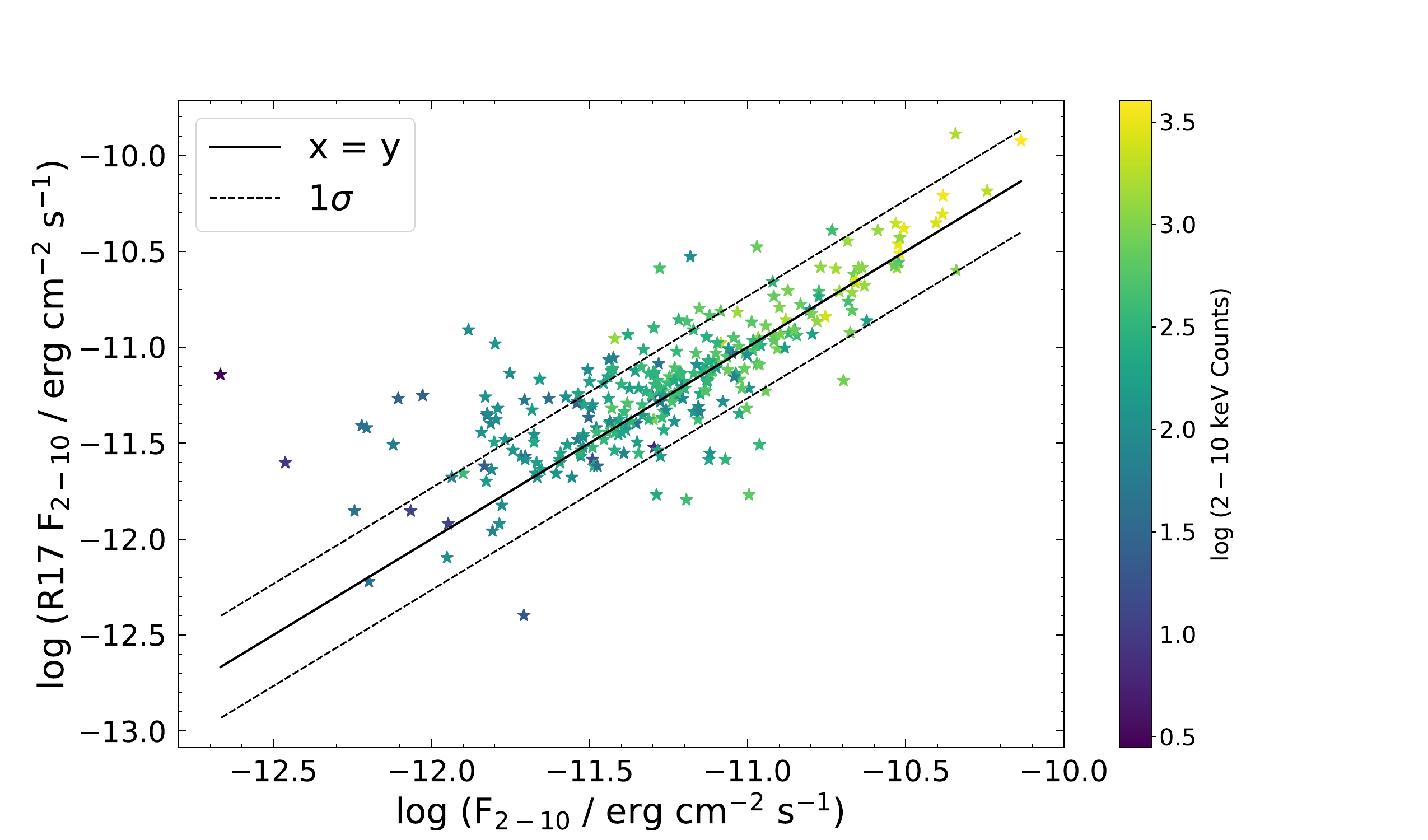} 
\caption{Comparison between the 2--10\,keV intrinsic flux we obtained and that obtained by \citetalias{2017ApJS..233...17R} following a detailed broadband X-ray spectral analysis. The points are color-coded with the counts in the 2--10\,keV energy range showing that the maximum offset with 1:1 line is observed in sources with least counts. The $x=y$ line is shown in black and the $1\sigma$ dispersion line is shown as black dashed. The typical uncertainties on $F_{2-10}$ calculated from our work are less than 5\%.}
\label{fig:xflux_comp}
\end{figure}


The spectra were fit using Cash statistics (C-stat; \citealp{1979ApJ...228..939C}). The best-fit values of all free parameters were checked using the \texttt{steppar} command in \textsc{XSPEC}. Additionally, the 90\% confidence intervals for all parameters were estimated using the \texttt{error} command in \textsc{XSPEC}. All the fit residuals were visually inspected and for $\sim$ 90\% of the sources we obtained good fits (see Table \ref{tab:xray} for goodness of each fit) using either model A1 or A2. For the remaining 10\% of the sources, we investigated the best-fit model and the fits individually. In some cases ($\sim7\%$
), a component to account for absorption by ionized gas, \texttt{ZXIPCF} (\citealp{2008MNRAS.385L.108R}), was added to the model to improve the fits. This component assumes a partial covering geometry of a partially ionized warm absorber. The photon index and normalization of the power law along with the three parameters (column density [$N_{\rm H}^{\rm W}$], ionization parameter [$\xi$], and the covering factor [$f_{\rm cov}^{\rm W}$] of the warm absorber) corresponding to the \texttt{ZXIPCF} component were allowed to vary during the fitting. After the fitting, the best-fit values of all the free parameters were checked using the \texttt{steppar} command to avoid local minima in the fitting. One of the sources, SWIFTJ1844.5-6221 (BAT ID = 995; more commonly known as Fairall 51), is known to show spectral variability on small timescales, due to the presence of a variable absorber, probably located in the broad line region (BLR; see \citealp{2015A&A...578A..96S} and references within). We found features of a possible changing-obscuration transition (e.g., \citealp{2023NatAs...7.1282R}) in its spectrum with a very low photon index ($\Gamma \sim 0.4$). A detailed analysis of the X-ray spectrum of this object is beyond the scope of this work, so we decided to remove it from the final sample (updated sample size $=$ 290). One source, SWIFTJ0335.4+1907 (BAT ID = 187), was fit using just a power law due to a low number of counts. Four sources (SWIFTJ0021.2-1909, SWIFTJ0945.6-1420, SWIFTJ1334.8-2328D2, and SWIFTJ1345.5+4139) showed signatures of strong obscuration in their spectra, corresponding to them being at the high end of the column density distribution of our sample. Hence, for these sources, we used model B1 from \citetalias{2017ApJS..233...17R}, which was specifically used to fit obscured sources. This model consists of additional components to account for the reprocessed primary X-ray emission and Thomson-scattered X-ray radiation (e.g., \citealp{2021MNRAS.504..428G}; \citealp{2023MNRAS.526.5072M}).

For all objects, we obtained the best-fit values for parameters including the photon index (Fig. \ref{fig:gamma}) and normalization of the primary power law, line-of-sight column density of the neutral material responsible for absorption, and temperature and normalization of the blackbody component (see Table \ref{tab:xray}), from the X-ray spectral fits. To ensure that our results were consistent with those reported by \citetalias{2017ApJS..233...17R}, we compared the best-fit values of $\Gamma$ of the primary power law and obtained a linear relation with a dispersion of 0.2 dex, expected due to variability. We calculated the observed and intrinsic (rest-frame and absorption-corrected) 2--10\,keV and 14--195\,keV fluxes (Fig. \ref{fig:xray_flux}) and luminosities for all our sources. All luminosities were calculated using the best estimated distances (including redshift-independent distances) listed in Table \ref{tab:src_list}. The upper and lower uncertainties in the flux were estimated using the \texttt{cflux} command in \textsc{XSPEC}.

In Fig. \ref{fig:xflux_comp}, we show a comparison between the 2--10\,keV fluxes from this work and those reported by \citetalias{2017ApJS..233...17R} to confirm that we get similar values. We obtain a 1:1 correlation (shown as a black line) with a 0.2 dex dispersion (similar to $\Gamma$; shown as a dashed black line), which could be an effect of the variability in the X-rays between the time of the observation used in our analysis and those used by \citetalias{2017ApJS..233...17R}. The visible scatter also emphasizes the need for a new X-ray spectral analysis that is contemporaneous with the optical/UV data for this specific work. The points are color-coded with the 2--10\,keV counts to show that the sources with maximum offset from the 1:1 line are indeed the ones with lower counts. The total X-ray luminosity ($L_{\rm x}$; to be later combined with the optical/UV disk emission, to finally estimate the bolometric luminosity) was computed from the 0.1 to 500\,keV energy range. The lower bound of this energy range was determined based on the X-ray spectrum generated by the thermal model \texttt{NTHCOMP} (\citealp{1996MNRAS.283..193Z}; \citealp{1999MNRAS.309..561Z}), which describes the X-ray corona. We assumed a photon index of 1.8 \citep{2018MNRAS.480.1819R}, coronal electron temperature ($kT_{\rm e}$) of 100\,keV \citep{2018MNRAS.480.1819R}, and seed photon temperature ($kT_{\rm bb}$) of 0.032\,keV based on the median black hole mass (${\rm log}\,[M_{\rm BH}/M_{\odot}] = 7.9$) of our sample. The X-ray spectrum thus created shows a sharp drop at 0.1\,keV, indicating the lack of significant X-ray emission beyond those energies.


\begin{table*}
\centering
\caption{Results of the X-ray spectral fitting (Sect. \ref{sect:xray_spec}).}
\begin{tabular}{cccccc}
\hline
\hline
\specialrule{0.1em}{0em}{0.5em}
\vspace{1mm}
BAT ID & Swift ID & $\Gamma$ & $F_{2-10}$ & Model & ${\rm Statistic/dof}$\\
\hline\\
\vspace{2mm}
2 & SWIFTJ0001.6-7701 & $1.55_{-0.64}^{+1.01}$ & $2.90_{-0.44}^{+0.48}$ & A1 $+$ \texttt{ZXIPCF} & $86.84/113$\\
\vspace{2mm}
3 & SWIFTJ0002.5+0323 & $1.97_{-0.16}^{+0.24}$ & $3.19_{-0.34}^{+0.37}$ & A2 & $134.4/148$\\
\vspace{2mm}
6 & SWIFTJ0006.2+2012 & $2.34_{-0.04}^{+0.03}$ & $10.8_{-0.21}^{+0.21}$ & A1 & $796.5/597$\\
\vspace{2mm}
10 & SWIFTJ0021.2-1909 & $1.60_{-0.18}^{+0.18}$ & $9.03_{-0.48}^{+0.50}$  & B1 & $360.5/445$\\
\vspace{2mm}
14 & SWIFTJ0026.5-5308 & $1.70_{-0.23}^{+0.35}$ & $1.97_{-0.23}^{+0.25}$ & A1 & $130.7/139$\\
\vspace{2mm}
16 & SWIFTJ0029.2+1319 & $2.20_{-0.17}^{+0.17}$ & ${5.44}_{-0.21}^{+0.22}$ & A2 & $335.2/397$\\
\vspace{2mm}
19 & SWIFTJ0034.5-7904 & $2.36_{-0.07}^{+0.11}$ & ${6.14}_{-0.21}^{+0.22}$ & A2 & $390.3/390$\\
\vspace{2mm}
34 & SWIFTJ0051.6+2928 & $1.99_{-0.23}^{+0.25}$ & ${1.71}_{-0.13}^{+0.13}$ & A2 & $197.3/253$\\
\vspace{2mm}
36 & SWIFTJ0051.9+1724 & $1.91_{-0.03}^{+0.10}$ & ${16.6}_{-0.40}^{+0.40}$ & A2 & $564.5/578$\\
\vspace{2mm}
39 & SWIFTJ0054.9+2524 & $1.97_{-0.15}^{+0.18}$ & ${5.58}_{-0.23}^{+0.24}$ & A2 & $328.9/403$\\
\hline
\end{tabular}
\tablefoot{We have reported the main outputs of the fitting here, including the best-fit values of the photon index, the 2--10\,keV intrinsic fluxes with errors, the model used to get the best fit, and the value of the statistic and the degrees of freedom of the fit. Flux units: $10^{-12}\,{\rm erg\,\,cm^{-2}\,s^{-1}}$. The table in its entirety is available at the CDS.}
\label{tab:xray}
\end{table*}


\section{AGN-galaxy image decomposition using GALFIT}\label{sect:galfit}

When estimating the optical/UV fluxes and magnitudes for AGN, one needs to correct for contamination due to the host galaxy light. Even though our sample consists of nearby unobscured AGN, where the central AGN is typically brighter than the host galaxy at optical/UV wavelengths, the fluxes calculated by selecting 5$"$ regions around the AGN can have a significant contribution from the host galaxy starlight. To obtain accurate measurements of the AGN flux devoid of any host galaxy light, we employed \textsc{GALFIT}\footnote{\url{https://users.obs.carnegiescience.edu/peng/work/galfit/galfit.html}}(\citealp{2002AJ....124..266P}; \citealp{2010AJ....139.2097P}) to fit the 2D surface brightness profiles of the sources and decompose them into AGN and galaxy light. \textsc{GALFIT} is an image analysis tool that can be used to model light profiles of galaxies by separating them into different parts such as disk, bulge, and point source, and using different radial distributions like a PSF, S\'ersic profile and Gaussian function, to fit them. For our analysis, we used a PSF profile to fit the AGN and a S\'ersic profile to reproduce the host galaxy emission, if and when needed. The size of the \textit{Swift}/UVOT PSF ranges between 2--3 arcseconds, which translates to physical scales of $\sim$ 0.04--4\,kpc for the redshift range of the sources in our sample. This is comparable to the typical sizes of bulge components of large galaxies (e.g., \citealp{2008IAUS..245....3F}), where most of the optical/UV emission can be attributed to the central AGN. For eight nearby sources with extended galaxy features, we adopted two S\'ersic profiles to better estimate the AGN flux, while in most of the faintest sources, \textsc{GALFIT} was not able to fit a S\'ersic profile to the images, and hence we only used the PSF profile to estimate their AGN flux. Due to the flux-limited nature of the \textit{Swift}/BAT AGN sample, the sources with the lowest fluxes are also the farthest and appear more point-like in the images, and hence are suitable to be fit with a PSF profile due to their comparatively small host galaxy contamination. In all cases, a uniform background component was also included in the fits.

\textsc{GALFIT} consists of a function-fitting algorithm and gives the best-fit results based on least-squares statistics. Therefore, the user is expected to provide the necessary inputs needed to do the fitting. \textsc{GALFIT} requires a template file that consists of all the details about the input image, including the name and location of the image file, size of the image region to fit, bad-pixel mask, zero-point magnitude, and plate scale. \textsc{GALFIT} needs some specific information for the input source image and looks for these in the image header corresponding to the keywords EXPTIME (the exposure time), GAIN (electrons/counts), and NCOMBINE (number of images combined to create the source image). Therefore, the input image file has to be edited to include these keywords and their values in the image header. There is also the option to add a bad-pixel mask, which is useful to discard very nearby sources such as bright stars and stray objects contaminating the source of interest or its immediate surroundings.

If the user wishes to fit a PSF, they have to provide the PSF image as well as the PSF fine sampling factor relative to the data in this template file. Additionally, all the functions or profiles corresponding to different parts of the galaxy that the user wants to fit along with the initial values of their fitting parameters are also defined in this template file. The user can also specify if a particular parameter is fixed or free to vary during the fitting. In the next section (Sect. \ref{sect:galfit_comp}), we describe in detail the different components we used for the image decomposition.


\subsection{Light profiles used in GALFIT}\label{sect:galfit_comp}

In our work, we use three different profiles to fit the source images. For most sources, this three-component model yields good fit residuals (fitting details provided in Sect. \ref{sect:galfiting}). Using more than three components typically results in overfitting of the images and over-subtracted residuals. The three components used for the fitting are:\\

\begin{itemize}
    \item \textit{Background sky}: The background in the \textit{Swift}/UVOT images is quite uniform for all the filters. Hence, we used a constant sky value to fit it. For the starting value, we used the background value as estimated by the Astropy (\citealp{astropy:2022}) package \texttt{photutils}\footnote{\url{https://photutils.readthedocs.io/en/stable/index.html}} (used for photometry; \citealp{2021zndo...4624996B}), using the sigma-clipping method. By using a source detection threshold specified by the user, the median background value is calculated by removing the sources with source counts above the threshold and recalculating the background. Since this is just an input estimation for the background to be better fit by \textsc{GALFIT}, we used a value of two sigma above the median for source clipping. As input, \textsc{GALFIT} requires the user to provide an image region of the background without any sources to fit the sky. Whenever possible, we used offset background regions as close as possible to the source of interest to reduce small deviations across the area of the detector. For the majority of sources, a region of $100\times100$ pixels\footnote{The plate scale of \textit{Swift}/UVOT is 0.502 arcseconds.} was selected, which was otherwise modified according to the availability of larger regions without any sources in the image.\\
    
    \item \textit{PSF profile for the AGN}: To fit a PSF profile in \textsc{GALFIT}, one needs to provide as input a PSF image, which is then convolved with the model to create the PSF function used to fit the point source in the image. For our analysis, we created the PSF images for each source in each filter using the \texttt{EPSFBuilder} task in the \texttt{photutils} Astropy package, which uses bright stars in the source image to create an effective PSF. This method involves a series of steps, starting from finding the appropriate stars to create the PSF, extracting them, and finally iterating over the selected stars to build the PSF. Firstly, a few stars within a certain range of the count rate of the source of interest are selected. We used the \texttt{find\_peaks} function to do this by specifying a threshold count rate slightly below that of the source, such that we have at least five to ten stars with count rates similar to that of the source to begin with. We then used the \texttt{extract\_stars} function to extract $25\times25$ pixel cutouts of these stars that were visually inspected for distortions, fake detections, saturated pixels, and bad quality, and to ensure that they were sufficiently isolated. Finally, three to six stars were selected to create the PSF using the \texttt{epsf\_builder} function with an oversampling factor of unity and ten iterations. The PSF image thus created was used as input in \textsc{GALFIT} along with the input source image and the bad-pixel mask. One needs to specify the region of the source image to be fit in the \textsc{GALFIT} template file. For most of the sources, we used a $100 \times 100$ pixel image region around the source, which is almost 25 times the size of the PSF and is sufficient to include the extended wings of the PSF. In the case of more extended sources, we used image regions large enough to include the source as well as some background. The PSF profile has only three free parameters in the \textsc{GALFIT} template file: the $x$ and $y$ coordinates and the total PSF magnitude. We specified the initial $x$ and $y$ coordinates based on the location of the center of the source and the input magnitude value as the one calculated for each source within the 5$"$ aperture using the \texttt{uvotsource} command of the \textit{Swift}/UVOT data reduction pipeline.\\
    
    \item \textit{S\'ersic profile for the host galaxy}: The host galaxy light in our sources, if and when needed, was fit using a S\'ersic profile. The free parameters for this profile include the $x$ and $y$ coordinates of the source, the integrated magnitude, the S\'ersic index,\footnote{In a S\'ersic light profile, the S\'ersic index ($n$) defines the curvature of the light profile with distance from the center. Larger values of $n$ indicate steeper inner light profiles and more extended outer wings, while low values of $n$ indicate flatter corelight profiles and sharper truncation at large radii.} the effective radius, the axis ratio, and the position angle (PA). In most cases, we left all the parameters free to vary, unless the fit did not converge due to one specific parameter going out of bounds, in which case we fixed that parameter to an appropriate value (see Sect. \ref{sect:galfiting}). For most of the sources, a single S\'ersic profile along with a PSF was sufficient to reproduce all the light coming from the galaxy. However, for some extended sources (8/290), we employed two S\'ersic profiles, one for the bulge ($n\sim 4$) and one for the disk ($n\sim1$), to fit the overall host galaxy light. On the contrary, for most of the distant sources ($z > 0.05$), the PSF was almost as large as the host galaxy, and hence we only used a PSF profile to fit the source.
\end{itemize}


\begin{table*}
\centering
\setlength{\tabcolsep}{0.3\tabcolsep}
\caption{Corrected AGN fluxes as estimated by \textsc{GALFIT} image decomposition (Sect. \ref{sect:galfit}).}
\begin{tabular}{ccccccccc}
\hline
\hline
\specialrule{0.1em}{0em}{0.5em}
\vspace{1mm}
BAT ID & Swift ID & $F_{\rm V}$ & $F_{\rm B}$ & $F_{\rm U}$ & $F_{\rm W1}$ & $F_{\rm M2}$ & $F_{\rm W2}$ & Flag\\
\hline\\
\vspace{1mm}
3 & SWIFTJ0002.5+0323 & \,\,\,$0.75\pm0.07$ & \,\,\,$0.82\pm0.06$ & \,\,\,$1.59\pm0.10$ & \,\,\,$1.59\pm0.14$& \,\,\,$2.17\pm0.11$ & \,\,\,$2.24\pm0.15$ & -\\
\vspace{1mm}
6 & SWIFTJ0006.2+2012 & \,\,\,\,\,$5.82\pm0.13$\tablefootmark{o} & \,\,\,\,\,$6.70\pm0.15$\tablefootmark{o} & $10.89\pm0.46$ & \,\,$19.63\pm1.14$\tablefootmark{o} & \,\,$22.33\pm0.76$\tablefootmark{o} & \,\,$24.60\pm1.19$\tablefootmark{o} & -\\
\vspace{1mm}
16 & SWIFTJ0029.2+1319 & \,\,\,\,\,$3.17\pm0.15$\tablefootmark{o} & \,\,\,\,\,$3.07\pm0.08$\tablefootmark{o} & \,\,\,\,\,$3.77\pm0.16$\tablefootmark{o} & \,\,\,\,\,$4.90\pm0.35$\tablefootmark{o} & \,\,\,\,\,$6.50\pm0.25$\tablefootmark{o} & \,\,\,\,\,$5.96\pm0.52$\tablefootmark{o} & c\\
\vspace{1mm}
19 & SWIFTJ0034.5-7904 & \,\,\,$0.89\pm0.07$ & \,\,\,$1.09\pm0.05$ & \,\,\,$2.53\pm0.21$ & \,\,$10.28\pm0.75$\tablefootmark{o} & \,\,$13.99\pm0.60$\tablefootmark{o} & \,\,$12.30\pm1.09$\tablefootmark{o} & c\\
\vspace{1mm}
34 & SWIFTJ0051.6+2928 & \,\,\,$1.17\pm0.06$ & \,\,\,$1.11\pm0.05$ & \,\,\,$1.16\pm0.07$ & \,\,\,$1.14\pm0.10$ & \,\,\,$1.00\pm0.06$ & \,\,\,\,\,$1.23\pm0.07$\tablefootmark{o} & -\\
\vspace{1mm}
36 & SWIFTJ0051.9+1724 & \,\,\,\,\,$3.45\pm0.24$\tablefootmark{o} & \,\,\,\,\,$3.01\pm0.07$\tablefootmark{o} & \,\,\,\,\,$4.55\pm0.20$\tablefootmark{o} & \,\,\,\,\,$6.91\pm0.44$\tablefootmark{o} & \,\,\,\,\,$7.60\pm0.31$\tablefootmark{o} & \,\,\,\,\,$6.26\pm0.58$\tablefootmark{o} & c\\
\vspace{1mm}
39 & SWIFTJ0054.9+2524 & \,\,\,$1.68\pm0.05$ & \,\,\,\,\,$2.80\pm0.06$\tablefootmark{o} & \,\,\,\,\,$4.76\pm0.21$\tablefootmark{o} & \,\,\,\,\,$5.40\pm0.38$\tablefootmark{o} & \,\,\,\,\,$7.24\pm0.38$\tablefootmark{o} & \,\,\,\,\,$7.60\pm0.66$\tablefootmark{o} & c\\
 \vspace{1mm}
43 & SWIFTJ0059.4+3150 & \,\,\,$1.17\pm0.07$ & \,\,\,$0.46\pm0.90$ & \,\,\,\,\,$0.97\pm0.14$\tablefootmark{o} & \,\,\,\,\,$4.00\pm0.30$\tablefootmark{p} & \,\,\,\,\,$3.64\pm0.16$\tablefootmark{o} & \,\,\,$3.55\pm0.17$ & -\\
 \vspace{1mm}
45 & SWIFTJ0101.5-0308 & \,\,\,$0.40\pm0.04$ & \,\,\,$0.14\pm0.04$ & \,\,\,\,\,$0.36\pm0.04$\tablefootmark{o} & \,\,\,\,\,$0.29\pm0.05$\tablefootmark{o} & \,\,\,\,\,$0.27\pm0.03$\tablefootmark{o} & \,\,\,\,\,$0.32\pm0.04$\tablefootmark{o} & -\\
 \vspace{1mm}
51 & SWIFTJ0105.7-1414 & \,\,\,$1.93\pm0.14$ & \,\,\,$2.50\pm0.21$ & \,\,\,$3.57\pm0.12$ & \,\,\,$6.08\pm0.43$ & $11.56\pm2.29$ & \,\,\,$8.87\pm0.54$ & c\\
\hline
\end{tabular}
\tablefoot{The last column shows, in the form of flags (c), if the GALFIT-estimated fluxes were corrected using corrections calculated in Sect. \ref{sect:galfit_corr} to create the final SEDs. All fluxes are in units of $10^{-15}\,{\rm erg\,\,cm^{-2}\,s^{-1}\,\AA^{-1}}$. The table in its entirety is available at the CDS.\\
\tablefoottext{o}{A sky $+$ PSF model was used to get the best \textsc{GALFIT} residual as \textsc{GALFIT} was unable to fit a S\'ersic profile and hence, the corresponding correction was employed (see Fig. \ref{fig:galfit_corr}), if and when needed.}\\
\tablefoottext{p}{UVOT pipeline-estimated flux was used in the final SED.}
}
\label{tab:galfit_flux}
\end{table*}


\subsection{GALFIT-ing procedure}\label{sect:galfiting}

For our sample of 290 unobscured AGN across the six \textit{Swift}/UVOT filters, we decomposed images into galaxy and AGN light profiles using \textsc{GALFIT}, following the procedure stated below:

\begin{enumerate}
    \item We first fit a $100\times100$ pixel region of the blank sky in the source image with a constant sky background.
    \item Next, we selected a $100\times100$ pixel (or larger if needed) region around the source of interest to fit the central AGN and the background together using a sky $+$ PSF profile. The input value for the sky was kept the same as the one obtained from the previous step. For the PSF profile, all the parameters were kept free during the fit.
    \item Finally, we fit the same source region as the previous step with a sky $+$ PSF $+$ S\'ersic profile. The best-fit values estimated in the second step were used as input values for the sky and PSF parameters in this step, while they were allowed to vary. This was mainly done because in almost all the cases the residuals in the second step showed over-fitting, especially in the central pixels. Hence, in this step we allowed the PSF and the S\'ersic profiles to adjust to each other and simultaneously fit the source light profile. For the S\'ersic profile, the starting values for the $x$ and $y$ coordinates and the integrated magnitude were kept the same as those of the PSF profile. The input value for the S\'ersic index ($n$) was set at 2.5, the axis ratio ($b/a$) at 1.0, the effective radius (half-light radius; $R_{\rm e}$) at 5.0 pix, and the PA at 0.0 deg. In most cases, if the fit does not converge, it was due to the S\'ersic index or the effective radius reaching their limits.\footnote{\textsc{GALFIT} allows the S\'ersic index to vary from 0.01 to 20 and the effective radius from 0.01 onward.} In the case of the S\'ersic index, we repeated the fits with different input values, going in the following order: $n$ = 2.0, 1.5, 1.0, 0.5, 3.0, 3.5, and 4.0. In the case of the effective radius, we repeated the fits with various values depending on the source but noticed that the input value of $R_{\rm e}$ does not affect the PSF (AGN) magnitude. Since this work aims to construct and fit AGN SEDs to obtain reliable estimates of the quantities describing AGN physics, we focused on extracting good measurements of AGN magnitudes through the image decomposition process. Hence, during the fitting, the S\'ersic profile was primarily used to apply constraints on the PSF profile, instead of exactly describing the host galaxy light profile. The input values used for parameters like $R_{\rm e}$, $b/a$, and PA were always kept the same as they do not affect the final estimates of the PSF magnitude. In this respect, in cases when the fit with sky $+$ PSF $+$ S\'ersic profile does not converge due to one of the S\'ersic profile parameters going out of bounds, we still get reliable estimates for the PSF magnitude that we can use to construct the SEDs (see Fig. \ref{fig:galfit_sim-noconv}). We have flagged such cases while reporting the best-fit values from the fitting.
\end{enumerate}

\begin{figure*}
  \begin{subfigure}[t]{0.52\textwidth}
    \centering
    \includegraphics[width=\textwidth]{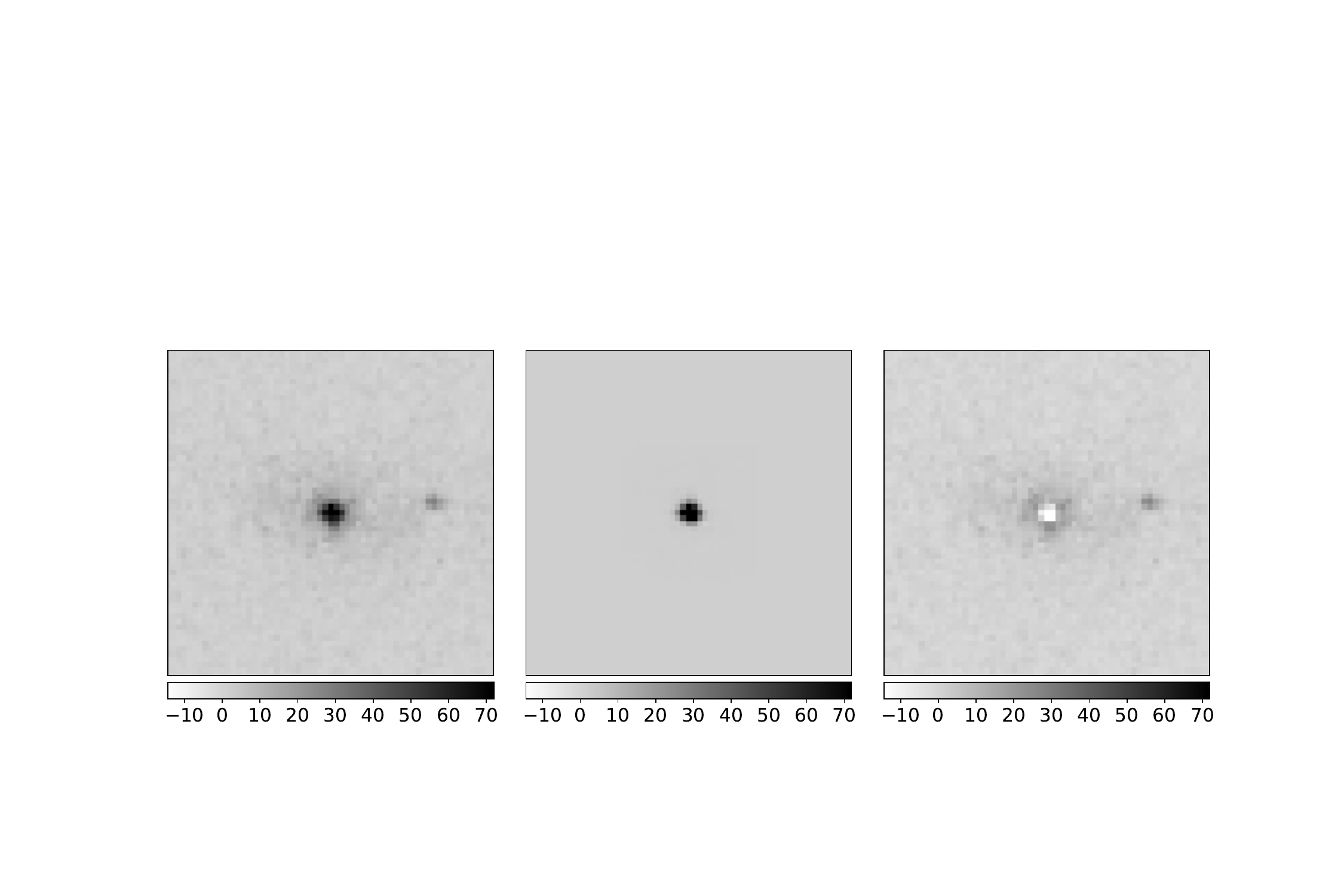} 
    \label{fig:galfit_eg11}
  \end{subfigure}
  \hspace{-1cm}
  \begin{subfigure}[t]{0.52\textwidth}
    \centering
    \includegraphics[width=\textwidth]{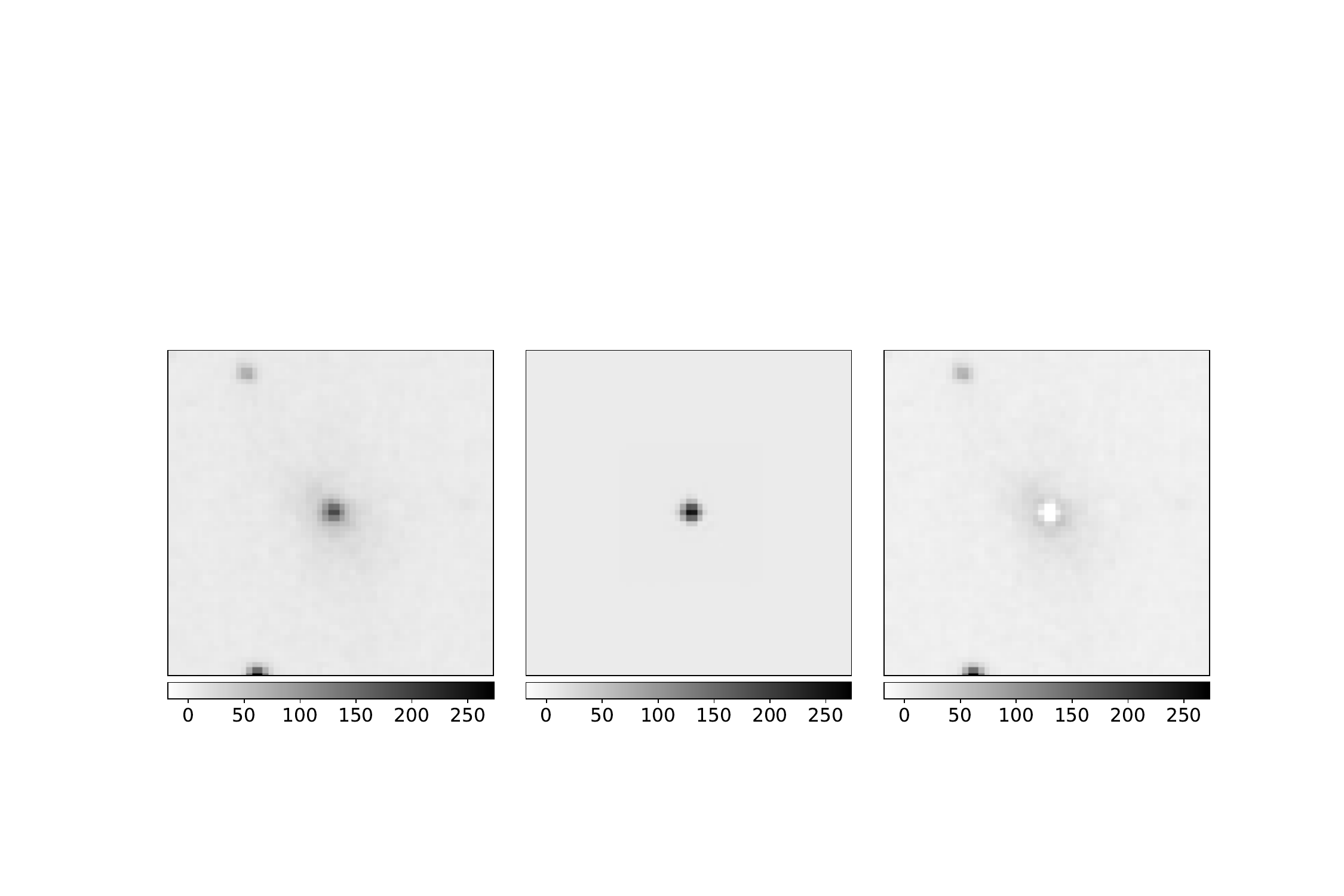} 
    \label{fig:galfit_eg21}
  \end{subfigure}
  
  \vspace{-3.5cm}
  
  \begin{subfigure}[t]{0.52\textwidth}
    \centering
    \includegraphics[width=\textwidth]{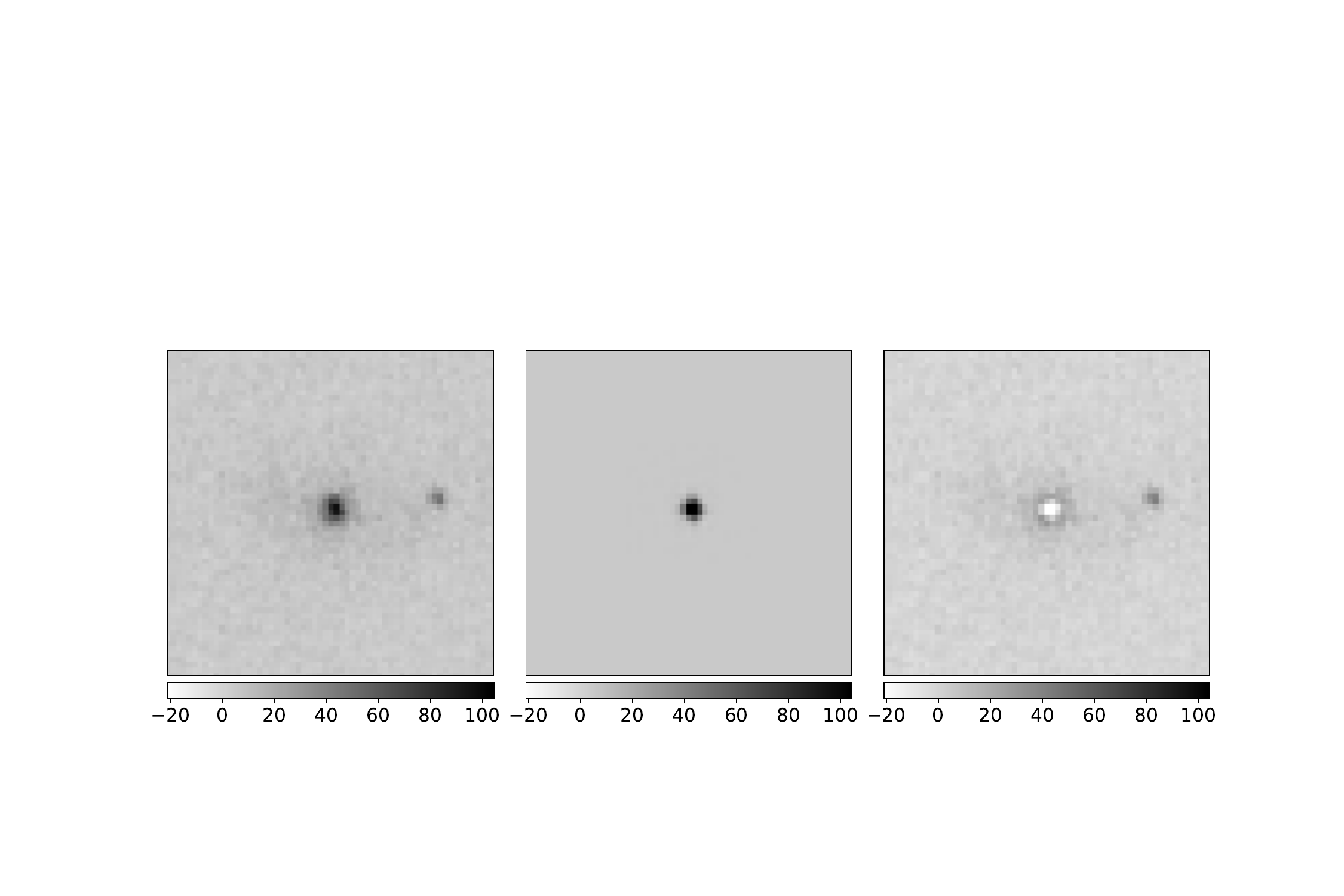}
    \label{fig:galfit_eg12}
  \end{subfigure}
  \hspace{-1cm}
  \begin{subfigure}[t]{0.52\textwidth}
    \centering
    \includegraphics[width=\textwidth]{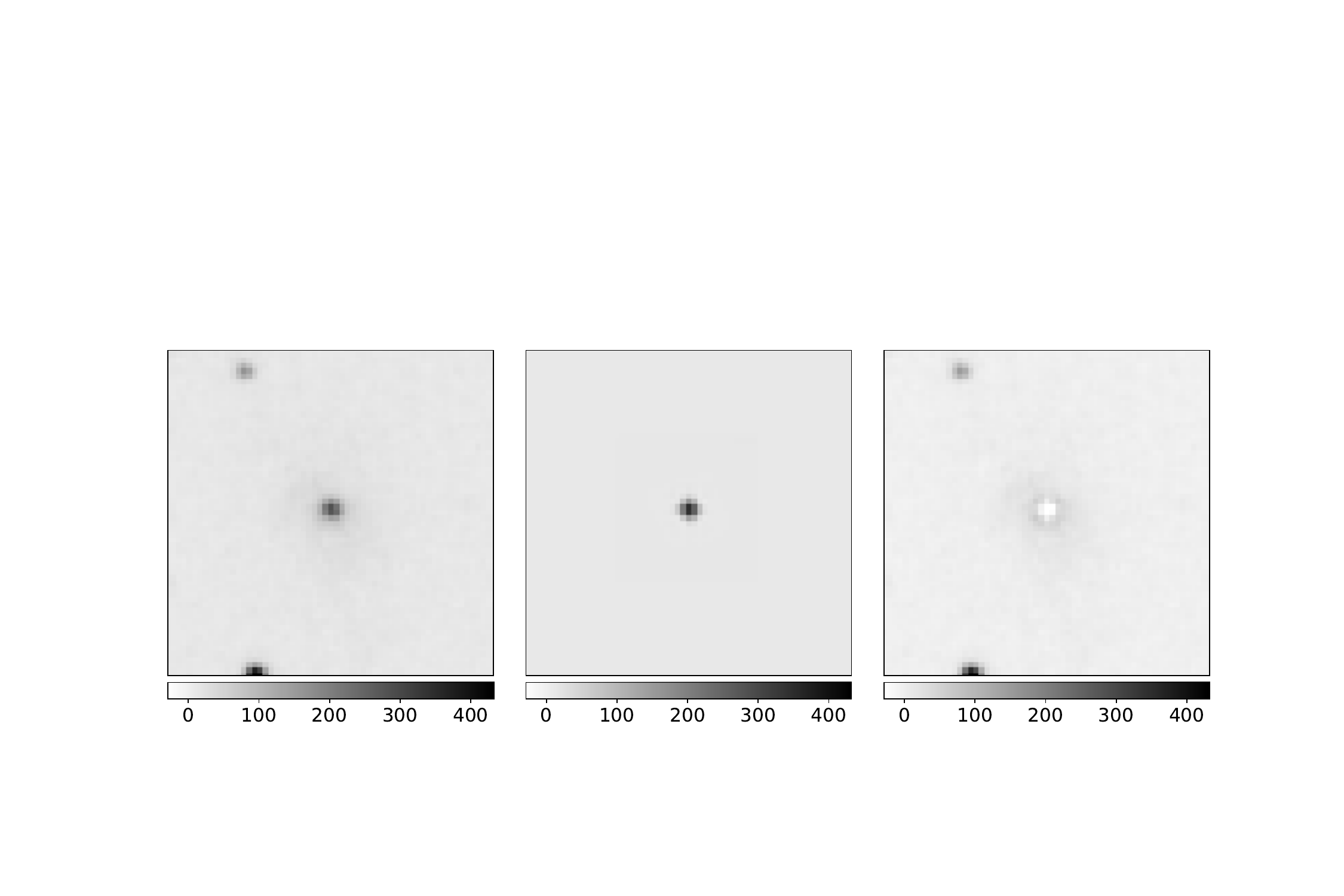} 
    \label{fig:galfit_eg22}
  \end{subfigure}
  
  \vspace{-3.5cm}
  
  \begin{subfigure}[t]{0.52\textwidth}
    \centering
    \includegraphics[width=\textwidth]{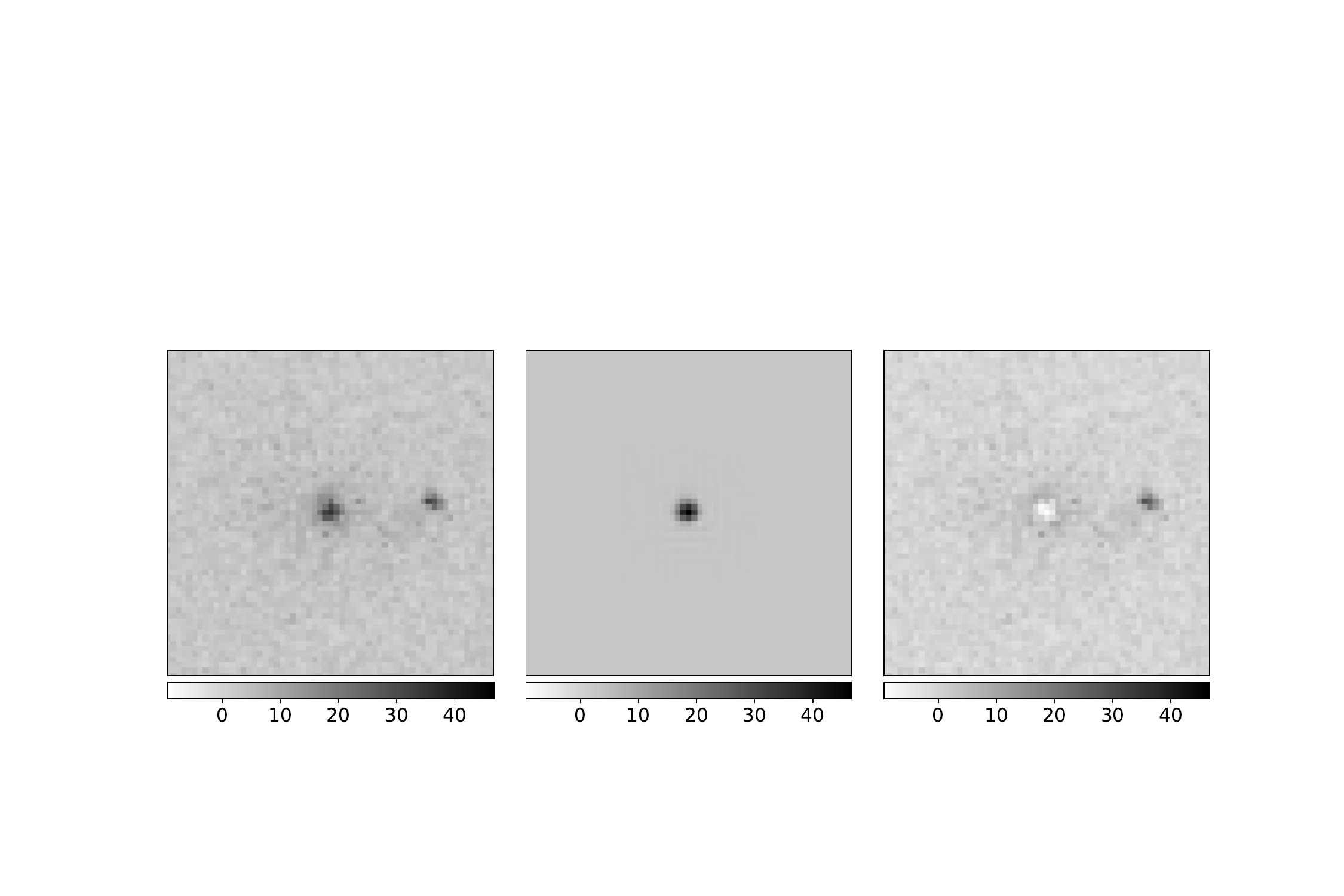}
    \label{fig:galfit_eg13}
  \end{subfigure}
 \hspace{-1cm}
  \begin{subfigure}[t]{0.52\textwidth}
    \centering
    \includegraphics[width=\textwidth]{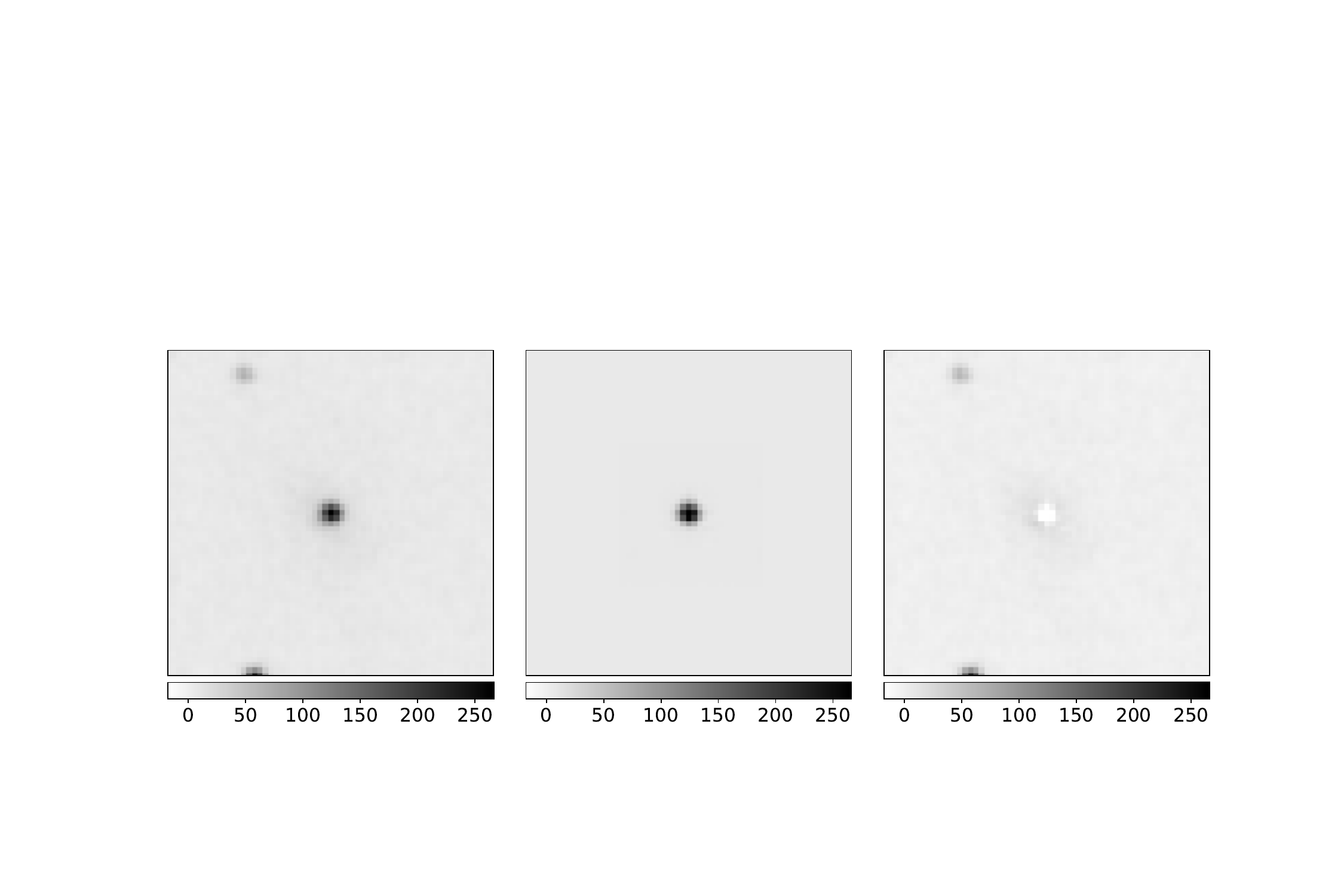} 
    \label{fig:galfit_eg23}
  \end{subfigure}
  
  \vspace{-3.5cm}
  
  \begin{subfigure}[t]{0.52\textwidth}
    \centering
    \includegraphics[width=\textwidth]{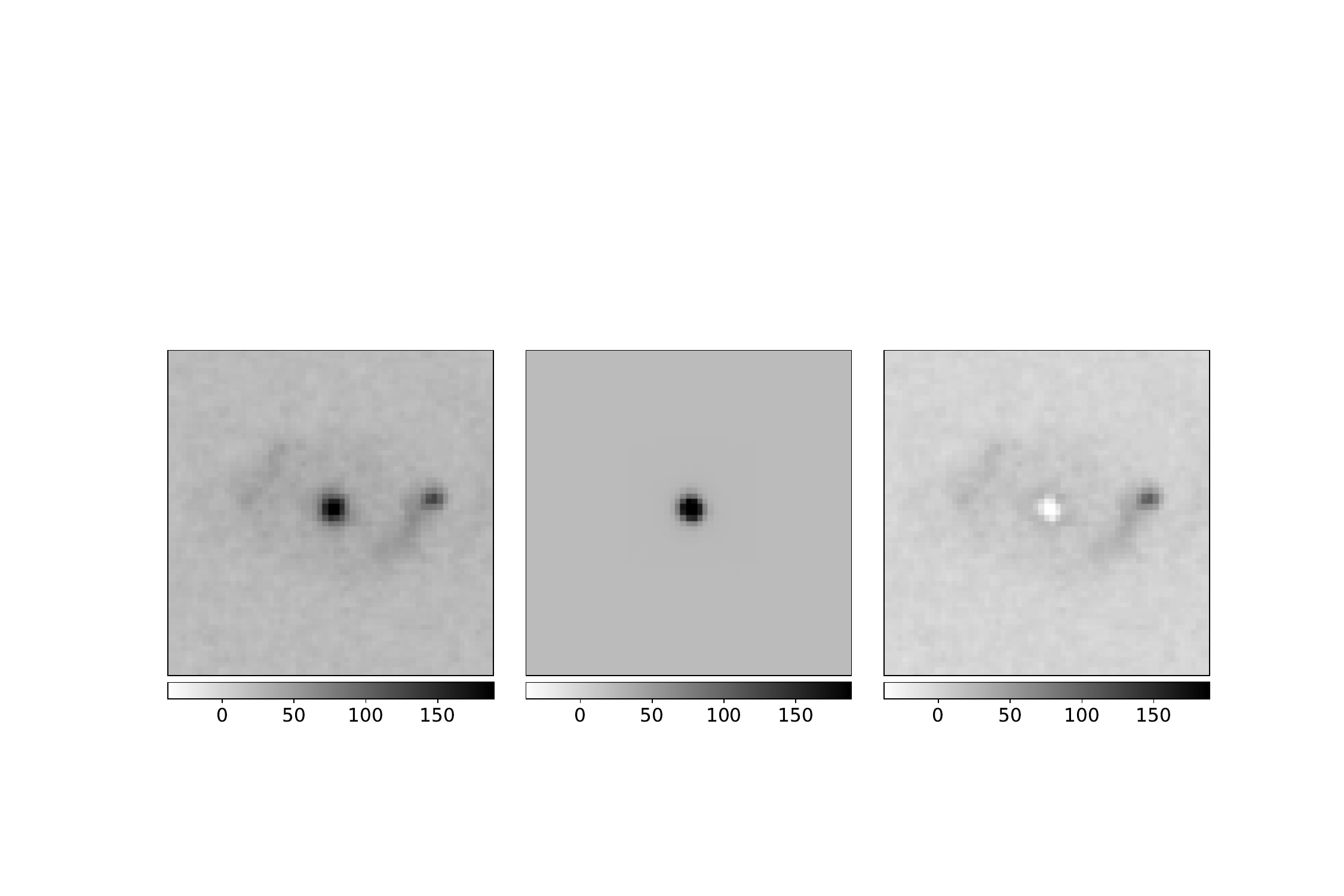}
    \label{fig:galfit_eg14}
  \end{subfigure}
  \hspace{-1cm}
  \begin{subfigure}[t]{0.52\textwidth}
    \centering
    \includegraphics[width=\textwidth]{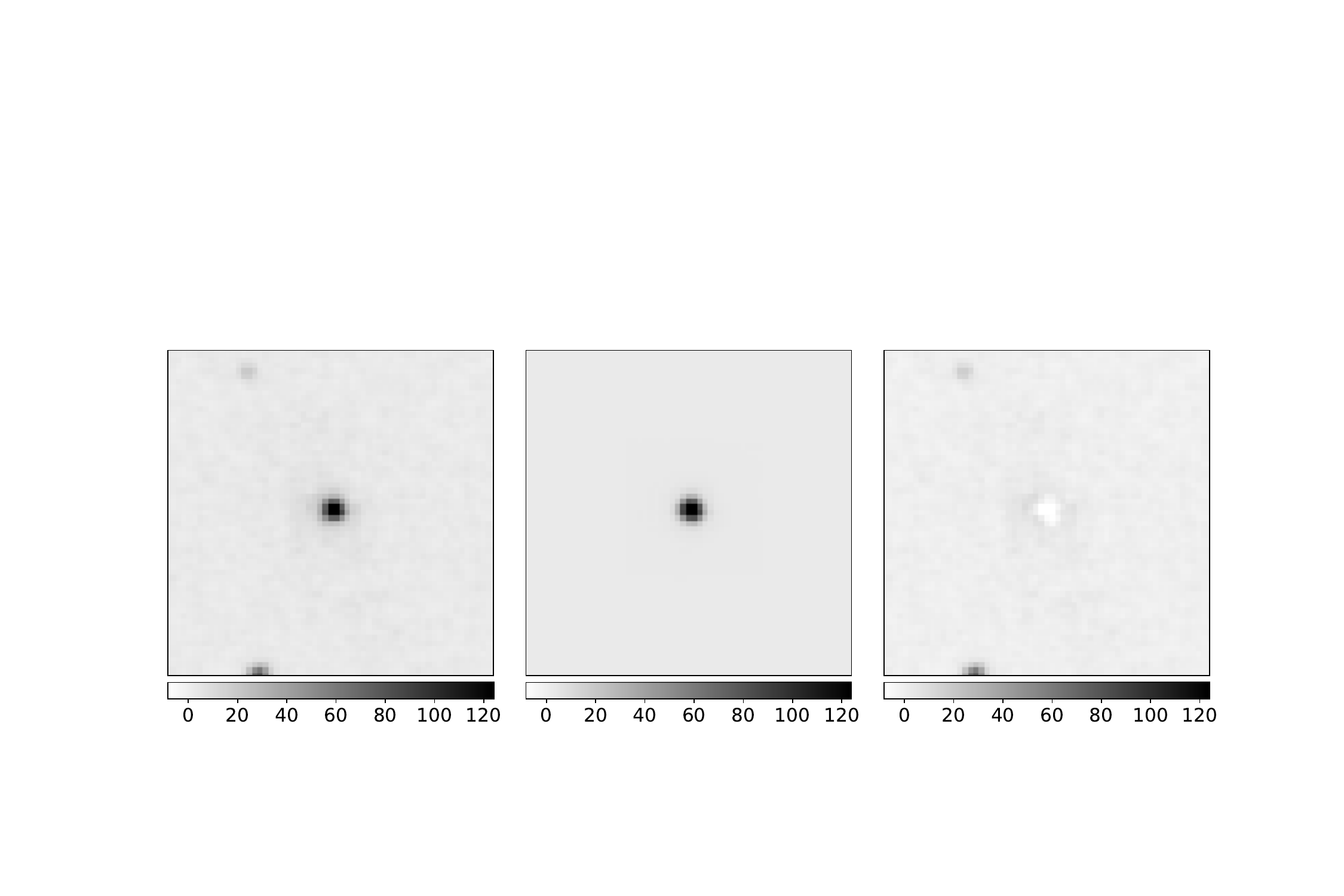} 
    \label{fig:galfit_eg24}
  \end{subfigure}
  
  \vspace{-3.5cm}
  
  \begin{subfigure}[t]{0.52\textwidth}
    \centering
    \includegraphics[width=\textwidth]{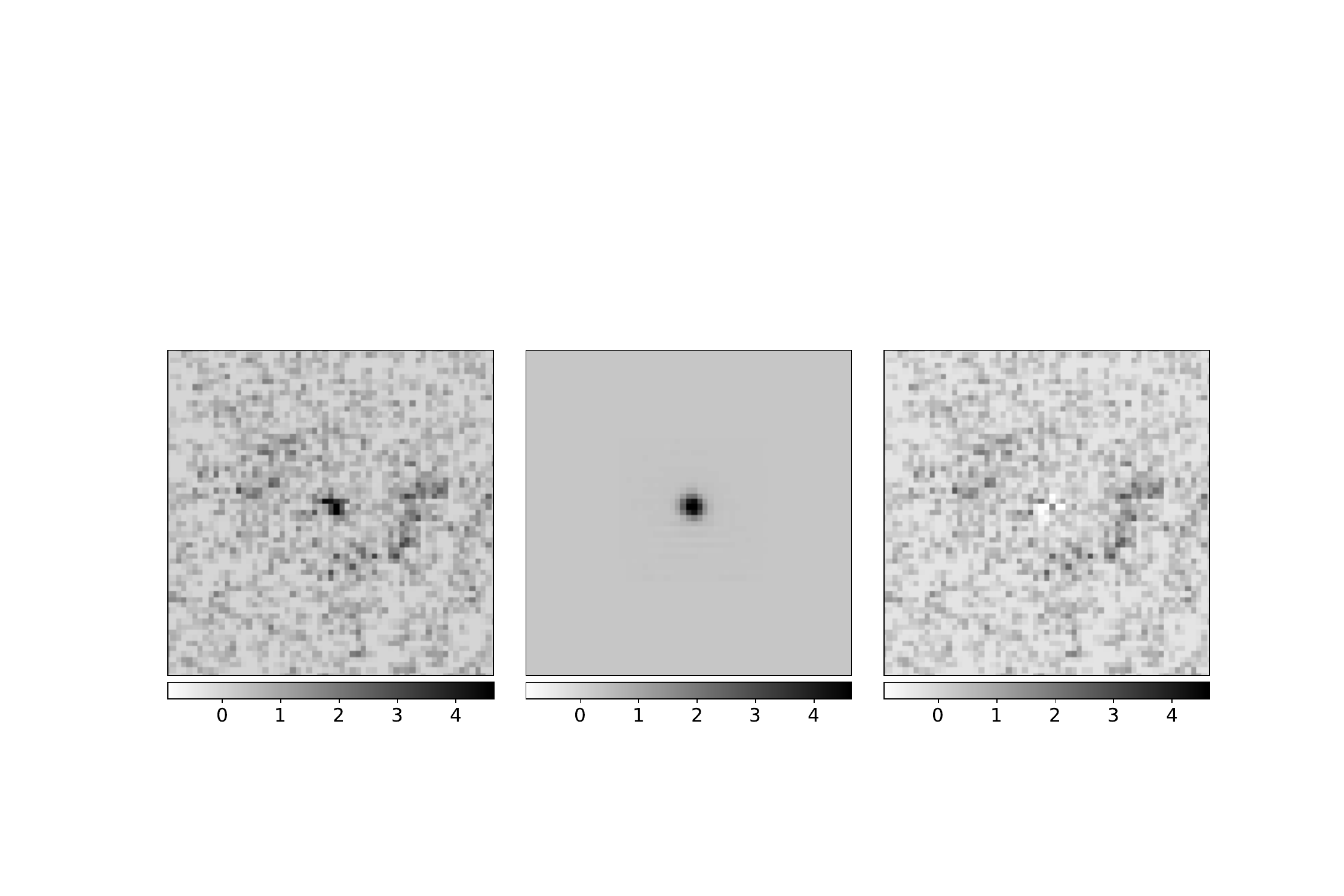}
    \label{fig:galfit_eg15}
  \end{subfigure}
  \hspace{-1cm}
  \begin{subfigure}[t]{0.52\textwidth}
    \centering
    \includegraphics[width=\textwidth]{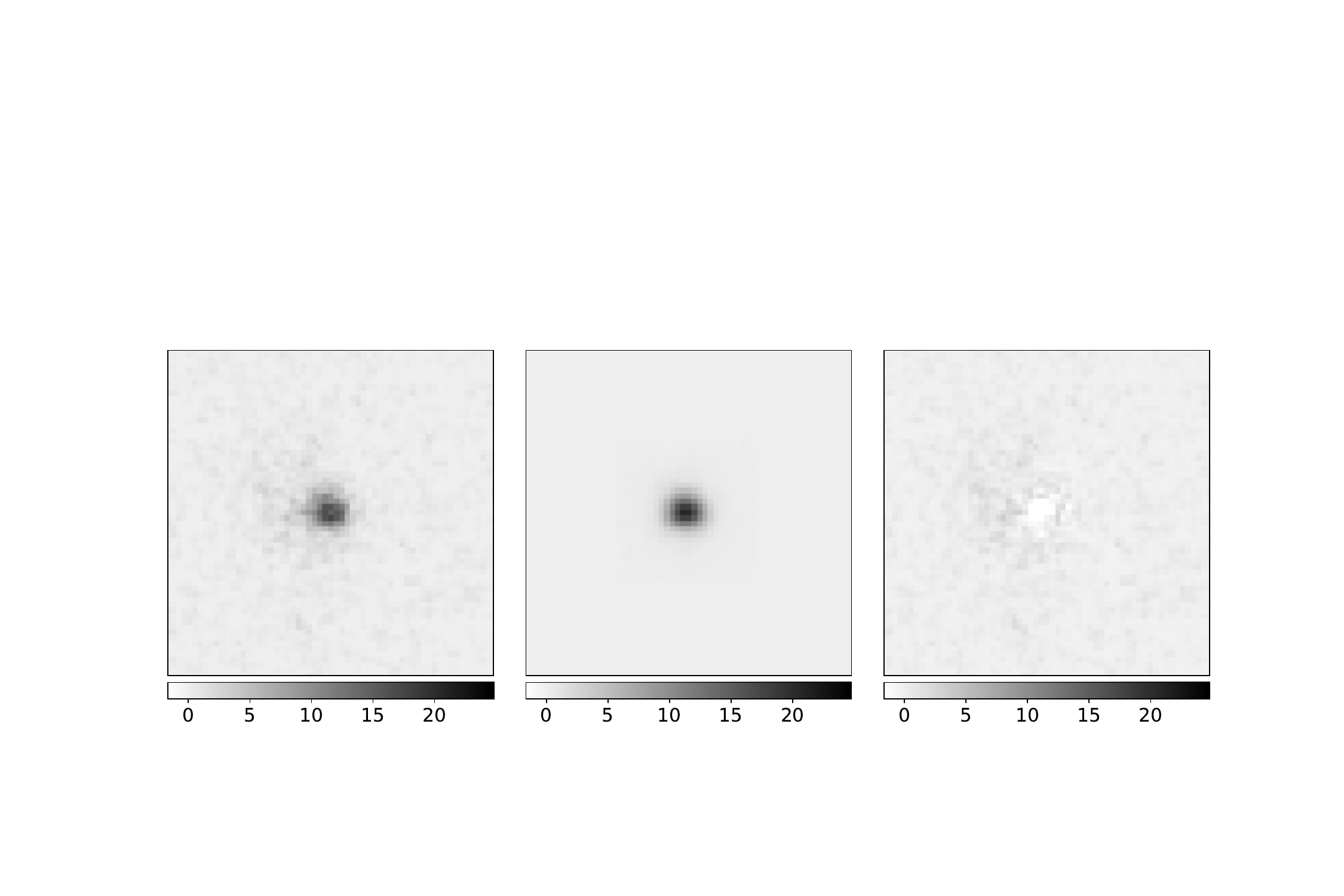} 
    \label{fig:galfit_eg25}
  \end{subfigure}
  
  \vspace{-3.5cm}
  
  \begin{subfigure}[t]{0.52\textwidth}
    \centering
    \includegraphics[width=\textwidth]{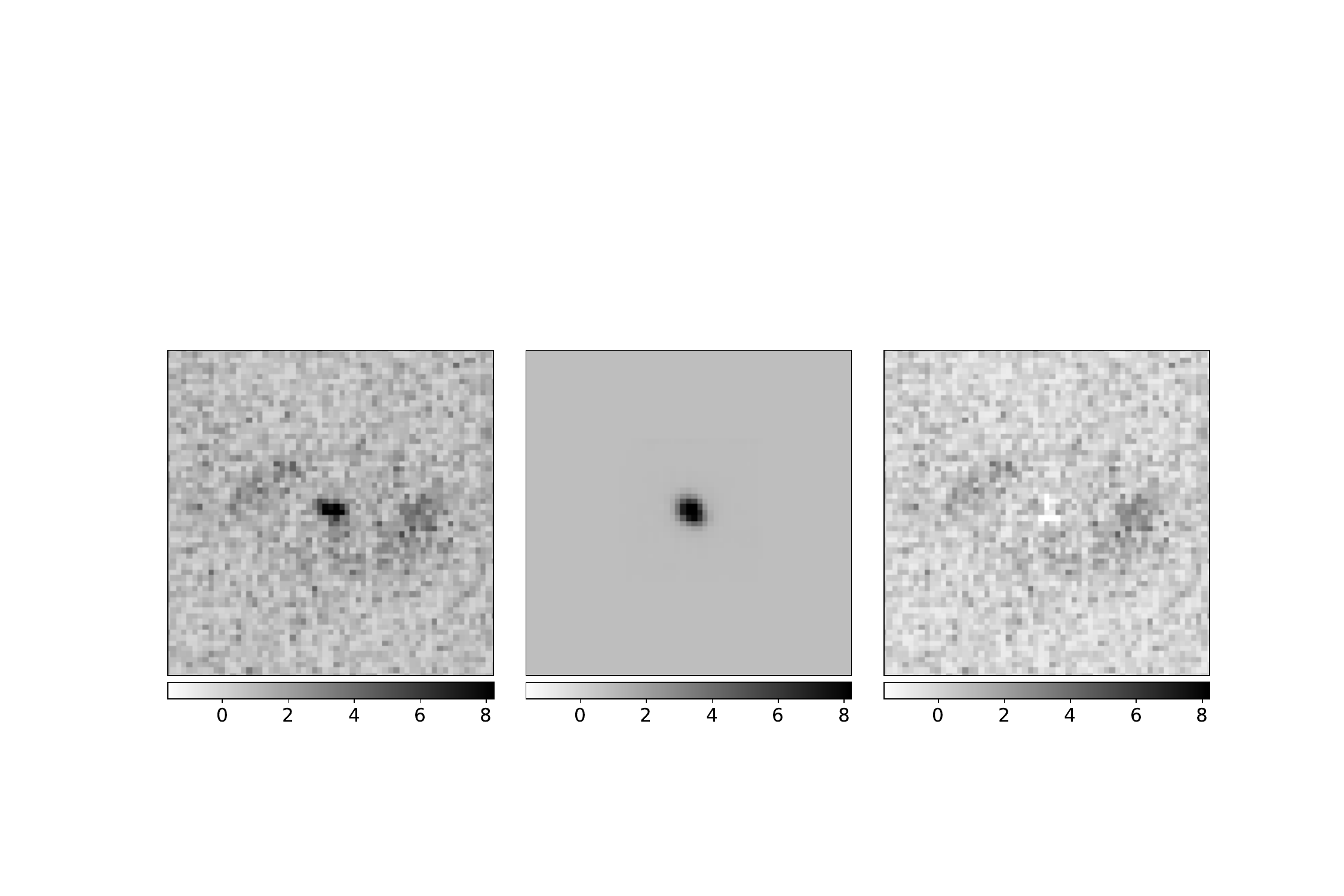}
    \label{fig:galfit_eg16}
    \vspace{-1.5cm}
    \caption{}
  \end{subfigure}
  \hspace{-1cm}
  \begin{subfigure}[t]{0.52\textwidth}
    \centering
    \includegraphics[width=\textwidth]{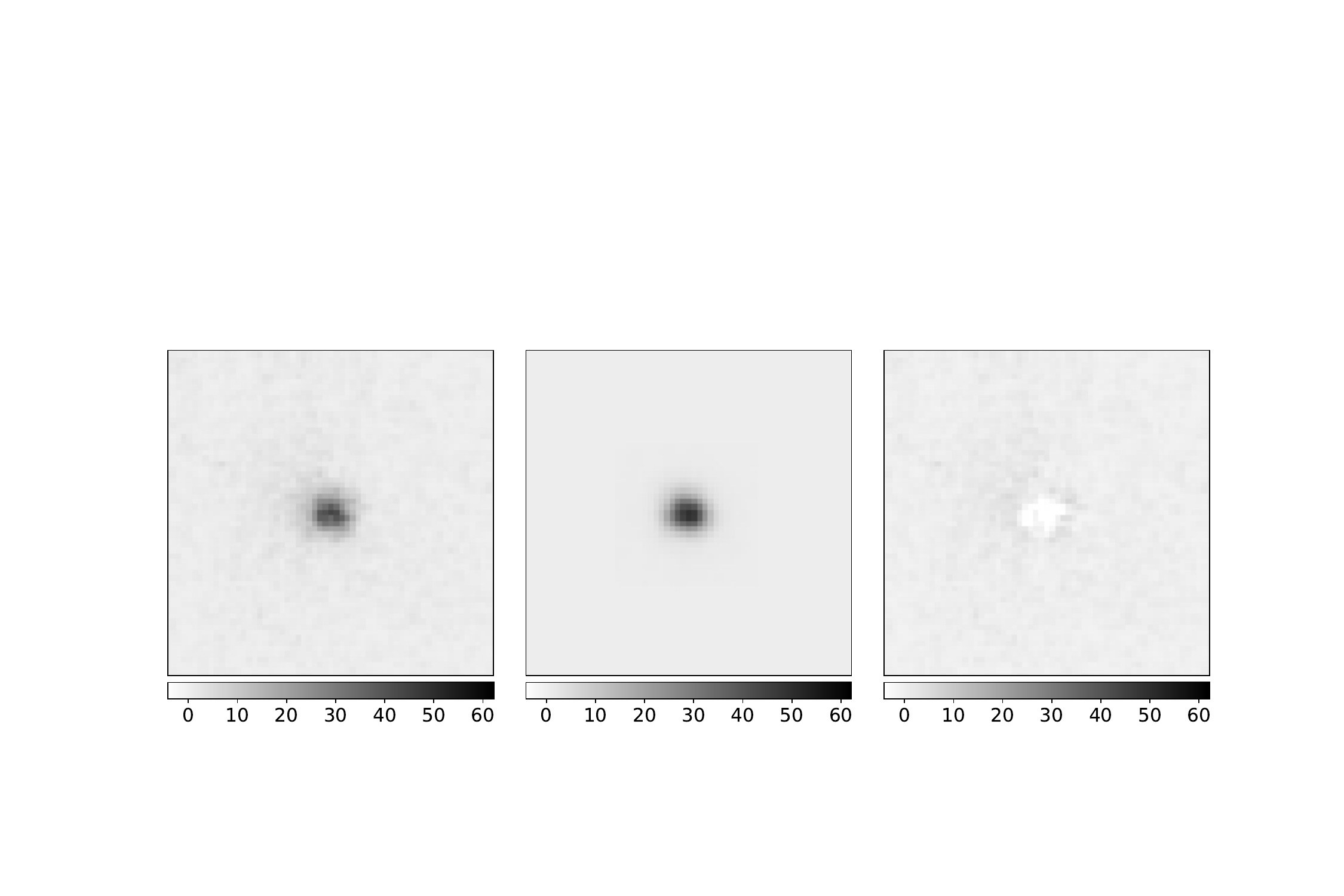} 
    \label{fig:galfit_eg26}
    \vspace{-1.5cm}
    \caption{}
  \end{subfigure}
\caption{Examples of source image fitting with \textsc{GALFIT} using a sky $+$ PSF $+$ S\'ersic profile (Sect. \ref{sect:galfit}): (a) SWIFTJ2226.8+3628 and (b) SWIFTJ0640.4-2554. In each panel, the left image is the source image, the middle one is the model, and the right one is the fit-residual. We show the fit in all six \textit{Swift}/UVOT filters from top to bottom in order (V, B, U, UVW1, UVM2, and UVW2). The colorbars represent the counts per pixel.}  
\label{fig:galfit_eg1}
\end{figure*}

\begin{figure*}
  \begin{subfigure}[t]{0.52\textwidth}
    \centering
    \includegraphics[width=\textwidth]{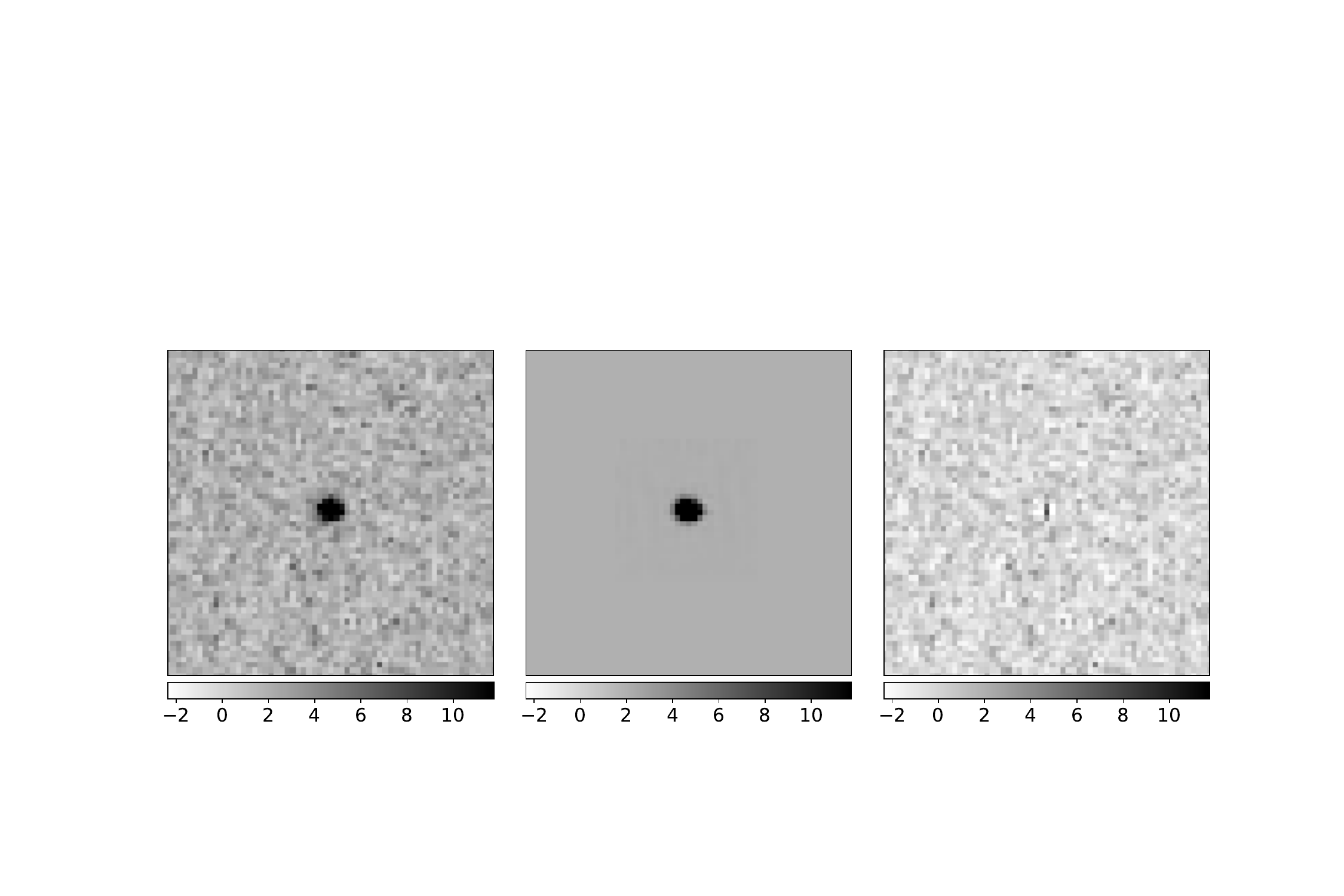} 
    \label{fig:galfit_eg31}
  \end{subfigure}
  \hspace{-1cm}
  \begin{subfigure}[t]{0.52\textwidth}
    \centering
    \includegraphics[width=\textwidth]{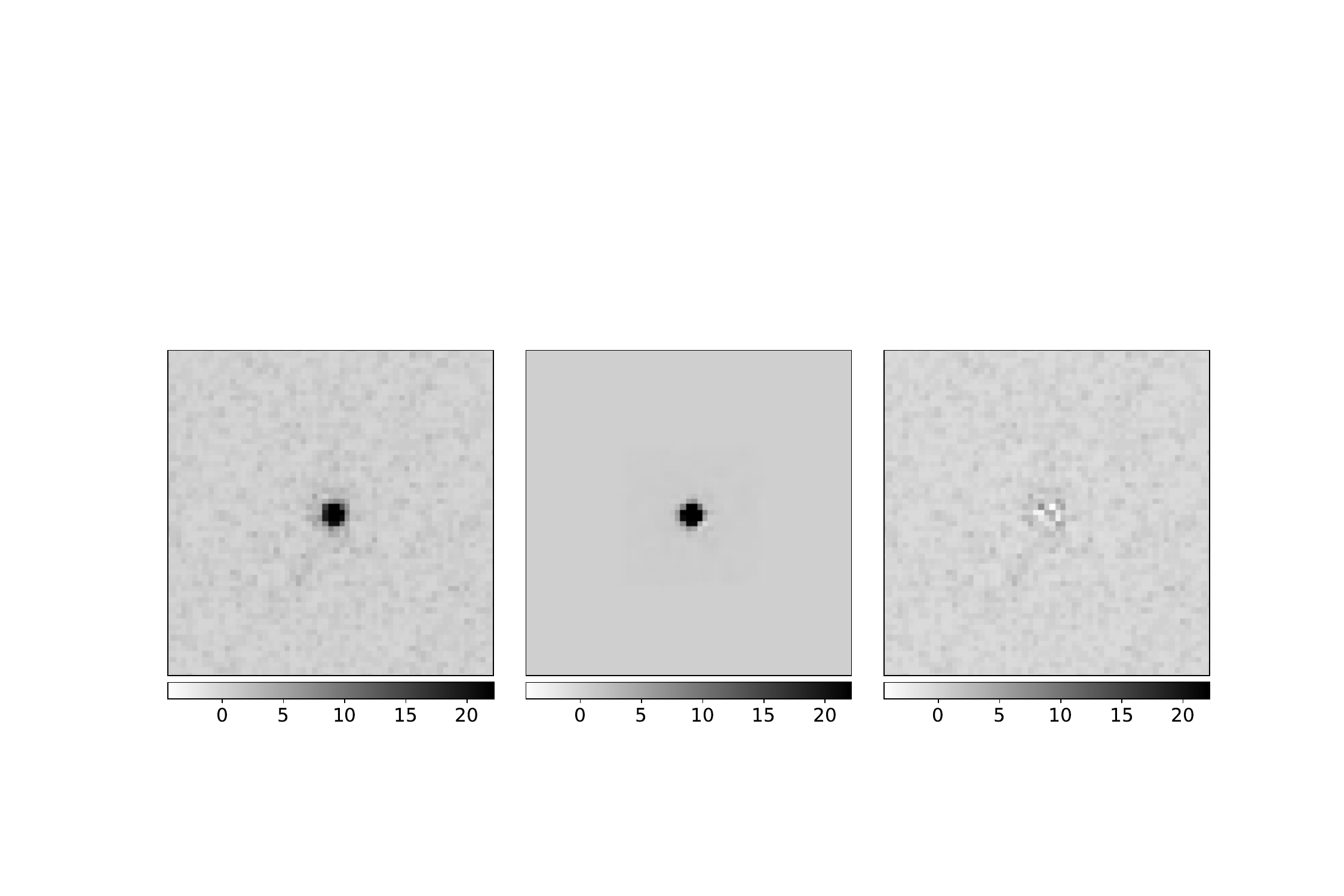} 
    \label{fig:galfit_eg41}
  \end{subfigure}
  
  \vspace{-3.5cm}
  
  \begin{subfigure}[t]{0.52\textwidth}
    \centering
    \includegraphics[width=\textwidth]{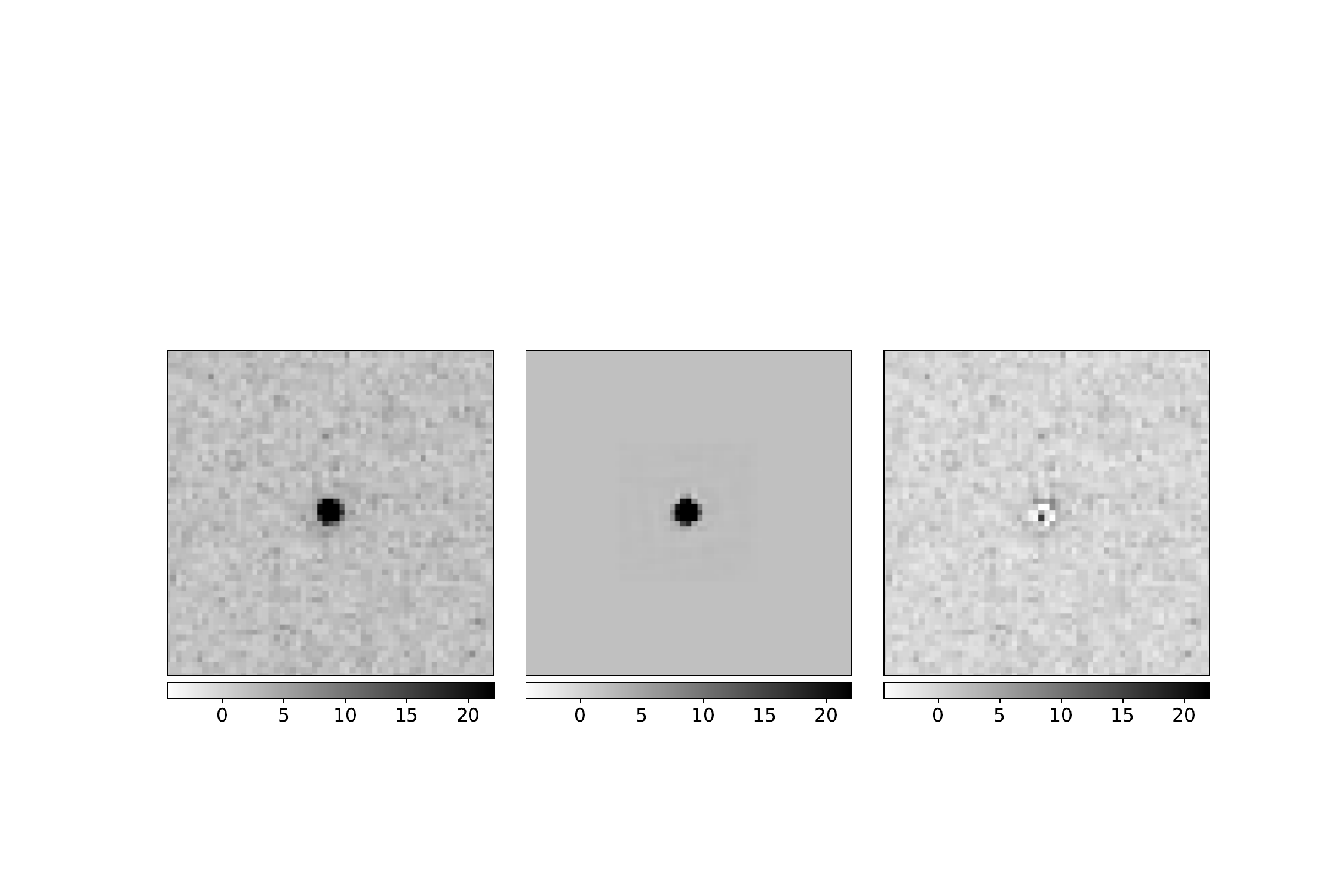}
    \label{fig:galfit_eg32}
  \end{subfigure}
  \hspace{-1cm}
  \begin{subfigure}[t]{0.52\textwidth}
    \centering
    \includegraphics[width=\textwidth]{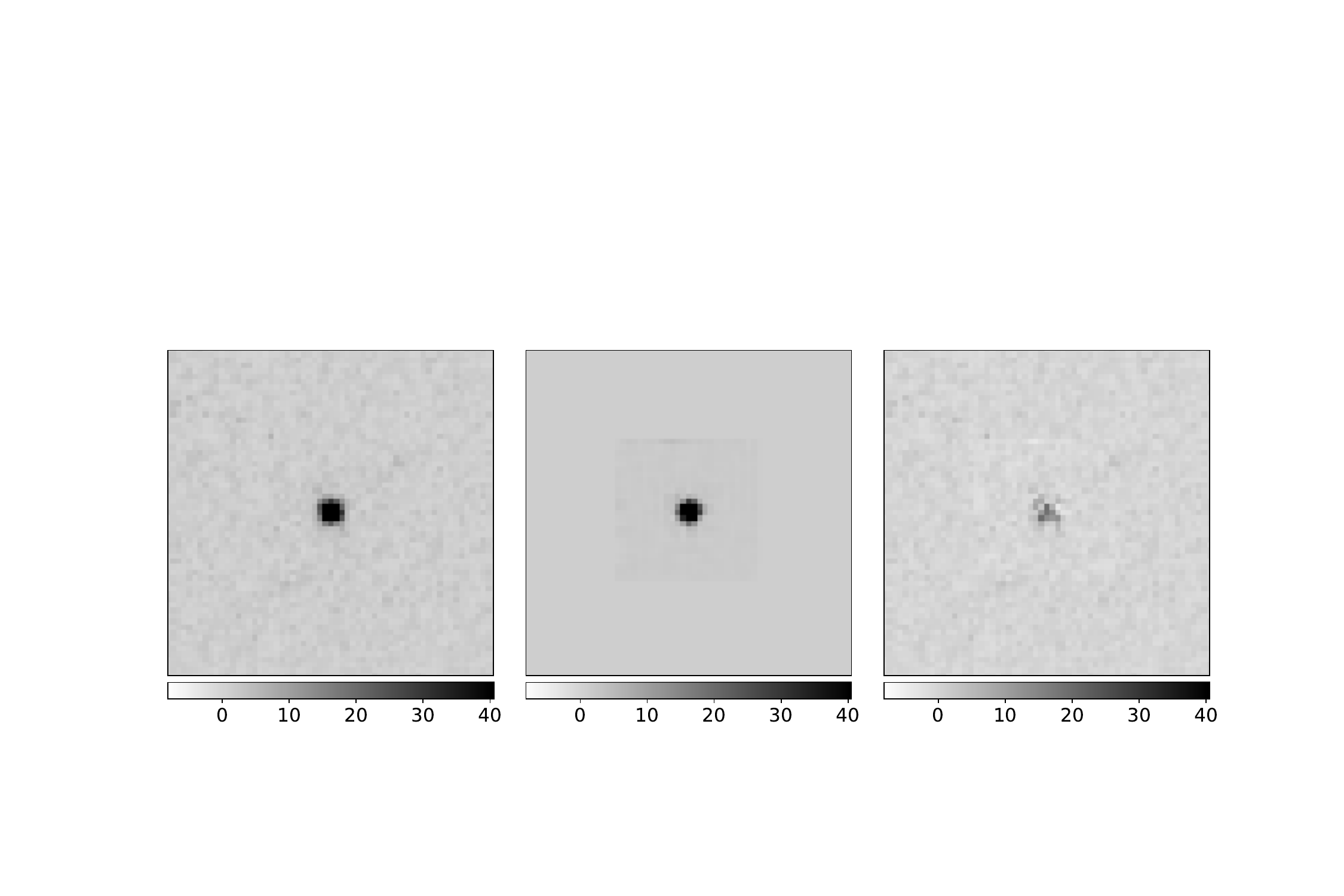} 
    \label{fig:galfit_eg42}
  \end{subfigure}
  
  \vspace{-3.5cm}
  
  \begin{subfigure}[t]{0.52\textwidth}
    \centering
    \includegraphics[width=\textwidth]{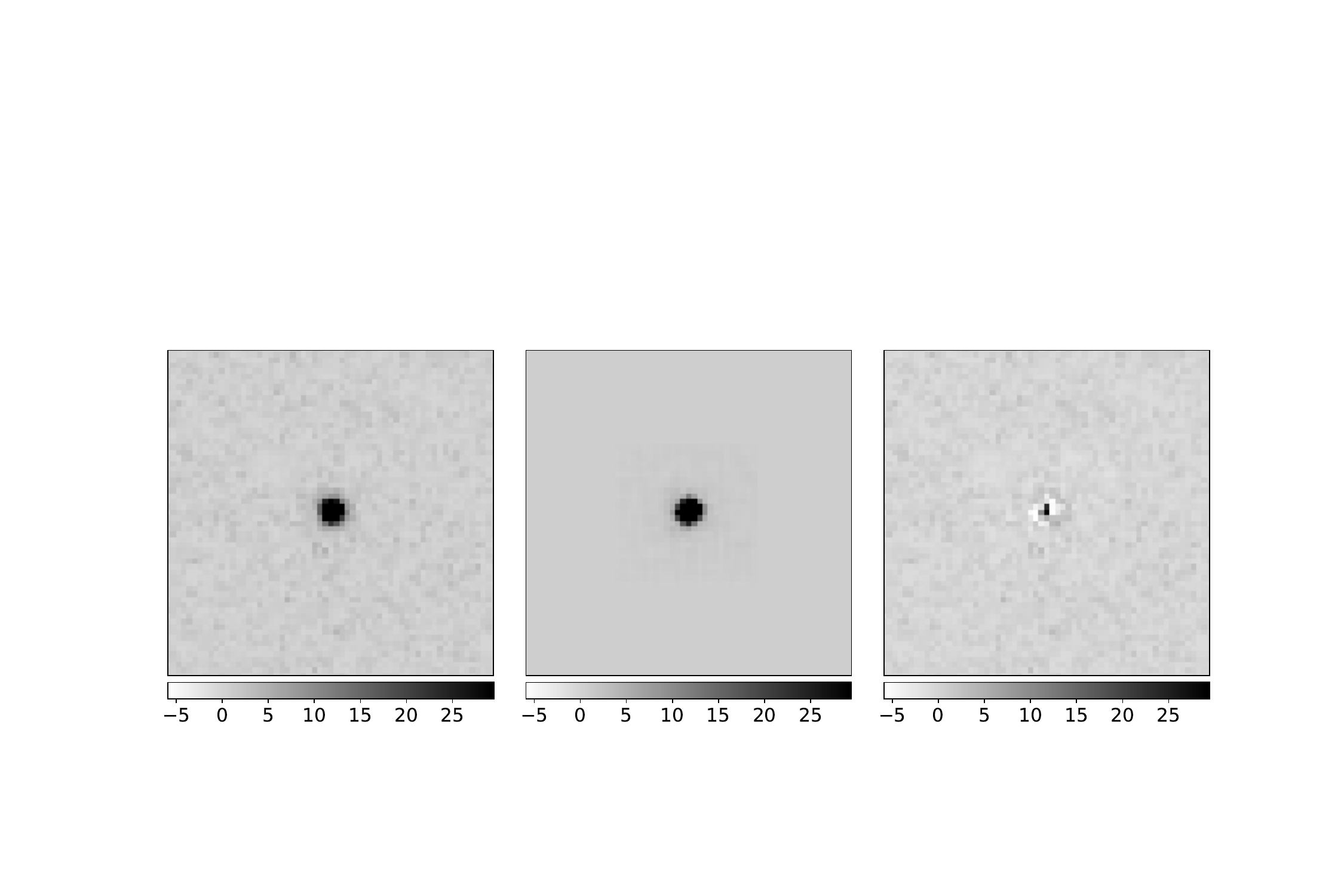}
    \label{fig:galfit_eg33}
  \end{subfigure}
 \hspace{-1cm}
  \begin{subfigure}[t]{0.52\textwidth}
    \centering
    \includegraphics[width=\textwidth]{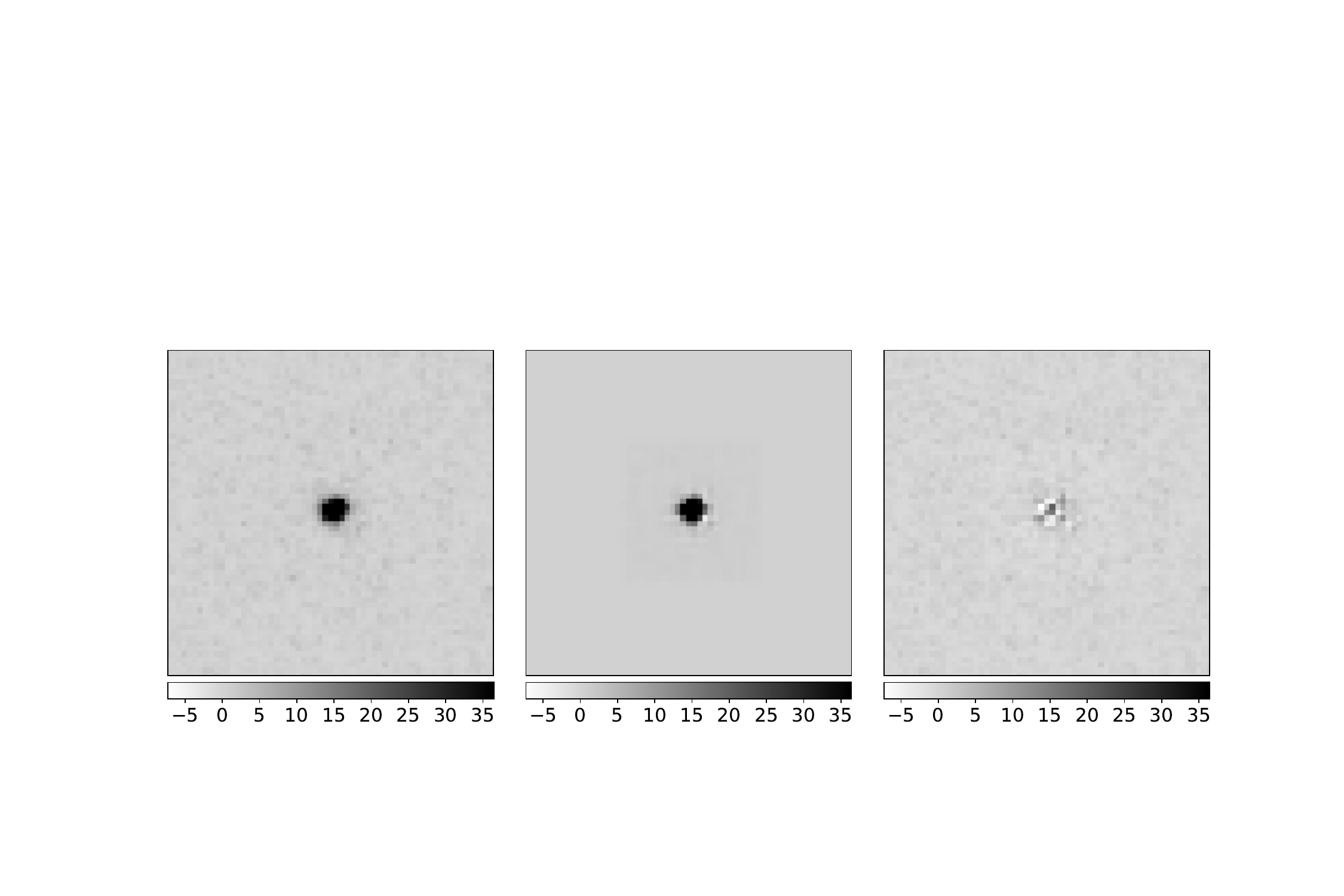} 
    \label{fig:galfit_eg43}
  \end{subfigure}
  
  \vspace{-3.5cm}
  
  \begin{subfigure}[t]{0.52\textwidth}
    \centering
    \includegraphics[width=\textwidth]{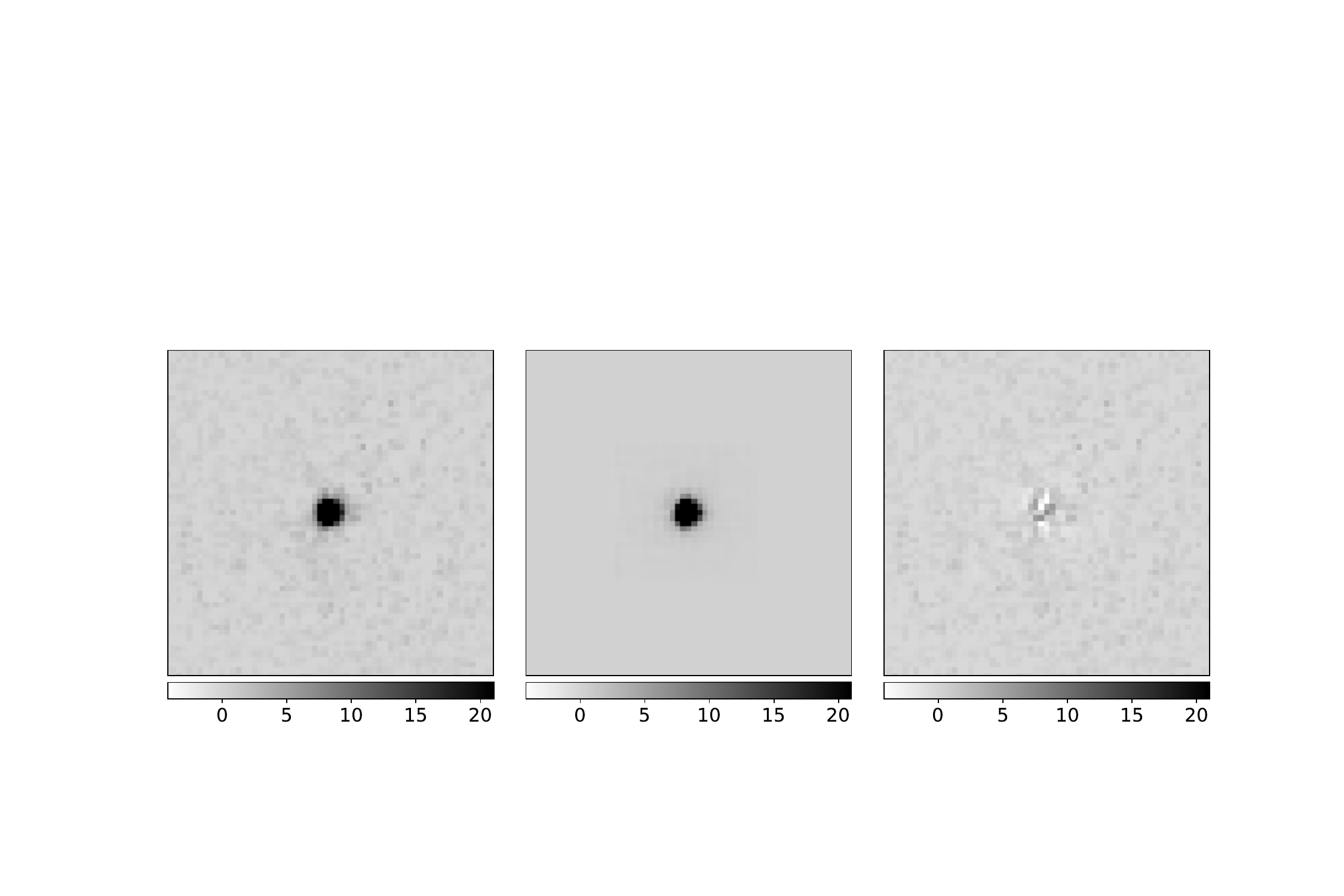}
    \label{fig:galfit_eg34}
  \end{subfigure}
  \hspace{-1cm}
  \begin{subfigure}[t]{0.52\textwidth}
    \centering
    \includegraphics[width=\textwidth]{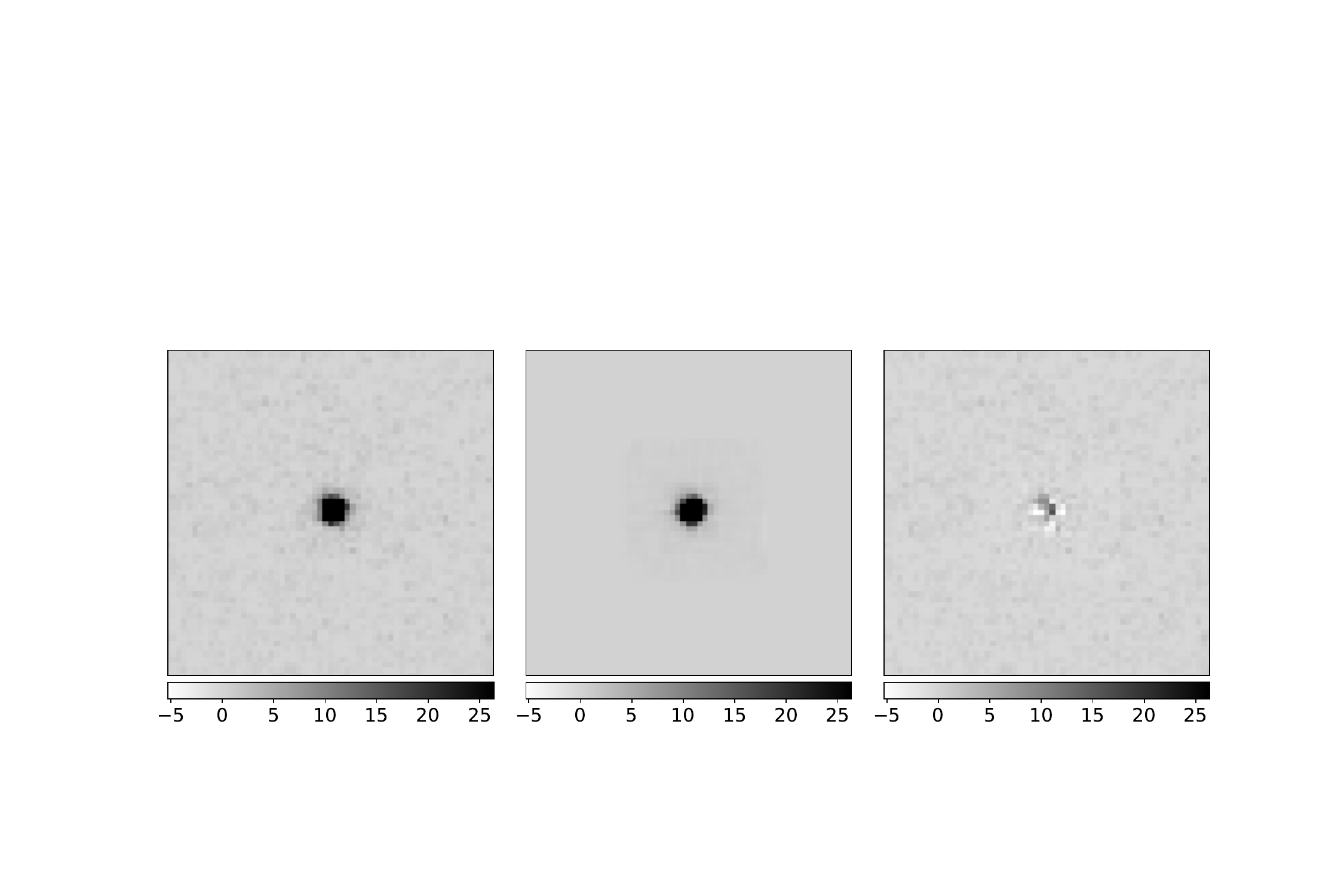} 
    \label{fig:galfit_eg44}
  \end{subfigure}
  
  \vspace{-3.5cm}
  
  \begin{subfigure}[t]{0.52\textwidth}
    \centering
    \includegraphics[width=\textwidth]{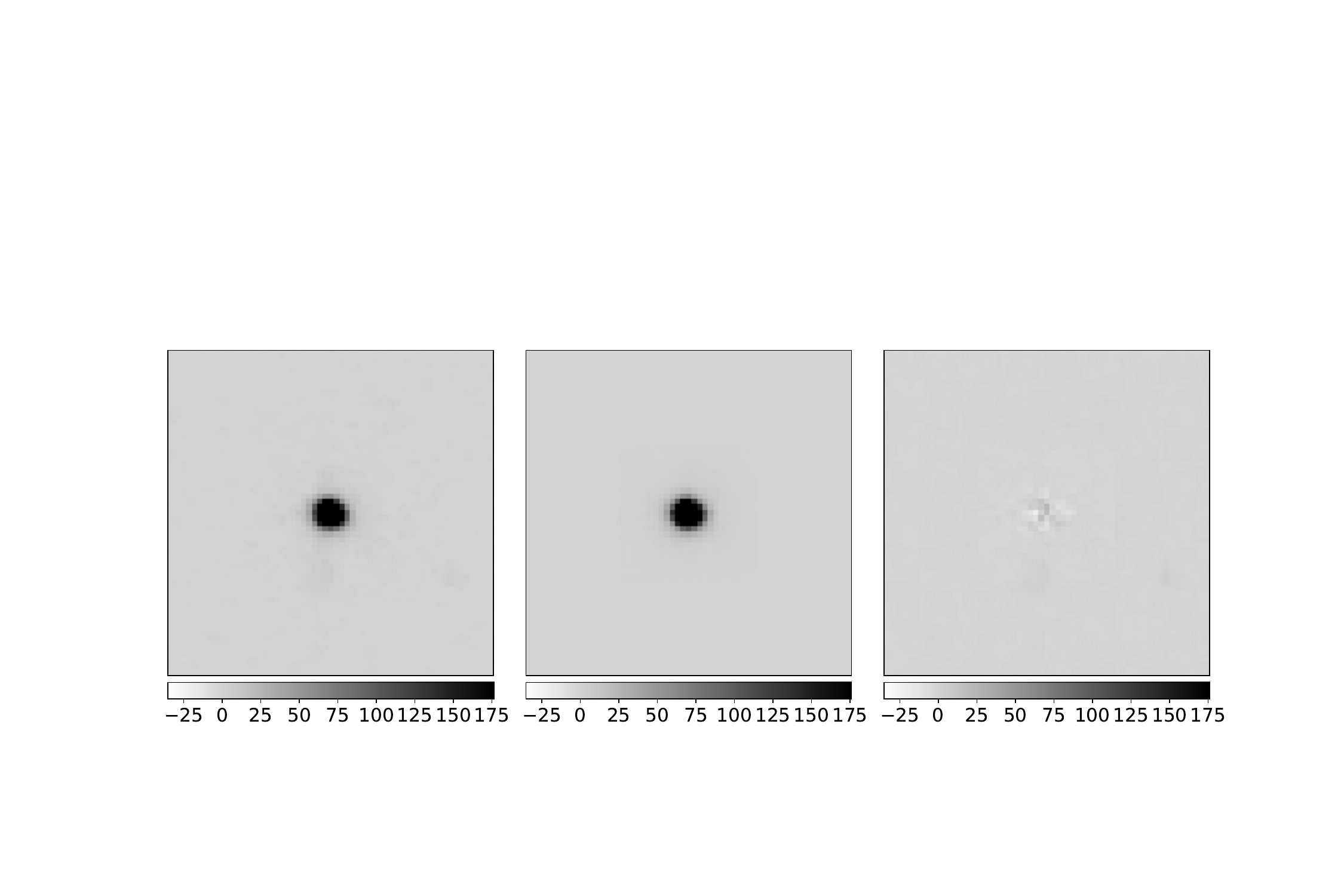}
    \label{fig:galfit_eg35}
  \end{subfigure}
  \hspace{-1cm}
  \begin{subfigure}[t]{0.52\textwidth}
    \centering
    \includegraphics[width=\textwidth]{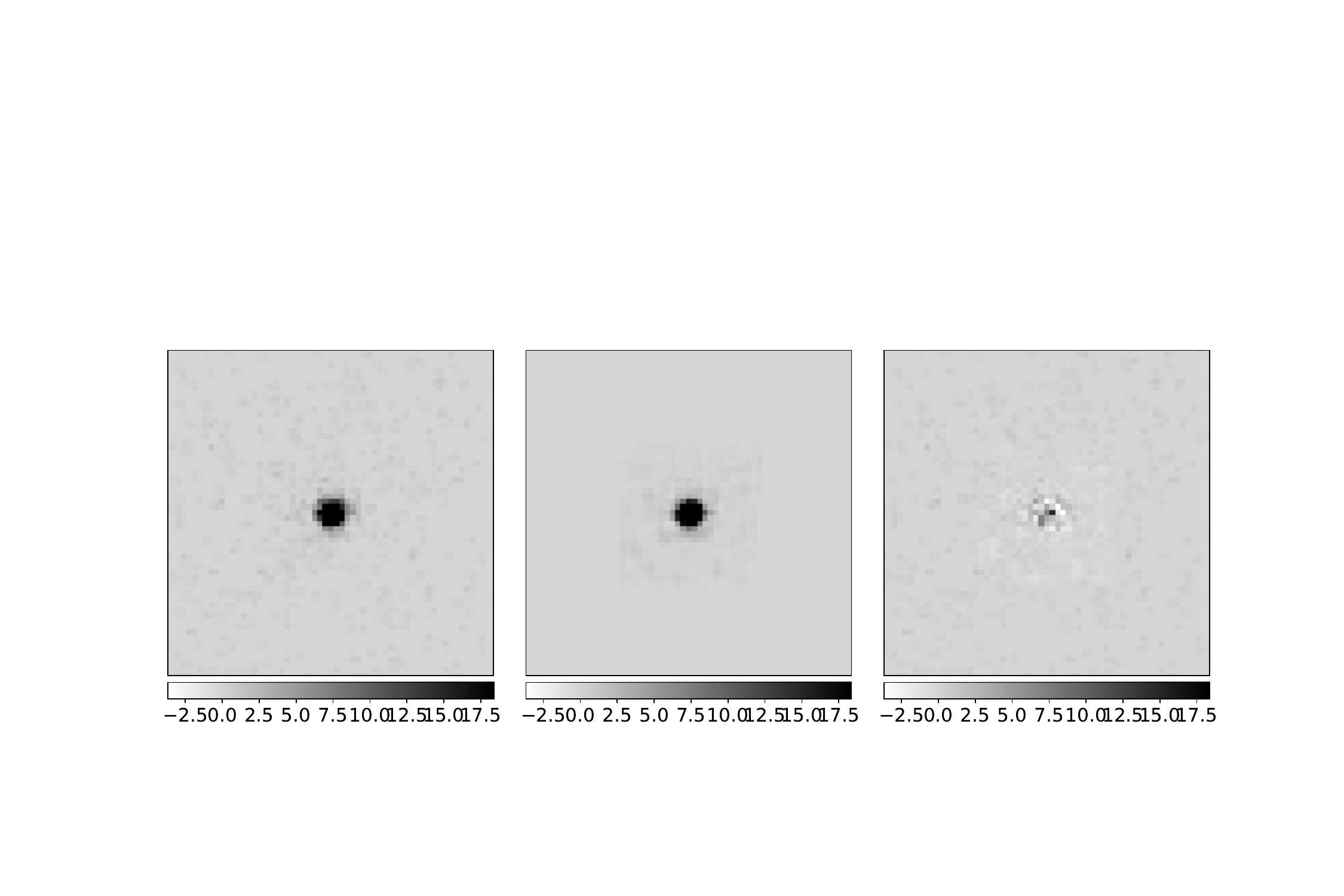} 
    \label{fig:galfit_eg45}
  \end{subfigure}
  
  \vspace{-3.5cm}
  
  \begin{subfigure}[t]{0.52\textwidth}
    \centering
    \includegraphics[width=\textwidth]{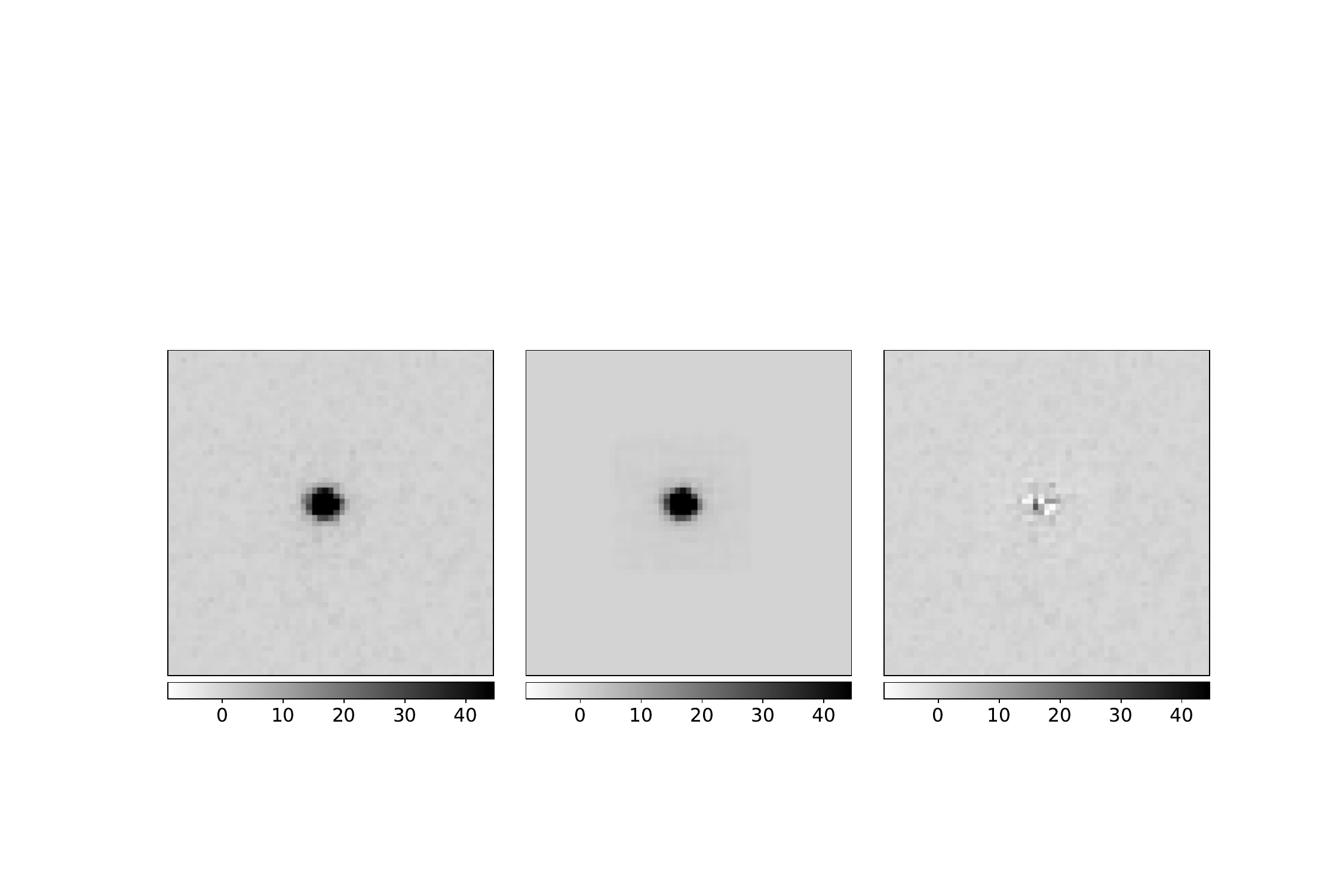}
    \label{fig:galfit_eg36}
    \vspace{-1.5cm}
    \caption{}
  \end{subfigure}
  \hspace{-1cm}
  \begin{subfigure}[t]{0.52\textwidth}
    \centering
    \includegraphics[width=\textwidth]{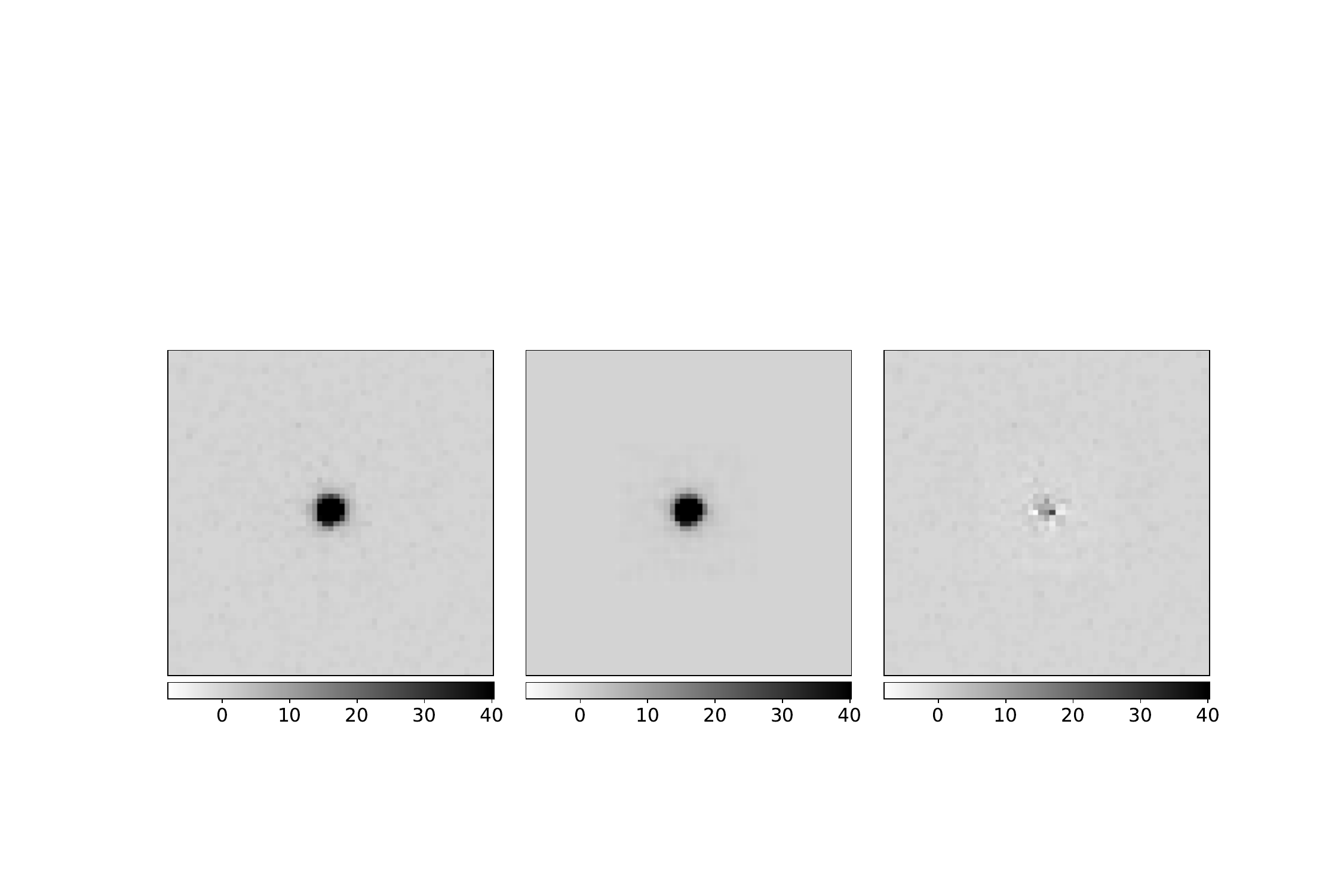} 
    \label{fig:galfit_eg46}
    \vspace{-1.5cm}
    \caption{}
  \end{subfigure}
\caption{Examples of source image fitting with \textsc{GALFIT} using a sky $+$ PSF profile for high redshift sources: (a) SWIFTJ1723.2+3418 and (b) SWIFTJ1421.4+4747. We corrected the GALFIT-estimated fluxes for such sources (Sect. \ref{sect:galfit_corr}) as \textsc{GALFIT} was unable to fit a S\'ersic profile to these sources and hence, missed some AGN flux. In each panel, the left image is the source, the middle one is the model, and the right one is the fit-residual. We show the fit in all six \textit{Swift}/UVOT filters from top to bottom in order (V, B, U, UVW1, UVM2, and UVW2). The colorbars represent the counts per pixel.}  
\label{fig:galfit_eg2}
\end{figure*}

We visually inspected the residuals of all fits. We show a few examples of the fitting with GALFIT and the obtained residuals in Figs. \ref{fig:galfit_eg1} and \ref{fig:galfit_eg2}. As expected, considering the wide variety of light profiles in our source sample in terms of spatial extension and brightness, there were instances when the fit residuals were not good or some parameters like the S\'ersic index or the effective radius were not constrained. There were cases when either the residuals still contained significant source emission that \textsc{GALFIT} was not able to model or the three component fit resulted in over-fitting of the source. For example, in the case of the farthest sources ($z > 0.1$) with very faint host galaxy profiles compared to the central AGN (predominately point source-like), the fit residuals with \textsc{GALFIT} were negative at the central pixels, showing over-subtraction from the fitting. In such cases ($\sim 14\%$), we adopted the magnitudes estimated using the \texttt{uvotsource} command of the \textit{Swift}/UVOT data reduction pipeline for an aperture of 5$"$ around the source, since within the defined region the flux was dominated by the central AGN. For the remaining sources, with a non-negligible contribution from the host galaxy, we wanted to be consistent in our image decomposition and quantify \textsc{GALFIT}'s response to \textit{Swift}/UVOT images. Hence, we carried out numerous simulations to check the reliability of PSF magnitudes estimated by \textsc{GALFIT}. As is described in detail in Appendix \ref{sect:appendixa}, we simulated a fake population of type\,I AGN by placing a point-source (star) with varying relative magnitudes at the center of a type\,II AGN (similar to the procedure followed by \citealp{2011ApJ...739...57K}). We then fit these fake type\,Is with \textsc{GALFIT} and compared the PSF magnitudes computed by \textsc{GALFIT} with the initial magnitude of the star. We did this for a total of ten sources spanning the redshift range of our sample and also including different galaxy morphologies, and with ten different magnitudes for the central star to achieve ten different AGN light to host light ratios (5\% to 95\%). Based on these simulations, we concluded that for sources at redshifts $> 0.05$ and with fluxes $< 10^{-15}\,{\rm erg\,\,cm^{-2}\,s^{-1}\,\AA^{-1}}$, the AGN magnitudes reported by \textsc{GALFIT} were brighter than the actual values (see Figs. \ref{fig:galfit_sim_W1}, \ref{fig:galfit_sim_M2}, and \ref{fig:galfit_sim_W2}). This was primarily attributed to the fact that \textsc{GALFIT} was not able to fit a S\'ersic profile for these sources and as a result, the PSF magnitudes and fluxes estimated by \textsc{GALFIT} also included contributions from the host galaxy. From these simulations, we calculated the necessary corrections needed to obtain improved estimates of PSF magnitudes for these sources. In Sect. \ref{sect:galfit_corr}, we describe in detail how we estimated and employed these corrections in this work.

We also did some tests to check how the \textsc{GALFIT}-estimated AGN magnitudes for our \textit{Swift}/UVOT images compare with those obtained by fitting the \textit{Hubble} Space Telescope (HST) images for the same sources. In Appendix \ref{sect:appendixb} we describe these tests in detail. Our main conclusion from these tests was that, even if the HST images show detailed and extended host galaxy profiles, if we just fit a small region around the nucleus with \textsc{GALFIT}, the estimated magnitudes are within 0.1 mag of those calculated using the lower resolution \textit{Swift}/UVOT images. This is acceptable, considering the difference in the central wavelengths of the overlapping HST and UVOT filters.


\subsection{Corrections to GALFIT-estimated PSF magnitudes}\label{sect:galfit_corr}

Based on the simulations (described in Appendix \ref{sect:appendixa}) performed to understand how \textsc{GALFIT} decomposes the \textit{Swift}/UVOT images and how reliable the estimated fluxes are, we reached some important conclusions. Firstly, for sources with very high counts ($>1000$ counts/pixel), \textsc{GALFIT} was unable to fit all the flux in the central pixels. This was primarily due to a poorly defined PSF. For these sources, the stars in the field of view that were used to create the PSF always had lower counts than the AGN itself. Therefore, the PSF was not suitable to fit such bright sources. However, the fit residuals in these cases were mostly $<$ 20\% of the source image, and therefore, the estimated PSF magnitudes were consistent with the expected AGN magnitudes within their error range. Secondly, in the case of more distant sources ($z > 0.05$) with lower fluxes ($F < 10^{-15}\,{\rm erg\,\,cm^{-2}\,s^{-1}\,\AA^{-1}}$), most of the time \textsc{GALFIT} was unable to converge while fitting a sky $+$ PSF $+$ S\'ersic profile to the source image. As a result, the estimated PSF fluxes were higher than expected. Even in cases when \textsc{GALFIT} successfully fit a S\'ersic profile to the source, the estimated PSF fluxes were higher than the expected values. We flagged such sources from our sample, to eventually be corrected before the final broadband SED fitting.

\begin{figure*}
  \begin{subfigure}[t]{0.54\textwidth}
    \centering
    \includegraphics[width=\textwidth]{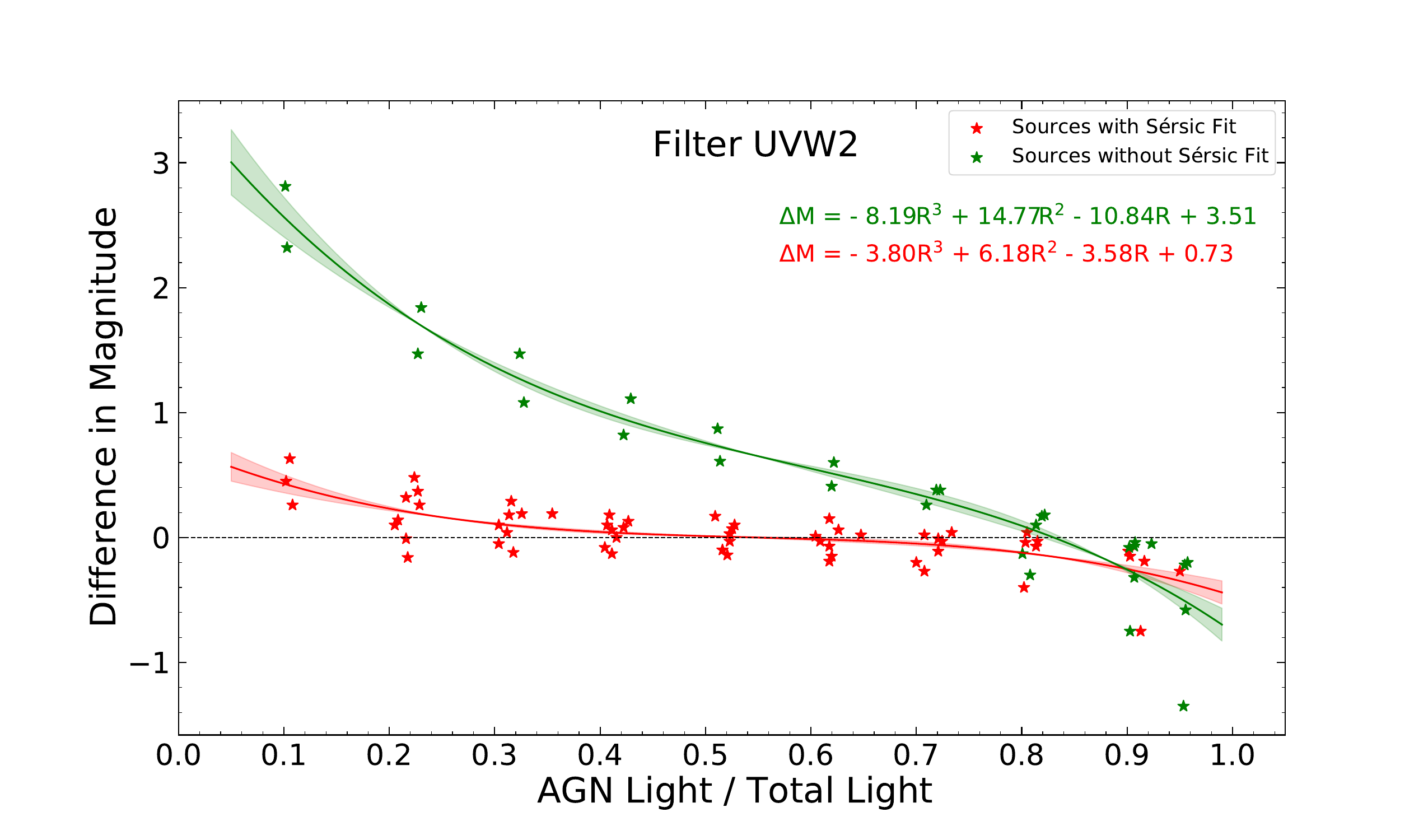}
    \vspace{-5mm}
    \caption{}
    \label{fig:galfit_corr_w2}
  \end{subfigure}
  \hspace{-1cm}
  \begin{subfigure}[t]{0.54\textwidth}
    \centering
    \includegraphics[width=\textwidth]{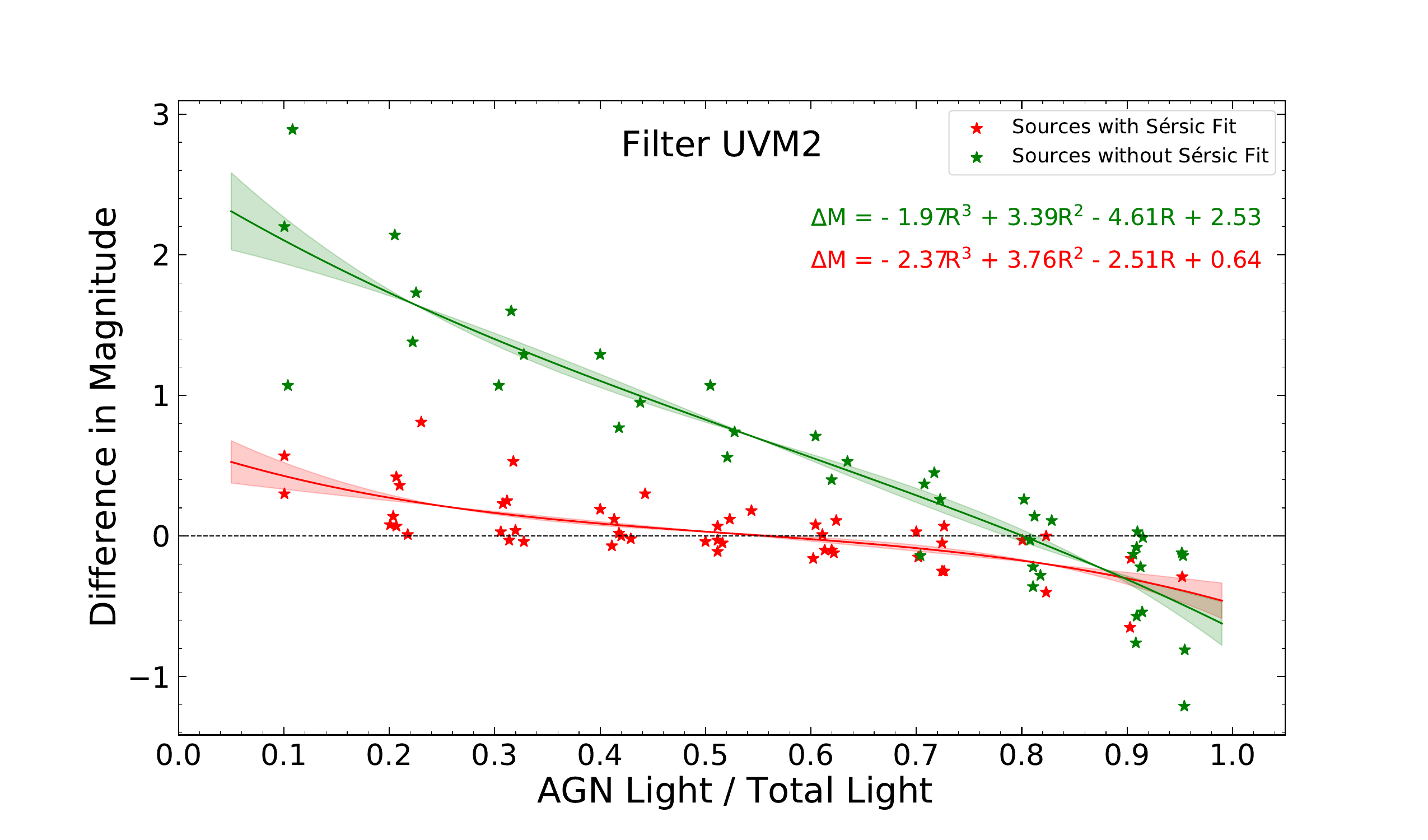}
    \vspace{-5mm}
    \caption{}
    \label{fig:galfit_corr_m2}
  \end{subfigure}

\vspace{-5mm}

  \begin{subfigure}[t]{0.54\textwidth}
    \centering
    \includegraphics[width=\textwidth]{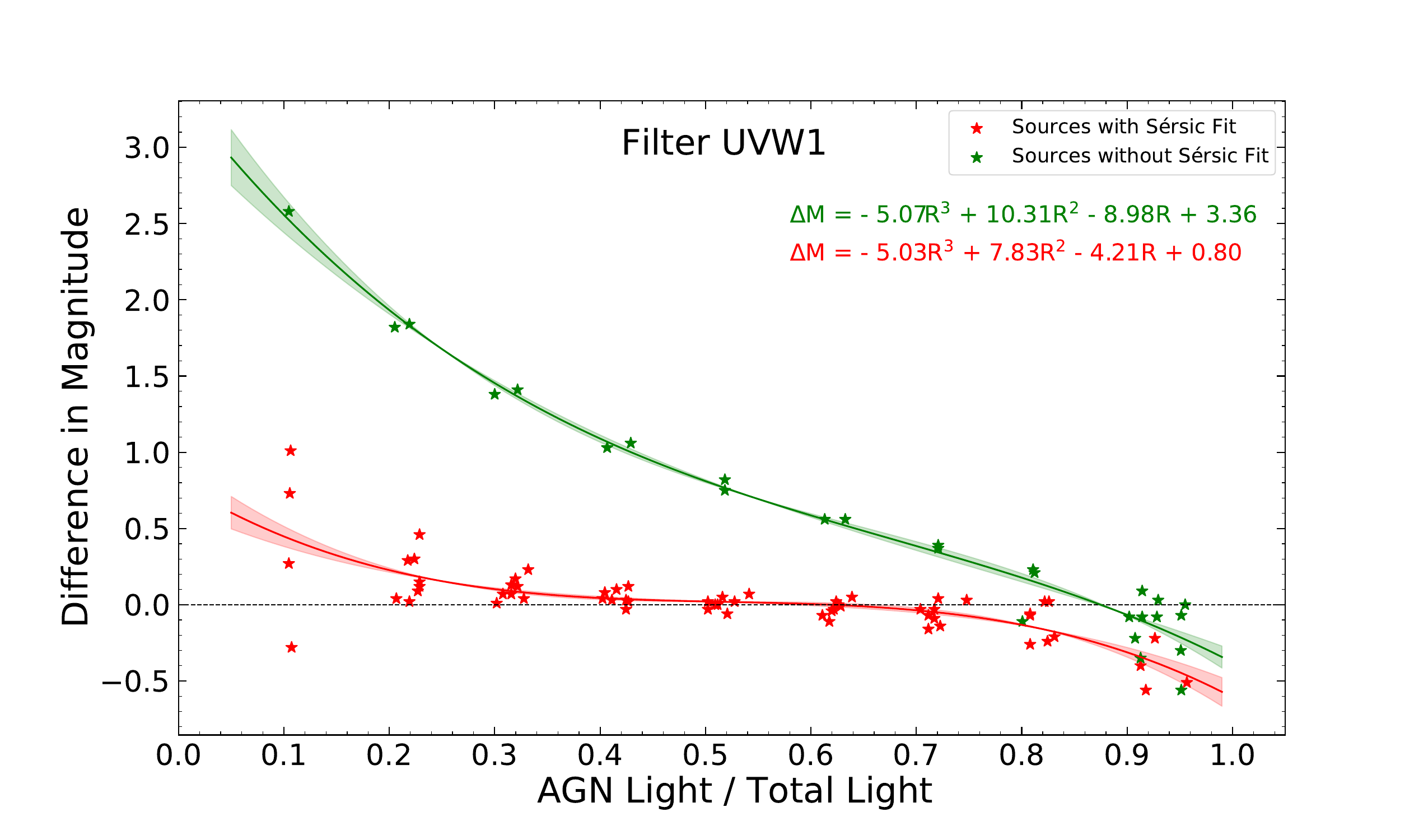}
    \vspace{-5mm}
    \caption{}
    \label{fig:galfit_corr_w1}
  \end{subfigure}
  \hspace{-1cm}
  \begin{subfigure}[t]{0.54\textwidth}
    \centering
    \includegraphics[width=\textwidth]{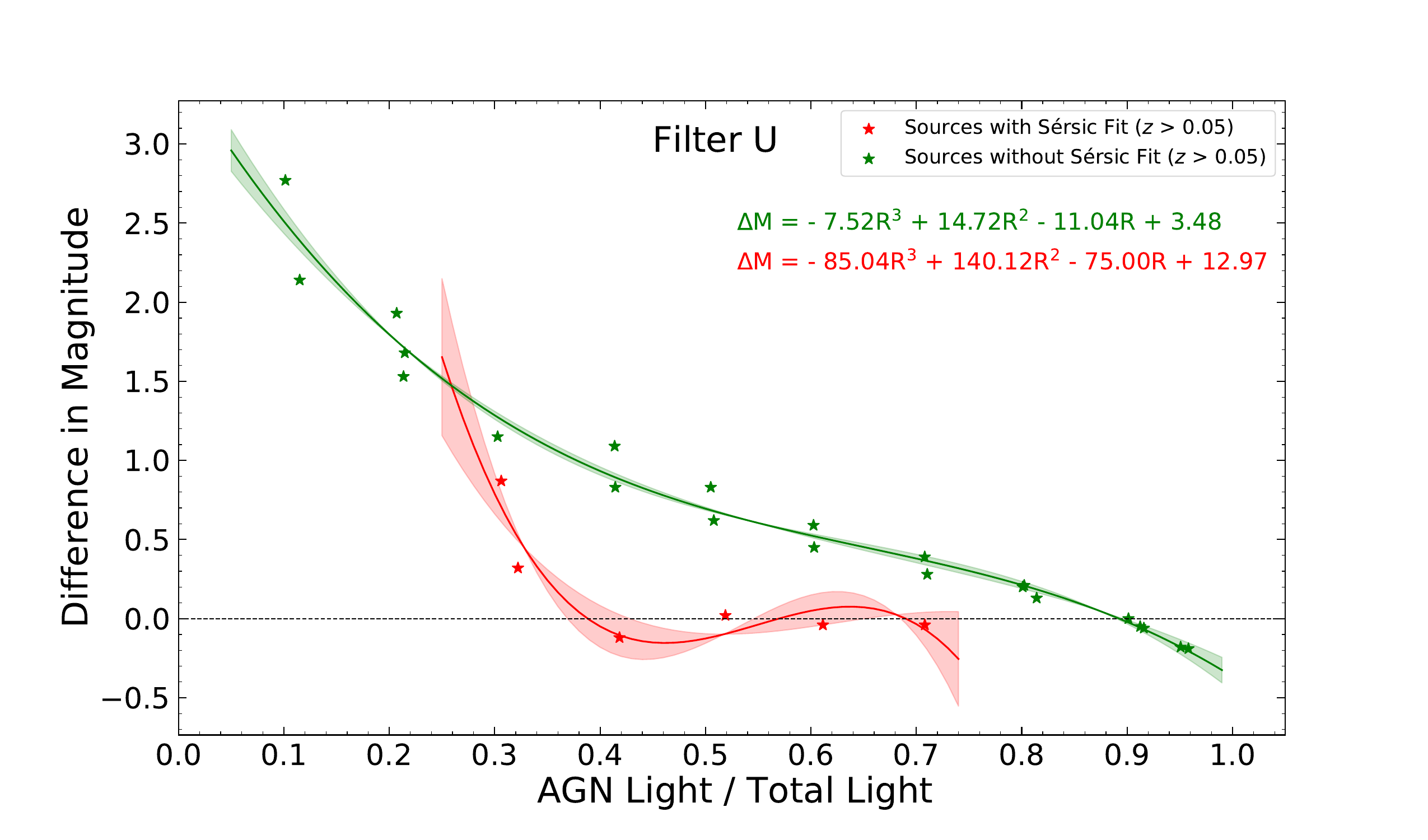}
    \vspace{-5mm}
    \caption{}
    \label{fig:galfit_corr_u}
  \end{subfigure}

\vspace{-5mm}

  \begin{subfigure}[t]{0.54\textwidth}
    \centering
    \includegraphics[width=\textwidth]{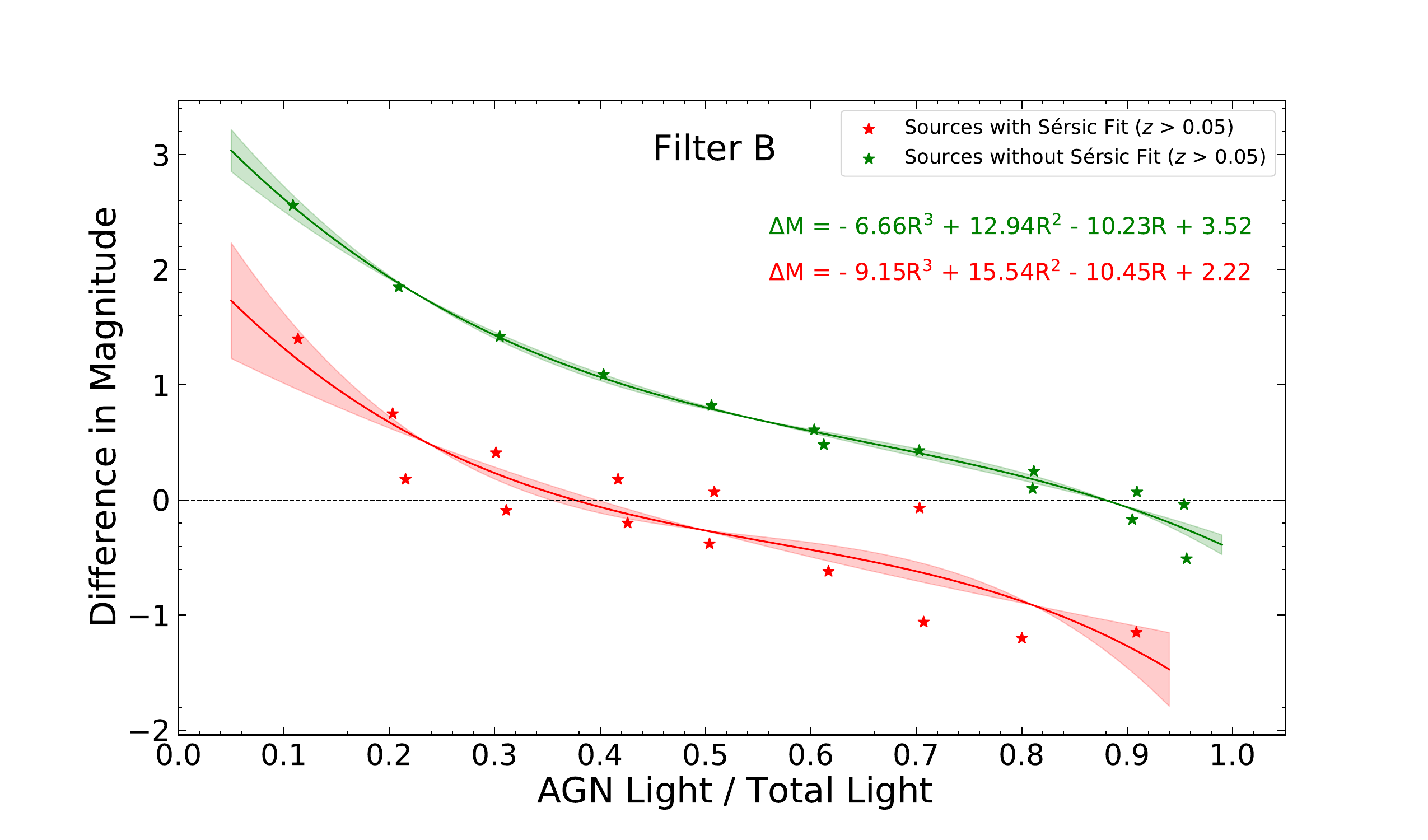}
    \vspace{-5mm}
    \caption{}
    \label{fig:galfit_corr_b}
  \end{subfigure}
  \hspace{-1cm}
  \begin{subfigure}[t]{0.54\textwidth}
    \centering
    \includegraphics[width=\textwidth]{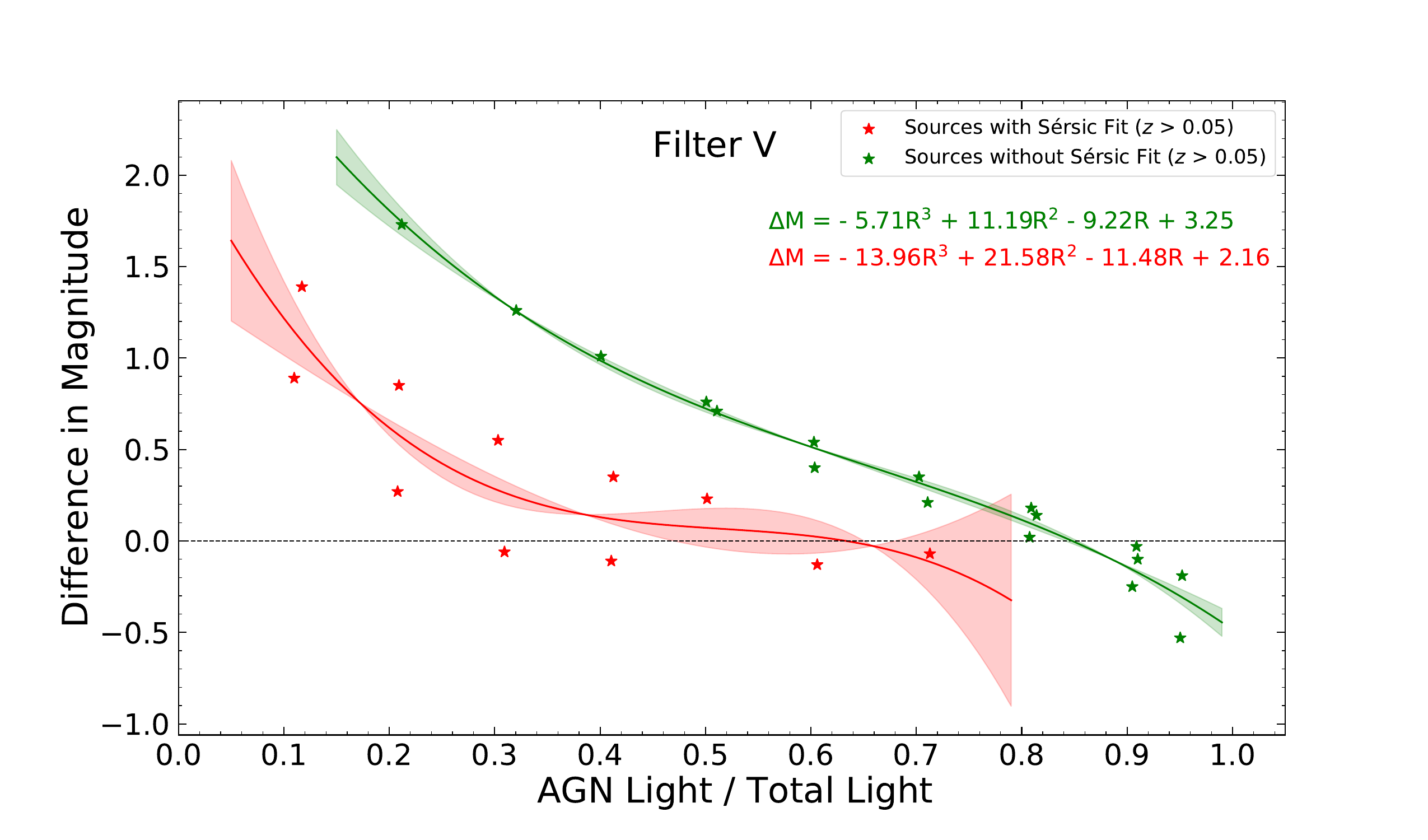}
    \vspace{-5mm}
    \caption{}
    \label{fig:galfit_corr_v}
  \end{subfigure}
\caption{Difference between the PSF magnitudes estimated by GALFIT and the actual magnitude of the point source used to replicate the AGN at the center of the simulated type\,I AGN population ($\Delta M$) vs. the AGN light ratio ($R_{\rm light}$) for all six \textit{Swift}/UVOT filters, (a) UVW2, (b) UVM2, (c) UVW1, (d) U, (e) B, and (f) V, based on simulations (described in Appendix \ref{sect:appendixa}). The plots also show the best-fit function and their equations for both cases: when a S\'ersic profile fit was obtained (in red) and when a fit with a S\'ersic profile fit was not possible (in green). The equations mentioned in each figure were then used to calculate the corresponding corrections to improve the PSF magnitudes estimated by \textsc{GALFIT} during the source image decomposition to obtain the AGN light devoid of host galaxy contamination (see Sect. \ref{sect:galfit_corr}).}  
\label{fig:galfit_corr}
\end{figure*}

To calculate the appropriate corrections for different sources in the various optical and UV filters, we first plotted the difference in the PSF magnitude estimated by \textsc{GALFIT} and the actual magnitude of the point source used to replicate the AGN at the center of the simulated type\,I AGN population ($\Delta M$) with respect to the AGN light ratio [$R_{\rm light} = {\rm AGN\,\,Light}/({\rm AGN\,\,Light\,+\, Host\,\,Galaxy\,\,Light}$)]. We use this particular representation ($R_{\rm light}$ vs. $\Delta M$) because the simulations performed previously showed that the reliability of \textsc{GALFIT}'s PSF flux prediction depends on the brightness of the AGN relative to the host galaxy (represented by $R_{\rm light}$). As is shown in Fig. \ref{fig:galfit_corr}, it is evident that the cases in which \textsc{GALFIT} was able to fit a S\'ersic profile occupy a different parameter space compared to those where a S\'ersic profile fit was not achieved. Therefore, we used two separate functions to fit these two trends for all six filters. Compared to a second-degree polynomial, a cubic polynomial provided a better fit, as it was able to better cover the sources at the higher end of $R_{\rm light}$. Based on these fits, we derived equations to calculate the value of $\Delta M$ for every filter corresponding to the two scenarios, whether a S\'ersic fit was achieved or not. We would like to point out that a definite caveat of this analysis is the systematic uncertainties introduced in the final AGN magnitudes of the sources for which these corrections are applied. Therefore, one needs to be careful when using these corrections for individual source analysis. We emphasize that in order to minimize the effect of the systematic uncertainties in the corrected AGN fluxes (especially, in the optical bands), one should only use them for large sample studies, like this one.


In our final sample, for 46 sources ($\sim$ 20\%) we used the corrections defined above to improve the PSF magnitudes estimated by \textsc{GALFIT} image decomposition. For 33 sources ($\sim$ 14\%), we used the magnitudes estimated during the data reduction process instead of the \textsc{GALFIT} estimated PSF magnitudes due to over-subtracted residuals in at least one of the optical/UV filters. All these sources were visually confirmed to be point-like with negligible contribution from the host galaxy. The final AGN magnitudes, fluxes, and errors are listed in Tables \ref{tab:galfit_flux} and \ref{tab:galfit_mag}. The sources for which \textsc{GALFIT} magnitudes were corrected and the sources for which we used the UVOT pipeline magnitudes are flagged accordingly in the tables. As is mentioned in Sect. \ref{sect:galfiting}, we do not report the values for the S\'ersic profile (host galaxy) parameters, as the focus of this work is solely to constrain the AGN light to construct the corresponding AGN SEDs.


\section{Broadband SED fitting}\label{sect:sed}

In this final part of the analysis, we use the optical and UV magnitudes calculated for each source to populate the optical/UV part of their SED. We start by using the PSF magnitudes estimated by \textsc{GALFIT} after the source image decomposition. As an initial step, we do not include the corrections computed on these magnitudes for the sources at $z>0.05$, to understand how the results change before and after applying the corrections. The original source PHA\footnote{\textsc{XSPEC}-compatible fits format} files for each source in each filter were edited to include the AGN magnitudes instead of the total source magnitudes estimated during the data reduction. The background PHA files were edited to have a value = 0.0 since the final magnitudes estimated by \textsc{GALFIT} are already corrected for the background in the first step of the decomposition (see Sect. \ref{sect:galfit}). The response files from \textit{Swift}/UVOT were edited to include the appropriate energy ranges (see Table \ref{tab:uvot}) for respective filters.

The optical/UV part of the SED is fit in \textsc{XSPEC} (\citealp{1996ASPC..101...17A}) using the \texttt{DISKPN} model, which is a thermal accretion disk model comprising multiple blackbody components (e.g., \citealp{1984PASJ...36..741M}; \citealp{1986ApJ...308..635M}). While it can successfully describe the optical/UV disk emission in many AGN sources (e.g., \citealp{1982ApJ...254...22M}; \citealp{1983ApJ...268..582M}), the multi-temperature accretion disk model, which is otherwise assumed to be in a quasi-steady state equilibrium (e.g., \citealp{1973A&A....24..337S}; \citealp{1974ApJ...191..507T}), is inadequate to explain the observed variability timescales in the UV (e.g., \citealp{1985ApJ...288..205A}; \citealp{2001ApJ...555..775C}; \citealp{2012MNRAS.423..451L}). We also acknowledge that by using a common disk model for all sources, we are making an assumption regarding the type of accretion flow in these sources. Numerous studies have shown that accretion in an AGN changes with its accretion rate, from possible ADAFs to standard thin disks to slim disks, as the accretion rate increases (see \citealp{1999ASPC..161..295B} for a review). However, at this point, we cannot assume or define specifically how the surrounding material is accreted onto the central SMBH. Hence, we fit all the SEDs with the widely acceptable thermal disk model, so that we can later evaluate our results in a consistent way. We also ignore any contribution from emission lines in our disk model. Multiple studies on continuum reverberation mapping have shown that emission lines can have a small but non-negligible contribution to the disk continuum emission (e.g., \citealp{2013ApJ...772....9C}; \citealp{2023ApJ...953..137M}). This contribution increases from $< 10\%$ to $\sim 30\%$  as we move to photometric bands at higher wavelengths (r and i bands). However, in the context of this work, we can safely ignore this contribution ($\sim 10\%$) up till our highest wavelength V band. SED models that include emission lines (e.g., \citealp{2021MNRAS.508..737T}) can be useful to study their effects on broadband SED fitting analysis, but it is beyond the scope of this work. In addition to the disk model, we include two components of the \texttt{ZDUST} model to take into account extinction due to dust grains (\citealp{1992ApJ...395..130P}) in the optical and UV bands, associated with the Milky Way (\texttt{ZDUST}\textsubscript{\texttt{MW}}) and the host galaxy of the AGN (\texttt{ZDUST}\textsubscript{\texttt{HG}}), respectively. We use the convolution model \texttt{ZASHIFT} to shift the disk spectrum to the redshift of the respective source. Hence, the final model used to fit the optical/UV SED is:

\begin{center}
  \texttt{ZDUST}\textsubscript{\texttt{MW}} $\times$ \texttt{ZDUST}\textsubscript{\texttt{HG}} $\times$ \texttt{ZASHIFT} $\times$ \texttt{DISKPN}.  
\end{center}
For the two reddening components, we adopt a Milky Way (MW) extinction curve (method = 1) and $R_{\rm v} = 3.08$. In the case of dust extinction due to our galaxy, we fix the $E(B-V)$ value to those estimated by \cite{1998ApJ...500..525S}, who combined the results of IRAS and COBE/DIRBE to create a 100-micron intensity map of the sky. The $E(B-V)$ for the AGN host galaxy is allowed to vary during the fitting. The redshifts for the two components are fixed to 0.0 and the redshift of the source, respectively.

\begin{figure*} 
  \begin{subfigure}[t]{0.54\textwidth}
    \centering
    \includegraphics[width=\textwidth]{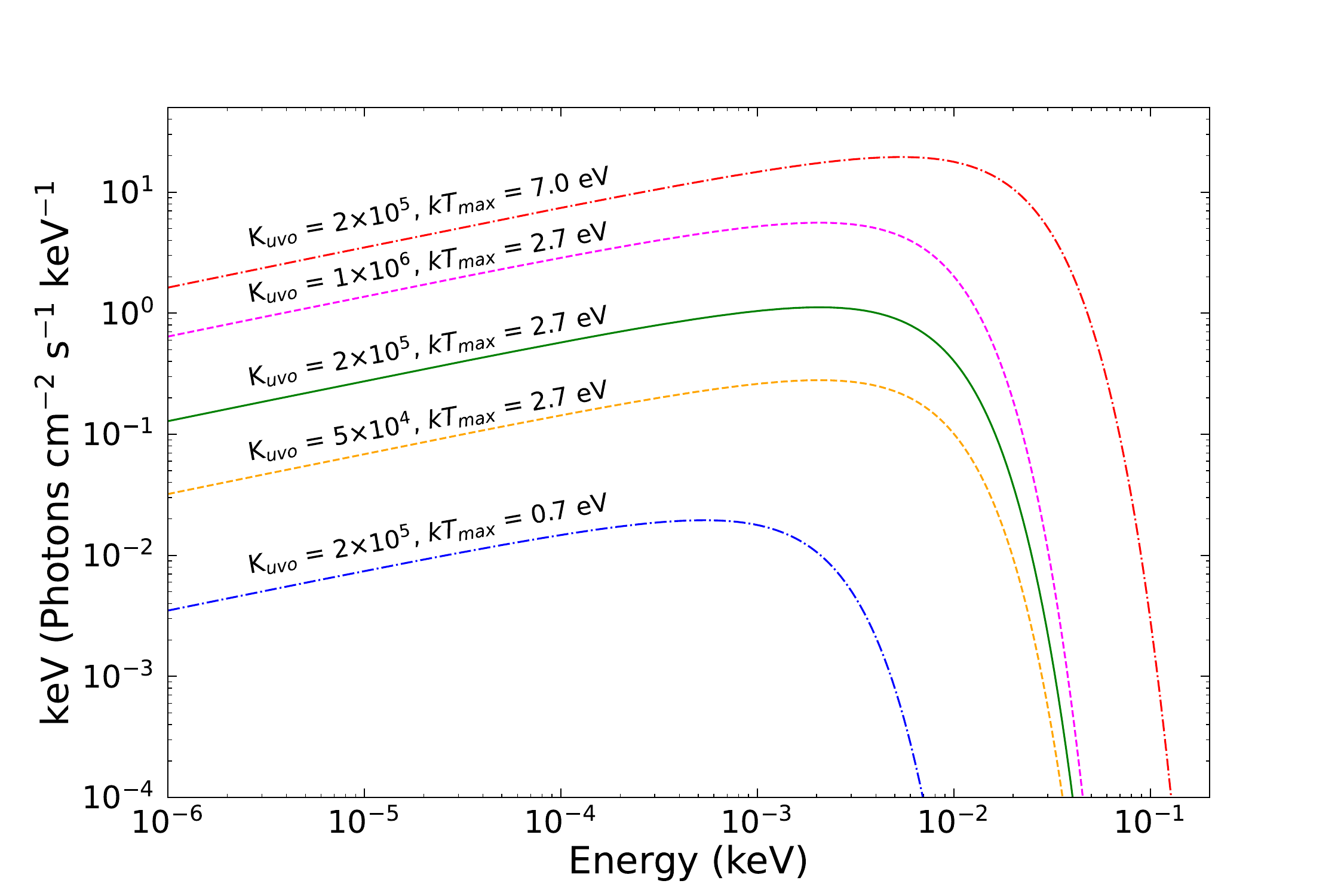}
    \caption{}
    \label{fig:diskpn}
  \end{subfigure}
  \hspace{-1.cm}
  \begin{subfigure}[t]{0.54\textwidth}
    \centering
    \includegraphics[width=\textwidth]{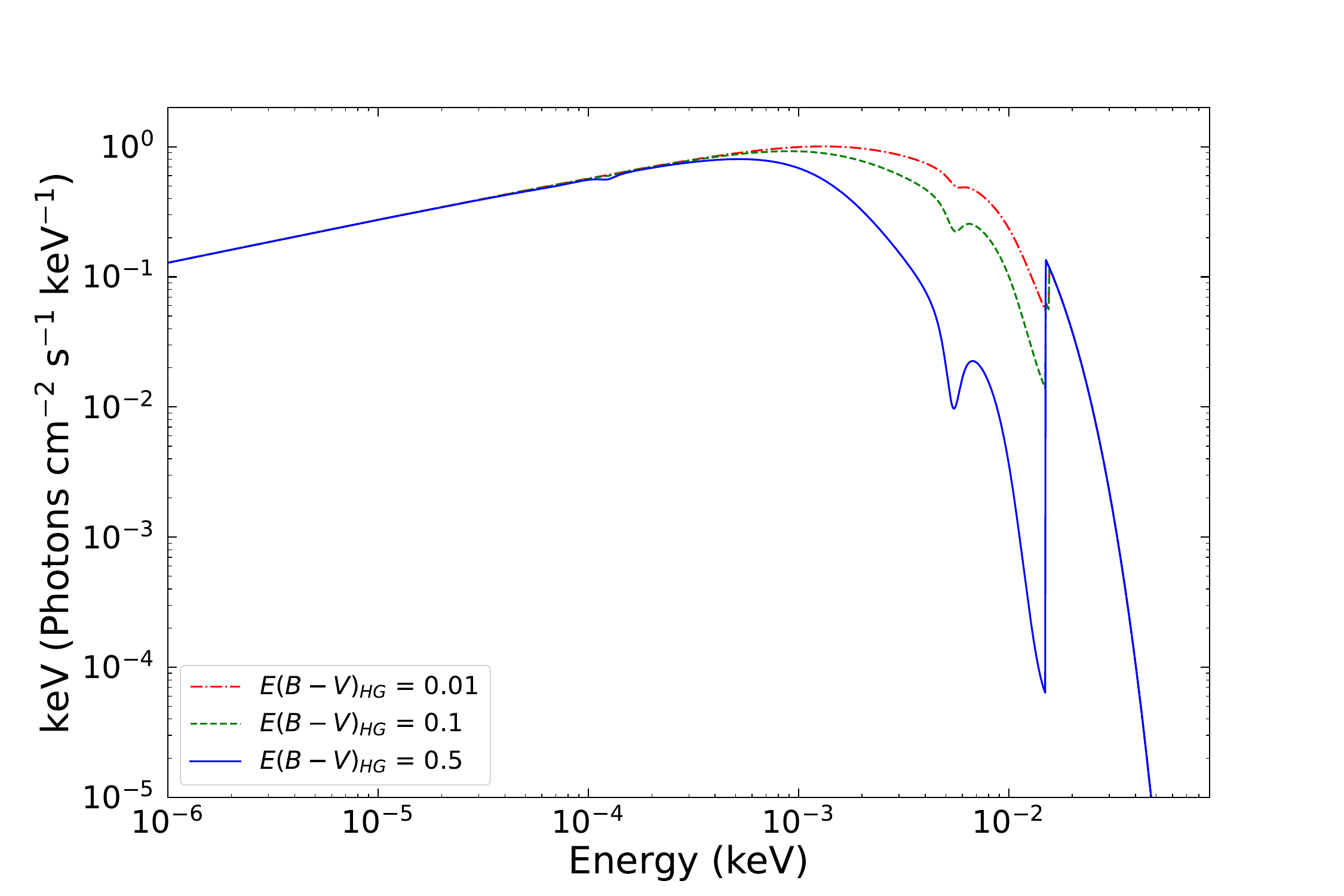}
    \caption{}
    \label{fig:ebvhg}
  \end{subfigure}
\caption{Optical/UV SED model adopted for this study (a) without absorption, for different values of $kT_{\rm max}$ (dash-dotted lines), keeping $K_{\rm uvo}$ constant at $2\times10^5$, and for different values of $K_{\rm uvo}$ (dashed lines), keeping $kT_{\rm max}$ constant at 2.7 eV and (b) with absorption, for different values of $E(B-V)_{\rm HG}$, with $kT_{\rm max} = 2.7\,{\rm eV}$, $K_{\rm uvo} = 2\times10^5$, and $E(B-V)_{\rm MW} = 0.05$. The former plot shows possible degeneracy between parameters $kT_{\rm max}$ and $K_{\rm uvo}$, while the latter shows effects of host galaxy dust extinction on the optical/UV SED model, almost negligible for $E(B-V)_{\rm HG}\leq0.01$.}
\label{fig:model}
\end{figure*}

For the \texttt{DISKPN}, the free parameters include the normalization ($K_{\rm uvo}$) and the maximum temperature of the disk ($kT_{\rm max}$; in units of keV). Here, the normalization is defined as:

\begin{equation}\label{eq:norm_uvo}
    K_{\rm uvo} = M^2 {\rm cos}(i)/(D^2 \times \beta^4),
\end{equation}\\
where $M$ is the mass of the central body (the SMBH in our case) in solar masses, $i$ is the inclination angle of the disk, $D$ is the distance to the source in kpc, and $\beta$ is the color to effective temperature ratio. The inner radius of the disk ($R_{\rm in}$; in units of $R_{\rm g} = GM/c^2$) is fixed to 6.0 gravitational radii\footnote{\url{https://heasarc.gsfc.nasa.gov/xanadu/xspec/manual/node169.html}}. In Fig. \ref{fig:model}, we show different realizations of our adopted model for the optical/UV SED based on various parameter values. In the left panel (\ref{fig:diskpn}), we have plotted the multi-temperature accretion disk model (without absorption) for: (i) three values of $kT_{\rm max}$ (0.7 eV, 2.7 eV, and 7.0 eV), keeping $K_{\rm uvo}$ constant at $2\times10^5$, assuming $M = 10^{7.9}\,M_{\odot}$ (median black hole mass), $i = 0^{\circ}$ (for face-on disks), $D = 176$ Mpc (median distance), and $\beta = 1$ (\citealp{2009MNRAS.392.1124V}) and (ii) for three values of $K_{\rm uvo}$ ($5\times10^4$, $2\times10^5$, and $1\times10^6$), keeping $kT_{\rm max}$ constant at 2.7 eV (median value). In the right panel (\ref{fig:ebvhg}), we show the effect of dust extinction due to the host galaxy on our SED models for a source at the median redshift of our sample ($z=0.04$), with $kT_{\rm max} = 2.7\,{\rm eV}$ (median value), $K_{\rm uvo} = 2\times10^5$, and $E(B-V)_{\rm MW} = 0.05$ (median value). We note that as we increase the maximum temperature of the disk, the optical/UV emission not only extends to higher energies but the amount of flux emitted also increases. The latter is also visible as an outcome of increasing the disk normalization. Hence, there is a clear degeneracy between the parameters, $kT_{\rm max}$ and $K_{\rm uvo}$. To achieve a good fit of the optical/UV photometric points of a high (low) flux source, the model could either increase (decrease) $kT_{\rm max}$ without changing $K_{\rm uvo}$ or vice versa. To ensure that this degeneracy does not affect our SED fitting results, we make sure that the best-fit values of $kT_{\rm max}$ are reasonable and do not exceed the upper limit of a few tens of eVs ($\sim10^{5.8}\,{\rm Kelvin}$), expected from typical \cite{1973A&A....24..337S} accretion disks around nonrotating Schwarzchild or rotating Kerr black holes (e.g., \citealp{2007ApJ...659..211B}; \citealp{2013ApJ...770...30B}). 


\begin{figure*}
  \begin{subfigure}[t]{\textwidth}
    \centering
    \includegraphics[width=1.05\textwidth]{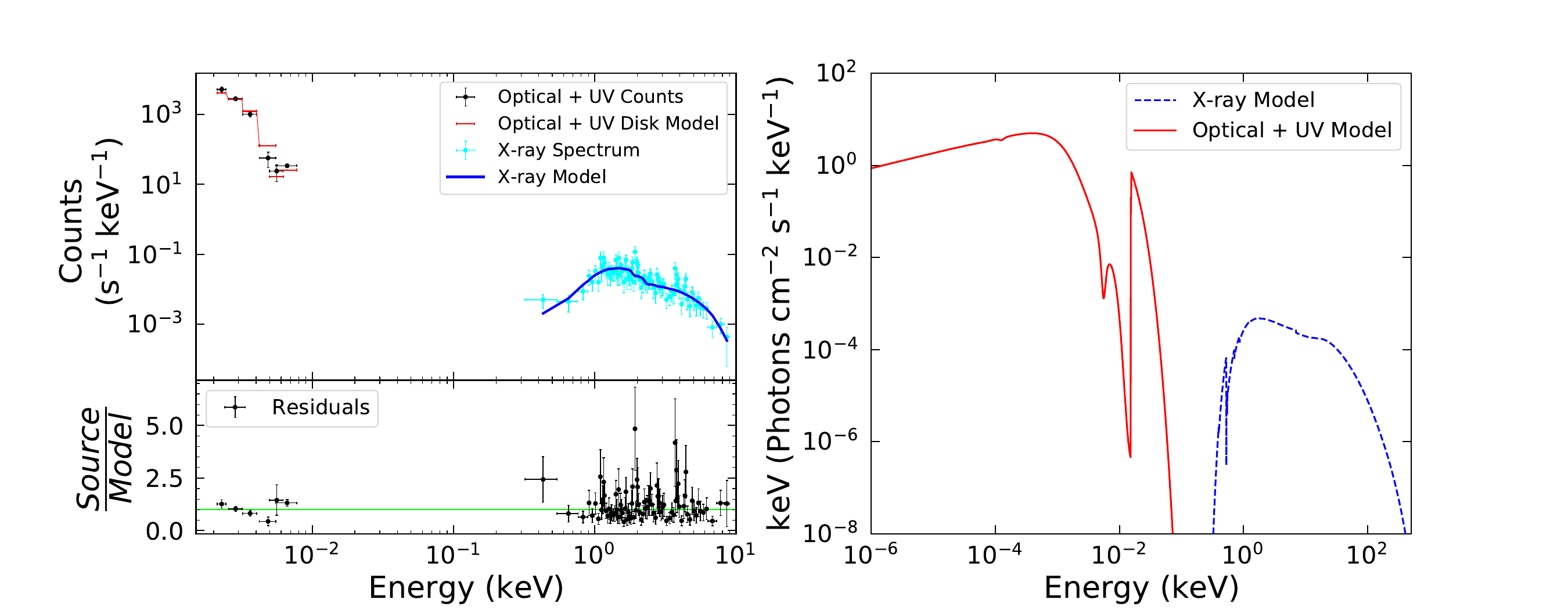} 
    \label{fig:sed_110}
  \end{subfigure}

\vspace{-0.5cm}

  \begin{subfigure}[t]{\textwidth}
    \centering
    \includegraphics[width=1.05\textwidth]{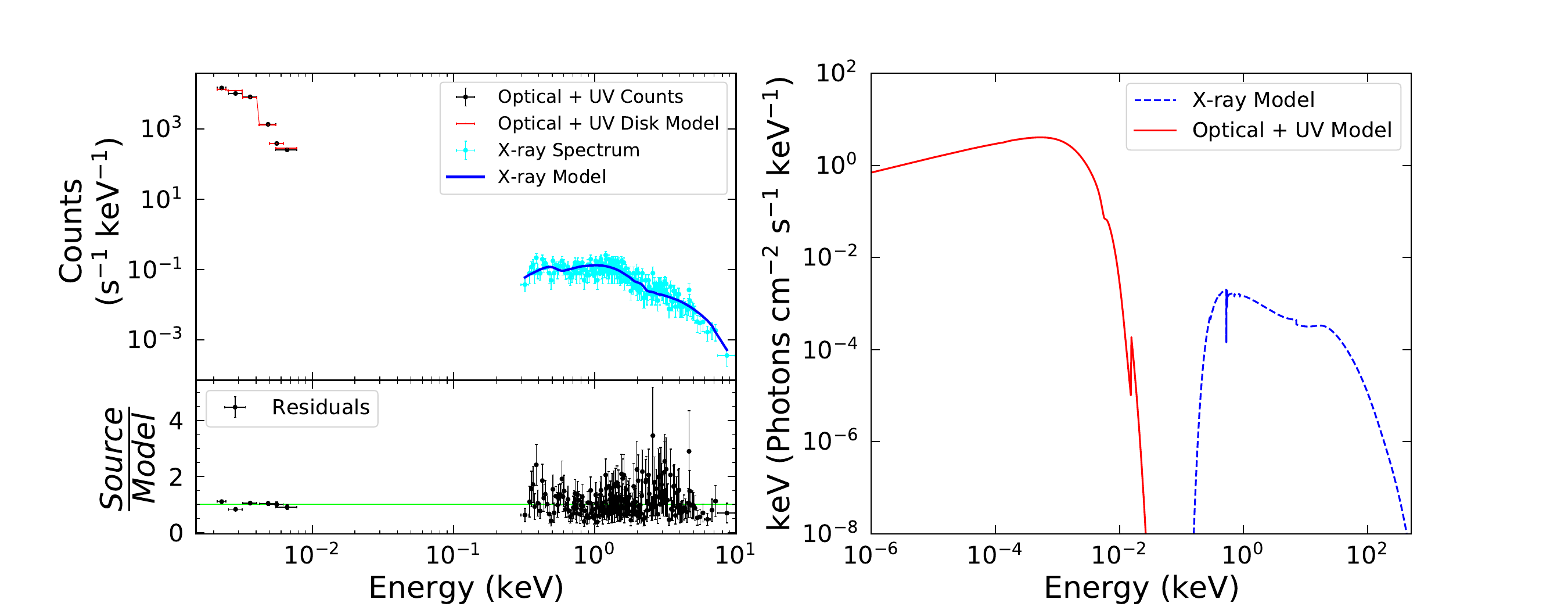}
    \label{fig:sed_335}
  \end{subfigure}

\vspace{-0.5cm}

  \begin{subfigure}[t]{\textwidth}
    \centering
    \includegraphics[width=1.05\textwidth]{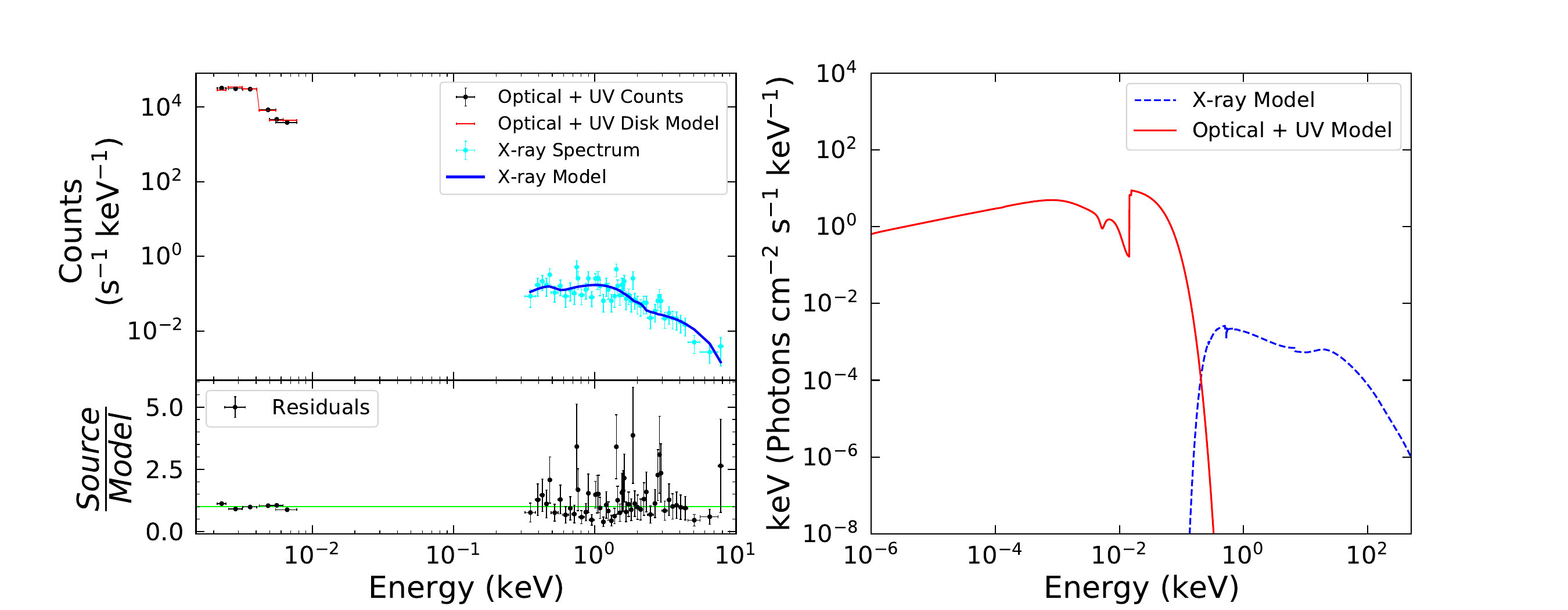}
    \label{fig:sed_722}
  \end{subfigure}
    \caption{Examples of the optical-to-X-ray SED fits. From top to bottom, the fits are for SWIFTJ0208.2+4452, SWIFTJ0630.7+6342, and SWIFTJ1421.4+4747. The left panel shows the counts in the optical/UV filters (black points) and the 0.3--10\,keV X-ray spectrum (cyan points). The figure also shows the fit obtained using the optical/UV disk model including dust extinction (in red) and the X-ray model (in blue), along with the fit residuals (bottom left panel). The right panel shows the best-fit SED model for each source.}
    \label{fig:sed1}
\end{figure*}

\begin{figure*}
  \begin{subfigure}[t]{\textwidth}
    \centering
    \includegraphics[width=1.05\textwidth]{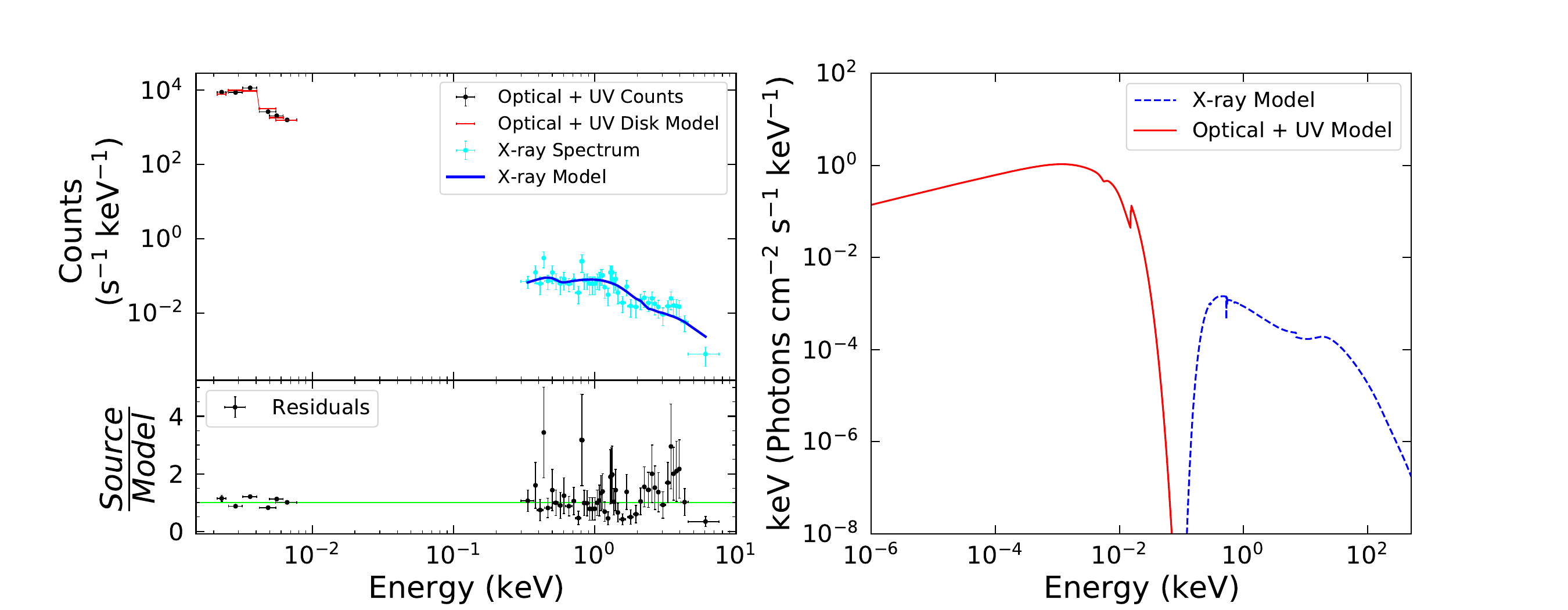} 
    \label{fig:sed_3}
  \end{subfigure}

\vspace{-0.5cm}

  \begin{subfigure}[t]{\textwidth}
    \centering
    \includegraphics[width=1.05\textwidth]{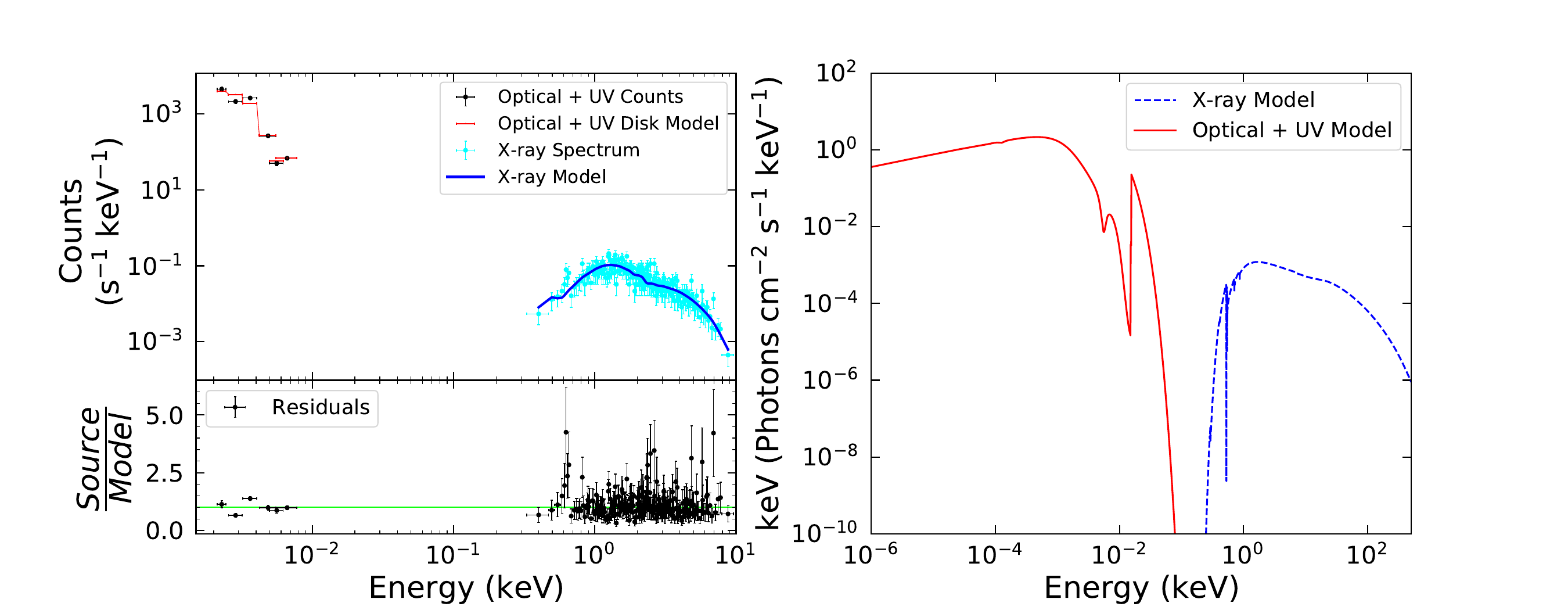}
    \label{fig:sed_348}
  \end{subfigure}

\vspace{-0.5cm}

  \begin{subfigure}[t]{\textwidth}
    \centering
    \includegraphics[width=1.05\textwidth]{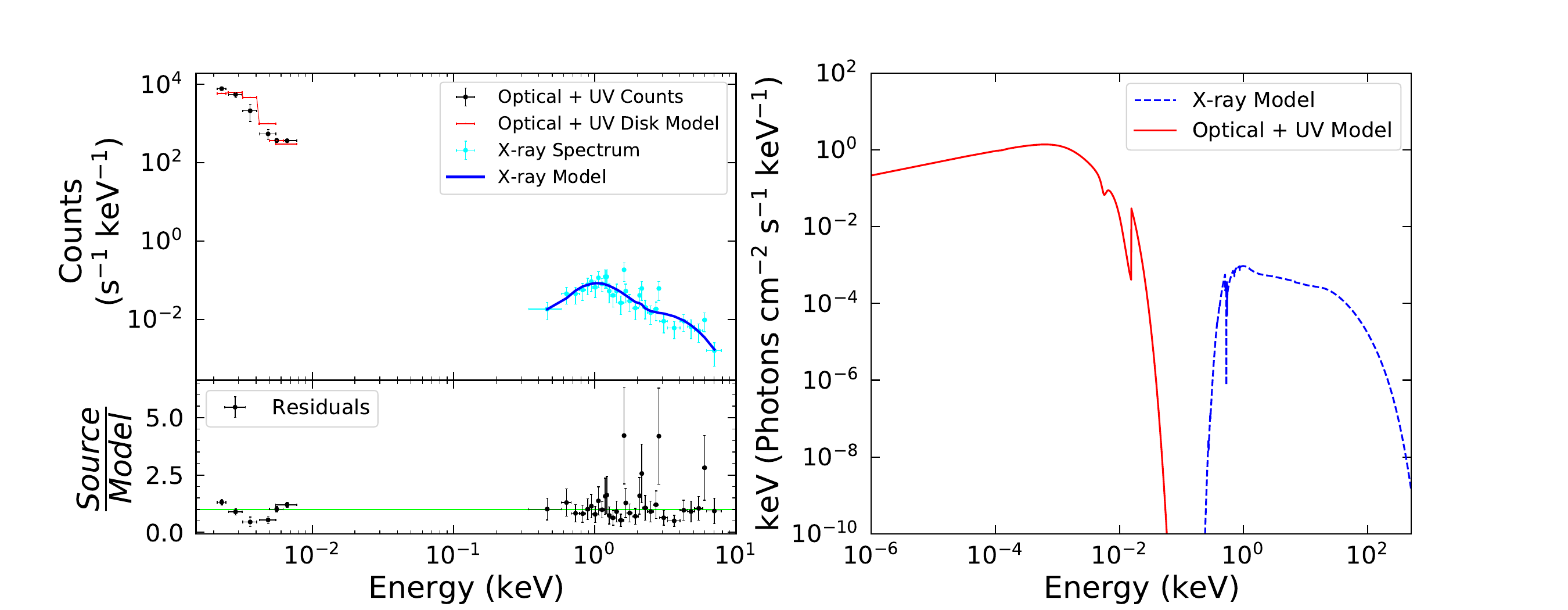}
    \label{fig:sed_667}
  \end{subfigure}
    \caption{Examples of the optical-to-X-ray SED fits. From top to bottom, the fits are for SWIFTJ0002.5+0323, SWIFTJ0654.6+0700, and SWIFTJ1316.9-7155. The left panel shows the counts in the optical/UV filters (black points) and the 0.3--10\,keV X-ray spectrum (cyan points). The figure also shows the fit obtained using the optical/UV disk model including dust extinction (in red) and the X-ray model (in blue), along with the fit residuals (bottom left panel). The right panel shows the best-fit SED model for each source.}
    \label{fig:sed2}
\end{figure*}

After compiling and preparing all the source PHA files, and defining the model to be used, we begin by fitting 223/290 ($\sim 77\%$) sources that have observations in all the six \textit{Swift}/UVOT filters. The best fit is obtained based on the reduced chi-square and visually inspecting the fit residuals. In cases when the host galaxy $E(B-V)$ was of the order $10^{-4}$ or less, or when it was not constrained, we repeated the fit by fixing it to zero, as effects of host galaxy dust extinction are almost negligible below $E(B-V)_{\rm HG}\sim 0.01$ (see Fig. \ref{fig:ebvhg}). The best-fit values of all free parameters were checked using the \texttt{steppar} command. The errors in all the free parameters ($E[B-V]_{\rm HG}$, $kT_{\rm max}$, and $K_{\rm uvo}$) were calculated using the \texttt{error} command. Next, we fit the sources with observations in at least four of the \textit{Swift}/UVOT filters (13/290 $\sim 4.5\%$ sources) following the same procedure as before. For the remaining 54/290 ($\sim 18.5\%$) sources, we had observations in fewer than four filters. Considering the three free parameters in the model used for fitting, it was not possible to fit these sources (with three or fewer data points), and to get reliable estimates of all the fit parameters. Therefore, we did not fit the optical/UV SEDs for these sources due to missing observations in more than two \textit{Swift}/UVOT filters. We show a few examples of our SED fits, along with the best-fit model and residuals, in Figs. \ref{fig:sed1} and \ref{fig:sed2}.

Out of the 236 sources for which we perform the optical/UV SED fits, the parameter corresponding to the maximum temperature of the accretion disk ($kT_{\rm max}$) was not constrained for 27 ($\sim 11\%$) sources. As was mentioned previously, these were high flux sources for which the model initially favored extremely high $kT_{\rm max}$ values ($\geq20\,{\rm eV}$; otherwise impossible for AGN disks) to obtain a suitable fit. In these objects, we fixed $kT_{\rm max}$ to its median value (2.7\,eV) obtained from the sources that were successfully fit while constraining the disk temperature. The final best fits were used to calculate the observed and intrinsic (rest-frame and absorption-corrected) model fluxes and luminosities in all the six \textit{Swift}/UVOT filters. The total intrinsic optical $+$ UV luminosity ($L_{\rm uvo}$) was estimated in the energy range from $10^{-7}$ to 0.1\,keV, since the standard \cite{1973A&A....24..337S} disk typically has very little to no emission in the X-rays. The luminosity thus computed will later be combined with $L_{\rm x}$ (see Sect. \ref{sect:xray_spec}) to estimate the total accretion luminosities ($L_{\rm bol}$) of the AGN in our sample. The upper and lower limits of these fluxes were calculated considering the 90\% confidence intervals of $K_{\rm uvo}$ and $E(B-V)_{\rm HG}$. The lower limit in the flux of each source corresponds to the best-fit model of that source with the highest $E(B-V)_{\rm HG}$ (most dust extinction) and lowest $K_{\rm uvo}$ as estimated during the fitting, and vice versa (see Table \ref{tab:sed}).

We repeat the optical/UV SED fitting, as stated above, for the more distant ($z>0.05$) sources after applying the corrections to their \textsc{GALFIT}-estimated PSF magnitudes (described in Sect. \ref{sect:galfit_corr}). For a total of 46 ($\sim 20\%$) sources, we redo the SED fitting after editing their source PHA files to include the corrected \textsc{GALFIT} PSF magnitudes. We report the best-fit values of $kT_{\rm max}$ and $E(B-V)_{\rm HG}$ along with the various flags regarding the SED fitting in Table \ref{tab:sed}. In the next section we discuss in detail the results we found for all three samples: (a) the total sample of 236 sources without applying corrections to the magnitudes necessary for distant sources, (b) a sub-sample of 190 sources, removing the sources that need to be corrected, and (c) the total sample of 236 sources including corrections to the magnitudes estimated by \textsc{GALFIT} for the distant sources.


\begin{table*}
\centering
\caption{Results of the optical/UV SED fitting (Sect. \ref{sect:sed}).}
\begin{tabular}{cccccc}
\hline
\hline
\specialrule{0.1em}{0em}{0.5em}
\vspace{1mm}
BAT ID & Swift ID & $kT_{\rm max}$ & $E(B-V)_{\rm HG}$ & $F_{\rm uvo}$ & Flag\\
\hline\\
\vspace{2mm}
3 & SWIFTJ0002.5+0323 & $2.8_{-0.8}^{+2.3}$ & $0.058_{-0.058}^{+0.070}$ & $1.85_{-0.65}^{+0.97}$ & -\\
\vspace{2mm}
6 & SWIFTJ0006.2+2012 & $3.9_{-0.3}^{+1.5}$ & $<0.014$ & $17.6_{-10.6}^{+5.05}$ & -\\
\vspace{2mm}
16 & SWIFTJ0029.2+1319 & $\,\,\,6.1_{-2.5}^{+12.1}$ & $0.117_{-0.050}^{+0.037}$ & ${24.0}_{-3.37}^{+5.24}$ & -\\
\vspace{2mm}
19 & SWIFTJ0034.5-7904 & $2.7$ & $<0.001$ & ${3.05}_{-0.15}^{+0.13}$ & t\\
\vspace{2mm}
34 & SWIFTJ0051.6+2928 & $3.7_{-1.3}^{+4.3}$ & $0.298_{-0.061}^{+0.065}$ & ${8.28}_{-1.55}^{+2.06}$ & -\\
\vspace{2mm}
36 & SWIFTJ0051.9+1724 & $2.9_{-0.2}^{+0.3}$ & $<0.012$ & ${6.13}_{-0.47}^{+0.28}$ & -\\
\vspace{2mm}
39 & SWIFTJ0054.9+2524 & $6.4_{-2.7}^{+2.7}$ & $0.058_{-0.046}^{+0.018}$ & ${15.7}_{-1.67}^{+3.81}$ & -\\
\vspace{2mm}
43 & SWIFTJ0059.4+3150 & $5.5_{-1.2}^{+2.5}$ & $<0.007$ & ${3.89}_{-0.92}^{+4.64}$ & -\\
\vspace{2mm}
45 & SWIFTJ0101.5-0308 & $2.8_{-1.2}^{+1.2}$ & $0.268_{-0.160}^{+0.154}$ & ${1.45}_{-0.70}^{+1.82}$ & -\\
\vspace{2mm}
51 & SWIFTJ0105.7-1414 & $4.0_{-0.6}^{+0.9}$ & \,\,\,\,\,0.000 & ${5.50}_{-0.13}^{+0.13}$ & e\\

\hline
\end{tabular}
\tablefoot{We have listed the best-fit values of the disk temperature ($kT_{\rm max}$ in eV), the extinction due to the host galaxy $E(B-V)_{\rm HG}$, and the total, intrinsic optical/UV flux (from $10^{-7}$--0.1\,keV) with errors. The last column shows, in the form of flags, if the disk temperature was fixed to its median value of 2.7\,eV (t) and if the host galaxy extinction was fixed to zero (e) during the fitting. Flux units: $10^{-11}\,{\rm erg\,\,cm^{-2}\,s^{-1}}$. The table in its entirety is available at the CDS.}
\label{tab:sed}
\end{table*}


\section{Results and discussion}\label{sect:discussion}

In the previous sections we explained in detail the various steps we followed to systematically construct and fit the optical-to-X-ray SEDs of a sample of $\sim$ 240 unobscured AGN. In this section we focus on the outputs of our SED fitting and explore some of the important scaling relations between them. All the results reported in the following sections are for the final sample of 236 unobscured AGN with GALFIT fluxes corrected as is explained in Sect. \ref{sect:galfit_corr} for the sources at $z>0.05$. However, we checked all our results for the sample without employing these corrections and did not find any qualitative differences. We also confirmed all our results to hold for a cleaner sample of 190 AGN excluding the distant sources that need to be corrected. In Sect. \ref{sect:med} we present the SED models for our sample of unobscured AGN. In Sect. \ref{sect:disk} we study the parameters that describe the optical/UV disk emission, such as the temperature of the accretion disk (Sect. \ref{sect:disktemp}) and the extinction due to host galaxy (Sect. \ref{sect:ebv}). And finally, in Sects. \ref{sect:aox} and \ref{sect:k} we present the optical-to-X-ray spectral index ($\alpha_{\rm ox}$) and various bolometric corrections ($\kappa_{\lambda}$), respectively, and explore their dependence on different physical properties of AGN, such as black hole mass, bolometric luminosity, and Eddington ratio.


\begin{figure*} 
  \begin{subfigure}[t]{0.54\textwidth}
    \centering
    \includegraphics[width=\textwidth]{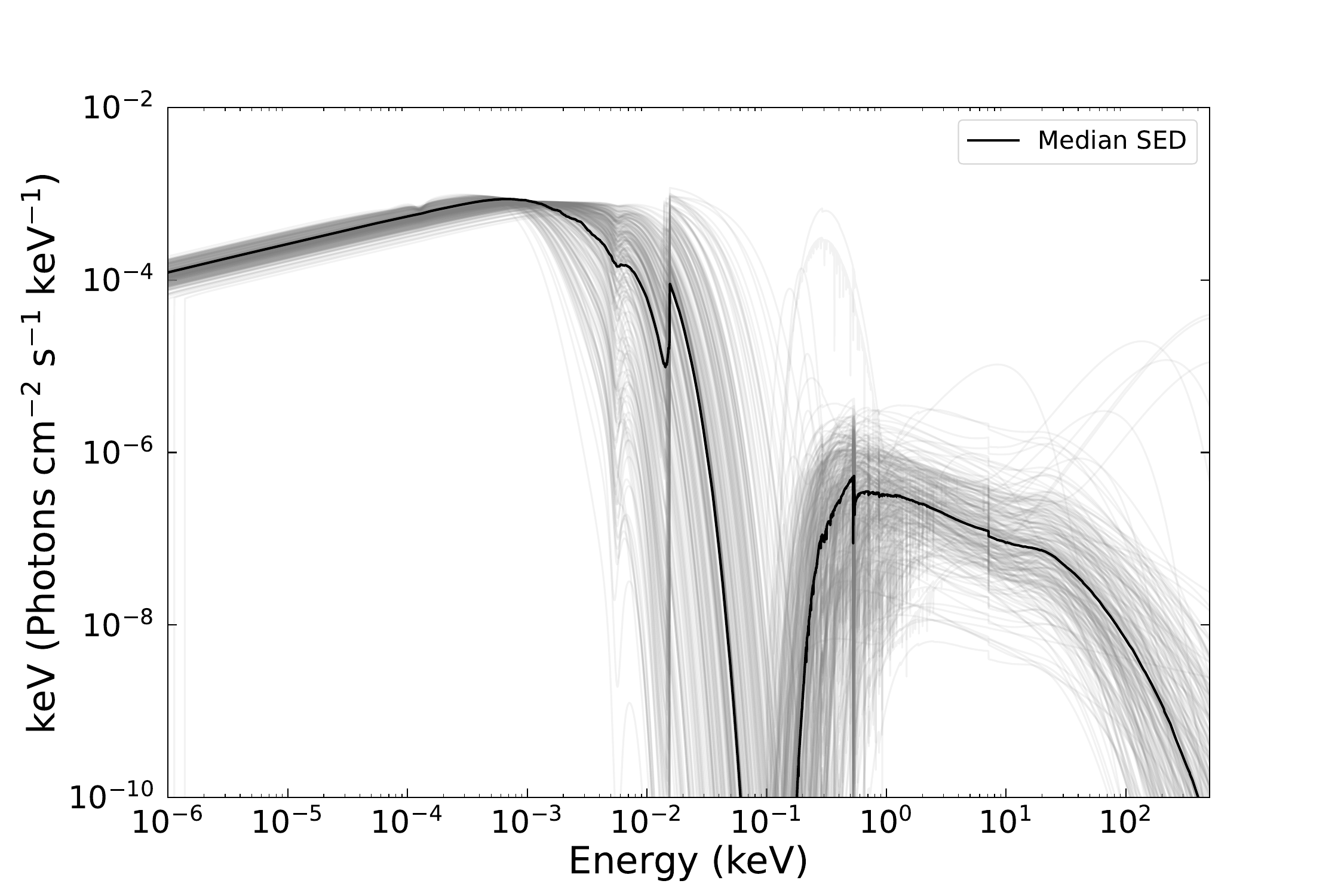}
    \caption{}
    \label{fig:med_abs}
  \end{subfigure}
  \hspace{-1.cm}
  \begin{subfigure}[t]{0.54\textwidth}
    \centering
    \includegraphics[width=\textwidth]{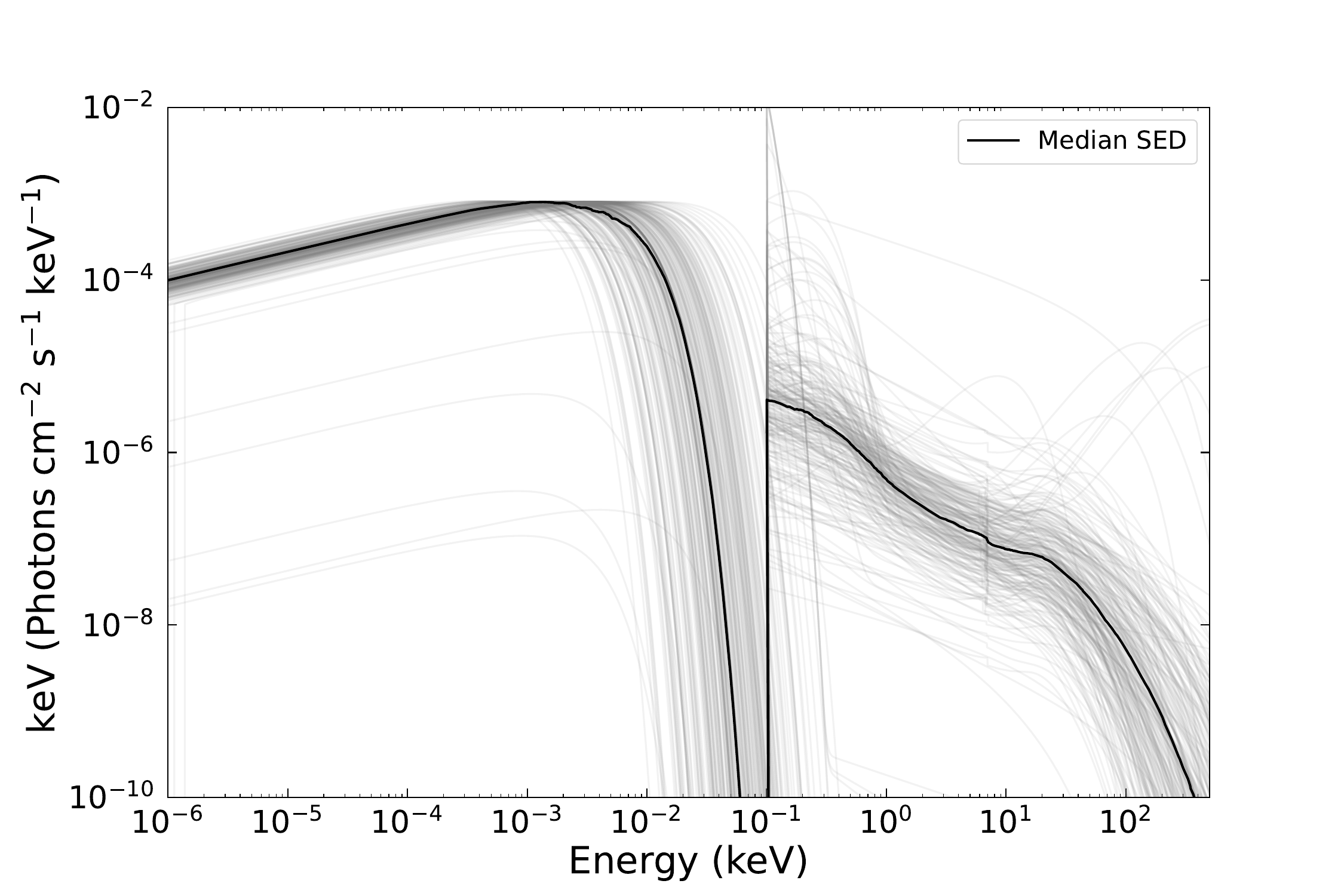}
    \caption{}
    \label{fig:med_intr}
  \end{subfigure}
\caption{Rest frame optical-to-X-ray best-fit SED models (a) with absorption and (b) without absorption, for the 236 unobscured, nearby AGN in our sample (in gray). The SEDs have been normalized by total flux. The median SED model is shown as a solid black line.}
\label{fig:med1}
\end{figure*}


\subsection{Diversity in SED models}\label{sect:med}

In Fig. \ref{fig:med_abs}, we show all the optical-to-X-ray SED models for our sample, as well as the rest-frame median SED model (in solid black). In the adjacent panel (Fig. \ref{fig:med_intr}), we show the intrinsic, absorption-corrected SED models for our sample along with the median, absorption-corrected SED (in solid black). All the SEDs have been normalized by the total flux. Comparing the two panels, we can clearly see the absorption features due to the host galaxy and the Milky Way in the optical/UV and soft X-rays. A reflection feature around 7\,keV is visible in both cases. As has been mentioned in many previous works on broadband SED fitting (e.g., \citealp{1994ApJS...95....1E}; \citealp{2006ApJS..166..470R}; \citealp{2009MNRAS.399.1553V}), and also visible in Fig. \ref{fig:med1}, there exists a large variety in the shapes of AGN SEDs that one needs to take into account to explore the full range of properties exhibited by the AGN population. A more detailed analysis of SED shapes of unobscured AGN based on their physical properties, such as black hole mass, luminosity, and Eddington ratio will be presented in another upcoming publication (\textcolor{blue}{Gupta et al. in prep.}). Additionally, a complete study of the X-ray-to-far-IR SEDs of all \textit{Swift}/BAT AGN is ongoing (\textcolor{blue}{Rojas et al. in prep.}).


\subsection{Optical/UV disk emission}\label{sect:disk}

The optical/UV part of the multiwavelength AGN SED corresponds to the emission that is believed to originate from the accretion disk, at different radii, as material spirals into the central SMBH (e.g., \citealp{1978Natur.272..706S}; \citealp{1982ApJ...254...22M}; 
\citealp{1987ApJ...315...74W}). This emission is expected to be the largest contributor to the total accretion luminosity of the AGN. In the following two sections, we look in detail at the parameters constrained from the optical/UV SED fitting, that is, the maximum disk temperature (Sect. \ref{sect:disktemp}), and the dust extinction due to the host galaxy (Sect. \ref{sect:ebv}).


\begin{figure*} 
  \begin{subfigure}[t]{0.54\textwidth}
    \centering
    \includegraphics[width=\textwidth]{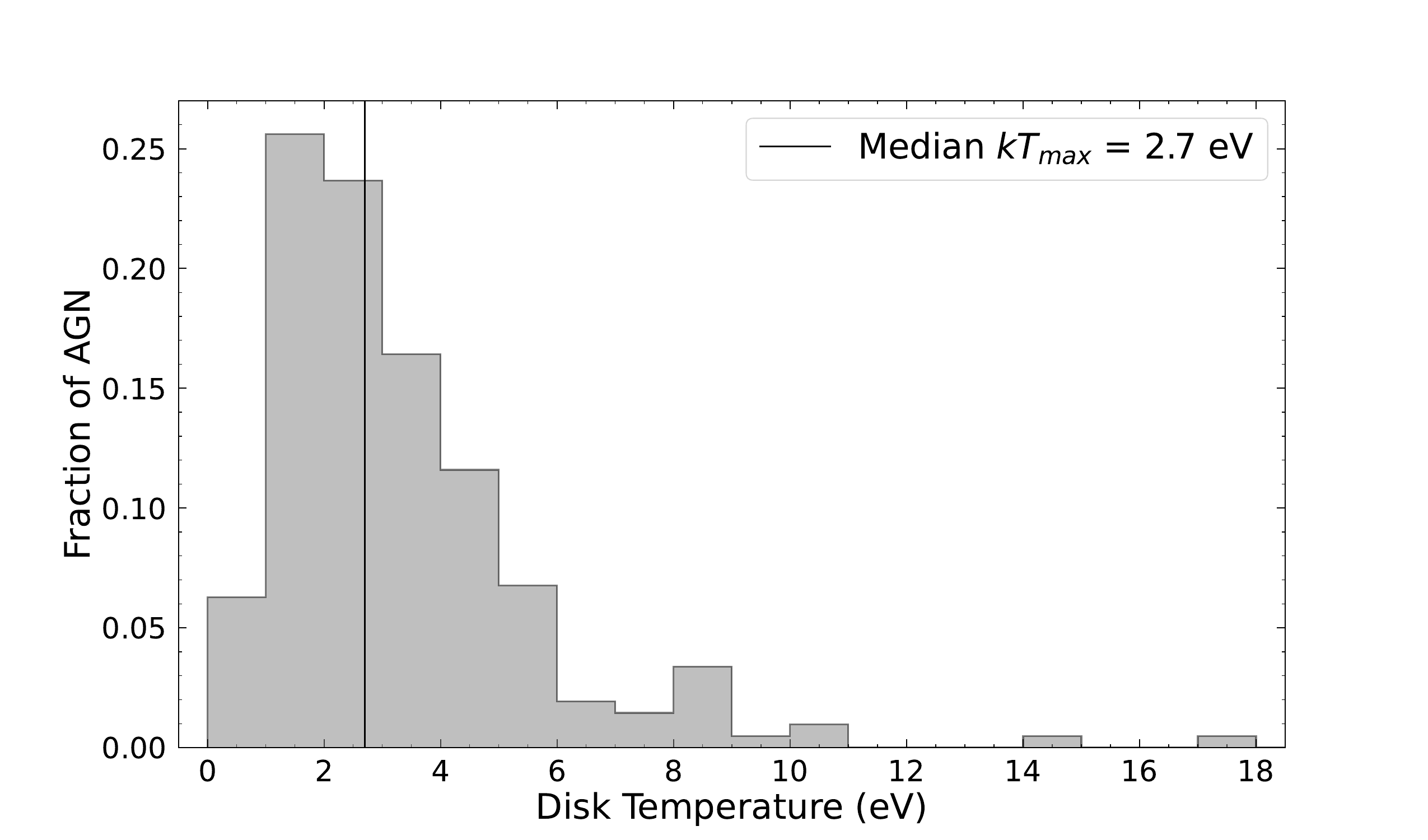}
    \caption{}
    \label{fig:temp}
  \end{subfigure}
  \hspace{-1.cm}
  \begin{subfigure}[t]{0.54\textwidth}
    \centering
    \includegraphics[width=\textwidth]{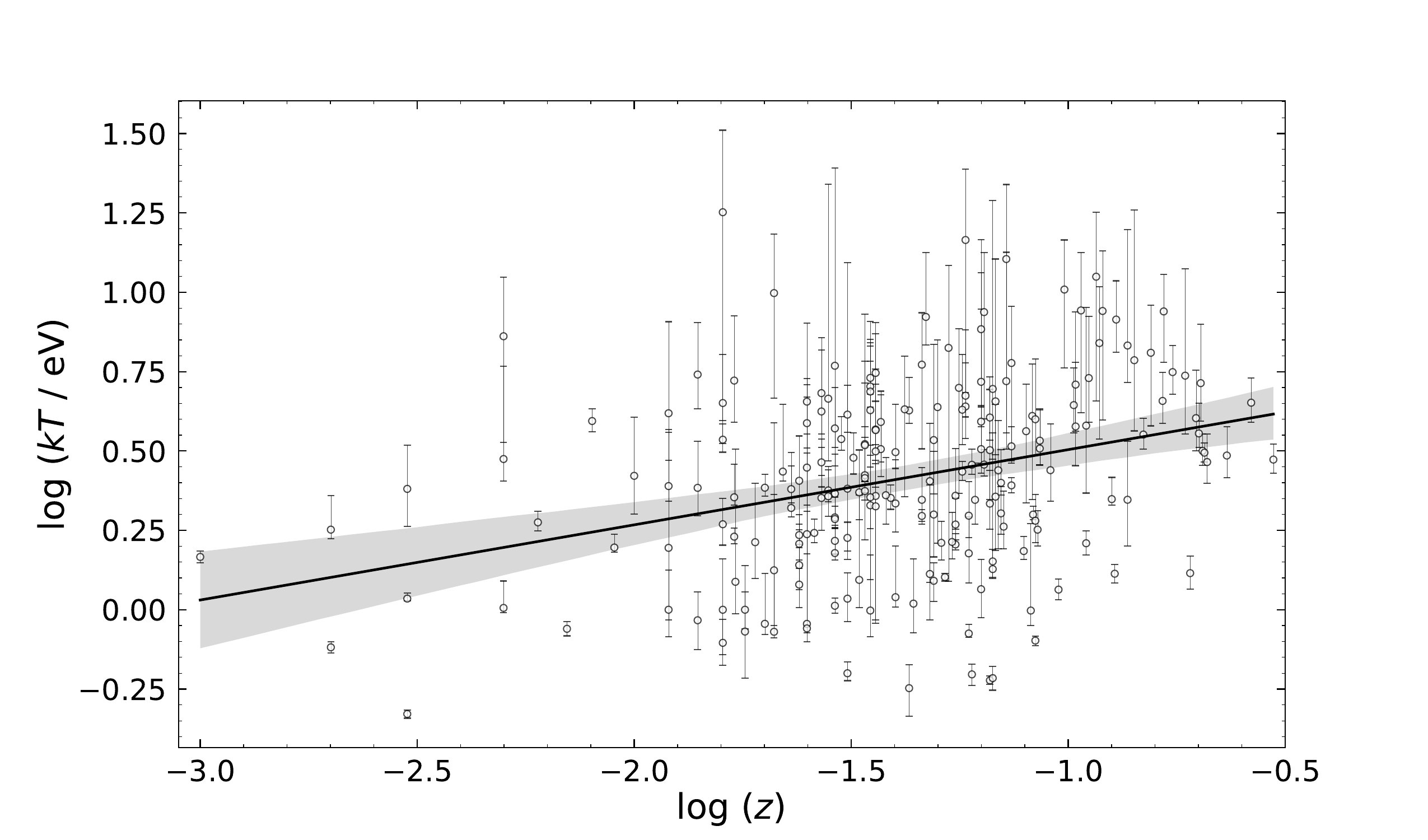}
    \caption{}
    \label{fig:kT_vs_z}
  \end{subfigure}
\caption{Maximum disk temperature ($kT_{\rm max}$) as obtained after fitting the optical/UV SEDs of our source sample with a dust-extinction-corrected thermal disk model (described in Sect. \ref{sect:sed}). (a) Distribution of the best-fit values of $kT_{\rm max}$. The median value of $kT_{\rm max}$ from this distribution is 2.7\,eV. (b) $kT_{\rm max}$ as a function of the redshift of the source. The solid black line shows the best-fit relation (Sect. \ref{sect:disktemp}, Table \ref{tab:kt}) and the shaded gray region displays the uncertainty on the regression.}  
\label{fig:kT}
\end{figure*}


\begin{figure*} 
  \begin{subfigure}[t]{0.54\textwidth}
    \centering
    \includegraphics[width=\textwidth]{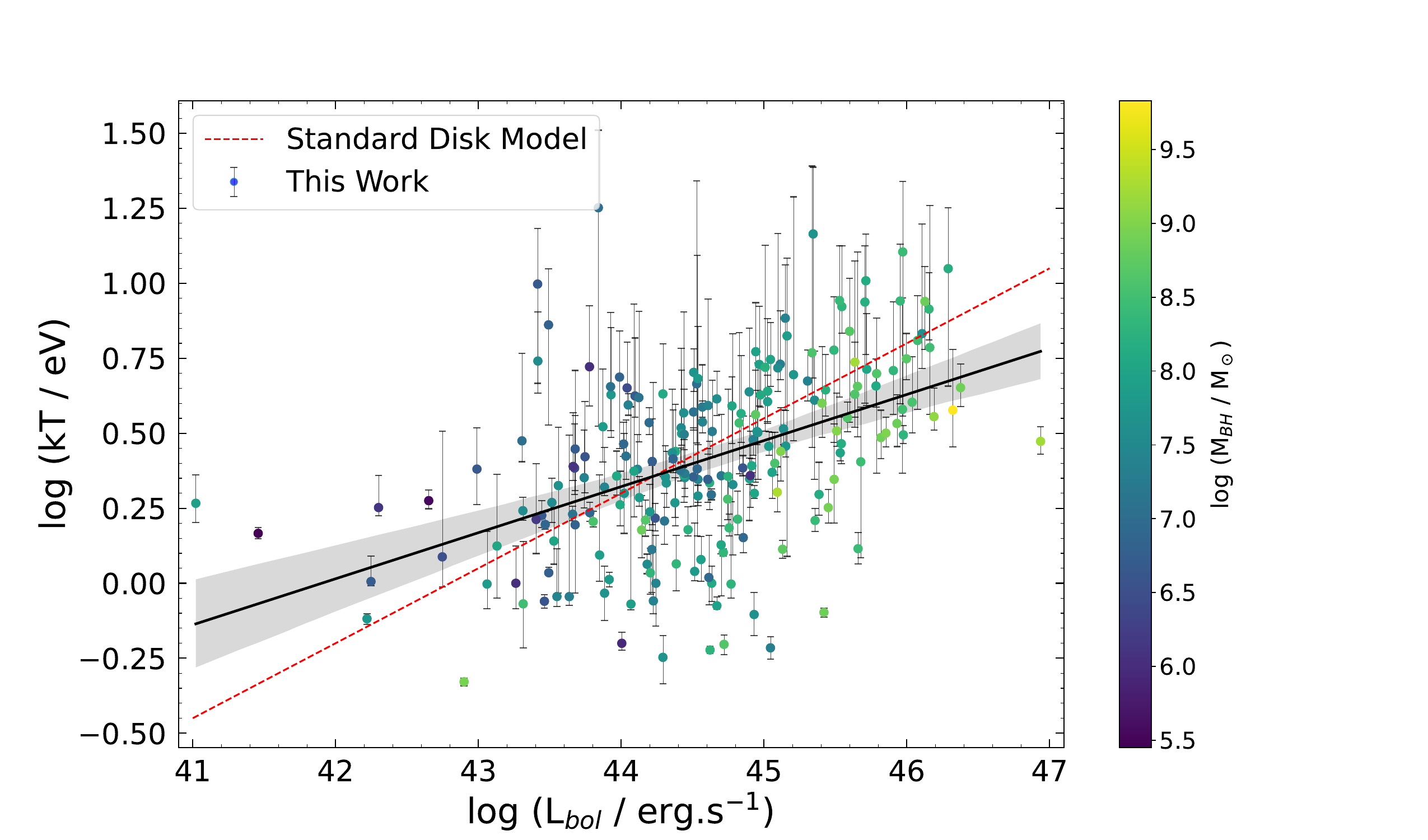}
    \caption{}
    \label{fig:kT_vs_lbol}
  \end{subfigure}
  \hspace{-1.cm}
  \begin{subfigure}[t]{0.54\textwidth}
    \centering
    \includegraphics[width=\textwidth]{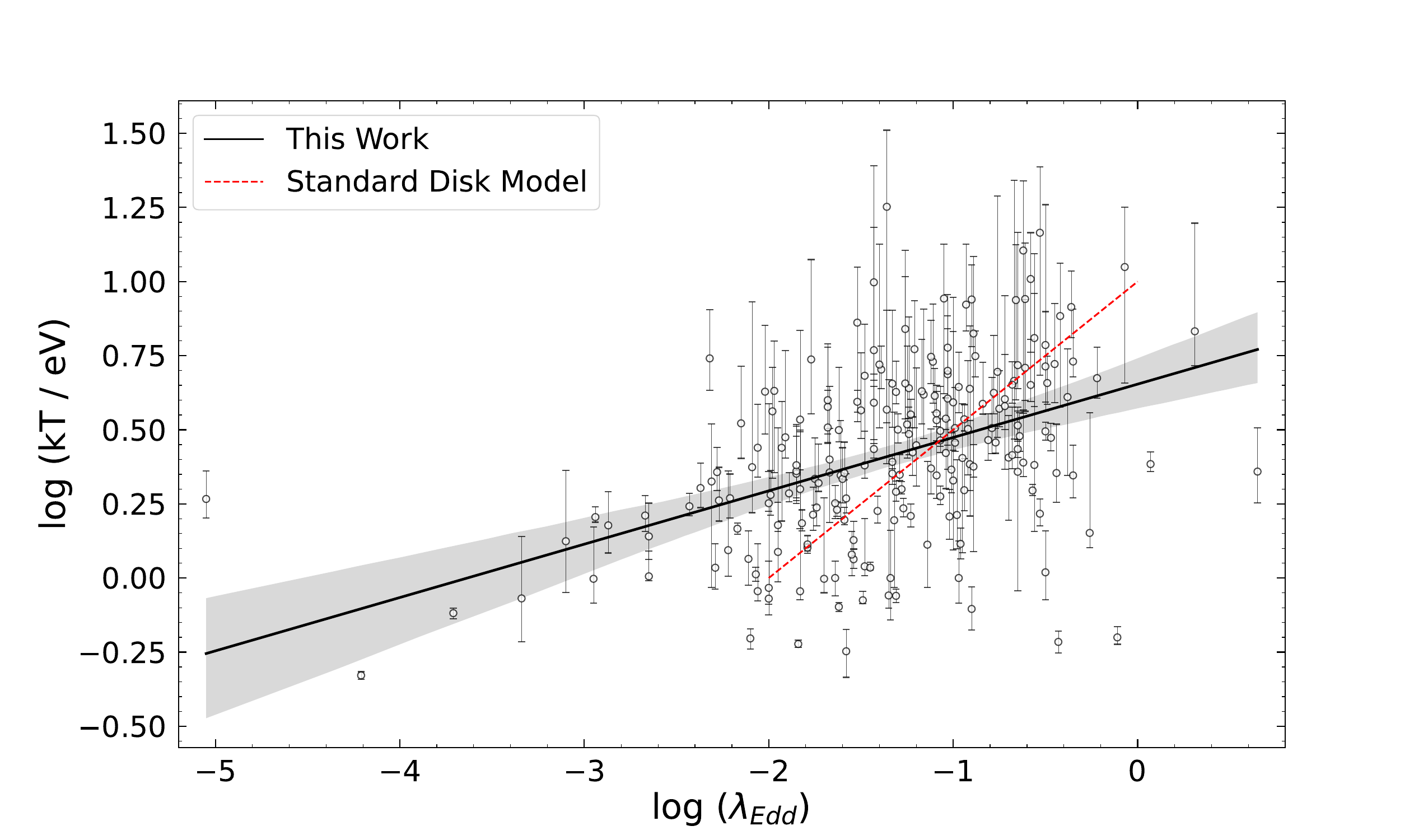}
    \caption{}
    \label{fig:kT_vs_er}
  \end{subfigure}
\caption{Maximum disk temperature ($kT_{\rm max}$) as a function of (a) the bolometric luminosity (color-coded with black hole mass), and (b) the Eddington ratio. The solid black line shows the best-fit relation (Sect. \ref{sect:disktemp}, Table \ref{tab:kt}) and the shaded gray region is the one sigma confidence interval. The dashed red line is the theoretical relation between (a) $kT_{\rm max}$ and $L_{\rm bol}$ ($T_{\rm max} \propto L_{\rm bol}^{1/4}$) and (b) $kT_{\rm max}$ and $\lambda_{\rm Edd}$ ($T_{\rm max} \propto \lambda_{\rm Edd}^{1/4}$), as is expected from the standard \citet{1973A&A....24..337S} disk model for a fixed black hole mass.}  
\label{fig:kT2}
\end{figure*}


\subsubsection{Disk temperature}\label{sect:disktemp}

The standard accretion disk model suggested by \cite{1973A&A....24..337S} considers a geometrically thin and optically thick structure for the disk with a temperature gradient, such that higher temperatures are expected at closer proximity to the SMBH, until the innermost stable circular orbit. Based on our SED fitting analysis using a multi-temperature accretion disk model, we obtained the maximum disk temperatures for our sample of unobscured AGN with a median $kT_{\rm max} = 2.70 \pm 0.01$\,eV (see Fig. \ref{fig:temp}). The range of $kT_{\rm max}$ values from this work are in agreement with those calculated using non-LTE (local thermodynamic equilibrium) AGN disk models (\citealp{2000ApJ...533..710H}) for a wide range of black hole masses and Eddington ratios (see \citealp{2007ApJ...659..211B} and \citealp{2013ApJ...770...30B}). When plotted against the redshift of the source, we see an increase in the value of $kT_{\rm max}$ with $z$ (as is shown in Fig. \ref{fig:kT_vs_z}). This positive correlation between $kT_{\rm max}$ and $z$ (see Table \ref{tab:kt}) does not necessarily imply that higher redshift sources have higher disk temperatures. In fact, it is an effect of the flux-limited nature of the \textit{Swift}/BAT AGN sample. Since more distant sources have to be brighter to be detected, this results in a sample where farther away sources are more luminous and have higher disk temperatures.

We also explore possible correlations of $kT_{\rm max}$ of the accretion disk with other properties of the AGN, such as bolometric luminosity, black hole mass, and Eddington ratio. From the broadband SED fitting, we use the X-ray flux from 0.1 to 500\,keV ($L_{\rm x}$) and the optical/UV flux from $10^{-7}$ to 0.1\,keV ($L_{\rm uvo}$) to estimate the total accretion luminosity or the bolometric luminosity ($L_{\rm bol} = L_{\rm x}+L_{\rm uvo}$) for our unobscured AGN sample (see Table \ref{tab:params}). All luminosities are calculated using the distances listed in Table \ref{tab:src_list}. The Eddington ratios for all our AGN (reported in Table \ref{tab:params}) are calculated using $\lambda_{\rm Edd} = L_{\rm bol}/L_{\rm Edd}$. To be consistent with the BASS DR2 (see \citealp{2022ApJS..261....2K}), we define the Eddington luminosity as:

\begin{equation}
    L_{\rm Edd} = 1.5 \times 10^{38} \times \frac{M_{\rm BH}}{M_\odot}\,{\rm erg\,s^{-1}}
\end{equation}
As part of the BASS DR2, \citet{2022ApJS..261....5M} calculated the black hole masses using the virial method and reverberation mapping for almost all \textit{Swift}/BAT type\,I AGN in the 70-month catalog. We have these black hole mass estimates for 235/236 sources in our sample and we use them throughout this work.

We find a strong positive correlation of the disk temperature with both $L_{\rm bol}$ and $\lambda_{\rm Edd}$ (Fig. \ref{fig:kT2}, Table \ref{tab:kt}). This is not surprising as under the assumption of the standard disk model \citep{1973A&A....24..337S} employed here to fit the optical/UV SED, we would expect sources with larger luminosities and Eddington ratios (up to the Eddington limit) to have hotter disks. This is a reasonable assumption for our sample since it does not probe high Eddington ratios ($\lambda_{\rm Edd} > 1$), and only has a few sources ($< 15\%$) extending to the low Eddington ratio regime of $\lambda_{\rm Edd} < 0.01$ where the standard disk model might fail (e.g., \citealp{1999PASP..111....1K}; \citealp{2005Ap&SS.300..177N}; \citealp{2014ARA&A..52..529Y}). Theoretically, the maximum value of the effective temperature of the \citet{1973A&A....24..337S} disk relates with the luminosity and the black hole mass as $T_{\rm max} \propto L_{\rm bol}^{1/4}\,M_{\rm BH}^{-1/2}$ (see review by \citealp{1973blho.conf..343N}). This relation can be further modified as $T_{\rm max}^4 \propto \lambda_{\rm Edd}\,M_{\rm BH}^{-1}$. However, we do not find any such dependence of $kT_{\rm max}$ on $M_{\rm BH}$. This is shown in Fig. \ref{fig:kT_vs_lbol}, where the points are color-coded with the black hole mass, but the only visible trend of $M_{\rm BH}$ is with $L_{\rm bol}$ (intrinsic to our sample). We show the theoretical relations (considering a fixed black hole mass) in Fig. \ref{fig:kT2} as a dashed red line, along with our best-fit regression lines (in black). The slopes of our relations ($0.15 \pm 0.02$ and $0.18\pm0.03$) are slightly flatter compared to the theoretical expectation (0.25), which could be due to the lack of any information on the accretion efficiency of our sources. Furthermore, this also adds on to the already established concerns on the applicability of the \citet{1973A&A....24..337S} disk model to a large AGN population such as ours, spanning wide ranges in luminosity, black hole mass, and Eddington ratio.


\begin{figure*} 
  \begin{subfigure}[t]{0.54\textwidth}
    \centering
    \includegraphics[width=\textwidth]{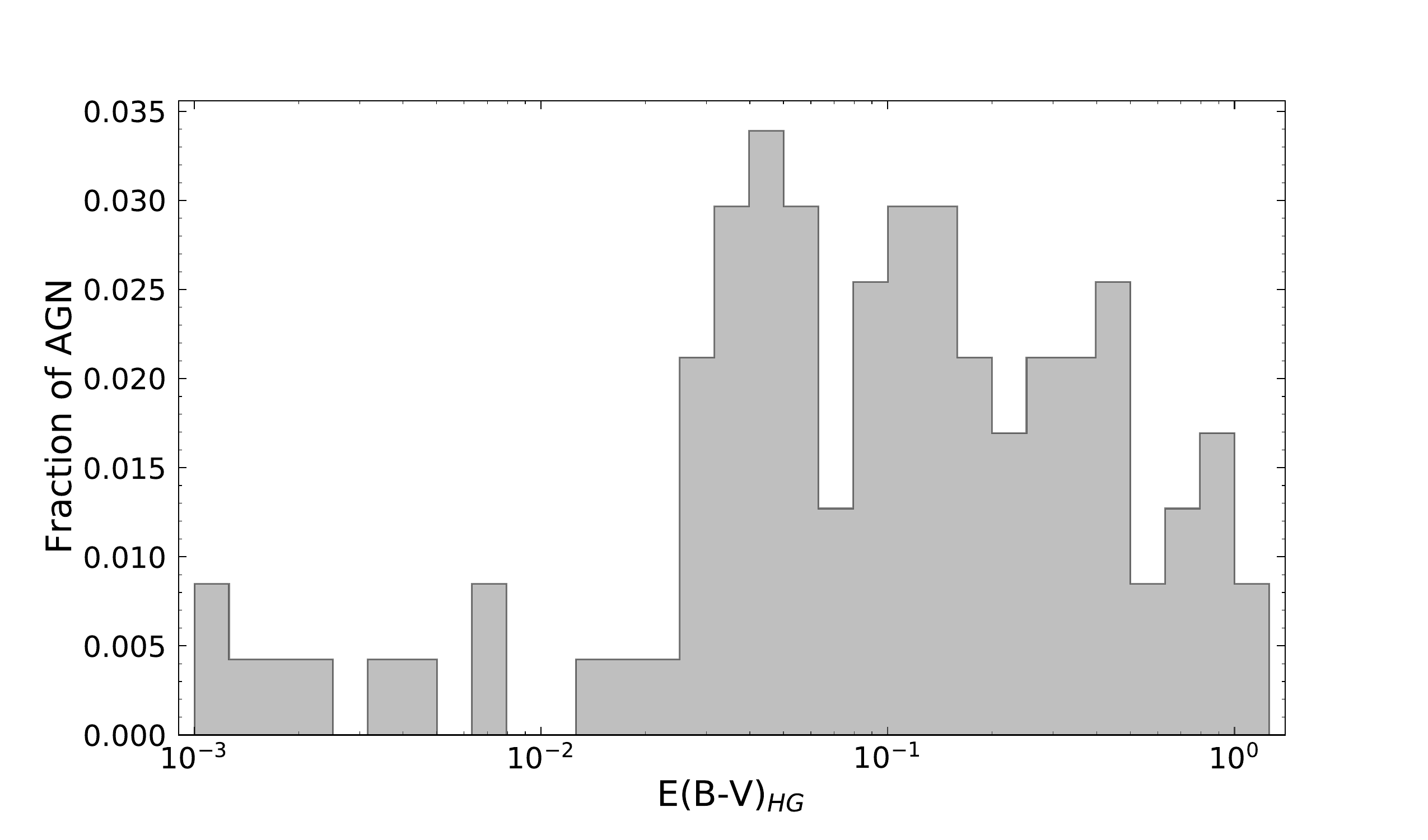}
    \caption{}
    \label{fig:ebv_log}
  \end{subfigure}
  \hspace{-1.cm}
  \begin{subfigure}[t]{0.54\textwidth}
    \centering
    \includegraphics[width=\textwidth]{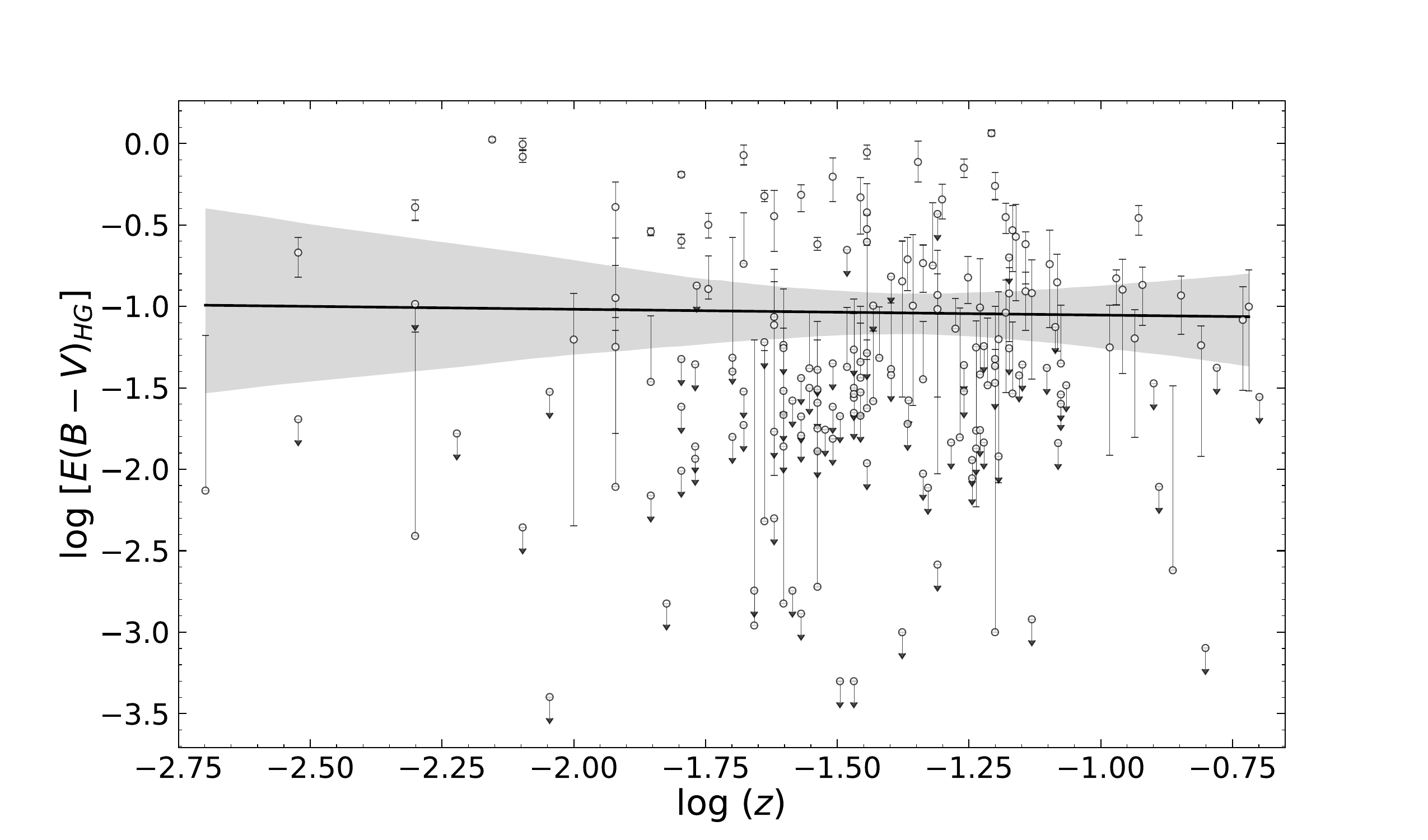}
    \caption{}
    \label{fig:ebv_vs_z}
  \end{subfigure}
\caption{The intrinsic dust extinction of the AGN emission in the optical/UV due to the host galaxy, quantified by the parameter $E(B-V)_{\rm HG}$, calculated from the optical/UV SED fitting (described in Sect. \ref{sect:sed}). (a) Distribution of the best-fit values of $E(B-V)_{\rm HG}$ and (b) $E(B-V)_{\rm HG}$ as a function of the redshift (upper limits shown as downward arrows). The solid black line shows the best-fit relation excluding the upper limits and the shaded gray region marks the one sigma confidence interval. We find a similar relation if we include the upper limits using survival analysis and hence, do not show it here.}  
\label{fig:ebv}
\end{figure*}


\begin{figure}
    \centering
    \includegraphics[width=0.52\textwidth]{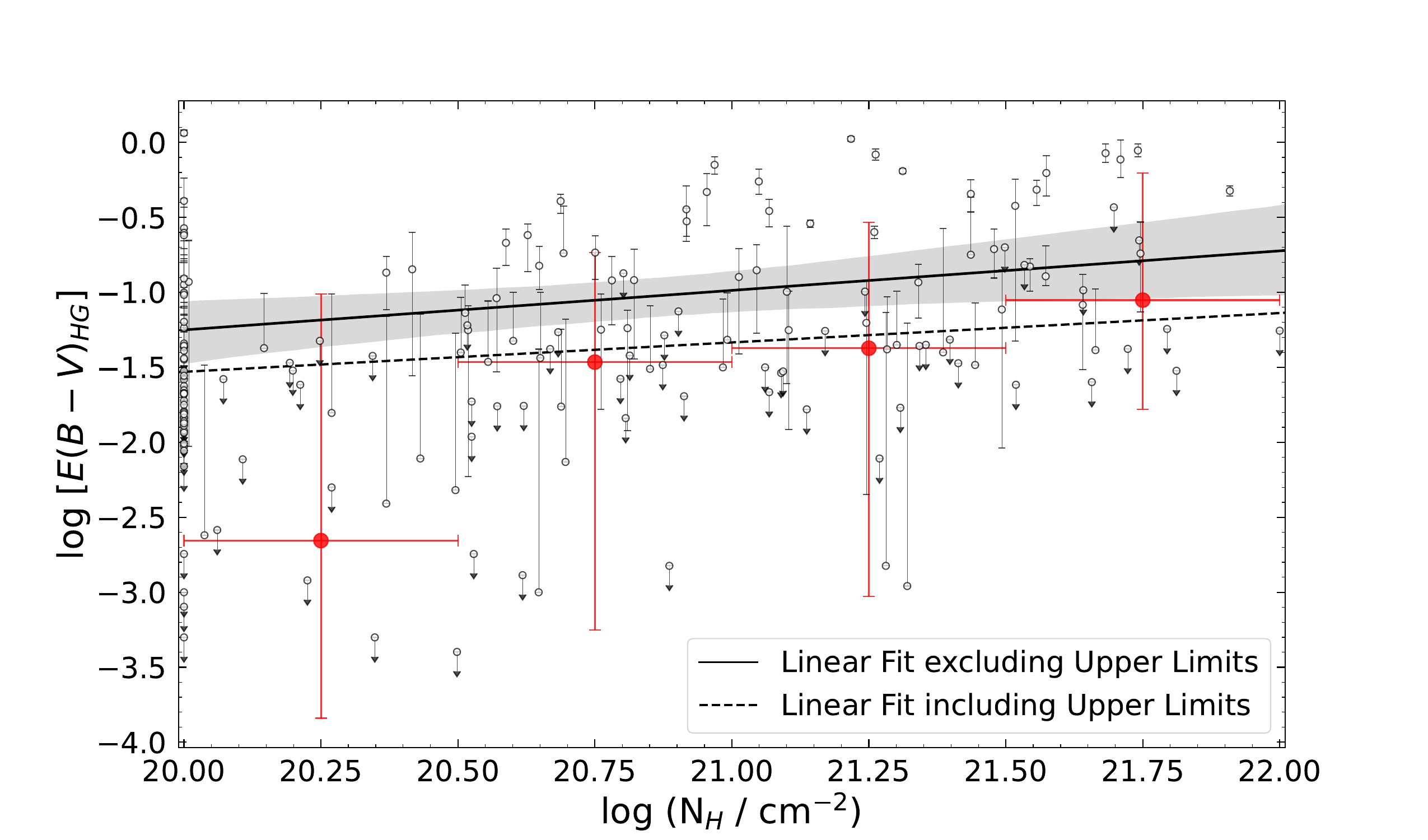}
    \caption{Host galaxy dust extinction in the optical/UV [$E(B-V)_{\rm HG}$] as a function of the X-ray column density ($N_{\rm H}$). Upper limits in $E(B-V)_{\rm HG}$ are shown as downward arrows. The solid black line shows the weak correlation between the two quantities if we exclude the upper limits (Sect. \ref{sect:ebv}, Table \ref{tab:kt}) and the shaded gray region represents the uncertainty on the regression relation. The red points are the median $E(B-V)_{\rm HG}$ values (including upper limits) in bins of $N_{\rm H}$ and the dashed black line is the linear fit using survival analysis to include the contribution from all upper limits.}
    \label{fig:ebv_vs_nh}
\end{figure}


\subsubsection{Host galaxy dust extinction} \label{sect:ebv}

One of the components (\texttt{ZDUST}) of our optical/UV SED model takes into account the intrinsic dust extinction of the source emission due to its host galaxy. This is quantified by the parameter $E(B-V)_{\rm HG}$. For 188 sources ($\sim 80\%$), our SED fitting was able to constrain the value of $E(B-V)_{\rm HG}$ (with 90 upper limits), while for the rest it was either obtained or fixed to be zero (see Sect. \ref{sect:sed} for details). We use survival analysis (e.g., \citealp{1985ApJ...293..192F}; \citealp{2017MNRAS.466.3161S}) to include the effects of upper limits while calculating the correlations in this section. In Fig. \ref{fig:ebv}, we show the distribution of $E(B-V)_{\rm HG}$ and also how it changes with redshift. We do not find any dependence of the intrinsic dust extinction experienced by the disk emission on the redshift. The Fig. \ref{fig:ebv_vs_z} only shows the linear relation (in black) obtained after excluding the upper limits, as their inclusion also gives a similar result. We also checked if $E(B-V)_{\rm HG}$ correlates with any other physical properties of the AGN, such as $L_{\rm bol}$, $M_{\rm BH}$, and $\lambda_{\rm Edd}$ but did not find any significant trends.

While in the optical/UV, $E(B-V)_{\rm HG}$ quantifies the extent of extinction undergone by the overall AGN disk emission, in the X-rays the line-of-sight column density ($N_{\rm H}$) describes the level of obscuration experienced by the more compact corona emission produced near the innermost portions of the accretion disk. To determine if there exists any link between the two wavelength-dependent forms of obscuration, we plot them against each other in Fig. \ref{fig:ebv_vs_nh} and obtain a weak positive correlation between them (see Table \ref{tab:kt}). We show both linear fits, with or without upper limits included, in the figure. It is worth mentioning here that even if both parameters quantify the level of obscuration of the emission at respective energies, considering that this emission originates in different regions of the AGN (at different physical scales), the material responsible for the extinction or obscuration could be different. Additionally, the lack of a strong correlation could also be rooted in the different spatial extents of the obscurers responsible in each waveband. The X-ray corona can be affected by individual BLR clouds (e.g., \citealp{2002ApJ...571..234R}; \citealp{2012AdAst2012E..17B}), dust clumps in the torus (e.g., \citealp{2014MNRAS.439.1403M}; \citealp{2017NatAs...1..679R}), in addition to the gas and dust surrounding the AGN, whereas the reddening of the disk requires more large-scale and extended structures. 


\begin{table}
\centering
\caption{Correlation results for Figs. \ref{fig:kT_vs_z}, \ref{fig:kT2}, \ref{fig:ebv_vs_z}, and \ref{fig:ebv_vs_nh}, described in Sect. \ref{sect:disk}.}
\begin{tabular}{cccc}
\hline
\hline
\specialrule{0.1em}{0em}{0.5em}
\vspace{1mm}
Correlation & R-Value\tablefootmark{a} & P-Value\tablefootmark{b} & Scatter\\
\hline\\
\vspace{2mm}
$kT_{\rm max}-z$ & \,\,\,\,0.32 & $1.76 \times 10^{-6}$ & 0.29\\
\vspace{2mm}
$kT_{\rm max}-L_{\rm bol}$ & \,\,\,\,0.47 & \,\,$4.32\times10^{-13}$ & 0.27\\
\vspace{2mm}
$kT_{\rm max}-\lambda_{\rm Edd}$ & \,\,\,\,0.45 & \,\,$9.95\times10^{-12}$ & 0.27\\
\vspace{2mm}
$E(B-V)_{\rm HG}-z$ & $-0.02$ & 0.85 & -\\
\vspace{2mm}
$E(B-V)_{\rm HG}-N_{\rm H}$ & \,\,\,\,0.25 & 0.02 & -\\

\hline
\end{tabular}
\tablefoot{We also report the one sigma dispersion (or scatter in dex) for parameters with significant correlations.\\
\tablefoottext{a}{The Pearson's correlation coefficient.}\\
\tablefoottext{b}{The probability of the data set appearing if the null hypothesis is correct.}}
\label{tab:kt}
\end{table}


\subsection{Optical-to-X-ray spectral index ($\alpha_{\rm ox}$)}\label{sect:aox}

The relation between the optical/UV and X-ray emission of AGN has long been an area of extensive investigation to better understand the energy generation mechanism in AGN (e.g., \citealp{1979ApJ...234L...9T}; \citealp{1982ApJ...262L..17A}; \citealp{1986ApJ...305...83A}; \citealp{1994ApJS...92...53W}; \citealp{1998A&A...334..498Y}). In this context, the optical-to-X-ray spectral index or $\alpha_{\rm ox}$ is widely used to explore how the optical/UV disk emission couples with the X-ray coronal emission of AGN (e.g., \citealp{2016ApJ...819..154L}). This parameter can be used to quantify the shape of the AGN SED and as a proxy for bolometric corrections, since it is relatively easy to calculate from limited data. $\alpha_{\rm ox}$ is defined as:

\begin{equation} \label{eq:aox}
    -0.3838\times{\rm log}\,[L_{\nu}(2500\,{\rm\AA})/L_{\nu}(\rm 2\,keV)]
\end{equation}\\
where $L_\nu$ is the luminosity at the specific wavelength or energy in units of ${\rm erg\,s^{-1}\,{\rm Hz}^{-1}}$ (e.g., \citealp{2005AJ....130..387S}; \citealp{2006AJ....131.2826S}). Over the years, many studies have been carried out focusing on how $\alpha_{\rm ox}$ evolves with redshift, and the AGN luminosity. Although no strong dependence has been reported with redshift (however, also refer to \citealp{2003ApJ...588..119B} and \citealp{2024MNRAS.527.9004R}), studies have shown that $\alpha_{\rm ox}$ has a significant anticorrelation with UV luminosity (e.g., \citealp{2005AJ....130..387S}; \citealp{2006AJ....131.2826S}; \citealp{2009MNRAS.392.1124V}; \citealp{2010A&A...512A..34L}; \citealp{2010ApJS..187...64G}; \citealp{2023MNRAS.523..646T}). Some studies have also investigated if the way we define $\alpha_{\rm ox}$ (using the 2500\,$\rm \AA$ and 2\,keV luminosities) is the most suitable, considering the purely historical choice of the two points (e.g., \citealp{2010ApJ...708.1388Y}; \citealp{2023A&A...676A.143S}; \citealp{2024MNRAS.527..356J}). We explore this issue briefly at the end of this section.

To calculate $\alpha_{\rm ox}$ using Eq. (\ref{eq:aox}), we estimate the monochromatic fluxes and luminosities at 2500\,$\rm \AA$ and 2\,keV using the best-fit models obtained from the broadband SED fitting. We first have a look at the well-established correlation between the two luminosities ($L_{2500\,{\rm \AA}}$ and $L_{\rm 2\,keV}$). Most studies have narrowed down the correlation between the UV and X-ray luminosity to $L_{\rm 2\,keV} \propto L_{\rm 2500\,\rm \AA}^\beta$, where $\beta$ ranges between 0.7 to 0.9 (e.g., \citealp{1986ApJ...305...83A}; \citealp{1994ApJS...92...53W}; \citealp{2005AJ....130..387S}; \citealp{2006AJ....131.2826S}; \citealp{2007ApJ...665.1004J}; \citealp{2010A&A...512A..34L}; \citealp{2010MNRAS.401..294S}; \citealp{2012A&A...539A..48M}). Slightly larger values for $\beta$ ($>0.9$) have been reported by some works (e.g., \citealp{2009ApJ...690..644G}; \citealp{2012MNRAS.425..907J}). This nonlinear correlation between the X-ray and the UV luminosity implies that optically bright AGN emit fewer X-rays (per UV luminosity) compared to optically faint sources (e.g., \citealp{1982ApJ...262L..17A}). This relation between the UV and X-ray luminosity is extremely useful to investigate possible links between the accretion disk and the hot corona, respectively. Additionally, the strong X-ray-UV luminosity relation is also fundamental for recent cosmological studies that intend to use quasars as standard candles and extend the Hubble diagram to higher redshifts (e.g., \citealp{2015ApJ...815...33R}; \citealp{2017A&A...602A..79L}; \citealp{2020A&A...642A.150L}; \citealp{2023A&A...676A.143S}).

\begin{figure}
    \centering
    \includegraphics[width=0.52\textwidth]{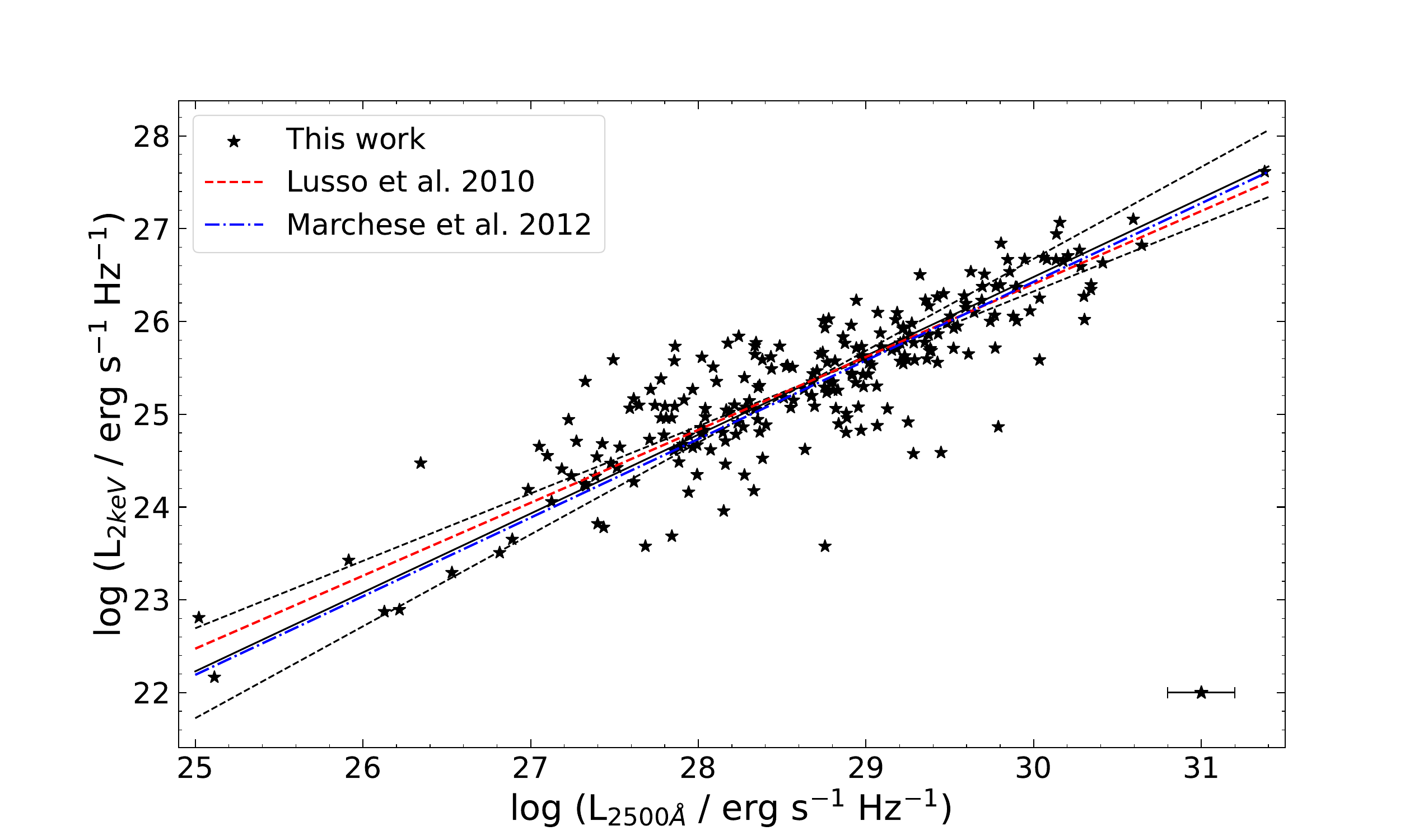}
    \caption{Monochromatic luminosity at 2500\,$\rm\AA$ vs. the monochromatic luminosity at 2\,keV. The two best-fit relations (Eqs. \ref{eq:l2keV_l2500_1} and \ref{eq:l2keV_l2500_2}) and the final bisector regression from our work (Eq. \ref{eq:l2keV_l2500}) are shown as dashed and solid black lines, respectively. We also show the relation reported by \citeauthor{2010A&A...512A..34L} (\citeyear{{2010A&A...512A..34L}}; for their photometrically selected sample) and \citet{2012A&A...539A..48M} as a dashed red line and a dash-dotted blue line, respectively. The typical uncertainties on $L_{\rm 2\,keV}$ are less than 5\% and those on $L_{\rm 2500\,\AA}$ are shown in the bottom right corner (also valid for subsequent plots including $L_{\rm 2\,keV}$ and $L_{\rm 2500\,\AA}$).}
    \label{fig:l2500_vs_l2keV}
\end{figure}

We estimate the best fit regression relation between $L_{2500\,\rm \AA}$ and $L_{\rm 2\,keV}$ using the Python package \texttt{linmix}, and by including the errors on the dependent variable. We first treat $L_{\rm 2500\,\AA}$ as the independent variable and get:

\begin{equation}\label{eq:l2keV_l2500_1}
    {\rm log}\,(L_{\rm 2\,keV}) = (0.726 \pm 0.029)\times {\rm log}\,(L_{\rm 2500\,\AA}) + (4.544 \pm 0.821)
\end{equation}\\
We then estimate the same relation by considering $L_{\rm 2\,keV}$ as the independent variable and find:

\begin{equation}\label{eq:l2keV_l2500_2}
    {\rm log}\,(L_{\rm 2\,keV}) = (0.990 \pm 0.039)\times {\rm log}\,(L_{\rm 2500\,\AA}) + (3.025 \pm 1.006)
\end{equation}\\
The final relation is the bisector of the above two regressions (as was described by \citealp{1990ApJ...364..104I}):

\begin{equation}\label{eq:l2keV_l2500}
    {\rm log}\,(L_{\rm 2\,keV}) = (0.850 \pm 0.013)\times {\rm log}\,(L_{\rm 2500\,\AA}) + 0.985
\end{equation}\\
In Fig. \ref{fig:l2500_vs_l2keV}, we plot the rest-frame monochromatic luminosity at 2\,keV as a function of the rest-frame monochromatic luminosity at 2500\,$\rm \AA$ along with the three best-fit relations mentioned above. When comparing our results with the literature, it is difficult to find an agreement with all previous works considering the wide range of values that have been reported for the slope of this correlation. We find that our results are the most consistent with \citeauthor{2010MNRAS.401..294S} (\citeyear{2010MNRAS.401..294S}; $\beta=0.870 \pm 0.001$) and \citeauthor{2012A&A...539A..48M} (\citeyear{2012A&A...539A..48M}; $\beta=0.847 \pm 0.036$) for their sample of X-ray-selected type I AGN. However, in comparison with other works that found $\beta < 0.8$ (e.g., 
\citealp{2006AJ....131.2826S}; $\beta=0.721 \pm 0.011$; \citealp{2007ApJ...665.1004J}; $\beta = 0.709 \pm 0.010$; \citealp{2010A&A...512A..34L}; $\beta = 0.760 \pm 0.022$), we find a steeper correlation between the X-ray and UV luminosity. This discrepancy in the computed slope could be attributed to several factors including selection effects, fitting techniques, variability, sample size, redshift range, etc. For example, our sample consists of hard-X-ray-selected AGN, further classified as unobscured based on their X-ray column density, whereas, most of the studies that obtained a shallower slope focused on optically selected AGN (e.g., \citealp{2006AJ....131.2826S}; \citealp{2007ApJ...665.1004J}). It has been discussed by \citet{1998A&A...334..498Y} and \citet{2007MNRAS.377.1113T} that intrinsic dispersions in the optical and X-ray luminosity can affect the overall correlation in optically selected, flux-limited samples. The slope quoted by \citeauthor{2009ApJ...690..644G} (\citeyear{2009ApJ...690..644G}; $\beta = 1.117\pm0.017$) is for their sample of optically selected quasars that have a low X-ray detection factor. However, for their sub-sample with a 100\% detection rate in the X-rays (zLxBox), they obtain a slope ($\beta=0.842\pm0.017$) consistent with ours. When the X-ray-selected sample studied by \citet{2010A&A...512A..34L} is further separated into spectroscopically and photometrically classified sources, they obtain a larger value for $\beta$ ($0.782\pm0.033$ and $0.786\pm0.033$), closer to our estimates. 


\begin{figure*} 
  \begin{subfigure}[t]{0.54\textwidth}
    \centering
    \includegraphics[width=\textwidth]{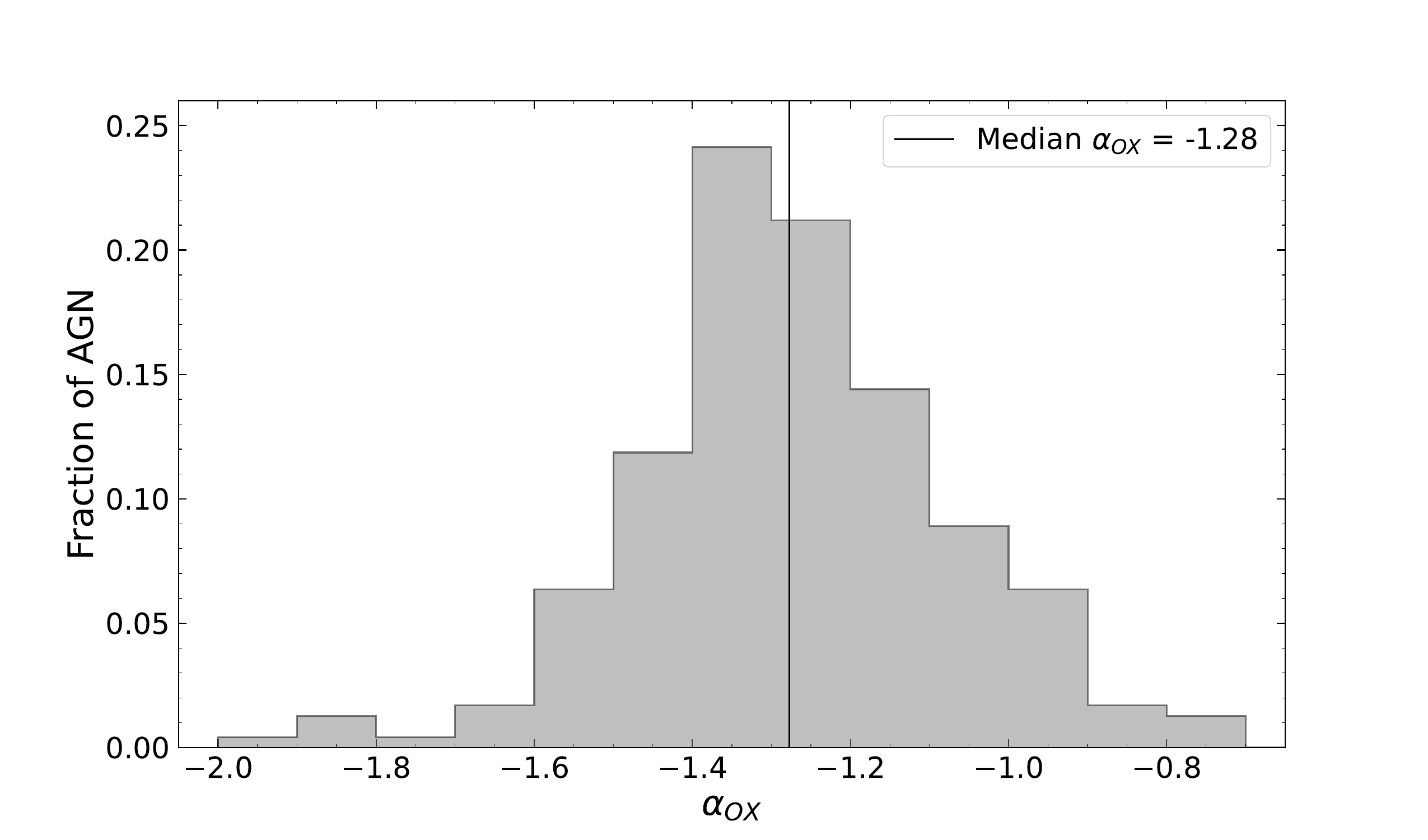}
    \caption{}
    \label{fig:aox}
  \end{subfigure}
  \hspace{-1.cm}
  \begin{subfigure}[t]{0.54\textwidth}
    \centering
    \includegraphics[width=\textwidth]{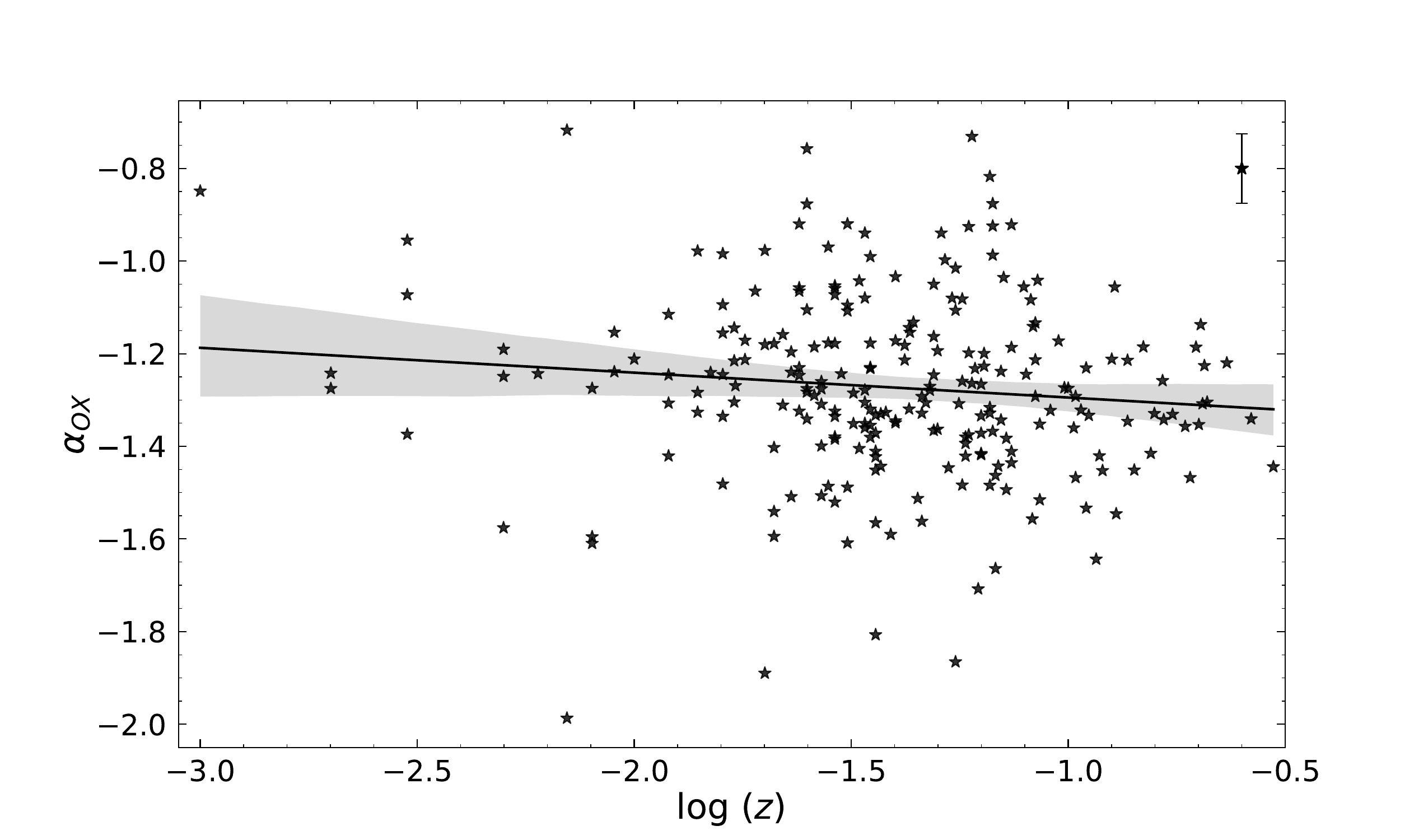}
    \caption{}
    \label{fig:aox_vs_z}
  \end{subfigure}
\caption{Optical-to-X-ray spectral index ($\alpha_{\rm ox}$) for our sample of hard-X-ray-selected nearby, unobscured AGN. (a) Distribution of $\alpha_{\rm ox}$ and (b) $\alpha_{\rm ox}$ as a function of redshift. The solid black line shows the best-fit linear relation to the data and the shaded gray region shows the uncertainty in the linear regression. We do not find any significant correlation between the two parameters. The typical errors on $\alpha_{\rm ox}$ are shown on the top right corner (also valid for subsequent plots including $\alpha_{\rm ox}$).}  
\label{fig:aox1}
\end{figure*}


\begin{figure*} 
  \begin{subfigure}[t]{0.54\textwidth}
    \centering
    \includegraphics[width=\textwidth]{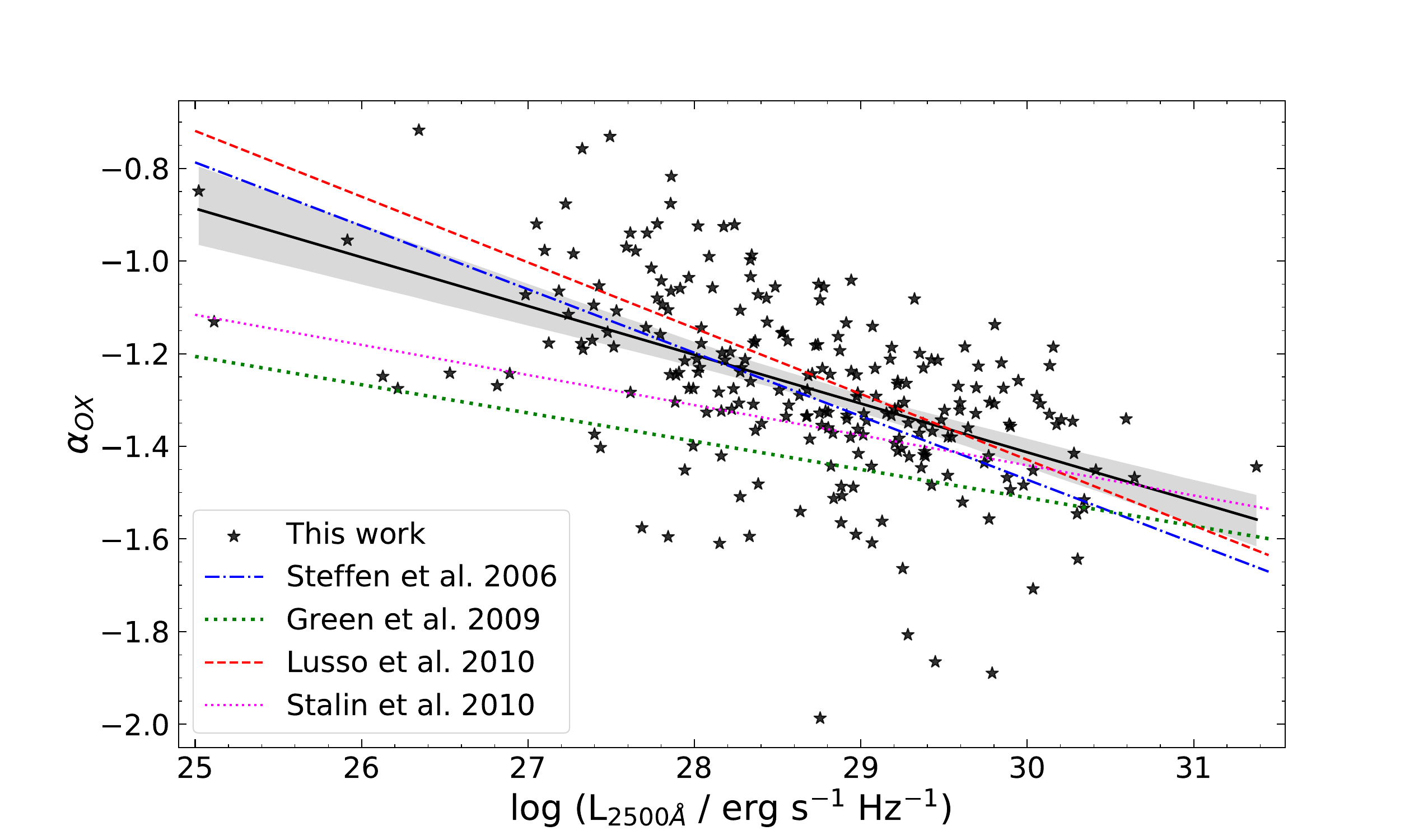}
    \caption{}
    \label{fig:aox_vs_l2500}
  \end{subfigure}
  \hspace{-1.cm}
  \begin{subfigure}[t]{0.54\textwidth}
    \centering
    \includegraphics[width=\textwidth]{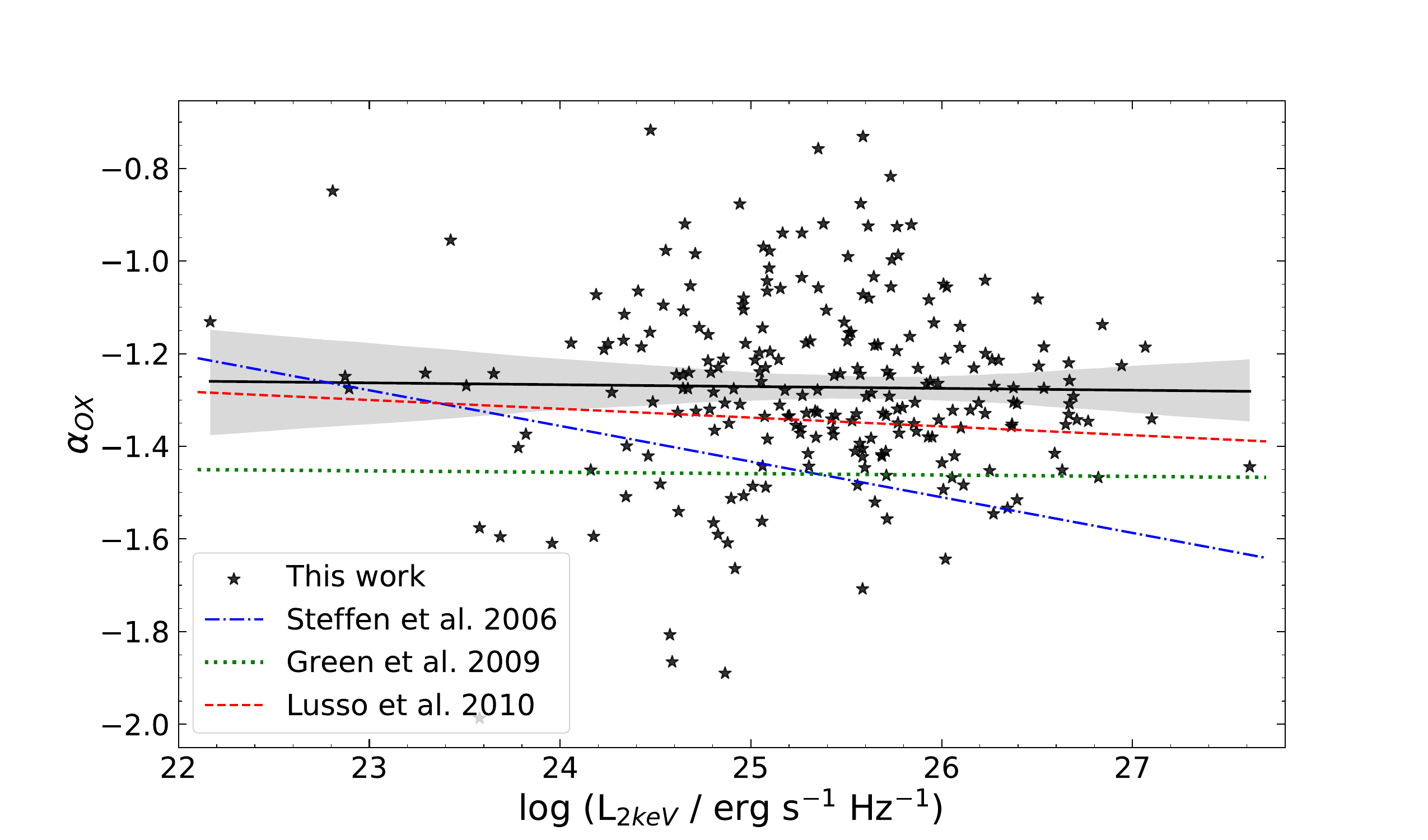}
    \caption{}
    \label{fig:aox_vs_l2keV}
  \end{subfigure}
\caption{$\alpha_{\rm ox}$ as a function of the rest-frame monochromatic luminosity at (a) 2500\,$\rm \AA$ (UV) and (b) 2\,keV (X-ray). We find a tight anticorrelation between $\alpha_{\rm ox}$ and $L_{\rm 2500\,\AA}$ (shown as the solid black line in the left panel) but no significant relation with $L_{\rm 2\,keV}$ (shown as a solid black line in the right panel). For comparison, we also plot the best-fit relations from \citeauthor{2006AJ....131.2826S} (\citeyear{2006AJ....131.2826S}; dash-dotted blue line), \citeauthor{2009ApJ...690..644G} (\citeyear{2009ApJ...690..644G}; thick dotted green line), \citeauthor{2010A&A...512A..34L} (\citeyear{2010A&A...512A..34L}; dashed red line), and \citeauthor{2010MNRAS.401..294S} (\citeyear{2010MNRAS.401..294S}; dotted magenta line). The shaded gray region shows the one sigma confidence interval for our best-fit linear regression.} 
\label{fig:aox2}
\end{figure*}


\begin{figure*} 
  \begin{subfigure}[t]{0.54\textwidth}
    \centering
    \includegraphics[width=\textwidth]{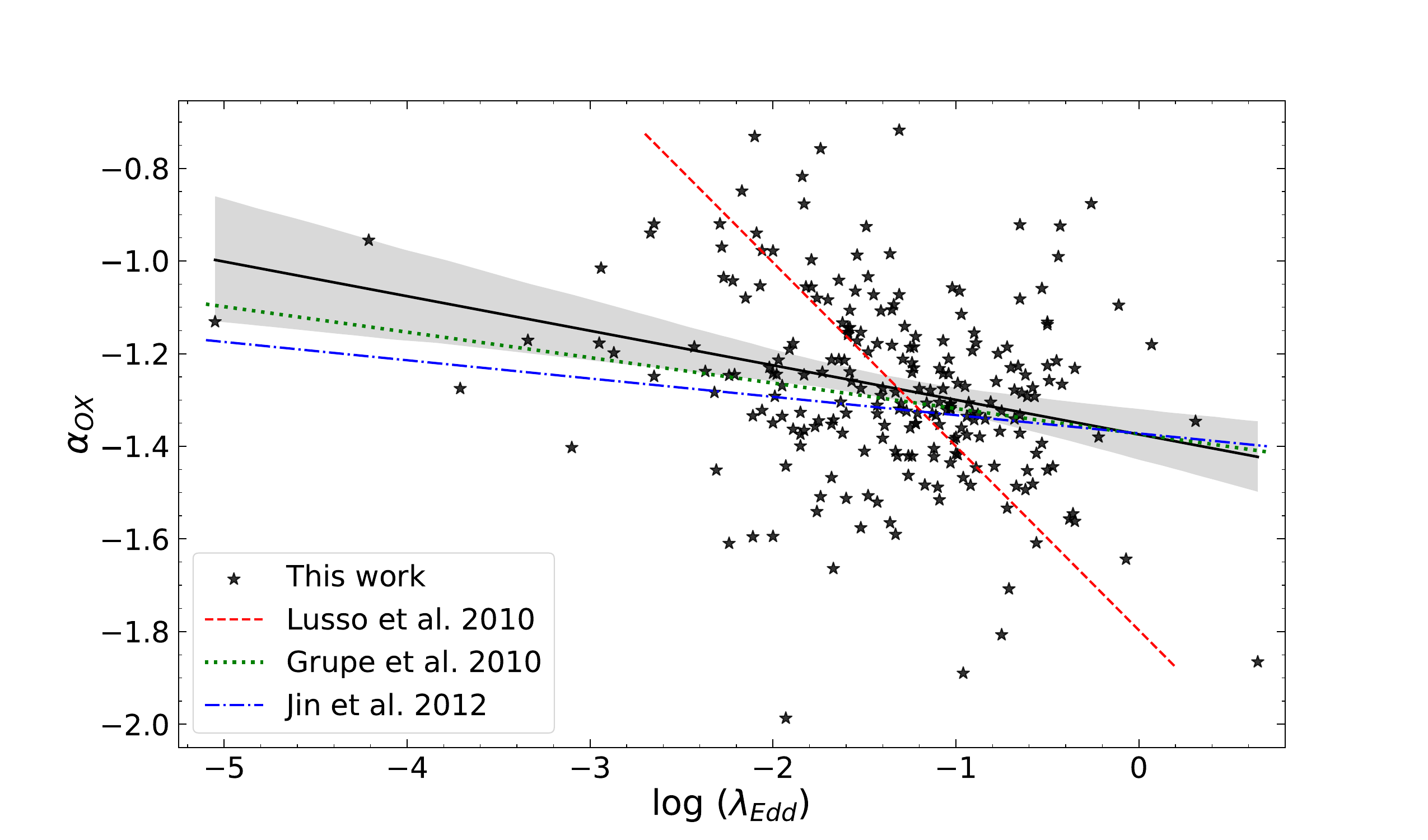}
    \caption{}
    \label{fig:aox_vs_ER}
  \end{subfigure}
  \hspace{-1.cm}
  \begin{subfigure}[t]{0.54\textwidth}
    \centering
    \includegraphics[width=\textwidth]{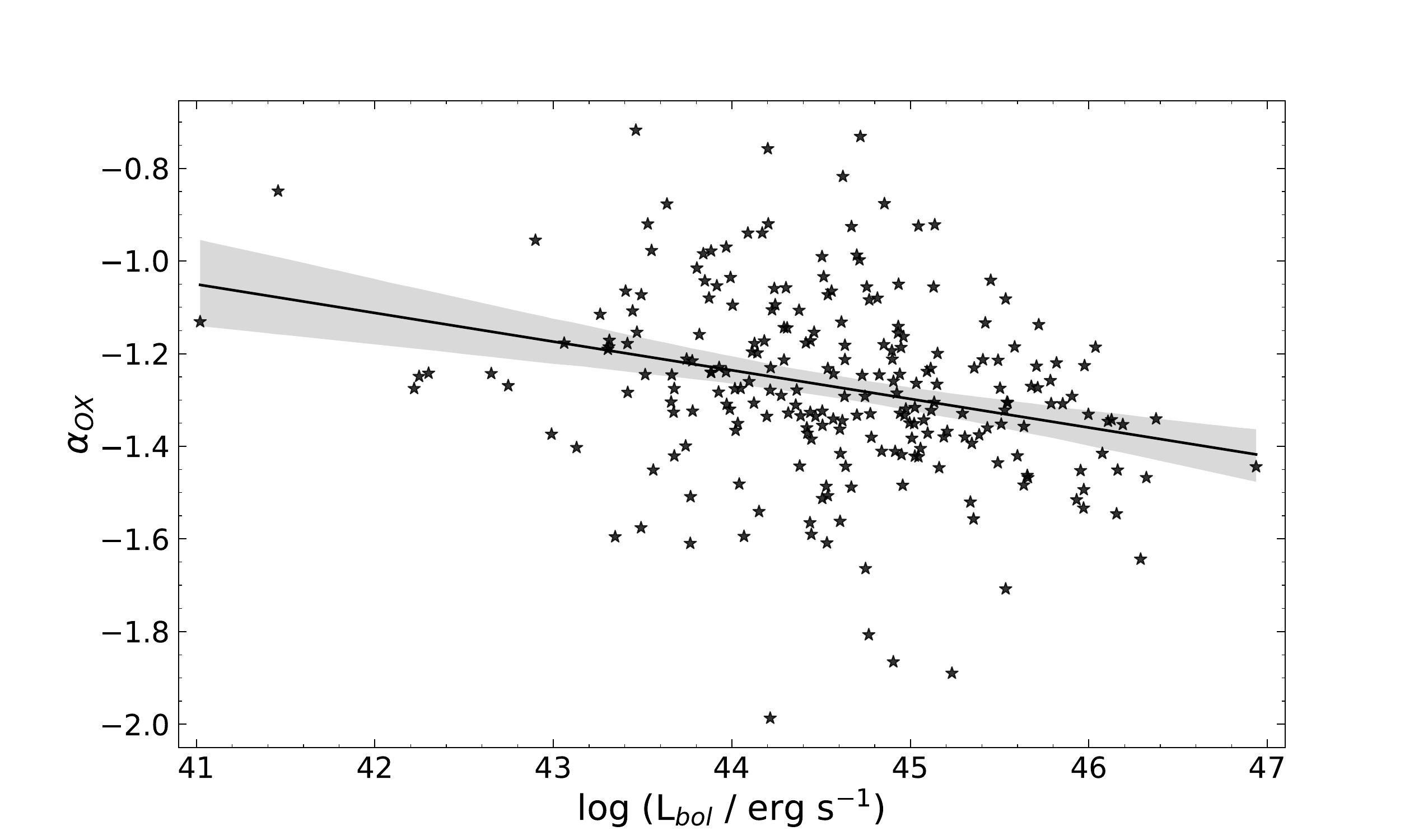}
    \caption{}
    \label{fig:aox_vs_lbol}
  \end{subfigure}
\caption{$\alpha_{\rm ox}$ as a function of (a) Eddington ratio and (b) bolometric luminosity. We find a strong anticorrelation of $\alpha_{\rm ox}$ with both the parameters (see Table \ref{tab:aox}). The solid black lines show our best-fit relations and the shaded gray regions are the one sigma confidence interval for each relation. Our linear relation between $\alpha_{\rm ox}$ and $\lambda_{\rm Edd}$ is flatter when compared to the one reported by \citeauthor{2010A&A...512A..34L} (\citeyear{2010A&A...512A..34L}; dashed red line), but consistent with \citeauthor{2010ApJS..187...64G} (\citeyear{2010ApJS..187...64G}; dotted green line) and \citeauthor{2012MNRAS.425..907J} (\citeyear{2012MNRAS.425..907J}; dash-dotted blue line).} 
\label{fig:aox3}
\end{figure*}


Next, we plot the distribution in $\alpha_{\rm ox}$ for our AGN sample in Fig. \ref{fig:aox}. We obtain $\alpha_{\rm ox}$ values ranging from $-2.0$ to $-0.7$, with a median at $-1.28$ (also see Table \ref{tab:params}). We do not see any significant trend in $\alpha_{\rm ox}$ with redshift (Fig. \ref{fig:aox_vs_z}, Table \ref{tab:aox}) as confirmed by most previous works (but see \citealp{1998A&A...334..498Y} and \citealp{2003ApJ...588..119B}). Following the strong, nonlinear relation established between the X-ray and UV luminosity, we expect the ratio of these luminosities to correlate with the UV luminosity. In Fig. \ref{fig:aox_vs_l2500}, we show $\alpha_{\rm ox}$ as a function of the monochromatic 2500\,$\rm \AA$ luminosity and the best-fit relation computed as follows:

\begin{equation}\label{eq:aox_l2500}
    \alpha_{\rm ox} = (-0.105 \pm 0.011) \times {\log}\,(L_{\rm 2500\,\AA}) + (1.745 \pm 0.315)
\end{equation}\\
\citet{2010ApJS..187...64G} found a similar slope ($-0.114\pm0.014$) for their sample of soft-X-ray-selected AGN with simultaneous optical-to-X-ray SEDs. This relation is, however, shallower compared to the one estimated by \citeauthor{2006AJ....131.2826S} (\citeyear{2006AJ....131.2826S}; $-0.137 \pm 0.008$) and \citeauthor{2010A&A...512A..34L} (\citeyear{2010A&A...512A..34L}; $-0.154 \pm 0.010$), and steeper compared to \citeauthor{2009ApJ...690..644G} (\citeyear{2009ApJ...690..644G}; $-0.061\pm0.009$) and \citeauthor{2010MNRAS.401..294S} (\citeyear{2010MNRAS.401..294S}; $-0.065\pm0.019$). Additionally, when compared with higher redshift samples (up to $z\sim5$), our slope is shallower than those reported by \citeauthor{2020MNRAS.492..719T}(\citeyear{2020MNRAS.492..719T}; $-0.199\pm0.011$) and \citeauthor{2021MNRAS.504.5556T}(\citeyear{2021MNRAS.504.5556T}; $-0.179\pm0.013$). Apart from sample selection and redshift effects mentioned previously, the difference in slopes could also be attributed to the fact that we cover lower luminosity regimes with our AGN sample. We have sources going down to $10^{25}\,{\rm erg\,s^{-1}\,Hz^{-1}}$, with $\sim 25\%$ sources below $10^{28}\,{\rm erg\,s^{-1}\,Hz^{-1}}$, which is almost three orders of magnitude below the luminosity range covered by most other works. In fact, we get a slightly shallower slope ($-0.097\pm0.018$), which is comparable with \citet{2010MNRAS.401..294S}, if we exclude the low luminosity sources ($< 10^{28}\,{\rm erg\,s^{-1}\,Hz^{-1}}$). \citet{2010A&A...512A..34L} also found a lower slope ($-0.142\pm0.012$) for their photometrically selected sample that extends to lower luminosity values. This effect of having low luminosity sources in our sample is also prominent in Fig. \ref{fig:aox_vs_l2keV}, where we plot $\alpha_{\rm ox}$ as a function of the 2\,keV monochromatic luminosity. In agreement with other works, we do not find any significant correlation between $\alpha_{\rm ox}$ and $L_{\rm 2\,keV}$:

\begin{equation}\label{eq:aox_l2keV}
    \alpha_{\rm ox} = (-0.004 \pm 0.015) \times {\rm log}\,(L_{\rm 2\,keV}) + (-1.172 \pm 0.388) 
\end{equation}\\
However, even in this case, we go down to luminosities of $10^{22}\,{\rm erg\,s^{-1}\,Hz^{-1}}$, which is up to two orders magnitudes lower when compared to \cite{2006AJ....131.2826S} and \cite{2010A&A...512A..34L}. This could explain the flatter slope we obtain for our best-fit regression relation compared to those inferred by \citeauthor{2006AJ....131.2826S} (\citeyear{2006AJ....131.2826S}; $-0.077 \pm 0.015$) and \citeauthor{2010A&A...512A..34L} (\citeyear{2010A&A...512A..34L}; $-0.019 \pm 0.013$). Our results are consistent with \citet{2009ApJ...690..644G}, who found a slope of $-0.003 \pm 0.010$ for their main sample, but with a different intercept value ($-1.384\pm0.261$).

Lastly, we check if there exists any dependence of $\alpha_{\rm ox}$ on the Eddington ratio. A correlation between $\alpha_{\rm ox}$ and $\lambda_{\rm Edd}$ was first reported by \citeauthor{2010A&A...512A..34L} (\citeyear{2010A&A...512A..34L}; although with a large dispersion), where they found that sources with higher $\lambda_{\rm Edd}$ showed higher ratios of UV to X-ray luminosity. This correlation was further confirmed by \citet{2010ApJS..187...64G} and \citet{2012MNRAS.425..907J}, but with a flatter slope and a higher significance. On the contrary, \cite{2009MNRAS.399.1553V} did not find any such relation for their sample of hard-X-ray-selected AGN. We illustrate the relation between $\alpha_{\rm ox}$ and $\lambda_{\rm Edd}$ for our sample in Fig. \ref{fig:aox_vs_ER}. We obtain a significant negative correlation (see Table \ref{tab:aox}) between the two quantities, where the best-fit linear regression is given as:

\begin{equation}\label{eq:aox_edd}
    \alpha_{\rm ox} = (-0.075 \pm 0.017) \times {\rm log}\,(\lambda_{\rm Edd}) + (-1.374 \pm 0.026)
\end{equation}\\
Our findings are in agreement with those of \citeauthor{2010ApJS..187...64G} (\citeyear{2010ApJS..187...64G}; $-0.11\pm0.02$) and \citeauthor{2012MNRAS.425..907J} (\citeyear{2012MNRAS.425..907J}; $-0.079\pm0.038$). We find a shallower slope for the correlation, compared to \citeauthor{2010A&A...512A..34L} (\citeyear{2010A&A...512A..34L}; $-0.133\pm0.023$), but with larger significance ($p$-value = $2.42 \times 10^{-5}$). A possible reason for this could be that our sample consists of many low UV luminosity sources (similar to \citealp{2010ApJS..187...64G} and \citealp{2012MNRAS.425..907J}), whereas the sample analyzed by \citet{2010A&A...512A..34L} are majorly bright in the UV. As a result, they have many sources with $\alpha_{\rm ox} < -1.5$ further increasing the dispersion in their relation.  When compared to \cite{2009MNRAS.399.1553V} who analyzed a sample of 26 low-absorption AGN, we have much better statistics ($\sim1$ dex improvement) that allows us to constrain this possible anticorrelation better. However, considering the large scatter visible in Fig. \ref{fig:aox_vs_ER} and also reported in previous works, one should be cautious when using the $\alpha_{\rm ox}-\lambda_{\rm Edd}$ relation (Eq. \ref{eq:aox_edd}) to estimate $\lambda_{\rm Edd}$. We also investigated whether $\alpha_{\rm ox}$ correlates with either the black hole mass ($M_{\rm BH}$) or the bolometric luminosity ($L_{\rm bol}$). We do not find any significant correlation of $\alpha_{\rm ox}$ with $M_{\rm BH}$. However, we obtain an anticorrelation between $\alpha_{\rm ox}$ and $L_{\rm bol}$ (Fig. \ref{fig:aox_vs_lbol}), with the best-fit relation as follows:

\begin{equation}\label{eq:aox_lbol}
    \alpha_{\rm ox} = (-0.062 \pm 0.014) \times {\rm log}\,(L_{\rm bol}) + (1.486 \pm 0.617)
\end{equation}\\
This correlation is similar to the one obtained between $\alpha_{\rm ox}$ and $\lambda_{\rm Edd}$ in terms of the slope (0.062 and 0.075), intercept (1.486 and 1.374), dispersion ($\sim 0.2$), and significance ($\sim 10^{-5}$), suggesting that the underlying physics responsible for these two relations could be the same and possibly related to the energy generation mechanism of AGN, where more luminous sources (with higher Eddington ratios, up to the Eddington limit) emit more in the UV relative to their X-ray emission.


\begin{figure}
    \centering
    \includegraphics[width=0.52\textwidth]{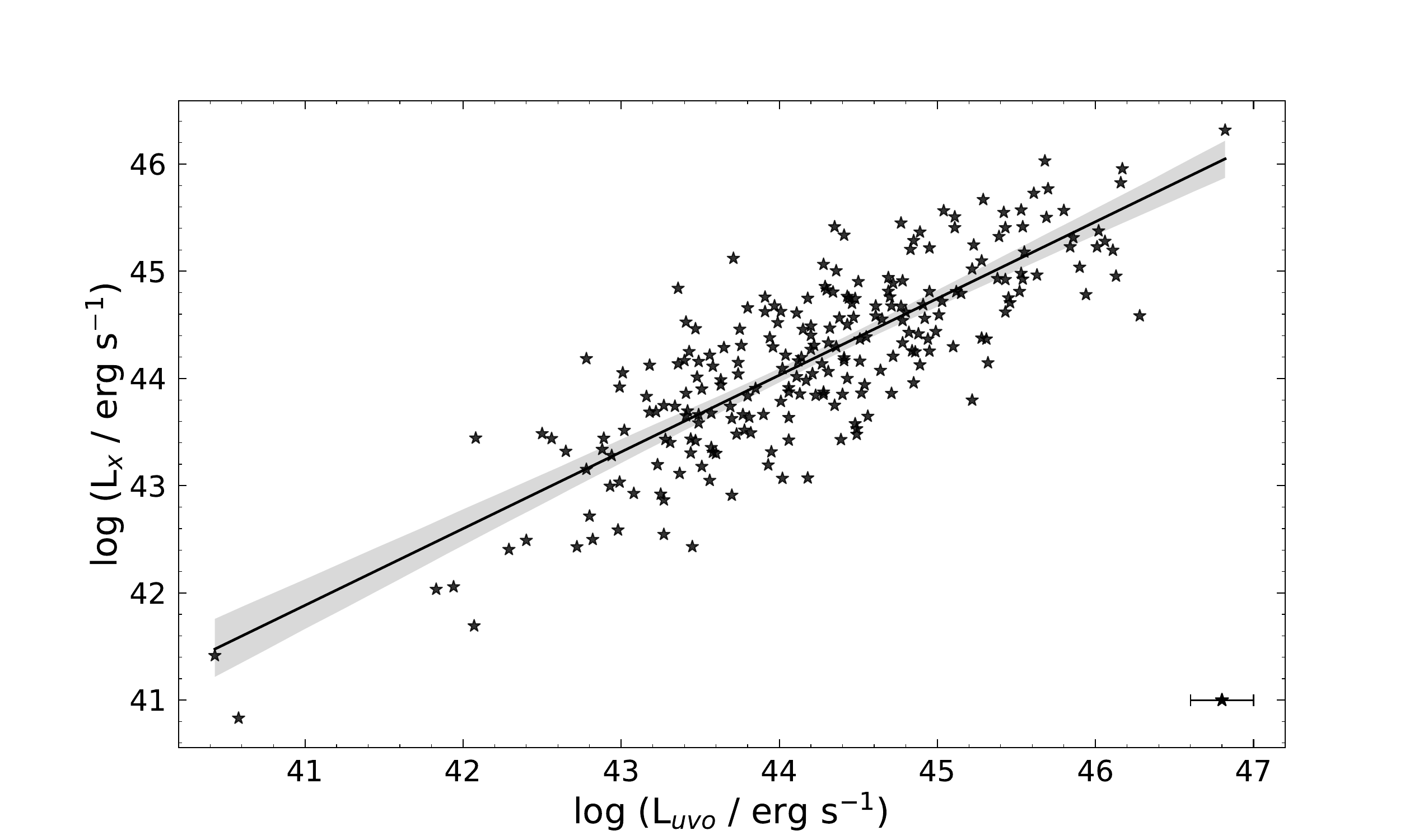}
    \caption{Optical/UV disk luminosity ($L_{\rm uvo}$) against the X-ray luminosity ($L_{\rm X}$). The solid black line shows our best-fit relation and the shaded gray region is the one sigma confidence interval. The typical errors on $L_{\rm x}$ are less than 5\% and those on $L_{\rm uvo}$ are shown in the bottom right corner.}
    \label{fig:luvo_vs_lx}
\end{figure}


The method of calculating $\alpha_{\rm ox}$ using the luminosities at 2500\,$\AA$ and 2\,keV as representing the disk and coronal emission, respectively goes back to the 1970s (\citealp{1968ApJ...151..393S}). However, over the years, we have managed to acquire observations over the entire electromagnetic spectrum, allowing us to constrain the total disk and coronal emission. Through this work, we provide a linear relation between the two major contributors of AGN emission as follows:

\begin{equation}\label{eq:luvo_lx}
    {\rm log}\,(L_{\rm x}) = (0.716 \pm 0.032) \times {\rm log}\,(L_{\rm uvo}) + (12.543 \pm 1.434)
\end{equation}\\
This equation (also shown in Fig. \ref{fig:luvo_vs_lx}) can be used to estimate either the disk luminosity from X-ray spectral fitting or the X-ray coronal luminosity if a fit disk model is available.


\begin{table}
\centering
\caption{Correlation results for Figs. \ref{fig:l2500_vs_l2keV}, \ref{fig:aox_vs_z}, \ref{fig:aox2}, \ref{fig:aox3}, and \ref{fig:luvo_vs_lx}, described in Sect. \ref{sect:aox}.}
\begin{tabular}{cccc}
\hline
\hline
\specialrule{0.1em}{0em}{0.5em}
\vspace{1mm}
Correlation & R-Value\tablefootmark{a} & P-Value\tablefootmark{b} & Scatter (dex)\\
\hline\\
\vspace{2mm}
$L_{\rm 2\,keV}-L_{\rm 2500\,\AA}$ & \,\,\,\,0.86 & \,\,$4.29 \times 10^{-69}$ & 0.45\\
\vspace{2mm}
$\alpha_{\rm ox}-z$ & $-0.11$ & 0.09 & -\\
\vspace{2mm}
$\alpha_{\rm ox}-L_{\rm 2500\,\AA}$ & $-0.53$ & \,\,$1.46 \times 10^{-18}$ & 0.15\\
\vspace{2mm}
$\alpha_{\rm ox}-L_{\rm 2\,keV}$ & $-0.02$ & 0.80 & -\\
\vspace{2mm}
$\alpha_{\rm ox}-\lambda_{\rm Edd}$ & $-0.27$ & $2.42\times10^{-5}$ & 0.17\\
\vspace{2mm}
$\alpha_{\rm ox}-L_{\rm bol}$ & $-0.28$ & $1.21 \times 10^{-5}$ & 0.17\\
\vspace{2mm}
$L_{\rm uvo}-L_{\rm x}$ & \,\,\,\,0.82 & \,\,$4.02 \times 10^{-59}$ & 0.50\\

\hline
\end{tabular}
\tablefoot{ We also report the one sigma dispersion (or scatter in dex) for parameters with significant correlations.\\
\tablefoottext{a}{The Pearson's correlation coefficient.}\\
\tablefoottext{b}{The probability of the data set appearing if the null hypothesis is correct.}
}
\label{tab:aox}
\end{table}


\begin{figure*} 
  \begin{subfigure}[t]{0.54\textwidth}
    \centering
    \includegraphics[width=\textwidth]{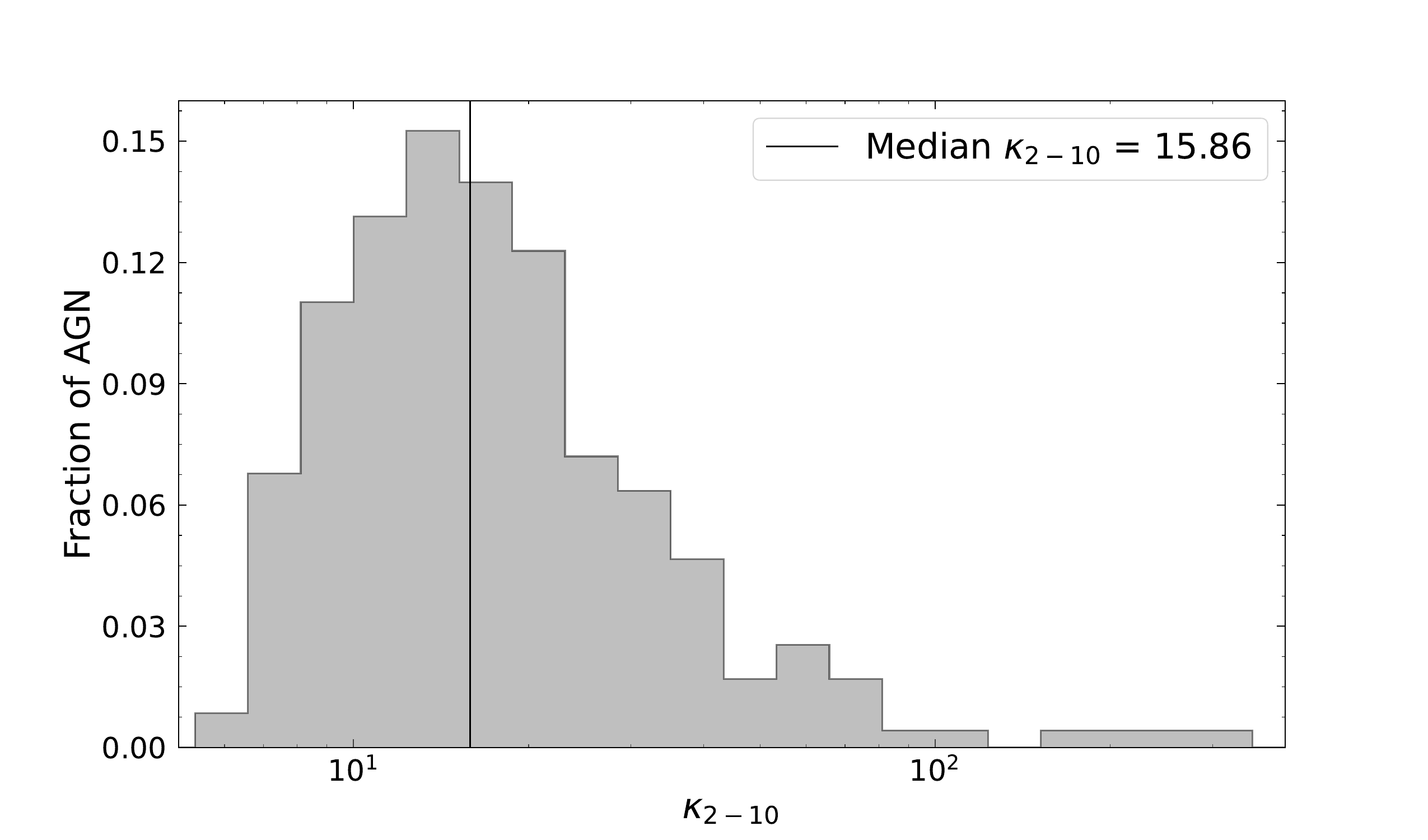}
    \caption{}
    \label{fig:k210_log}
  \end{subfigure}
  \hspace{-1.cm}
  \begin{subfigure}[t]{0.54\textwidth}
    \centering
    \includegraphics[width=\textwidth]{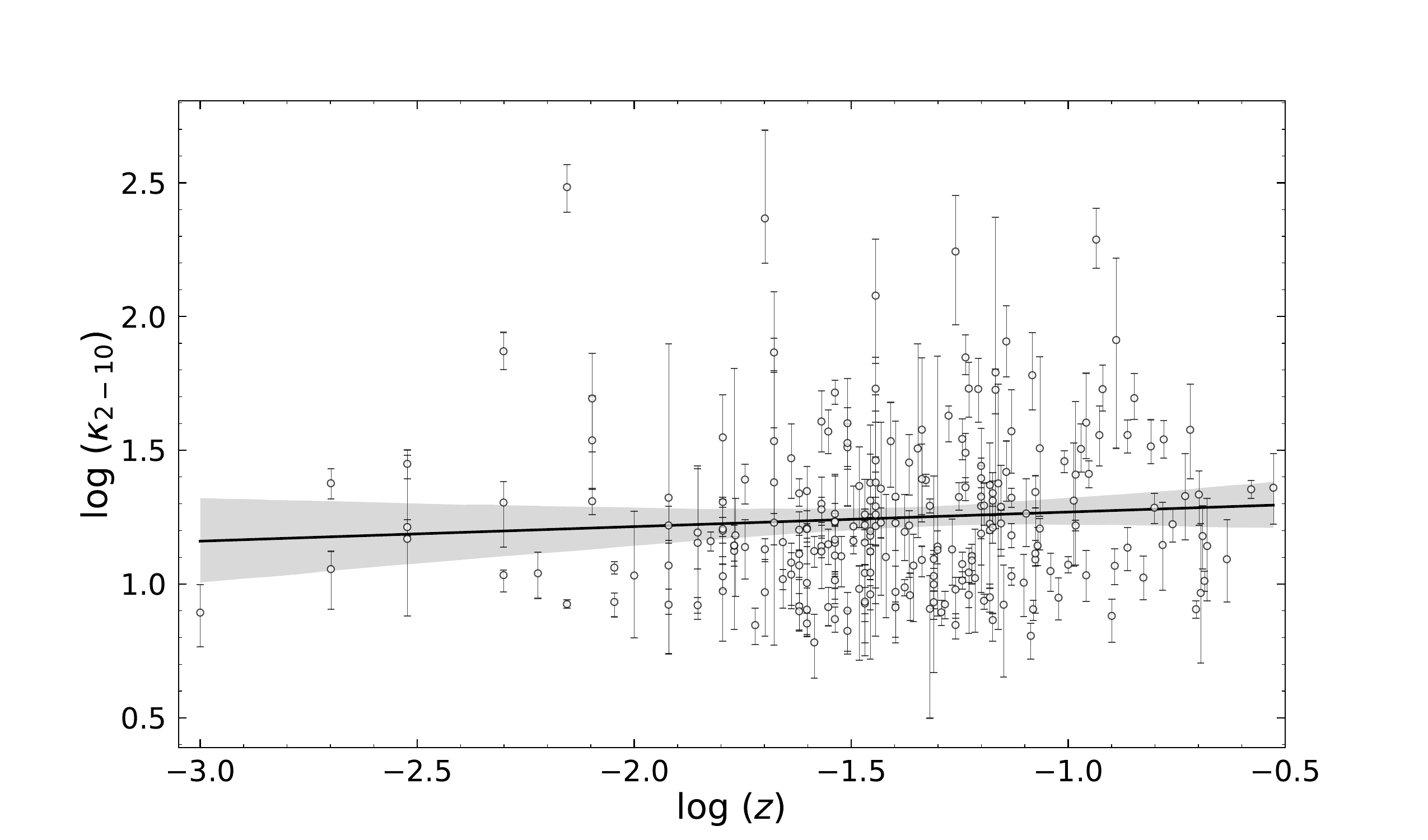}
    \caption{}
    \label{fig:k210_vs_z}
  \end{subfigure}
\caption{Soft X-ray bolometric correction ($\kappa_{2-10}$) for our sample of hard-X-ray-selected nearby, unobscured AGN. (a) Distribution of $\kappa_{2-10}$ and (b) $\kappa_{2-10}$ as a function of the redshift. The solid black line shows the best-fit relation to the data and the shaded gray region represents the uncertainty in the linear fit. We do not find any significant dependence of $\kappa_{2-10}$ on redshift.} 
\label{fig:k210}
\end{figure*}


\begin{figure}
    \centering
    \includegraphics[width=0.52\textwidth]{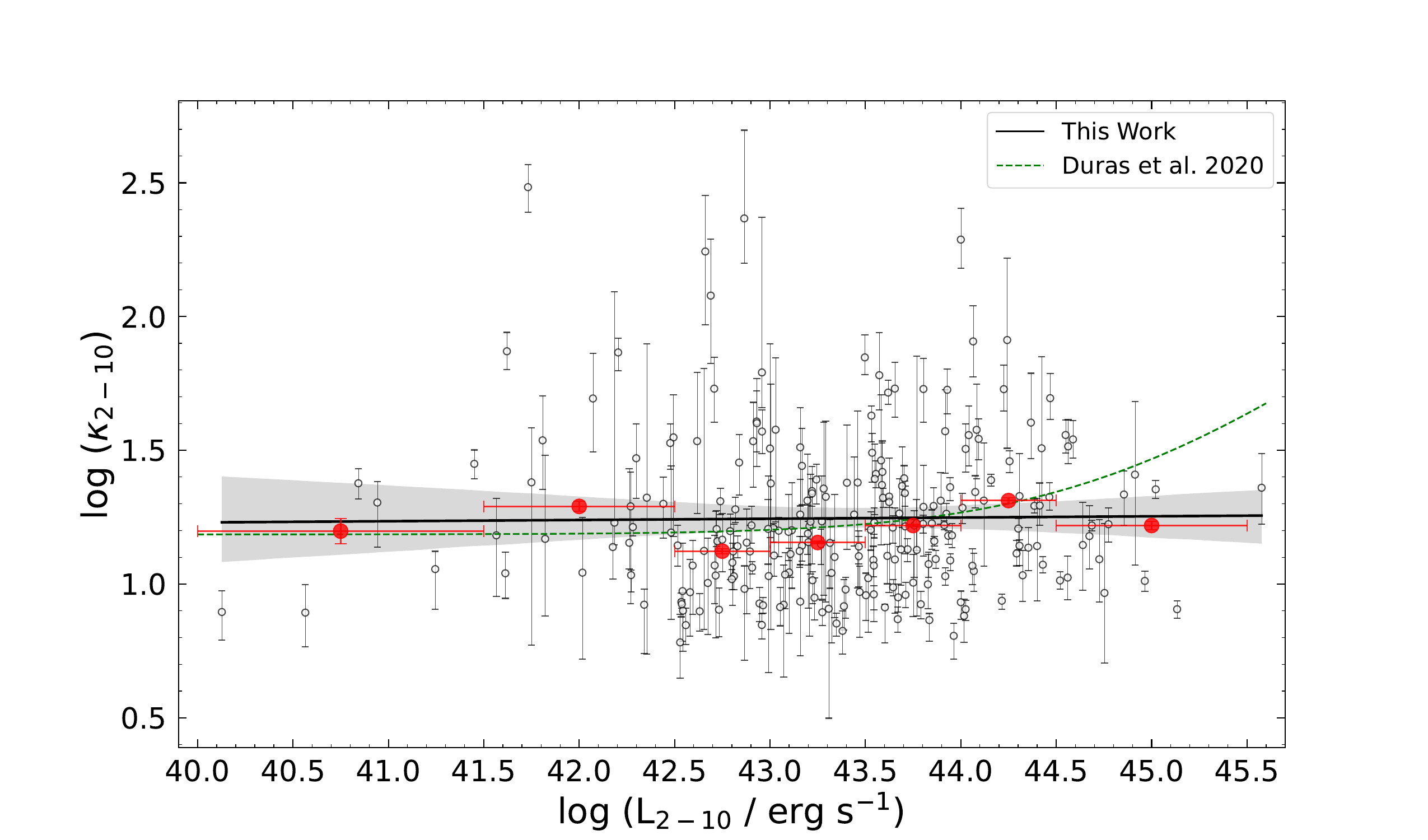}
    \caption{2--10\,keV bolometric correction ($\kappa_{2-10}$) against the intrinsic 2--10\,keV luminosity. The red points show the median value of $\kappa_{2-10}$ in bins of $L_{2-10}$. The solid black line shows our best-fit relation and the shaded gray region is the one sigma confidence interval. The dashed green line shows the relation reported by \citet{2020A&A...636A..73D}, which is consistent with our results up to $L_{2-10}\sim10^{45}\,{\rm erg/s}$.}
    \label{fig:k210_vs_l210}
\end{figure}


\begin{figure}
    \centering
    \includegraphics[width=0.52\textwidth]{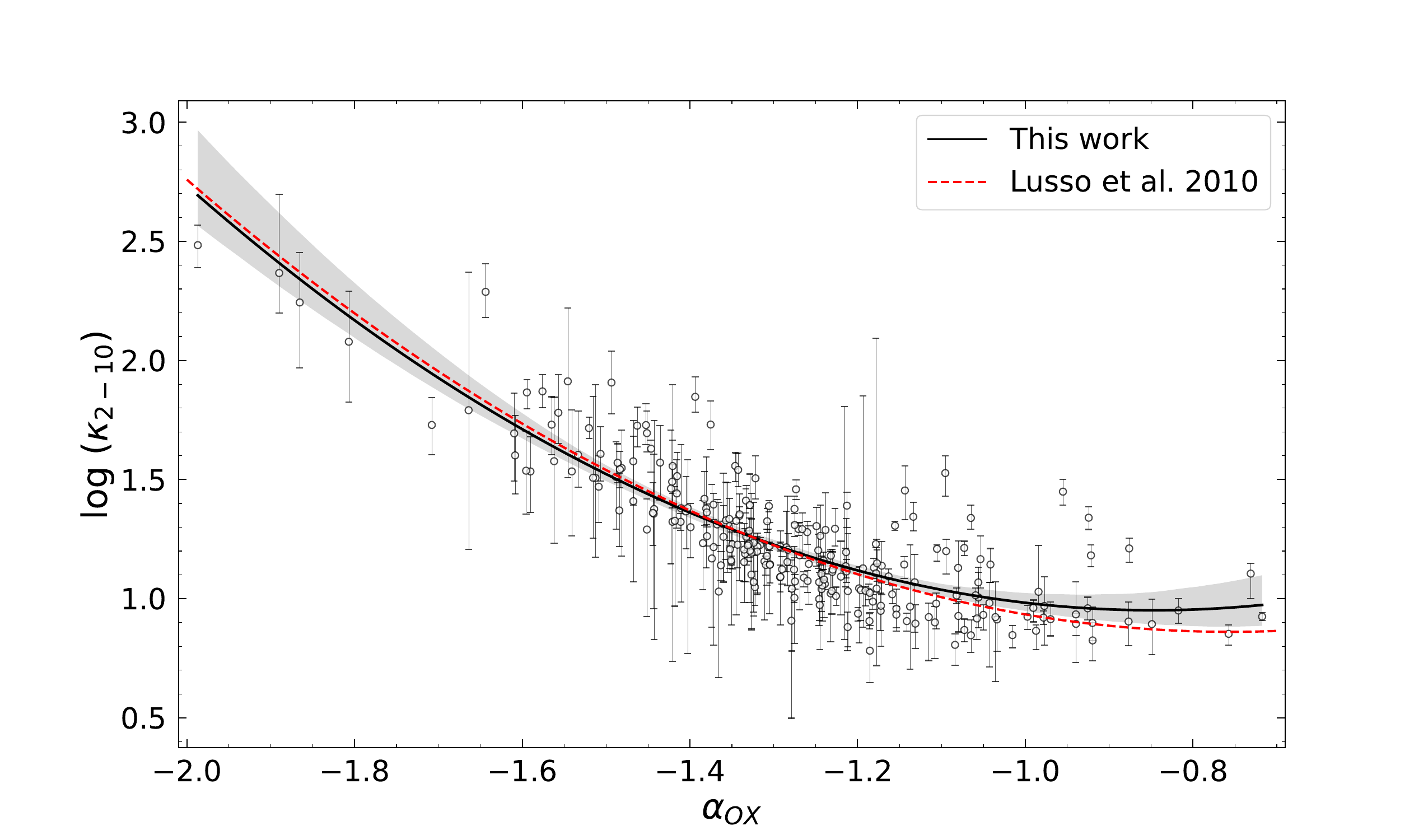}
    \caption{2--10\,keV bolometric correction ($\kappa_{2-10}$) against $\alpha_{\rm ox}$. The solid black line shows the best-fit relation we obtained (Sect. \ref{sect:kx}, Table \ref{tab:ktrends}), which is in good agreement with the one estimated by \citeauthor{2010A&A...512A..34L} (\citeyear{2010A&A...512A..34L}; dashed red line). The shaded gray region shows the one sigma confidence interval for our best-fit relation.}
    \label{fig:aox_vs_k210}
\end{figure}


\subsection{Bolometric corrections}\label{sect:k}

We calculated the bolometric corrections in different energy bands by taking a ratio of the bolometric luminosity ($L_{\rm bol} = L_{\rm x}+L_{\rm uvo}$) and the intrinsic luminosity in the specific energy range ($\kappa_\lambda = L_{\rm bol}/L_\lambda$). 


\subsubsection{X-ray}\label{sect:kx}

For the X-rays, we report bolometric corrections in three energy bins, the 0.2--2\,keV ($\kappa_{0.2-2}$), 2--10\,keV ($\kappa_{2-10}$), and the 14--195\,keV ($\kappa_{14-195}$). In this section we focus on $\kappa_{2-10}$, while the other two X-ray bolometric corrections are discussed in Sect. \ref{sect:k_add}. 

In Fig. \ref{fig:k210_log}, we show the distribution in the values of $\kappa_{2-10}$ for our sample (also see Table \ref{tab:params}). We do not find any dependence of $\kappa_{2-10}$ on the redshift (Fig. \ref{fig:k210_vs_z}). We obtain a median value of $\kappa_{2-10} = 15.86 \pm 0.17$, albeit with scatter up to 0.3 dex. Compared to \citet{2009MNRAS.392.1124V}, who reported a median $\kappa_{2-10} = 20$, our median $\kappa_{2-10}$ is slightly lower. A possible reason behind this could be the difference in the fitting techniques employed while fitting the optical/UV part of the SED. Even though we used the same disk model, in their analysis, \citet{2009MNRAS.392.1124V} fixed the parameter corresponding to the normalization of the disk emission (given by Eq. \ref{eq:norm_uvo}). For their models, they calculated $K_{\rm uvo}$ by substituting $M$ with the black hole mass estimates from reverberation mapping, using the luminosity distances for $D$, and assuming the inclination angle to be $0\deg$ (for face-on disks of low-absorption AGN), and the color-to-effective temperature ratio to be unity. Fixing the normalization parameter can drastically affect the overall SED fitting as the only parameter left to vary is the disk temperature (discussed in Sect. \ref{sect:disktemp}). We further checked the implication of fixing the normalization by following the approach taken by \citet{2009MNRAS.392.1124V} for our AGN sample and obtained higher values for $\kappa_{2-10}$, consistent with those reported by \citet{2009MNRAS.392.1124V}. However, once again, we do not recommend fixing the normalization in such disk models as that can strongly alter the SED fitting. Moreover, assuming a common inclination angle and color-to-temperature ratio for the entire sample could work for small samples like those analyzed by \citet{2009MNRAS.392.1124V} and \citet{2009MNRAS.399.1553V}. But doing the same for a sample as large as ours cannot be justified and can severely bias the final results.

We also plot $\kappa_{2-10}$ as a function of the intrinsic 2--10\,keV luminosity in Fig. \ref{fig:k210_vs_l210}. Across five orders of magnitude in the X-ray luminosity, we do not see any significant trend in the value of $\kappa_{2-10}$. This is in agreement with \citet{2020A&A...636A..73D}, who estimated $\kappa_{2-10}\sim$ 15--18 for ${\rm log}\,(L_{2-10}/{\rm erg\,s^{-1}})\sim$ 40--44 and only find a significant increase in $\kappa_{2-10}$ up to 30--70 for ${\rm log}\,(L_{2-10}/{\rm erg\,s^{-1}})>45$. Now although one might be tempted to just use the median value of $\kappa_{2-10}$ to estimate the bolometric luminosity from the intrinsic 2--10\,keV luminosity, we would like to emphasize that doing so could be problematic, owing to the visible scatter of $\sim$ 0.3 dex in $\kappa_{2-10}$. We confirm that this scatter is not due to sources with high Galactic extinction ($\sim$ 20 sources have $E[B-V]>0.3$). Moreover, the scatter cannot be due to variability as our analysis uses contemporaneous observations in the multiple bands. Hence, the scatter in $\kappa_{2-10}$ is intrinsic to our AGN sample and could be due to either inclination angle or black hole spin, quantities not explicitly constrained in our SED fitting (see Sect. \ref{sect:sed}). Additionally, $\kappa_{2-10}$ changes significantly with the bolometric luminosity and the Eddington ratio of the source, also confirmed by previous works (e.g., \citealp{2009MNRAS.399.1553V}; \citealp{2010A&A...512A..34L}; \citealp{2020A&A...636A..73D}). A detailed study of the main parameters regulating $\kappa_{2-10}$ and how it correlates with various physical properties of AGN will be presented in \textcolor{blue}{Gupta et al. in prep.}.

\cite{2010A&A...512A..34L} reported a strong correlation between $\kappa_{2-10}$ and $\alpha_{\rm ox}$. This relation could be extremely useful considering that independent estimates of bolometric luminosities ($\kappa_\lambda \times L_{\lambda}$) require one to have access to a wealth of multiband data, whereas $\alpha_{\rm ox}$ can be calculated from just the monochromatic luminosity at $2500\,{\rm\AA}$ and 2\,keV. Furthermore, the relation of $\kappa_{2-10}$ with $\alpha_{\rm ox}$ has a smaller scatter (0.15 dex) compared to the one with $L_{2-10}$ mentioned above. In Fig. \ref{fig:aox_vs_k210}, we show this $\alpha_{\rm ox}-\kappa_{2-10}$ relation obtained from our work, which is in excellent agreement with \cite{2010A&A...512A..34L}. The best-fit relation (shown as a solid black line in Fig. \ref{fig:aox_vs_k210}) between $\alpha_{\rm ox}$ and $\kappa_{2-10}$ is represented by a second-degree polynomial as follows:

\begin{equation}\label{eq:k210_aox}
    \begin{aligned}
        {\rm log}\,(\kappa_{2-10}) =& \,\,(1.339 \pm 0.143) \times \alpha_{\rm ox}^2 + (2.267 \pm 0.370) \times \alpha_{\rm ox}\\
        & \,\,+ (1.911 \pm 0.239)
    \end{aligned}
\end{equation}


\begin{figure*} 
  \begin{subfigure}[t]{0.54\textwidth}
    \centering
    \includegraphics[width=\textwidth]{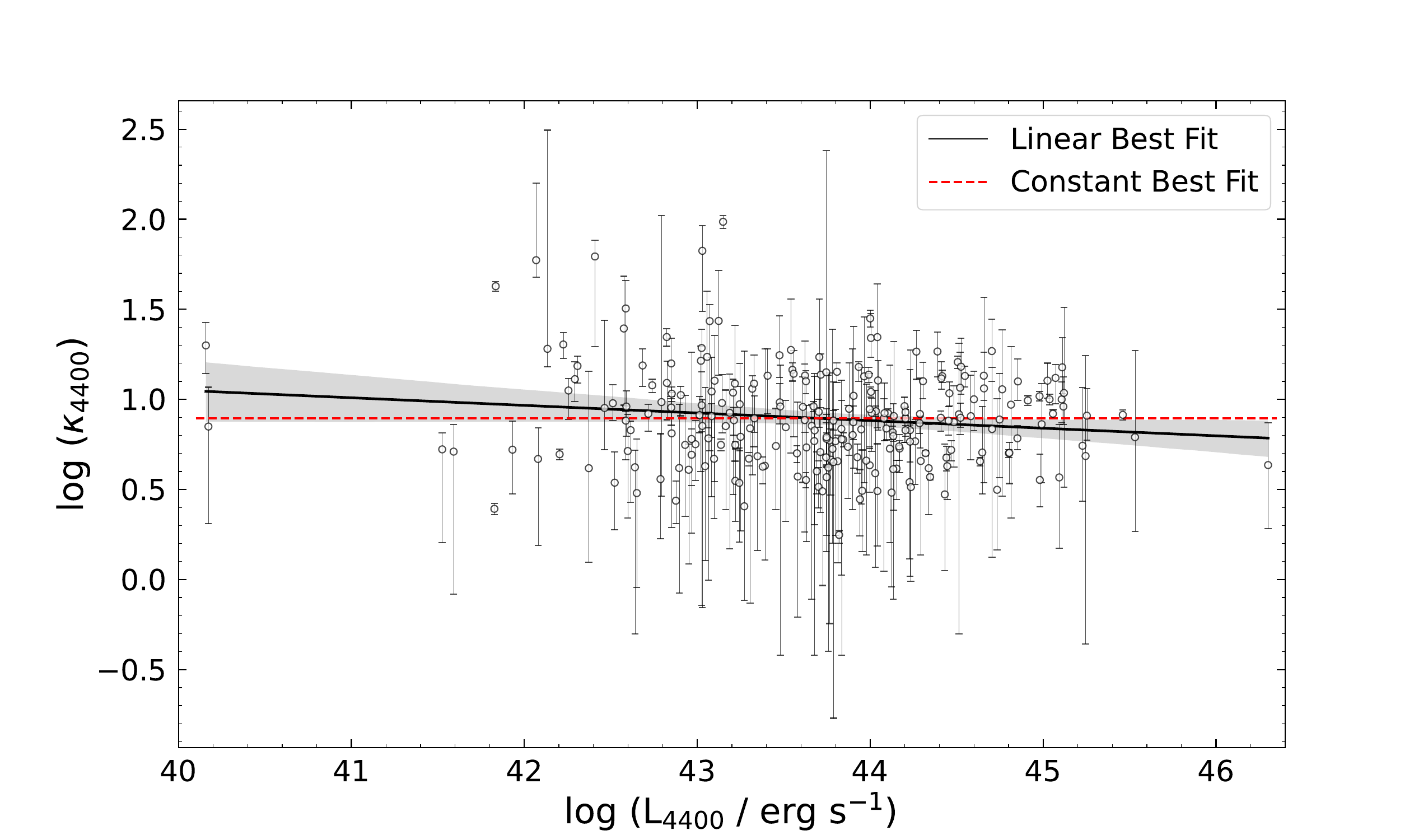}
    \caption{}
    \label{fig:k4400_vs_l4400}
  \end{subfigure}
  \hspace{-1.cm}
  \begin{subfigure}[t]{0.54\textwidth}
    \centering
    \includegraphics[width=\textwidth]{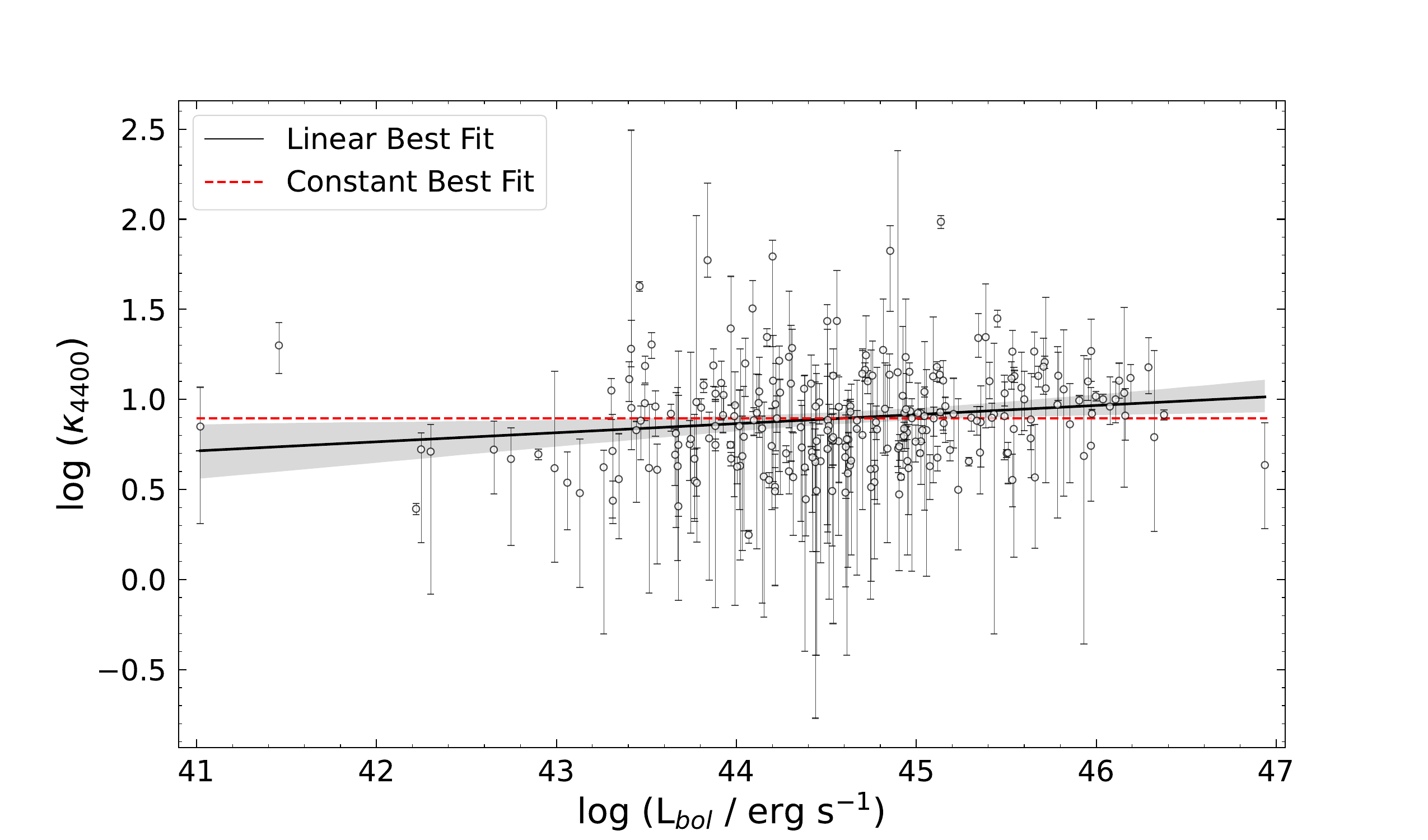}
    \caption{}
    \label{fig:k4400_vs_lbol}
  \end{subfigure}
\caption{4400 $\rm \AA$ bolometric correction as a function of (a) the 4400 $\rm \AA$ monochromatic luminosity and (b) the bolometric luminosity. We find a very weak correlation in both cases (Sect. \ref{sect:kuvo}, Table \ref{tab:ktrends}) shown as the black line. Hence, we show the best fit with a constant value as well (dashed red line). For both cases, we get a value of $\kappa_{\rm 4400\,\AA}$ = 7.9 $\pm$ 0.3.} 
\label{fig:k4400}
\end{figure*}


\begin{figure*}
    \centering
    \includegraphics[width=1.05\textwidth]{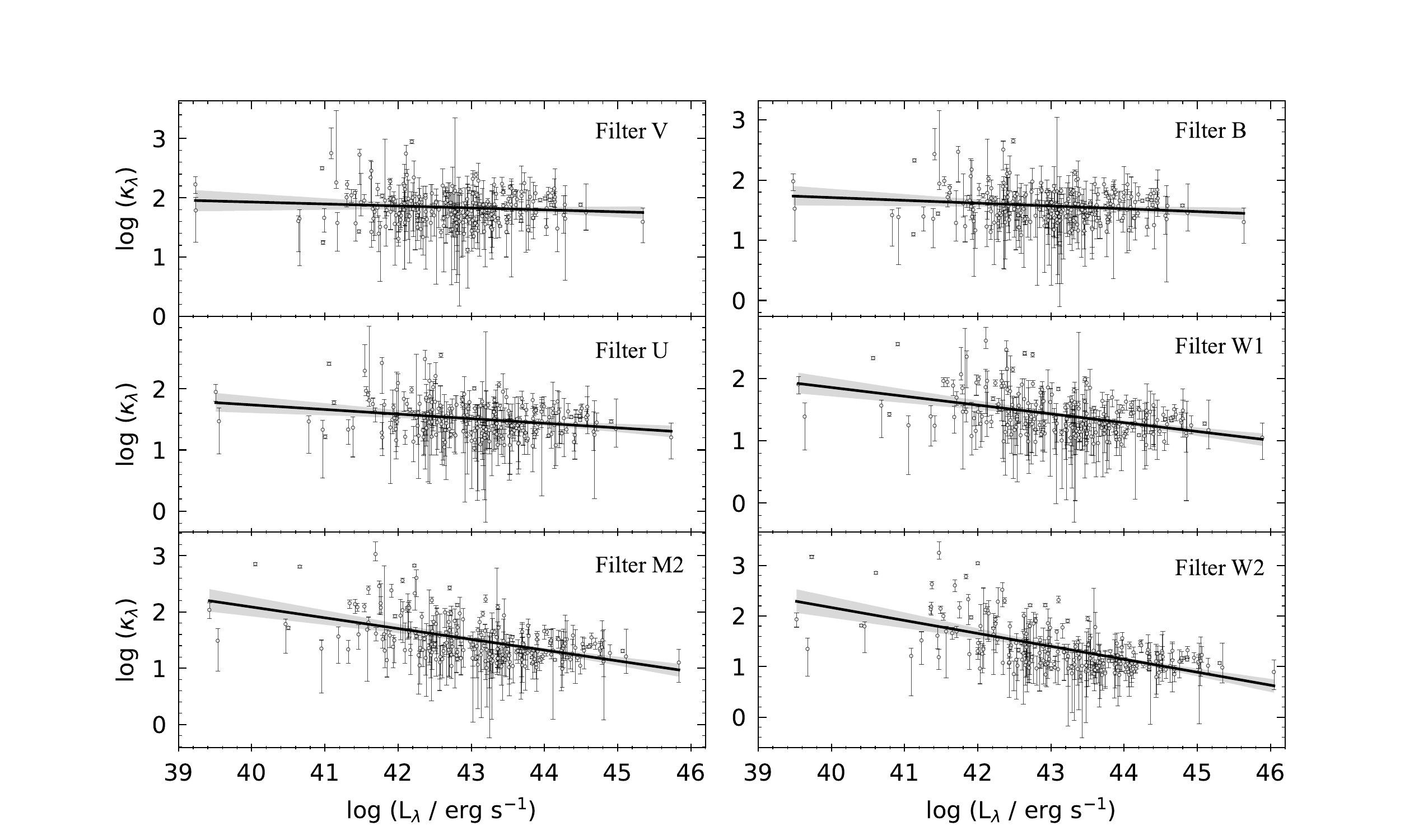}
    \caption{Six optical/UV bolometric corrections as a function of the intrinsic luminosity in their respective filters. The solid lines show our best-fit relations. A significant anticorrelation with luminosity is observed for bolometric corrections in the filters U, W1, M2, and W2. This dependence is stronger as we go to higher energies (Sect. \ref{sect:kuvo}, Table \ref{tab:ktrends}).}
    \label{fig:kall_vs_l}
\end{figure*}


\begin{figure*}
    \centering
    \includegraphics[width=1.05\textwidth]{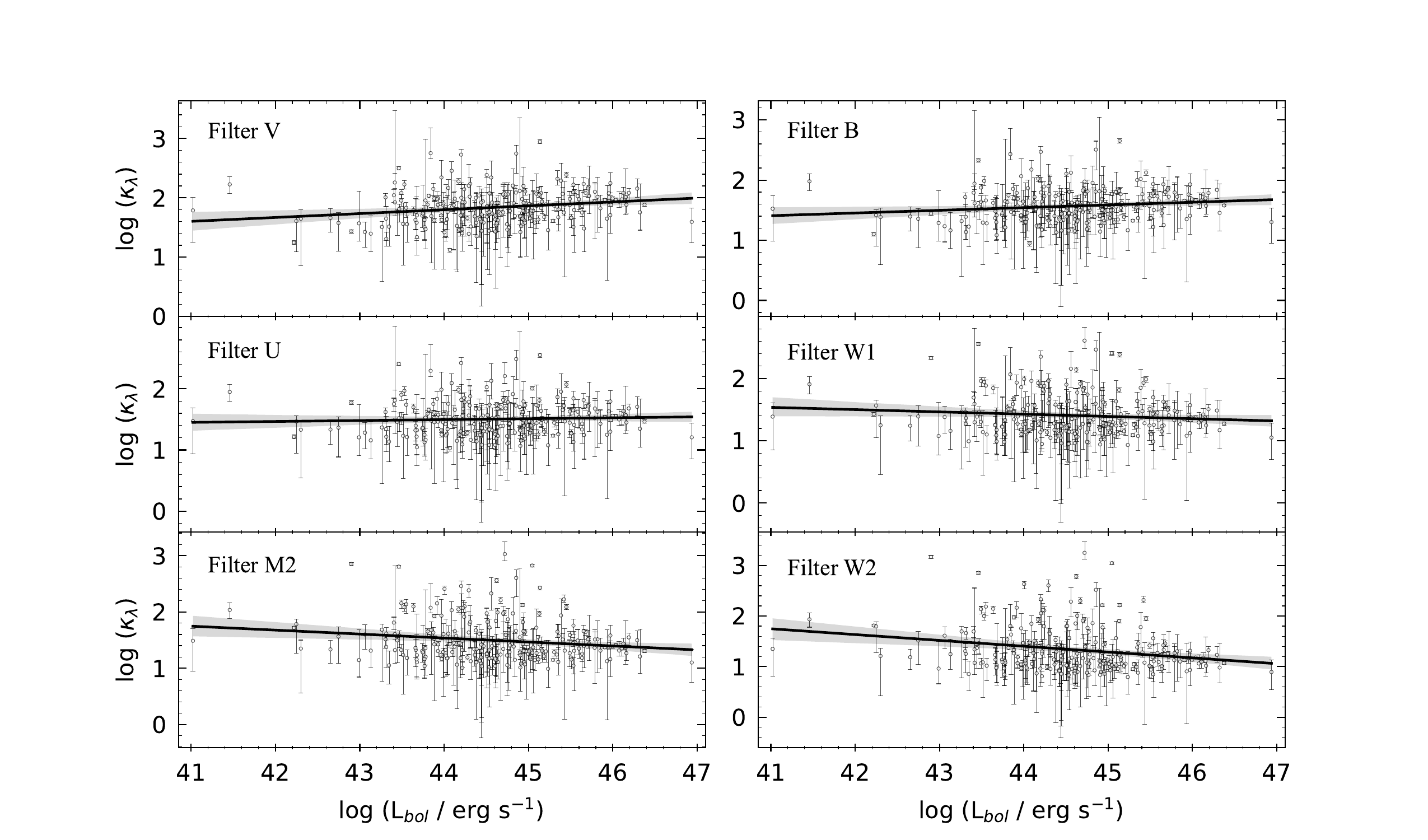}
    \caption{Six optical/UV bolometric corrections as a function of the bolometric luminosity. The solid lines show the best-fit relations. We only find a significant correlation with $L_{\rm bol}$ for the highest-energy UV filter W2 (Sect. \ref{sect:kuvo}, Table \ref{tab:ktrends}).}
    \label{fig:kall_vs_lbol}
\end{figure*}


\begin{figure*}
    \centering
    \includegraphics[width=1.05\textwidth]{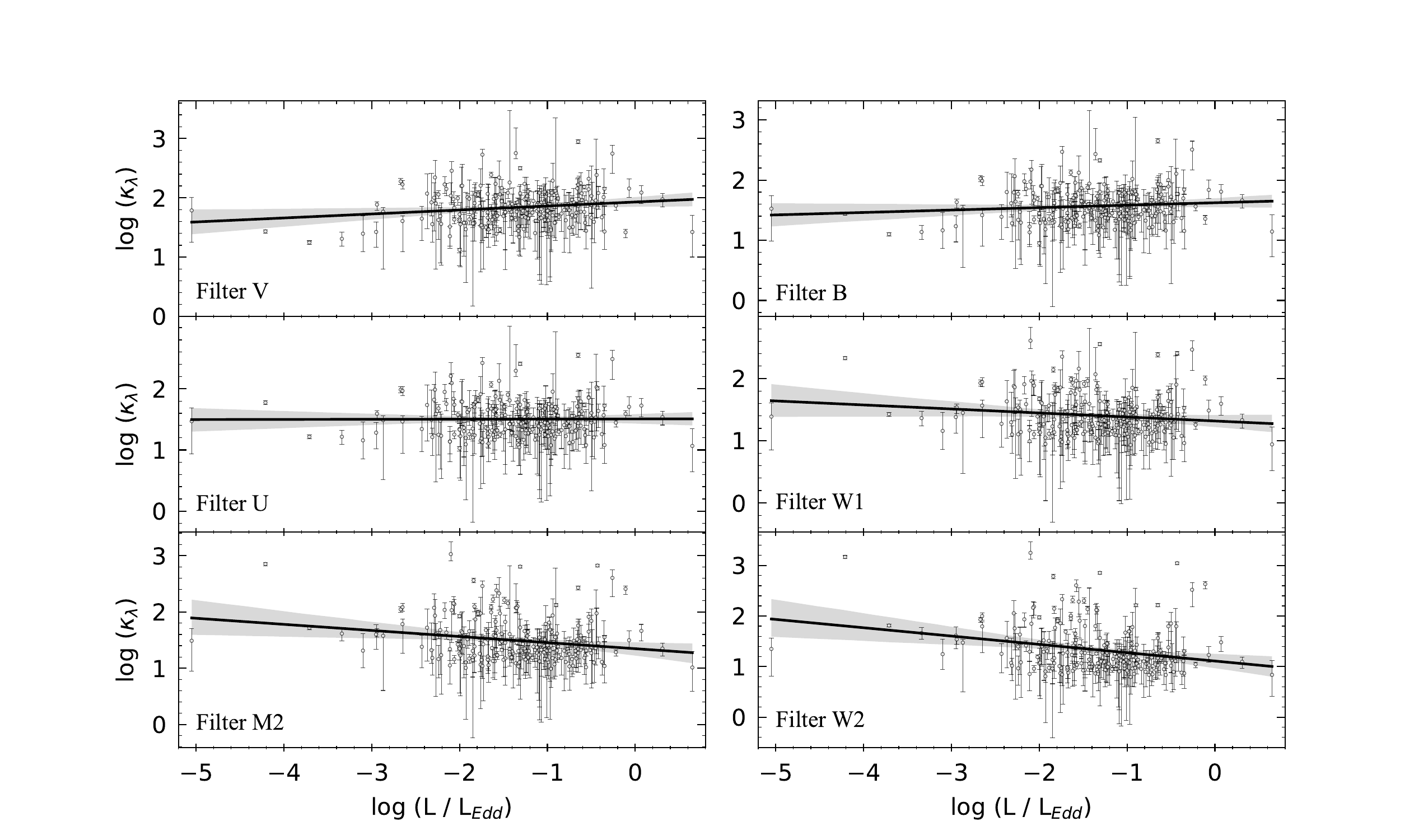}
    \caption{Six optical/UV bolometric corrections as a function of the Eddington ratio. The solid lines show the best-fit relations. We only find a significant correlation with $\lambda_{\rm Edd}$ for the highest-energy UV filters M2 and W2 (Sect. \ref{sect:kuvo}, Table \ref{tab:ktrends}).}
\label{fig:kall_vs_er}
\end{figure*}


\subsubsection{Optical/UV}\label{sect:kuvo}

Based on the optical/UV wavelength regime of the broadband SED, we also calculated bolometric corrections in the optical and UV energy bands. In Table \ref{tab:kbol}, we list the optical/UV bolometric corrections in the six \textit{Swift}/UVOT filters (V, B, U, UVW1, UVM2, and UVW2). These bolometric corrections were calculated by taking a ratio of the bolometric luminosity and the intrinsic luminosity in the corresponding \textit{Swift}/UVOT waveband ($L_{\lambda}$; wavelength range for each band is listed in Table \ref{tab:uvot}). We also report the $4400\rm\,\AA$ bolometric correction ($\kappa_{\rm 4400\,\AA} = L_{\rm bol}/L_{\rm 4400\,\AA}$; in the optical B-band) as has been done by many previous works (e.g., \citealp{2007ApJ...654..731H}; \citealp{2012MNRAS.425..623L}; \citealp{2020MNRAS.495.3252S}; \citealp{2020A&A...636A..73D}). First, we explore if these bolometric corrections change with redshift and do not find any significant dependence of $\kappa_\lambda$ on $z$. Since we do not expect the energy generation mechanism of AGN to change over such a small redshift range as that of our sample ($0.001< z < 0.3$), absence of a correlation between $\kappa_\lambda$ and $z$ further confirms that our SED analysis is consistent throughout the redshift range of our sample. Even though the \textit{Swift}/BAT AGN sample is flux-limited, due to which the farthest sources are the faintest and we had to apply corrections to the \textsc{GALFIT}-estimated AGN fluxes (described in Sect. \ref{sect:galfit_corr}) for the sources above $z=0.05$, our results are not influenced by it. This also verifies that the corrections that we calculated for \textsc{GALFIT} fluxes did not introduce any biases to our analysis.

Next, we check if any of the optical/UV bolometric correction factors show a dependence on $L_{\lambda}$, $L_{\rm bol}$, $M_{\rm BH}$, or $\lambda_{\rm Edd}$. We first focus on $\kappa_{\rm 4400\,\AA}$ so as to compare our results with other similar studies from the literature. In Fig. \ref{fig:k4400}, we show $\kappa_{\rm 4400\,\AA}$ as a function of $L_{\rm 4400\,\AA}$ and $L_{\rm bol}$. In agreement with past studies (e.g., \citealp{2012MNRAS.425..623L}; \citealp{2020A&A...636A..73D}), we do not find any significant correlation of $\kappa_{\rm 4400\,\AA}$ with either the intrinsic monochromatic luminosity at $4400\,{\rm \AA}$ or the bolometric luminosity (see Table \ref{tab:ktrends}). \citet{2020MNRAS.495.3252S}, however, reported a luminosity dependent $\kappa_{\rm 4400\,\AA}$, as they found higher values of $\kappa_{\rm 4400\,\AA}$ for lower luminosity sources ($L_{\rm bol}< 10^{44}\,{\rm erg/s}$). This dependence can be attributed to the fact that they used the observed luminosity (instead of the intrinsic luminosity) at $4400\,{\rm \AA}$ to calculate $\kappa_{\rm 4400\,\AA}$, due to which the lower luminosity sources that are affected by possible extinction from the surrounding material would have higher $\kappa_{\rm 4400\,\AA}$ due to lower values of observed $L_{\rm 4400\,\AA}$. Whereas, the higher luminosity sources ($L_{\rm bol} > 10^{44}\,{\rm erg/s}$) have $\kappa_{\rm 4400\,\AA}$ values similar to ours. Hence, we recommend the constant value fit that gives a value of $\kappa_{\rm 4400\,\AA}=7.9 \pm 0.3$. Additionally, $\kappa_{\rm 4400\,\AA}$ also does not show any dependence on $M_{\rm BH}$ or $\lambda_{\rm Edd}$. Lastly, we investigate the same trends for the bolometric correction factors in the six \textit{Swift}/UVOT filters ($\kappa_{\rm V}$, $\kappa_{\rm B}$, $\kappa_{\rm U}$, $\kappa_{\rm W1}$, $\kappa_{\rm M2}$, and $\kappa_{\rm W2}$). In the following part of the discussion, we first report all our findings and then later discuss their physical implications.

In Fig. \ref{fig:kall_vs_l}, we show the optical/UV bolometric corrections against the intrinsic luminosity in their respective filters and in Figs. \ref{fig:kall_vs_lbol} and \ref{fig:kall_vs_er}, we show their dependence on the bolometric luminosity and the Eddington ratio, respectively. Similar to $\kappa_{\rm 4400\,\AA}$, $\kappa_{\rm B}$ ($\lambda_{\rm B} = 4392\AA$) does not show any correlation with $L_{\rm B}$, $L_{\rm bol}$, $M_{\rm BH}$, or $\lambda_{\rm Edd}$ (top right panel in Figs. \ref{fig:kall_vs_l}, \ref{fig:kall_vs_lbol}, and \ref{fig:kall_vs_er}). Bolometric corrections of the highest-energy optical filter U and the lowest-energy UV filter W1 show a strong negative correlation with the intrinsic luminosities in their respective bands (middle panel in Fig. \ref{fig:kall_vs_l}, Tables \ref{tab:ktrends} and \ref{tab:keq}). They do not show dependence on any other properties of the AGN. The V filter bolometric correction factor shows a weak positive correlation with the bolometric luminosity and the Eddington ratio (top left panel in Figs. \ref{fig:kall_vs_lbol} and \ref{fig:kall_vs_er}). The bolometric corrections for the high-energy UV filters, $\kappa_{\rm M2}$ and $\kappa_{\rm W2}$, show significant anticorrelations with the intrinsic luminosities in their respective energy bands (bottom panel in Fig. \ref{fig:kall_vs_l}, Tables \ref{tab:ktrends} and \ref{tab:keq}) and with $\lambda_{\rm Edd}$ (bottom panel in Fig. \ref{fig:kall_vs_er}, Tables \ref{tab:ktrends} and \ref{tab:keq}), while $\kappa_{\rm W2}$ also shows a strong dependence on $L_{\rm bol}$ (see lower right panel in Fig. \ref{fig:kall_vs_lbol} and Tables \ref{tab:ktrends} and \ref{tab:keq}). They do not show any dependence on $M_{\rm BH}$.

Based on the trends mentioned above, we can draw some important conclusions. Firstly, we should pay attention to how the lowest-energy bolometric corrections ($\kappa_{\rm V}$ and $\kappa_{\rm B}$) do not correlate with any physical properties of the AGN. While those at the highest energies ($\kappa_{\rm M2}$ and $\kappa_{\rm W2}$) show strong correlations with our parameters of interest, except black hole mass. Studies have shown that compared to a disk model based on single-temperature blackbody emission, a multi-temperature accretion disk model better describes the optical/UV spectrum of AGN (e.g., \citealp{1978Natur.272..706S}; \citealp{1983ApJ...268..582M}). Assuming the disk to have a radial temperature profile [$T_{\rm eff}(R) \propto R^{-p}$; $p = 0.75$ for the standard \citet{1973A&A....24..337S} disk], with smaller radii being at higher temperatures, we naturally expect the highest-energy UV photons to originate mainly in the inner regions of the accretion disk, much closer to the central SMBH. Whereas, the cooler outer regions of the disk would mainly emit the lower-energy optical photons (e.g., \citealp{1998ApJ...500..162C}; \citealp{2005ApJ...622..129S}). This effect of the gradient in the disk temperature with distance from the central black hole on the wavelength of the emitted photons has been discussed in detail by several studies on reverberation mapping, where time lag measurements between different continuum wavelengths (wavelength-dependent lags) have been used to estimate the size of the accretion disk and other inner regions of AGN (e.g., \citealp{2007MNRAS.380..669C}; also see \citealp{2021iSci...24j2557C} for a review). Many such studies estimated accretion disk sizes to be larger than what is expected from the \citet{1973A&A....24..337S} disk model and some of them argued that a possible reason for this could be the diffused emission from the larger BLR contributing to the photometric measurements, especially in the optical (e.g., \citealp{2018ApJ...857...53C}; \citealp{2019NatAs...3..251C}; \citealp{2022ApJ...925...29C}). In this situation, we cannot be certain that the lower energy band of the broadband SED is solely dominated by emission arising in the disk without substantial contamination from the more extended regions. Hence explaining why the corresponding bolometric corrections are not strongly correlated with properties of the AGN, such as their total energy output (from accretion) or their mass normalized accretion rate. On the contrary, the higher-energy UV bands predominately trace the accretion disk emission and therefore, the corresponding $\kappa_{\lambda}$ correlate with quantities like $\lambda_{\rm Edd}$ (proxy of the accretion rate) and $L_{\rm bol}$ (total emission of the AGN). The observed negative correlation suggests that sources with larger accretion luminosities have more emission coming from the low wavelength, high-energy disk photons produced at shorter radii, and this is evident in the lower values of $\kappa_{\lambda}$ for these sources.

Secondly, we consistently do not find a dependence on $M_{\rm BH}$ for any of the optical/UV (or X-ray) bolometric corrections. This indicates that the mass of the central SMBH does not control or influence the total radiation emitted by the accreting material at different wavelengths probed here. The main driving factor here is in fact the Eddington ratio. This emphasizes the importance and requirement of detailed studies of AGN physics in different regimes of $\lambda_{\rm Edd}$. Finally, a significant scatter is visible in all optical/UV bolometric corrections ($\sim$ 0.1--1\,dex; see Table \ref{tab:kbol_med}). Similar to the X-ray bolometric correction ($\kappa_{2-10}$), we confirm that this scatter is not due to sources with high Galactic extinction and would like to reiterate the importance of including the respective scatters while calculating $L_{\rm bol}$ from $\kappa_{\lambda}$ for individual sources, to avoid miscalculations. We report the median values for all the optical/UV bolometric corrections, with uncertainties and dispersions, in Table \ref{tab:kbol_med}. We also provide scaling relations of all the optical/UV bolometric corrections with $\alpha_{\rm ox}$ in Table \ref{tab:aox_rel}.


\begin{figure}
    \centering
    \includegraphics[width=0.52\textwidth]{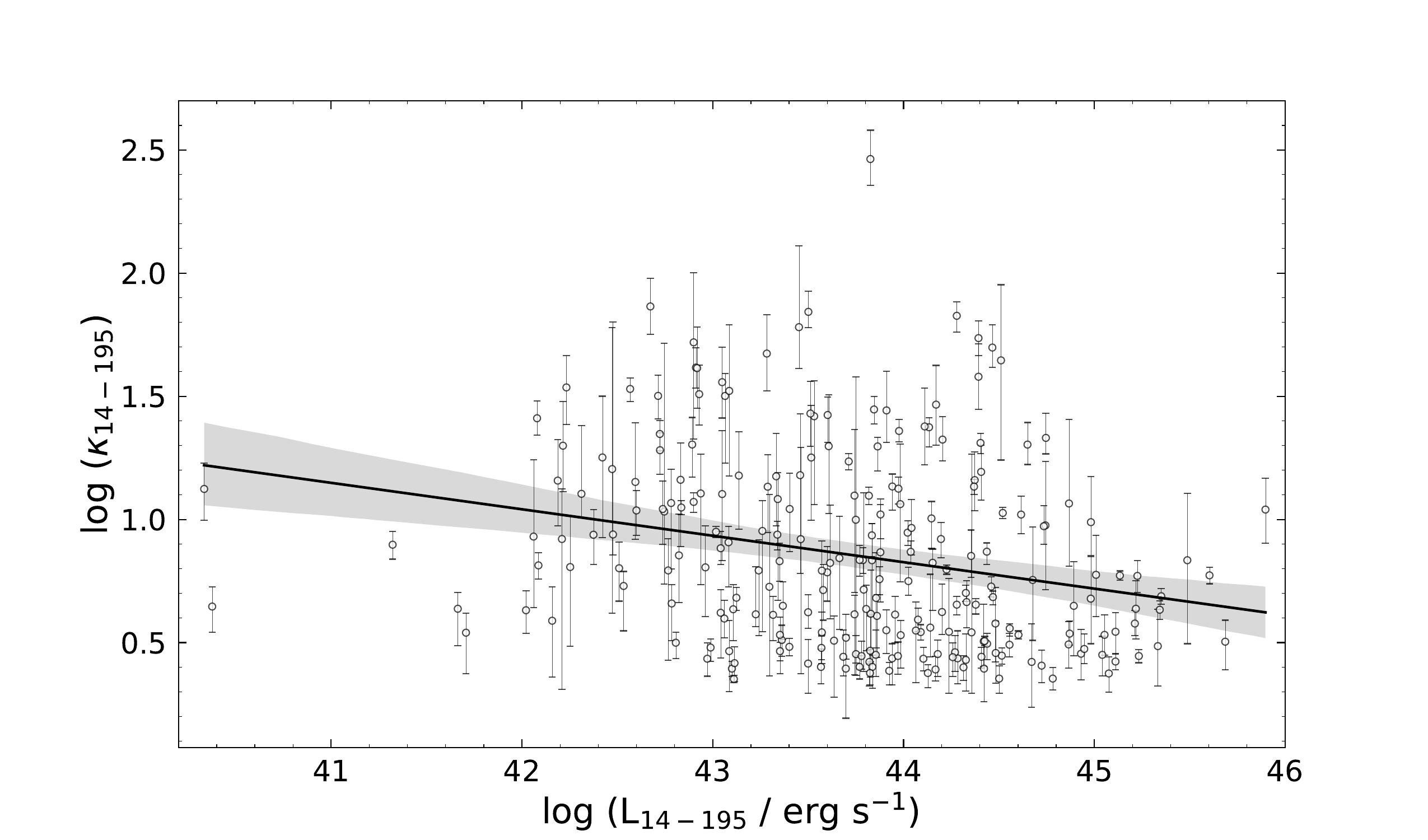}
    \caption{14--195\,keV bolometric correction ($\kappa_{14-195}$) against the intrinsic luminosity in the 14--195\,keV energy bin ($L_{14-195}$). The solid black line shows the best-fit relation we obtained (Sect. \ref{sect:k_add}, Table \ref{tab:ktrends}). The shaded gray region shows the one sigma confidence interval for our best-fit relation.}
    \label{fig:k14195_vs_l14195}
\end{figure}


\begin{figure*} 
  \begin{subfigure}[t]{0.54\textwidth}
    \centering
    \includegraphics[width=\textwidth]{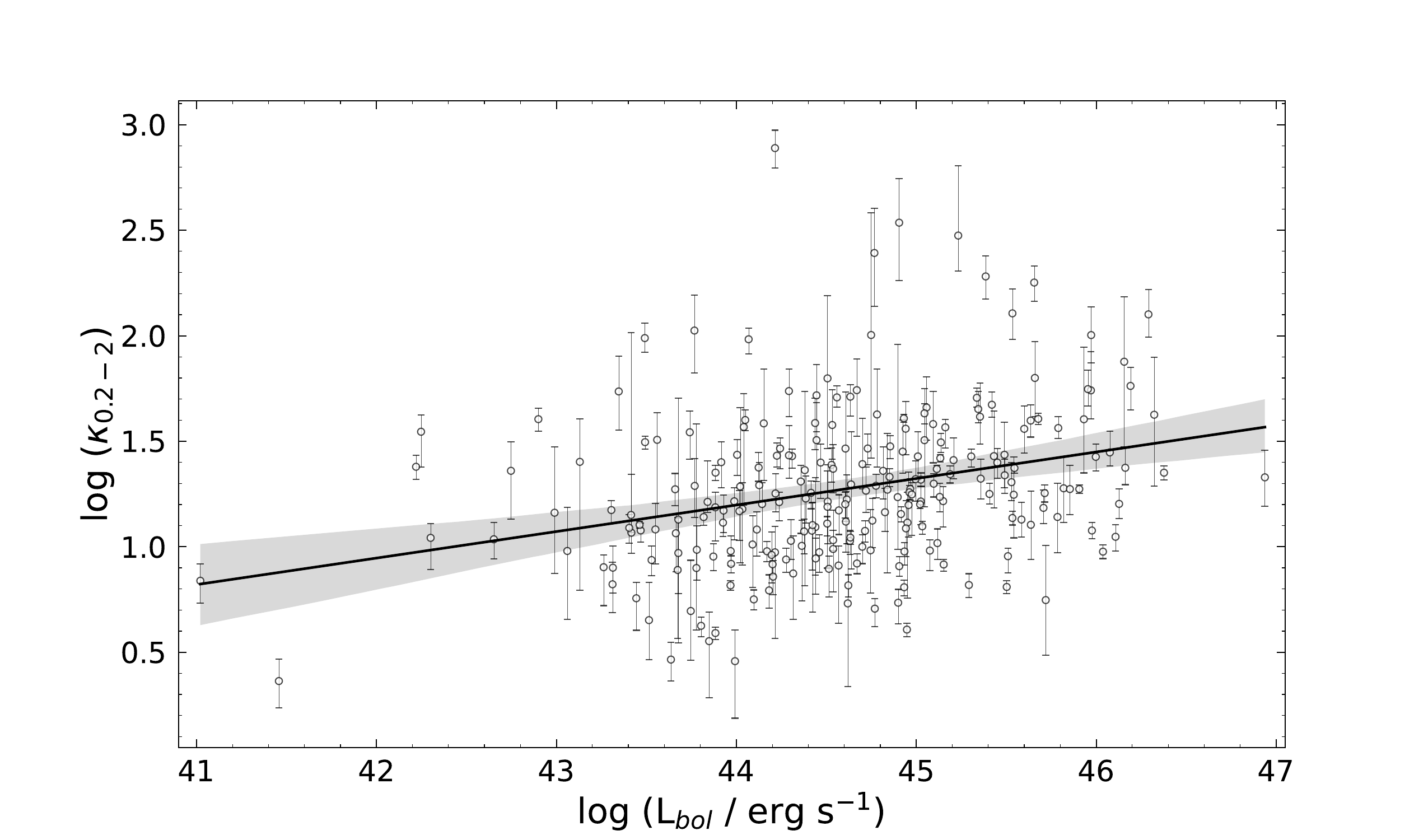}
    \caption{}
    \label{fig:k0.22_vs_lb}
  \end{subfigure}
  \hspace{-1.cm}
  \begin{subfigure}[t]{0.54\textwidth}
    \centering
    \includegraphics[width=\textwidth]{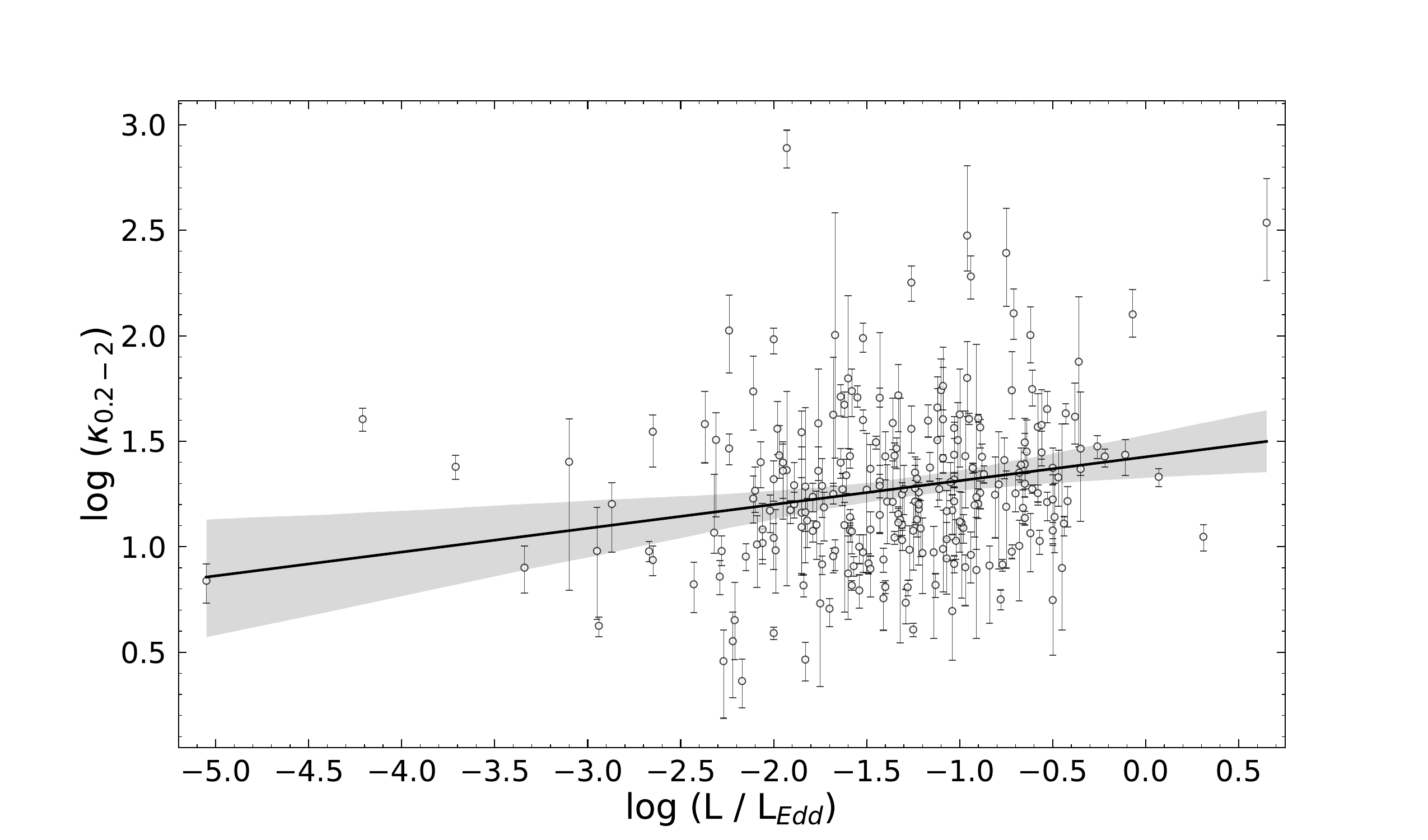}
    \caption{}
    \label{fig:k0.22_vs_er}
  \end{subfigure}
\caption{0.2--2\,keV bolometric correction as a function of (a) the bolometric luminosity and (b) the Eddington ratio. We find a significant correlation in both cases (Sect. \ref{sect:k_add}, Table \ref{tab:ktrends}) shown as the black line. The shaded region is the one sigma confidence interval for the linear regression line.} 
\label{fig:k0.22}
\end{figure*}


\begin{figure*} 
  \begin{subfigure}[t]{0.54\textwidth}
    \centering
    \includegraphics[width=\textwidth]{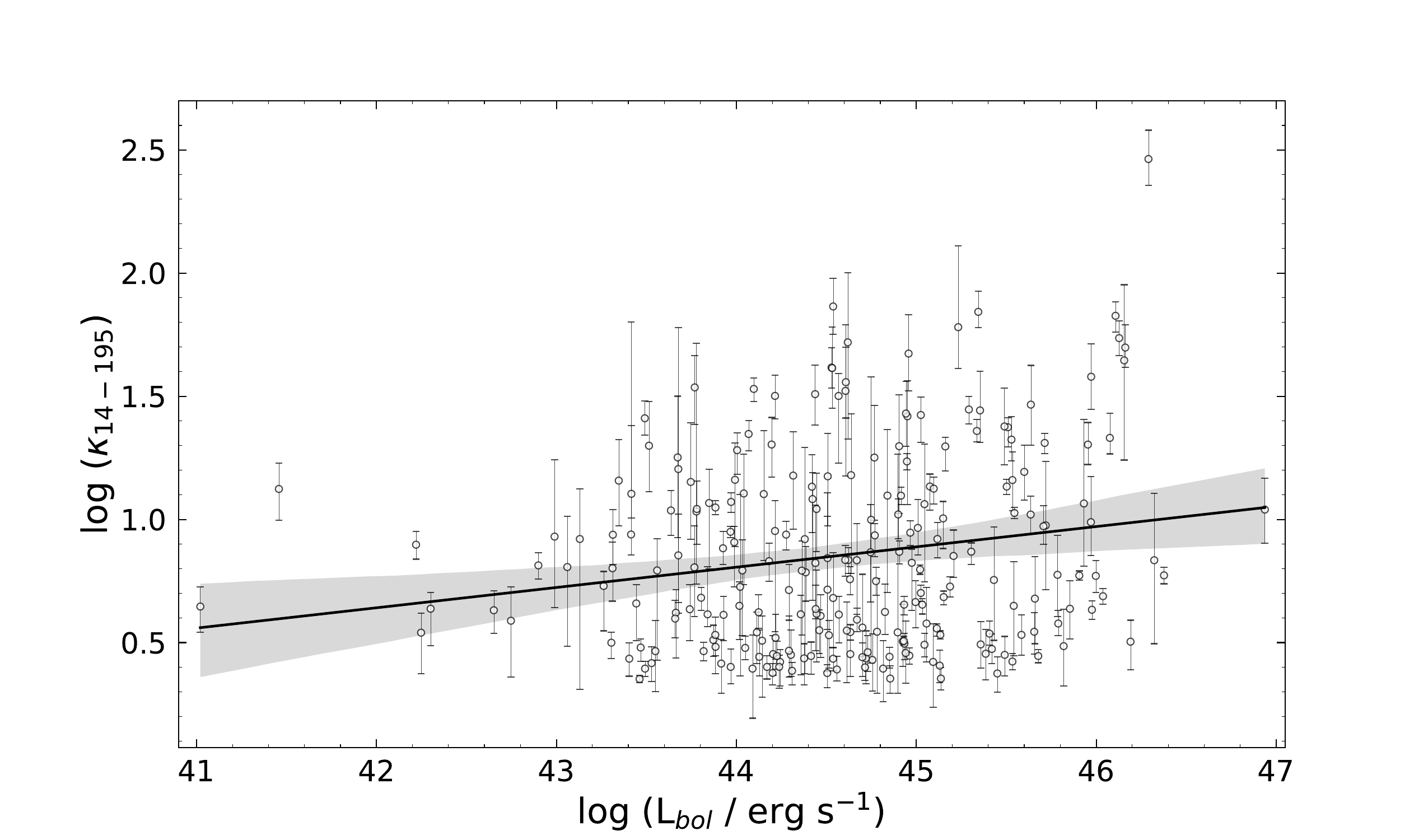}
    \caption{}
    \label{fig:k14195_vs_lb}
  \end{subfigure}
  \hspace{-1.cm}
  \begin{subfigure}[t]{0.54\textwidth}
    \centering
    \includegraphics[width=\textwidth]{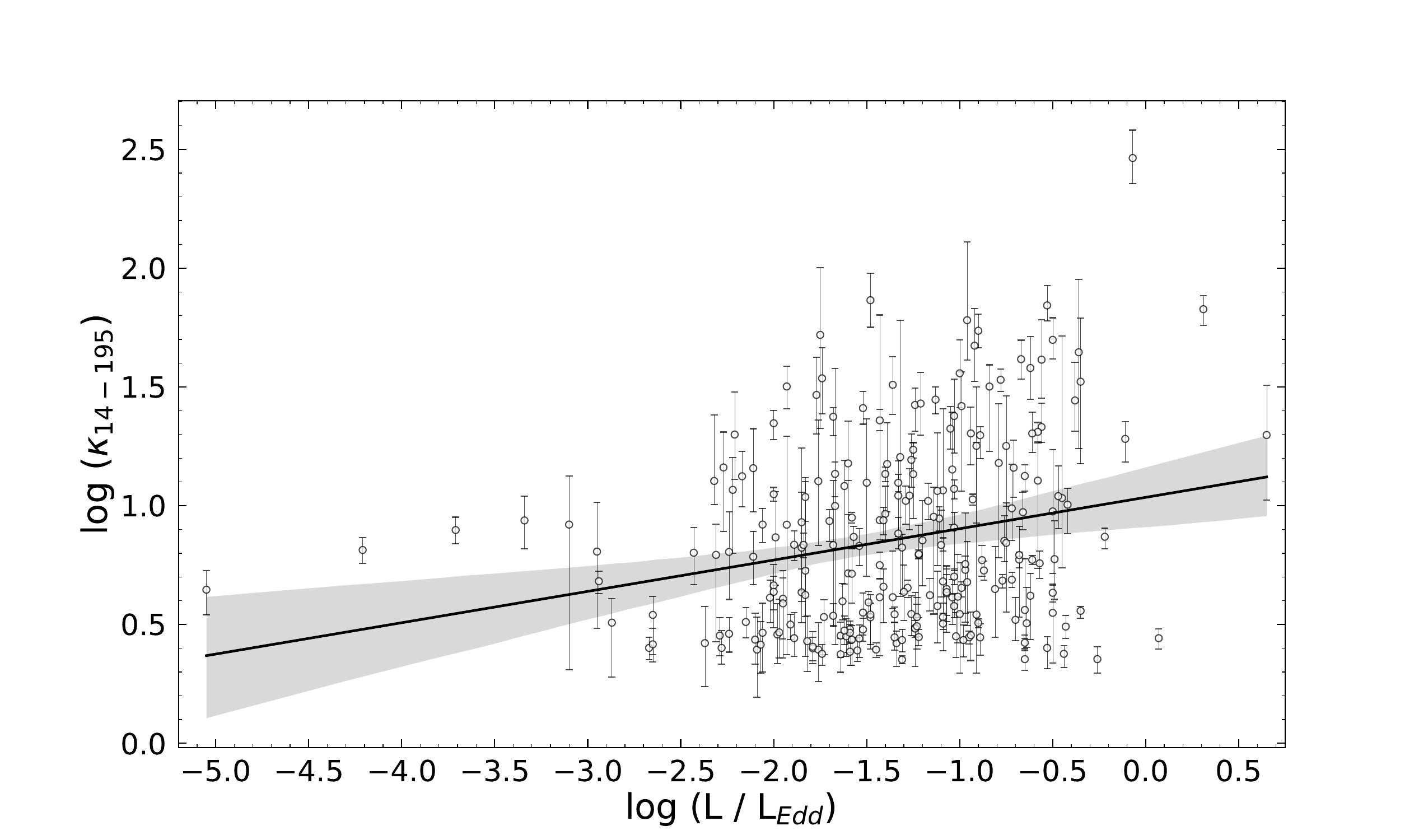}
    \caption{}
    \label{fig:k14195_vs_er}
  \end{subfigure}
\caption{14--195\,keV bolometric correction as a function of (a) the bolometric luminosity and (b) the Eddington ratio. We find a significant correlation in both cases (Sect. \ref{sect:k_add}, Table \ref{tab:ktrends}) shown as the black line. The shaded region is the one sigma confidence interval for the linear regression line.} 
\label{fig:k14195}
\end{figure*}


\begin{table*}
\centering
\caption{Useful outputs from the broadband SED fitting.}
\begin{tabular}{cccccccccc}
\hline
\hline
\specialrule{0.1em}{0em}{0.5em}
\vspace{1mm}
BAT ID & Swift ID & $L_{2500\,{\rm \AA}}$ & $L_{\rm 2\,keV}$ & $L_{\rm bol}$ & $M_{\rm BH}$ & $\alpha_{\rm ox}$ & $\kappa_{\rm 4400 \AA}$ & $\kappa_{2-10}$ & $\lambda_{\rm Edd}$\\
\hline\\
\vspace{2mm}
3 & SWIFTJ0002.5+0323 & 43.07 & 42.35 & 43.68 & 6.70 & $-1.27^{+0.10}_{-0.14}$ & $5.59^{+4.92}_{-3.35}$ & $10.09^{+4.73}_{-3.60}$ & $-1.20$\\
\vspace{2mm}
6 & SWIFTJ0006.2+2012 & 43.99 & 43.11 & 44.57 & 7.23 & $-1.34^{+0.14}_{-0.07}$ & $5.86^{+2.94}_{-4.10}$ & \,\,\,$22.26^{+10.38}_{-5.22}$ & $-0.84$\\
\vspace{2mm}
16 & SWIFTJ0029.2+1319 & 45.49 & 44.32 & 46.16 & 8.49 & $-1.45^{+0.04}_{-0.06}$ & $8.12^{+3.38}_{-2.19}$ & \,\,\,$49.53^{+8.32}_{-11.80}$ & $-0.50$\\
\vspace{2mm}
19 & SWIFTJ0034.5-7904 & 44.26 & 43.78 & 44.95 & 8.02 & $-1.19^{+0.02}_{-0.02}$ & $6.56^{+0.54}_{-0.59}$ & $10.70^{+0.81}_{-0.80}$ & $-1.25$\\
\vspace{2mm}
34 & SWIFTJ0051.6+2928 & 43.96 & 42.49 & 44.44 & 7.62 & $-1.56^{+0.07}_{-0.09}$ & $4.63^{+2.22}_{-1.68}$ & \,\,\,$53.75^{+13.46}_{-16.65}$ & $-1.36$\\
\vspace{2mm}
36 & SWIFTJ0051.9+1724 & 44.43 & 43.91 & 45.15 & 7.75 & $-1.20^{+0.02}_{-0.01}$ & $7.38^{+0.59}_{-0.91}$ & \,\,\,$8.66^{+0.61}_{-0.50}$ & $-0.77$\\
\vspace{2mm}
39 & SWIFTJ0054.9+2524 & 45.36 & 44.28 & 46.08 & 8.46 & $-1.41^{+0.04}_{-0.07}$ & $9.13^{+4.19}_{-1.86}$ & $32.70^{+4.55}_{-8.44}$ & $-0.56$\\
\vspace{2mm}
43 & SWIFTJ0059.4+3150 & 42.69 & 41.96 & 43.42 & 7.56 & $-1.28^{+0.06}_{-0.30}$ & \,\,$8.96^{+18.44}_{-3.69}$ & \,\,$14.26^{+2.90}_{-12.71}$ & $-2.32$\\
\vspace{2mm}
45 & SWIFTJ0101.5-0308 & 43.90 & 42.75 & 44.38 & 8.13 & $-1.44^{+0.19}_{-0.39}$ & $4.19^{+9.42}_{-3.79}$ & \,\,\,$23.78^{+17.03}_{-32.21}$ & $-1.93$\\
\vspace{2mm}
51 & SWIFTJ0105.7-1414 & 44.30 & 43.48 & 45.03 & 7.88 & $-1.32^{+0.01}_{-0.01}$ & $8.37^{+0.48}_{-0.47}$ & $16.79^{+1.26}_{-1.29}$ & $-1.03$\\

\hline
\end{tabular}
\tablefoot{We have reported the 2500\,${\rm \AA}$, 2\,keV, and the bolometric luminosities with their errors. We also present the black hole masses ($M_{\rm BH}$) of our sources as estimated by \citet{2022ApJS..261....5M} in units of ${\rm log}\,(M_{\rm BH}/{\rm M_{\odot}}$). Finally, we list the estimated value of the optical-to-X-ray spectral indices ($\alpha_{\rm ox}$; Sect. \ref{sect:aox}), the 2--10\,keV X-ray bolometric corrections ($\kappa_{2-10}$; Sect. \ref{sect:kx}), the 4400 ${\rm \AA}$ optical bolometric corrections ($\kappa_{\rm 4400 \AA}$; Sect. \ref{sect:kuvo}), and the Eddington ratios ($\lambda_{\rm Edd}$ in log). Luminosity units: $10^{-42}\,{\rm erg\,s^{-1}}$. The table in its entirety is available at the CDS.}
\label{tab:params}
\end{table*}


\begin{table*}
\centering
\caption{Estimated values of the optical and UV bolometric corrections (described in Sect. \ref{sect:kuvo}).}
\begin{tabular}{cccccccc}
\hline
\hline
\specialrule{0.1em}{0em}{0.5em}
\vspace{1mm}
BAT ID & Swift ID & $\kappa_{\rm V}$ & $\kappa_{\rm B}$ & $\kappa_{\rm U}$ & $\kappa_{\rm W1}$ & $\kappa_{\rm M2}$ & $\kappa_{\rm W2}$\\
\hline\\
\vspace{2mm}
3 & SWIFTJ0002.5+0323 & 50.67 & 26.02 & 21.04 & 14.94 & 16.98 & 10.68\\
\vspace{2mm}
6 & SWIFTJ0006.2+2012 & 54.23 & 27.10 & 21.09 & 14.00 & 15.22 & \,\,\,8.98\\
\vspace{2mm}
16 & SWIFTJ0029.2+1319 & 76.37 & 37.31 & 28.12 & 17.68 & 18.57 & 10.40\\
\vspace{2mm}
19 & SWIFTJ0034.5-7904 & 59.30 & 30.56 & 24.84 & 17.82 & 20.36 & 12.93\\
\vspace{2mm}
34 & SWIFTJ0051.6+2928 & 42.70 & 21.41 & 16.75 & 11.22 & 12.27 & \,\,\,7.30\\
\vspace{2mm}
36 & SWIFTJ0051.9+1724 & 67.06 & 34.31 & 27.61 & 19.44 & 21.96 & 13.70\\
\vspace{2mm}
39 & SWIFTJ0054.9+2524 & 85.93 & 41.90 & 31.51 & 19.72 & 20.66 & 11.53\\
\vspace{2mm}
43 & SWIFTJ0059.4+3150 & 83.93 & 41.17 & 31.22 & 19.83 & 20.95 & 11.85\\
\vspace{2mm}
45 & SWIFTJ0101.5-0308 & 38.02 & 19.52 & 15.78 & 11.21 & 12.74 & \,\,\,8.01\\
\vspace{2mm}
51 & SWIFTJ0105.7-1414 & 77.48 & 38.66 & 30.01 & 19.85 & 21.52 & 12.65\\

\hline
\end{tabular}
\tablefoot{The table in its entirety is available at the CDS.}
\label{tab:kbol}
\end{table*}


\begin{table}
\centering
\caption{Correlation results for Figs. \ref{fig:k210_vs_z}, \ref{fig:k210_vs_l210}, \ref{fig:aox_vs_k210}, \ref{fig:k4400}, \ref{fig:kall_vs_l}, \ref{fig:kall_vs_lbol}, \ref{fig:kall_vs_er}, \ref{fig:k14195_vs_l14195}, \ref{fig:k0.22}, and \ref{fig:k14195}, described in Sect. \ref{sect:k}.}
\begin{tabular}{cccc}
\hline
\hline
\specialrule{0.1em}{0em}{0.5em}
\vspace{1mm}
Correlation & R-Value\tablefootmark{a} & P-Value\tablefootmark{b} & Scatter (dex)\\
\hline\\
\vspace{2mm}
$\kappa_{2-10}-z$ & \,\,\,\,0.08 & 0.25 & -\\
\vspace{2mm}
$\kappa_{2-10}-L_{2-10}$ & \,\,\,\,0.13 & 0.84 & -\\
\vspace{2mm}
$\alpha_{\rm ox}-\kappa_{2-10}$ & $-0.86$ & \,\,$6.51 \times 10^{-69}$ & \,\,0.15\tablefootmark{c}\\
\vspace{2mm}
$\kappa_{\rm 4400\,\AA}-L_{\rm 4400\,\AA}$ & $-0.14$ & 0.03 & -\\
\vspace{2mm}
$\kappa_{\rm 4400\,\AA}-L_{\rm bol}$ & \,\,\,\,0.15 & 0.02 & -\\
\vspace{2mm}
$\kappa_{\rm U}-L_{\rm U}$ & $-0.25$ & $9.50\times10^{-5}$ & 0.27\\
\vspace{2mm}
$\kappa_{\rm W1}-L_{\rm W1}$ & $-0.42$ & \,\,$9.86\times10^{-12}$ & 0.30\\
\vspace{2mm}
$\kappa_{\rm M2}-L_{\rm M2}$ & $-0.52$ & \,\,$2.24\times10^{-17}$ & 0.33\\
\vspace{2mm}
$\kappa_{\rm W2}-L_{\rm W2}$ & $-0.61$ & \,\,$6.09\times10^{-25}$ & 0.37\\
\vspace{2mm}
$\kappa_{\rm W2}-L_{\rm bol}$ &$-0.22$ & $5.80\times10^{-4}$ & 0.45\\
\vspace{2mm}
$\kappa_{\rm M2}-\lambda_{\rm Edd}$ & $-0.20$ & $1.80\times10^{-3}$ & 0.38\\
\vspace{2mm}
$\kappa_{\rm W2}-\lambda_{\rm Edd}$ & $-0.26$ & $7.68\times10^{-5}$ & 0.45\\
\hline\\
\vspace{2mm}
$\kappa_{0.2-2}-L_{\rm bol}$ & \,\,\,\,0.30 & $2.56\times10^{-6}$ & 0.36\\
\vspace{2mm}
$\kappa_{0.2-2}-\lambda_{\rm Edd}$ & \,\,\,\,0.22 & $8.43\times10^{-4}$ & 0.37\\
\vspace{2mm}
$\kappa_{14-195}-L_{14-195}$ & $-0.25$ & $1.19\times10^{-4}$ & 0.38\\
\vspace{2mm}
$\kappa_{14-195}-L_{\rm bol}$ & \,\,\,\,0.19 & $3.80\times10^{-3}$ & 0.39\\
\vspace{2mm}
$\kappa_{14-195}-\lambda_{\rm Edd}$ & \,\,\,\,0.24 & $1.91\times10^{-4}$ & 0.38\\

\hline
\end{tabular}
\tablefoot{We also report the one sigma dispersion (or scatter in dex) for parameters with significant correlations.\\
\tablefoottext{a}{The Pearson's correlation coefficient.}\\
\tablefoottext{b}{The probability of the data set appearing if the null hypothesis is correct.}\\
\tablefoottext{c}{Best-fit achieved using a second-order polynomial.}
}
\label{tab:ktrends}
\end{table}


\begin{table}
\centering
\caption{Coefficients for the best-fit relation of various optical/UV bolometric corrections ($\kappa_{\lambda}$) with the intrinsic luminosity in their respective bands ($L_{\lambda}$), the bolometric luminosity ($L_{\rm bol}$), and the Eddington ratio ($\lambda_{\rm Edd}$), described in Sect. \ref{sect:kuvo}.}
\begin{tabular}{ccc}
\hline
\hline
\specialrule{0.1em}{0em}{0.5em}
\vspace{1mm}
Correlation & A & B\\
\hline\\
\vspace{2mm}
$\kappa_{\rm U}-L_{\rm U}$ & $-0.07\pm0.02$ & \,\,\,$4.74\pm0.81$\\
\vspace{2mm}
$\kappa_{\rm W1}-L_{\rm W1}$ & $-0.14\pm0.02$ & \,\,\,$7.52\pm0.85$\\
\vspace{2mm}
$\kappa_{\rm M2}-L_{\rm M2}$ & $-0.19\pm0.02$ & \,\,\,$9.75\pm0.90$\\
\vspace{2mm}
$\kappa_{\rm W2}-L_{\rm W2}$ & $-0.26\pm0.02$ & $12.41\pm0.95$\\
\vspace{2mm}
$\kappa_{\rm W2}-L_{\rm bol}$ & $-0.12\pm0.03$ & \,\,\,$6.47\pm1.47$\\
\vspace{2mm}
$\kappa_{\rm M2}-\lambda_{\rm Edd}$ & $-0.11\pm0.03$ & \,\,\,$1.35\pm0.05$\\
\vspace{2mm}
$\kappa_{\rm W2}-\lambda_{\rm Edd}$ & $-0.16\pm0.04$ & \,\,\,$1.11\pm0.06$\\

\hline
\end{tabular}
\tablefoot{The format for the linear relations is: ${\rm log}(y) = {\rm A}\times {\rm log}(x) + {\rm B}$, where $y$ is $\kappa_{\lambda}$ and $x$ is either $L_{\lambda}$, $L_{\rm bol}$, or $\lambda_{\rm Edd}$.}
\label{tab:keq}
\end{table}


\begin{table}
\centering
\caption{Median values of all the X-ray and optical/UV bolometric corrections with respective uncertainties and one sigma dispersions (Sect. \ref{sect:k}).}
\begin{tabular}{cccc}
\hline
\hline
\specialrule{0.1em}{0em}{0.5em}
\vspace{1mm}
$\kappa_{\lambda}$ & Median Value & Errors & Dispersion (dex)\\
\hline\\
\vspace{2mm}
$\kappa_{2-10}$ & 15.9 & 0.16 & 0.30\\
\vspace{2mm}
$\kappa_{\rm 4400\,\AA}$ & \,\,\,7.9 & 0.32 & 0.10\\
\vspace{2mm}
$\kappa_{\rm V}$ & 65.4 & 0.47 & 0.14\\
\vspace{2mm}
$\kappa_{\rm B}$ & 35.5 & 0.25 & 0.12\\
\vspace{2mm}
$\kappa_{\rm U}$ & 29.3 & 0.22 & 0.16\\
\vspace{2mm}
$\kappa_{\rm W1}$ & 21.4 & 0.27 & 0.38\\
\vspace{2mm}
$\kappa_{\rm M2}$ & 24.0 & 0.60 & 0.67\\
\vspace{2mm}
$\kappa_{\rm W2}$ & 15.2 & 0.96 & 1.07\\
\hline\\
\vspace{2mm}
$\kappa_{0.2-2}$ & 17.4 & 0.33 & 0.56\\
\vspace{2mm}
$\kappa_{14-195}$ & \,\,\,6.0 & 0.12 & 0.57\\
\vspace{2mm}
$\kappa_{\rm 2500\,\AA}$ & \,\,\,5.7 & 0.09 & 0.45\\
\vspace{2mm}
$\kappa_{\rm 3000\,\AA}$ & \,\,\,6.0 & 0.05 & 0.22\\
\vspace{2mm}
$\kappa_{\rm 5100\,\AA}$ & \,\,\,8.8 & 0.06 & 0.13\\

\hline
\end{tabular}
\tablefoot{Using the median bolometric corrections without appropriately including the dispersions in their values could lead to errors in calculations. The dispersion in dex lists the typical errors in estimating $L_{\rm bol}$ when using a monochromatic indicator ($\kappa_{\lambda}$) for a single source. Kindly refer to Sect. \ref{sect:k} for more details.}
\label{tab:kbol_med}
\end{table}


\begin{table}
\centering
\caption{Coefficients for the best-fit relation of various bolometric corrections with the optical-to-X-ray spectral index ($\alpha_{\rm ox}$) described in Sect. \ref{sect:k}.}
\begin{tabular}{cccc}
\hline
\hline
\specialrule{0.1em}{0em}{0.5em}
\vspace{1mm}
$\kappa_{\lambda}$ & A & B & C\\
\hline\\
\vspace{2mm}
$\kappa_{2-10}$ & \,\,\,\,$1.34\pm0.14$ & \,\,\,\,$2.27\pm0.37$ & $1.91\pm0.24$\\
\vspace{2mm}
$\kappa_{\rm 4400\,\AA}$ & \,\,\,\,$0.78\pm0.07$ & \,\,\,\,$1.89\pm0.09$ & -\\
\vspace{2mm}
$\kappa_{\rm V}$ & \,\,\,\,$0.68\pm0.08$ & \,\,\,\,$2.71\pm0.10$ & -\\
\vspace{2mm}
$\kappa_{\rm B}$ & \,\,\,\,$0.82\pm0.07$ & \,\,\,\,$2.61\pm0.09$ & -\\
\vspace{2mm}
$\kappa_{\rm U}$ & \,\,\,\,$0.87\pm0.18$ & \,\,\,\,$3.26\pm0.45$ & $4.20\pm0.29$\\
\vspace{2mm}
$\kappa_{\rm W1}$ & \,\,\,\,$1.44\pm0.16$ & \,\,\,\,$5.04\pm0.42$ & $5.44\pm0.27$\\
\vspace{2mm}
$\kappa_{\rm M2}$ & \,\,\,\,$1.81\pm0.18$ & \,\,\,\,$6.21\pm0.48$ & $6.40\pm0.31$\\
\vspace{2mm}
$\kappa_{\rm W2}$ & \,\,\,\,$2.27\pm0.24$ & \,\,\,\,$7.67\pm0.61$ & $7.34\pm0.39$\\
\hline\\
\vspace{2mm}
$\kappa_{0.2-2}$ & \,\,\,\,$1.86\pm0.25$ & \,\,\,\,$3.58\pm0.66$ & $2.75\pm0.43$\\
\vspace{2mm}
$\kappa_{14-195}$ & $-1.22\pm0.10$ & $-0.70\pm0.13$ & -\\
\vspace{2mm}
$\kappa_{\rm 2500\,\AA}$ & \,\,\,\,$1.54\pm0.17$ & \,\,\,\,$5.37\pm0.43$ & $5.12\pm0.28$\\
\vspace{2mm}
$\kappa_{\rm 3000\,\AA}$ & \,\,\,\,$1.09\pm0.16$ & \,\,\,\,$3.94\pm0.42$ & $4.04\pm0.27$\\
\vspace{2mm}
$\kappa_{\rm 5100\,\AA}$ & \,\,\,\,$0.72\pm0.08$ & \,\,\,\,$1.88\pm0.10$ & -\\

\hline
\end{tabular}
\tablefoot{The format for linear relations is: ${\rm log}(\kappa_{\lambda}) = {\rm A}\times\alpha_{\rm ox} + {\rm B}$ and the format for a second-degree best-fit relation is: ${\rm log}(\kappa_{\lambda}) = {\rm A}\times\alpha_{\rm ox}^2 + {\rm B}\times{\alpha_{\rm ox}} + {\rm C}$.}
\label{tab:aox_rel}
\end{table}


\subsubsection{Additional}\label{sect:k_add}

In this section we list some additional bolometric corrections calculated in this work that could be useful to estimate bolometric luminosities depending on the data at hand. In the X-rays, we report the values of bolometric correction in the 0.2--2\,keV (soft X-rays: $\kappa_{0.2-2}$) and the 14--195\,keV (hard X-rays: $\kappa_{14-195}$) energy bands. We obtain a median value of $\kappa_{0.2-2} = 17.4 \pm 0.3$ and $\kappa_{14-195} = 6.0 \pm 0.1$. Similar to $\kappa_{2-10}$ (see Sect. \ref{sect:kx}), $\kappa_{0.2-2}$ stays constant with the 0.2--2\,keV intrinsic luminosity across five orders of magnitude ($10^{40} < L_{0.2-2}/{\rm erg\,s^{-1}} < 10^{45}$). However, for the hard X-ray bolometric correction, an anticorrelation is recovered between $\kappa_{14-195}$ and $L_{14-195}$ (see Fig. \ref{fig:k14195_vs_l14195} and Table \ref{tab:ktrends}). This could probably just be an obvious consequence of how we define bolometric corrections and because of the large range of this specific energy band. The best-fit relation (shown as a black line in Fig. \ref{fig:k14195_vs_l14195}) is defined as follows:

\begin{equation}\label{eq:k14195l14195}
    {\rm log}\,(\kappa_{14-195}) = (-0.11 \pm 0.03) \times {\rm log}\,(L_{14-195}) + (5.55 \pm 1.20)
\end{equation}\\
We checked if $\kappa_{\rm 0.2-2}$ and $\kappa_{14-195}$ correlate with any other physical properties of AGN, such as $L_{\rm bol}$, $M_{\rm BH}$, and $\lambda_{\rm Edd}$. We found a significant dependence of both $\kappa_{\rm 0.2-2}$ and $\kappa_{14-195}$ on $L_{\rm bol}$ and $\lambda_{\rm Edd}$ (shown in Figs. \ref{fig:k0.22} and \ref{fig:k14195}, respectively and Table \ref{tab:ktrends}). The best-fit linear regressions (shown as black lines) are as follows:

\begin{equation}\label{eq:k0.22}
    {\rm log}\,(\kappa_{0.2-2}) = (0.13 \pm 0.03) \times {\rm log}\,(L_{\rm bol}) + (-4.33 \pm 1.16)
\end{equation}
\begin{equation}\label{eq:k0.22er}
    {\rm log}\,(\kappa_{0.2-2}) = (0.11 \pm 0.03) \times {\rm log}\,(\lambda_{\rm Edd}) + (1.43 \pm 0.05)
\end{equation}
\begin{equation}\label{eq:k14195}
    {\rm log}\,(\kappa_{14-195}) = (0.08 \pm 0.03) \times {\rm log}\,(L_{\rm bol}) + (-2.82 \pm 1.26)
\end{equation}
\begin{equation}\label{eq:k14195_er}
    {\rm log}\,(\kappa_{14-195}) = (0.13 \pm 0.04) \times {\rm log}\,(\lambda_{\rm Edd}) + (1.04 \pm 0.05)
\end{equation}\\
We also provide scaling relations of $\alpha_{\rm ox}$ with both $\kappa_{0.2-2}$ and $\kappa_{14-195}$ in Table \ref{tab:aox_rel}.

In the optical/UV wavelength bands, we report three additional bolometric corrections at $2500\AA$ ($\kappa_{\rm 2500\AA}$), $3000\AA$ ($\kappa_{\rm 3000\AA}$), and $5100\AA$ ($\kappa_{\rm 5100\AA}$). The median values with errors and dispersions are listed in Table \ref{tab:kbol_med}. We also list the corresponding coefficients for the best-fit relation with $\alpha_{\rm ox}$ in Table \ref{tab:aox_rel}.


\section{Summary and conclusion}\label{sect:summary}

In this paper, we present one of the largest studies to date of optical-to-X-ray wavelength SEDs for a sample of 236 unobscured, hard-X-ray-selected, nearby AGN from the 70-month \textit{Swift}/BAT catalog. We compiled simultaneous optical, UV, and X-ray observations for all sources to mitigate any issues arising due to differences in the variability timescales at different wavelengths. We extracted the 0.3--10\,keV X-ray spectra and magnitudes and fluxes in the six \textit{Swift}/UVOT filters: V, B, U, UVW1, UVM2, and UVW2 for all our sources. We corrected the optical/UV fluxes for galactic dust extinction as well as host galaxy contamination. The latter was done by fitting the 2D surface brightness of the source images using a PSF and a S\'ersic profile with \textsc{GALFIT} to fit the AGN and the host galaxy, respectively. We also performed numerous simulations by creating fake populations of type\,I AGN and fitting them using \textsc{GALFIT}, to quantitatively inspect the reliability of the AGN magnitudes calculated by \textsc{GALFIT}. As a result of these simulations, we were able to calculate correction factors in each of the six \textit{Swift}/UVOT filters, which can be used to correct the \textsc{GALFIT}-estimated PSF fluxes. We constructed and fit the multiwavelength AGN SEDs ($10^{-7}$ to 500\,keV) using a dust-reddened multi-temperature blackbody to model the optical/UV accretion disk and a variety of X-ray models consisting of multiple components to account for the primary X-ray emission and the reprocessed secondary emission. We then used the best-fit SED models to calculate important quantities used to describe AGN properties, such as the total accretion luminosity or bolometric luminosity of our sources, intrinsic optical, UV, and X-ray luminosities, optical-to-X-ray spectral indices, multiwavelength bolometric corrections, and Eddington ratios. Compared to similar past studies (e.g., \citealp{2009MNRAS.392.1124V}; \citealp{2009MNRAS.399.1553V}), we have improved the sample size by almost one dex, and hence expanded the range of luminosities and Eddington ratios that can be probed (both span five orders of magnitude), adopted improved SED fitting techniques, and implemented more detailed procedures to remove contamination due to the host galaxy to AGN light in the optical/UV using \textsc{GALFIT}. The main results of this study are summarized as follows:

\begin{itemize}
    \item We report total aperture optical/UV source magnitudes (Table \ref{tab:UVOT_mag}) and fluxes (Table \ref{tab:UVOT_flux}), along with errors, in the six \textit{Swift}/UVOT filters (Sect. \ref{sect:uvot}) for all 292 unobscured AGN in our original sample. For the final sample of 236 sources, we report AGN magnitudes (Table \ref{tab:galfit_mag}) and fluxes (Table \ref{tab:galfit_flux}) devoid of the host galaxy light, estimated from the image decomposition and three-component fitting done using \textsc{GALFIT} (Sect. \ref{sect:galfit}). The results of our X-ray spectral fitting in the 0.3--10\,keV band (Sect. \ref{sect:xray_spec}) are summarized in Table \ref{tab:xray}, listing values of the photon index ($\Gamma$) and the intrinsic 2--10\,keV X-ray flux for all AGN in our initial sample. Best-fit values of parameters from the optical/UV SED fitting (Sect. \ref{sect:sed}), such as the maximum disk temperature, host galaxy extinction, and the intrinsic flux in the optical $+$ UV energy range ($10^{-7}$ to 0.1\,keV) are presented in Table \ref{tab:sed}.
    \item Some important quantities, including monochromatic luminosities at $2500\,{\rm \AA}$ and 2\,keV, the optical-to-X-ray spectral index ($\alpha_{\rm ox}$), bolometric luminosity, black hole mass, and Eddington ratio, are reported for the 236 unobscured AGN in our sample in Table \ref{tab:params}. Eight different multiwavelength bolometric corrections are provided for all sources in Tables \ref{tab:params} and \ref{tab:kbol}, and their median values with uncertainties are also given in Table \ref{tab:kbol_med}.
    \item We confirm the strong correlation observed between the X-ray and UV monochromatic luminosity at 2\,keV and 2500\,$\rm \AA$ for AGN, given by $L_{\rm 2\,keV} \propto L_{\rm 2500\,\AA}^{\beta}$ with $\beta = 0.850 \pm 0.013$ (Eq. \ref{eq:l2keV_l2500}, Fig. \ref{fig:l2500_vs_l2keV}). 
    \item We calculated the optical-to-X-ray spectral index for our AGN sample, finding a median $\alpha_{\rm ox} = -1.28$. We find no dependence of $\alpha_{\rm ox}$ on redshift or $L_{\rm 2\,keV}$ (Figs. \ref{fig:aox_vs_z} and \ref{fig:aox_vs_l2keV}), and confirm the strong anticorrelation between $\alpha_{\rm ox}$ and $L_{\rm 2500\,\AA}$ (Eq. \ref{eq:aox_l2500}, Fig. \ref{fig:aox_vs_l2500}). We find weak anticorrelations between $\alpha_{\rm ox}$ and both $\lambda_{\rm Edd}$ and $L_{\rm bol}$ (Eqs. \ref{eq:aox_edd} and \ref{eq:aox_lbol}, Fig. \ref{fig:aox3}), and no correlation with $M_{\rm BH}$.
    \item We computed 2--10\,keV X-ray bolometric corrections for our unobscured AGN sample, finding a median $\kappa_{2-10} = 15.9 \pm 0.2$. We obtained a scatter of 0.3 dex at one sigma in the value of $\kappa_{2-10}$ for our AGN sample (Fig. \ref{fig:k210_vs_l210}), which could be due to quantities such as the inclination angle and/or black hole spin that were not constrained in our SED fitting. More findings on how $\kappa_{2-10}$ relates to AGN properties will be presented in a forthcoming dedicated paper.
    \item We present a tight (scatter = 0.15 dex) second-order relation between $\kappa_{2-10}$ and $\alpha_{\rm ox}$, which can be useful to calculate $L_{\rm bol}$ in the absence of multiband data, solely from the intrinsic, monochromatic luminosities at 2\,keV and $2500\,\rm\AA$ (Eq. \ref{eq:k210_aox}, Fig. \ref{fig:aox_vs_k210}).
    \item For the optical/UV energies, we list seven different bolometric corrections: the $4400\,\rm\AA$ optical bolometric correction and one for each of the six \textit{Swift}/UVOT filters. None of these $\kappa_{\lambda}$ show any dependence on redshift, which negates any biases due to redshift in our SED fitting analysis. We explore if and how these optical/UV bolometric corrections correlate with different physical properties of AGN. Our main finding is that the highest-energy UV bolometric corrections ($\kappa_{\rm M2}$ and $\kappa_{\rm W2}$) show significant negative correlations with both $L_{\rm bol}$ and $\lambda_{\rm Edd}$, while the lowest-energy ones ($\kappa_{\rm V}$, $\kappa_{\rm B}$, and $\kappa_{\rm 4400\,\AA}$) show no clear dependence on any of the AGN accretion properties (Figs. \ref{fig:kall_vs_lbol} and \ref{fig:kall_vs_er}, Tables \ref{tab:ktrends} and \ref{tab:keq}). This result can be explained if we assume the following: (a) the standard \citet{1973A&A....24..337S} disk model with a multi-temperature blackbody emission, and (b) possible contamination in the lower energy optical photometric bands from the diffused emission produced in the BLR (Sect. \ref{sect:kuvo}). Furthermore, the lack of any kind of dependence of $\kappa_{\lambda}$ on black hole mass implies that it is not in any way a controlling factor for the amount of radiation emitted via accretion among the wavelength ranges probed here.
    \item Our results show significant scatter ($\sim$ 0.1--1\,dex) in all bolometric correction factors (Table \ref{tab:kbol_med}). This scatter cannot be attributed either to sources with high Galactic extinction or to different variability timescales between the multiwavelength emission of AGN. We therefore strongly recommend the inclusion of the reported dispersion values while calculating $L_{\rm bol}$ from $\kappa_{\lambda}$ for individual sources and the use of appropriate scaling relations of $\kappa_{\lambda}$ when carrying out population studies.
\end{itemize}

Through this large multiwavelength study, we have attempted to perform a comprehensive broadband SED fitting analysis of a representative sample of AGN in the local Universe, to provide the community estimates of some useful quantities, such as bolometric luminosities, Eddington ratios, optical-to-X-ray spectral indices, and multiband bolometric corrections, as well as some important relations between these quantities. Given the time and resources, we have made an effort to keep this study as exhaustive as possible by including appropriate tools to take into account possible sources of errors that might affect our analysis, including AGN variability, dust-extinction, and host galaxy contamination. We have also tried to demonstrate here the important role such studies can play in improving our knowledge about the accretion and emission mechanisms in AGN. With the wealth of available archival data and by employing more sophisticated theoretical and computational models, we can also extract information about the host galaxy. To give a few examples, there exist SED fitting codes such as \textsc{CIGALE}/\textsc{X-CIGALE} (\citealp{2009A&A...507.1793N}; \citealp{2019A&A...622A.103B}; \citealp{2020MNRAS.491..740Y}; \citealp{2022ApJ...927..192Y}), \textsc{AGNfitter} (\citealp{2016ApJ...833...98C}; \citealp{2024A&A...688A..46M}), or \textsc{FortesFit} \citep{2018MNRAS.473.5658R}, AGN-galaxy templates provided by \citet{2006ApJS..166..470R}, \citet{2010ApJ...713..970A} or \citet{2021MNRAS.508..737T}, and theoretical accretion disk models (e.g., \citealp{2012MNRAS.420.1848D}; \citealp{2012MNRAS.426..656S}) that can simultaneously constrain the star formation history of the host galaxy and the accretion properties of the AGN (e.g., accretion rates and efficiency). An extensive effort in this direction could prove to be an interesting milestone in our understanding of the AGN-galaxy coevolution scenario. Two such studies are ongoing: one is focused on modeling the X-ray-to-far-IR SEDs of the BASS DR2 AGN using CIGALE (\textcolor{blue}{Rojas et al. in prep.}) and the other on modeling our optical-to-X-ray SEDs using \citet{2012MNRAS.420.1848D} models (\textcolor{blue}{Kallov\'a et al. in prep.}).\\


\section{Data availability}

All the optical-to-X-ray SED fits generated in this work are available in the online version of the paper.\\


\begin{acknowledgements}

We would like to thank the anonymous referee for their useful suggestions that helped improve this manuscript. This work made use of data from the NASA/IPAC Infrared Science Archive and NASA/IPAC Extragalactic Database (NED), which are operated by the Jet Propulsion Laboratory, California Institute of Technology, under contract with the National Aeronautics and Space Administration. This research has made use of data and/or software provided by the High Energy Astrophysics Science Archive Research Center (HEASARC), which is a service of the Astrophysics Science Division at NASA/GSFC and the High Energy Astrophysics Division of the Smithsonian Astrophysical Observatory. This research also made use of \texttt{photutils}, an Astropy package for the detection and photometry of astronomical sources \citep{2021zndo...4624996B}. We acknowledge financial support from: a 2018 grant from the ESO-Government of Chile Joint Committee (KKG); the Belgian Federal Science Policy Office (BELSPO) in the framework of the PRODEX Programme of the European Space Agency (KKG); ANID CATA-BASAL project FB210003 (CR, RJA, FEB); ANID FONDECYT Regular grant \#1230345 (CR), \#1231718 (RJA), and \#1200495 (FEB); NASA ADAP award 80NSSC19K0749 (MJK); ANID Millennium Science Initiative Program - ICN12\_009 (FEB); ANID FONDECYT Postdoctorado grant \#3220516 (MJT), \#3210157 (AFR), and \#3230310 (YD); the European Research Council (ERC) under the European Union’s Horizon 2020 research and innovation program, grant agreement \#950533 (BT); the Israel Science Foundation, grant \#1849/19 (BT); the Korea Astronomy and Space Science Institute under the R\&D program supervised by the Ministry of Science and ICT, Project \#2024-1-831-01 (KO); the National Research Foundation of Korea, NRF-2020R1C1C1005462 (KO).

\end{acknowledgements}


\bibliographystyle{aa}
\bibliography{References}

\begin{thebibliography}{153}
\expandafter\ifx\csname natexlab\endcsname\relax\def\natexlab#1{#1}\fi

\bibitem[{{Alloin} {et~al.}(1985){Alloin}, {Pelat}, {Phillips}, \& {Whittle}}]{1985ApJ...288..205A}
{Alloin}, D., {Pelat}, D., {Phillips}, M., \& {Whittle}, M. 1985, \apj, 288, 205

\bibitem[{{Ananna} {et~al.}(2020){Ananna}, {Urry}, {Treister}, {Hickox}, {Shankar}, {Ricci}, {Cappelluti}, {Marchesi}, \& {Turner}}]{2020ApJ...903...85A}
{Ananna}, T.~T., {Urry}, C.~M., {Treister}, E., {et~al.} 2020, \apj, 903, 85

\bibitem[{{Antonucci}(1993)}]{1993ARA&A..31..473A}
{Antonucci}, R. 1993, \araa, 31, 473

\bibitem[{{Arnaud}(1996)}]{1996ASPC..101...17A}
{Arnaud}, K.~A. 1996, Astronomical Society of the Pacific Conference Series, Vol. 101, {XSPEC: The First Ten Years}, ed. G.~H. {Jacoby} \& J.~{Barnes}, 17

\bibitem[{{Assef} {et~al.}(2010){Assef}, {Kochanek}, {Brodwin}, {Cool}, {Forman}, {Gonzalez}, {Hickox}, {Jones}, {Le Floc'h}, {Moustakas}, {Murray}, \& {Stern}}]{2010ApJ...713..970A}
{Assef}, R.~J., {Kochanek}, C.~S., {Brodwin}, M., {et~al.} 2010, \apj, 713, 970

\bibitem[{{Astropy Collaboration} {et~al.}(2022){Astropy Collaboration}, {Price-Whelan}, {Lim}, {Earl}, {Starkman}, {Bradley}, {Shupe}, {Patil}, {Corrales}, {Brasseur}, {N{"o}the}, {Donath}, {Tollerud}, {Morris}, {Ginsburg}, {Vaher}, {Weaver}, {Tocknell}, {Jamieson}, {van Kerkwijk}, {Robitaille}, {Merry}, {Bachetti}, {G{"u}nther}, {Aldcroft}, {Alvarado-Montes}, {Archibald}, {B{'o}di}, {Bapat}, {Barentsen}, {Baz{'a}n}, {Biswas}, {Boquien}, {Burke}, {Cara}, {Cara}, {Conroy}, {Conseil}, {Craig}, {Cross}, {Cruz}, {D'Eugenio}, {Dencheva}, {Devillepoix}, {Dietrich}, {Eigenbrot}, {Erben}, {Ferreira}, {Foreman-Mackey}, {Fox}, {Freij}, {Garg}, {Geda}, {Glattly}, {Gondhalekar}, {Gordon}, {Grant}, {Greenfield}, {Groener}, {Guest}, {Gurovich}, {Handberg}, {Hart}, {Hatfield-Dodds}, {Homeier}, {Hosseinzadeh}, {Jenness}, {Jones}, {Joseph}, {Kalmbach}, {Karamehmetoglu}, {Ka{l}uszy{'n}ski}, {Kelley}, {Kern}, {Kerzendorf}, {Koch}, {Kulumani}, {Lee}, {Ly}, {Ma}, {MacBride}, {Maljaars}, {Muna}, {Murphy}, {Norman}, {O'Steen},
  {Oman}, {Pacifici}, {Pascual}, {Pascual-Granado}, {Patil}, {Perren}, {Pickering}, {Rastogi}, {Roulston}, {Ryan}, {Rykoff}, {Sabater}, {Sakurikar}, {Salgado}, {Sanghi}, {Saunders}, {Savchenko}, {Schwardt}, {Seifert-Eckert}, {Shih}, {Jain}, {Shukla}, {Sick}, {Simpson}, {Singanamalla}, {Singer}, {Singhal}, {Sinha}, {Sip{H{o}}cz}, {Spitler}, {Stansby}, {Streicher}, {{{S}}umak}, {Swinbank}, {Taranu}, {Tewary}, {Tremblay}, {Val-Borro}, {Van Kooten}, {Vasovi{'c}}, {Verma}, {de Miranda Cardoso}, {Williams}, {Wilson}, {Winkel}, {Wood-Vasey}, {Xue}, {Yoachim}, {Zhang}, {Zonca}, \& {Astropy Project Contributors}}]{astropy:2022}
{Astropy Collaboration}, {Price-Whelan}, A.~M., {Lim}, P.~L., {et~al.} 2022, apj, 935, 167

\bibitem[{{Avni} \& {Tananbaum}(1982)}]{1982ApJ...262L..17A}
{Avni}, Y. \& {Tananbaum}, H. 1982, \apjl, 262, L17

\bibitem[{{Avni} \& {Tananbaum}(1986)}]{1986ApJ...305...83A}
{Avni}, Y. \& {Tananbaum}, H. 1986, \apj, 305, 83

\bibitem[{{Baek} {et~al.}(2019){Baek}, {Chung}, {Schawinski}, {Oh}, {Wong}, {Koss}, {Ricci}, {Trakhtenbrot}, {Smith}, \& {Ueda}}]{2019MNRAS.488.4317B}
{Baek}, J., {Chung}, A., {Schawinski}, K., {et~al.} 2019, \mnras, 488, 4317

\bibitem[{{Barthelmy} {et~al.}(2005){Barthelmy}, {Barbier}, {Cummings}, {Fenimore}, {Gehrels}, {Hullinger}, {Krimm}, {Markwardt}, {Palmer}, {Parsons}, {Sato}, {Suzuki}, {Takahashi}, {Tashiro}, \& {Tueller}}]{2005SSRv..120..143B}
{Barthelmy}, S.~D., {Barbier}, L.~M., {Cummings}, J.~R., {et~al.} 2005, \ssr, 120, 143

\bibitem[{{Baumgartner} {et~al.}(2013){Baumgartner}, {Tueller}, {Markwardt}, {Skinner}, {Barthelmy}, {Mushotzky}, {Evans}, \& {Gehrels}}]{2013ApJS..207...19B}
{Baumgartner}, W.~H., {Tueller}, J., {Markwardt}, C.~B., {et~al.} 2013, \apjs, 207, 19

\bibitem[{{Bechtold} {et~al.}(2003){Bechtold}, {Siemiginowska}, {Shields}, {Czerny}, {Janiuk}, {Hamann}, {Aldcroft}, {Elvis}, \& {Dobrzycki}}]{2003ApJ...588..119B}
{Bechtold}, J., {Siemiginowska}, A., {Shields}, J., {et~al.} 2003, \apj, 588, 119

\bibitem[{{Beloborodov}(1999)}]{1999ASPC..161..295B}
{Beloborodov}, A.~M. 1999, in Astronomical Society of the Pacific Conference Series, Vol. 161, High Energy Processes in Accreting Black Holes, ed. J.~{Poutanen} \& R.~{Svensson}, 295

\bibitem[{{Bianchi} {et~al.}(2012){Bianchi}, {Maiolino}, \& {Risaliti}}]{2012AdAst2012E..17B}
{Bianchi}, S., {Maiolino}, R., \& {Risaliti}, G. 2012, Advances in Astronomy, 2012, 782030

\bibitem[{{Bonning} {et~al.}(2007){Bonning}, {Cheng}, {Shields}, {Salviander}, \& {Gebhardt}}]{2007ApJ...659..211B}
{Bonning}, E.~W., {Cheng}, L., {Shields}, G.~A., {Salviander}, S., \& {Gebhardt}, K. 2007, \apj, 659, 211

\bibitem[{{Bonning} {et~al.}(2013){Bonning}, {Shields}, {Stevens}, \& {Salviander}}]{2013ApJ...770...30B}
{Bonning}, E.~W., {Shields}, G.~A., {Stevens}, A.~C., \& {Salviander}, S. 2013, \apj, 770, 30

\bibitem[{{Boquien} {et~al.}(2019){Boquien}, {Burgarella}, {Roehlly}, {Buat}, {Ciesla}, {Corre}, {Inoue}, \& {Salas}}]{2019A&A...622A.103B}
{Boquien}, M., {Burgarella}, D., {Roehlly}, Y., {et~al.} 2019, \aap, 622, A103

\bibitem[{{Bradley} {et~al.}(2021){Bradley}, {Sip{\H{o}}cz}, {Robitaille}, {Tollerud}, {Vin{\'\i}cius}, {Deil}, {Barbary}, {Wilson}, {Busko}, {Donath}, {G{\"u}nther}, {Cara}, {Conseil}, {Bostroem}, {Droettboom}, {Bray}, {Krachyon}, {Lim}, {Andersen Bratholm}, {Barentsen}, {Craig}, {Rathi}, {Pascual}, {Perren}, {Georgiev}, {De Val-Borro}, {Kerzendorf}, {Bach}, {Quint}, \& {Souchereau}}]{2021zndo...4624996B}
{Bradley}, L., {Sip{\H{o}}cz}, B., {Robitaille}, T., {et~al.} 2021, {astropy/photutils: 1.1.0}, Zenodo

\bibitem[{{Breeveld} {et~al.}(2010){Breeveld}, {Curran}, {Hoversten}, {Koch}, {Landsman}, {Marshall}, {Page}, {Poole}, {Roming}, {Smith}, {Still}, {Yershov}, {Blustin}, {Brown}, {Gronwall}, {Holland}, {Kuin}, {McGowan}, {Rosen}, {Boyd}, {Broos}, {Carter}, {Chester}, {Hancock}, {Huckle}, {Immler}, {Ivanushkina}, {Kennedy}, {Mason}, {Morgan}, {Oates}, {de Pasquale}, {Schady}, {Siegel}, \& {vand en Berk}}]{2010MNRAS.406.1687B}
{Breeveld}, A.~A., {Curran}, P.~A., {Hoversten}, E.~A., {et~al.} 2010, \mnras, 406, 1687

\bibitem[{{Brightman} {et~al.}(2017){Brightman}, {Balokovi{\'c}}, {Ballantyne}, {Bauer}, {Boorman}, {Buchner}, {Brandt}, {Comastri}, {Del Moro}, {Farrah}, {Gandhi}, {Harrison}, {Koss}, {Lanz}, {Masini}, {Ricci}, {Stern}, {Vasudevan}, \& {Walton}}]{2017ApJ...844...10B}
{Brightman}, M., {Balokovi{\'c}}, M., {Ballantyne}, D.~R., {et~al.} 2017, \apj, 844, 10

\bibitem[{{Burrows} {et~al.}(2005){Burrows}, {Hill}, {Nousek}, {Kennea}, {Wells}, {Osborne}, {Abbey}, {Beardmore}, {Mukerjee}, {Short}, {Chincarini}, {Campana}, {Citterio}, {Moretti}, {Pagani}, {Tagliaferri}, {Giommi}, {Capalbi}, {Tamburelli}, {Angelini}, {Cusumano}, {Br{\"a}uninger}, {Burkert}, \& {Hartner}}]{2005SSRv..120..165B}
{Burrows}, D.~N., {Hill}, J.~E., {Nousek}, J.~A., {et~al.} 2005, \ssr, 120, 165

\bibitem[{{Cackett} {et~al.}(2021){Cackett}, {Bentz}, \& {Kara}}]{2021iSci...24j2557C}
{Cackett}, E.~M., {Bentz}, M.~C., \& {Kara}, E. 2021, iScience, 24, 102557

\bibitem[{{Cackett} {et~al.}(2018){Cackett}, {Chiang}, {McHardy}, {Edelson}, {Goad}, {Horne}, \& {Korista}}]{2018ApJ...857...53C}
{Cackett}, E.~M., {Chiang}, C.-Y., {McHardy}, I., {et~al.} 2018, \apj, 857, 53

\bibitem[{{Cackett} {et~al.}(2007){Cackett}, {Horne}, \& {Winkler}}]{2007MNRAS.380..669C}
{Cackett}, E.~M., {Horne}, K., \& {Winkler}, H. 2007, \mnras, 380, 669

\bibitem[{{Cackett} {et~al.}(2022){Cackett}, {Zoghbi}, \& {Ulrich}}]{2022ApJ...925...29C}
{Cackett}, E.~M., {Zoghbi}, A., \& {Ulrich}, O. 2022, \apj, 925, 29

\bibitem[{{Calistro Rivera} {et~al.}(2016){Calistro Rivera}, {Lusso}, {Hennawi}, \& {Hogg}}]{2016ApJ...833...98C}
{Calistro Rivera}, G., {Lusso}, E., {Hennawi}, J.~F., \& {Hogg}, D.~W. 2016, \apj, 833, 98

\bibitem[{{Calzetti} {et~al.}(2000){Calzetti}, {Armus}, {Bohlin}, {Kinney}, {Koornneef}, \& {Storchi-Bergmann}}]{2000ApJ...533..682C}
{Calzetti}, D., {Armus}, L., {Bohlin}, R.~C., {et~al.} 2000, \apj, 533, 682

\bibitem[{{Cardelli} {et~al.}(1989){Cardelli}, {Clayton}, \& {Mathis}}]{1989ApJ...345..245C}
{Cardelli}, J.~A., {Clayton}, G.~C., \& {Mathis}, J.~S. 1989, \apj, 345, 245

\bibitem[{{Cash}(1979)}]{1979ApJ...228..939C}
{Cash}, W. 1979, \apj, 228, 939

\bibitem[{{Chelouche}(2013)}]{2013ApJ...772....9C}
{Chelouche}, D. 2013, \apj, 772, 9

\bibitem[{{Chelouche} {et~al.}(2019){Chelouche}, {Pozo Nu{\~n}ez}, \& {Kaspi}}]{2019NatAs...3..251C}
{Chelouche}, D., {Pozo Nu{\~n}ez}, F., \& {Kaspi}, S. 2019, Nature Astronomy, 3, 251

\bibitem[{{Collier} \& {Peterson}(2001)}]{2001ApJ...555..775C}
{Collier}, S. \& {Peterson}, B.~M. 2001, \apj, 555, 775

\bibitem[{{Collier} {et~al.}(1998){Collier}, {Horne}, {Kaspi}, {Netzer}, {Peterson}, {Wanders}, {Alexander}, {Bertram}, {Comastri}, {Gaskell}, {Malkov}, {Maoz}, {Mignoli}, {Pogge}, {Pronik}, {Sergeev}, {Snedden}, {Stirpe}, {Bochkarev}, {Burenkov}, {Shapovalova}, \& {Wagner}}]{1998ApJ...500..162C}
{Collier}, S.~J., {Horne}, K., {Kaspi}, S., {et~al.} 1998, \apj, 500, 162

\bibitem[{{den Brok} {et~al.}(2022){den Brok}, {Koss}, {Trakhtenbrot}, {Stern}, {Cantalupo}, {Lamperti}, {Ricci}, {Ricci}, {Oh}, {Bauer}, {Riffel}, {Rodr{\'\i}guez-Ardila}, {B{\"a}r}, {Harrison}, {Ichikawa}, {Mej{\'\i}a-Restrepo}, {Mushotzky}, {Powell}, {Boissay-Malaquin}, {Stalevski}, {Treister}, {Urry}, \& {Veilleux}}]{2022ApJS..261....7D}
{den Brok}, J.~S., {Koss}, M.~J., {Trakhtenbrot}, B., {et~al.} 2022, \apjs, 261, 7

\bibitem[{{Done} {et~al.}(2012){Done}, {Davis}, {Jin}, {Blaes}, \& {Ward}}]{2012MNRAS.420.1848D}
{Done}, C., {Davis}, S.~W., {Jin}, C., {Blaes}, O., \& {Ward}, M. 2012, \mnras, 420, 1848

\bibitem[{{Duras} {et~al.}(2020){Duras}, {Bongiorno}, {Ricci}, {Piconcelli}, {Shankar}, {Lusso}, {Bianchi}, {Fiore}, {Maiolino}, {Marconi}, {Onori}, {Sani}, {Schneider}, {Vignali}, \& {La Franca}}]{2020A&A...636A..73D}
{Duras}, F., {Bongiorno}, A., {Ricci}, F., {et~al.} 2020, \aap, 636, A73

\bibitem[{{Elvis} {et~al.}(1994){Elvis}, {Wilkes}, {McDowell}, {Green}, {Bechtold}, {Willner}, {Oey}, {Polomski}, \& {Cutri}}]{1994ApJS...95....1E}
{Elvis}, M., {Wilkes}, B.~J., {McDowell}, J.~C., {et~al.} 1994, \apjs, 95, 1

\bibitem[{{Feigelson} \& {Nelson}(1985)}]{1985ApJ...293..192F}
{Feigelson}, E.~D. \& {Nelson}, P.~I. 1985, \apj, 293, 192

\bibitem[{{Freeman}(2008)}]{2008IAUS..245....3F}
{Freeman}, K.~C. 2008, in Formation and Evolution of Galaxy Bulges, ed. M.~{Bureau}, E.~{Athanassoula}, \& B.~{Barbuy}, Vol. 245, 3--10

\bibitem[{{Green} {et~al.}(2009){Green}, {Aldcroft}, {Richards}, {Barkhouse}, {Constantin}, {Haggard}, {Karovska}, {Kim}, {Kim}, {Vikhlinin}, {Anderson}, {Mossman}, {Kashyap}, {Myers}, {Silverman}, {Wilkes}, \& {Tananbaum}}]{2009ApJ...690..644G}
{Green}, P.~J., {Aldcroft}, T.~L., {Richards}, G.~T., {et~al.} 2009, \apj, 690, 644

\bibitem[{{Grupe} {et~al.}(2010){Grupe}, {Komossa}, {Leighly}, \& {Page}}]{2010ApJS..187...64G}
{Grupe}, D., {Komossa}, S., {Leighly}, K.~M., \& {Page}, K.~L. 2010, \apjs, 187, 64

\bibitem[{{Gupta} {et~al.}(2021){Gupta}, {Ricci}, {Tortosa}, {Ueda}, {Kawamuro}, {Koss}, {Trakhtenbrot}, {Oh}, {Bauer}, {Ricci}, {Privon}, {Zappacosta}, {Stern}, {Kakkad}, {Piconcelli}, {Veilleux}, {Mushotzky}, {Caglar}, {Ichikawa}, {Elagali}, {Powell}, {Urry}, \& {Harrison}}]{2021MNRAS.504..428G}
{Gupta}, K.~K., {Ricci}, C., {Tortosa}, A., {et~al.} 2021, \mnras, 504, 428

\bibitem[{{Haardt} \& {Maraschi}(1991)}]{1991ApJ...380L..51H}
{Haardt}, F. \& {Maraschi}, L. 1991, \apjl, 380, L51

\bibitem[{{HI4PI Collaboration} {et~al.}(2016){HI4PI Collaboration}, {Ben Bekhti}, {Fl{\"o}er}, {Keller}, {Kerp}, {Lenz}, {Winkel}, {Bailin}, {Calabretta}, {Dedes}, {Ford}, {Gibson}, {Haud}, {Janowiecki}, {Kalberla}, {Lockman}, {McClure-Griffiths}, {Murphy}, {Nakanishi}, {Pisano}, \& {Staveley-Smith}}]{2016A&A...594A.116H}
{HI4PI Collaboration}, {Ben Bekhti}, N., {Fl{\"o}er}, L., {et~al.} 2016, \aap, 594, A116

\bibitem[{{Ho}(2008)}]{2008ARA&A..46..475H}
{Ho}, L.~C. 2008, \araa, 46, 475

\bibitem[{{Hopkins} {et~al.}(2007){Hopkins}, {Richards}, \& {Hernquist}}]{2007ApJ...654..731H}
{Hopkins}, P.~F., {Richards}, G.~T., \& {Hernquist}, L. 2007, \apj, 654, 731

\bibitem[{{Hubeny} {et~al.}(2000){Hubeny}, {Agol}, {Blaes}, \& {Krolik}}]{2000ApJ...533..710H}
{Hubeny}, I., {Agol}, E., {Blaes}, O., \& {Krolik}, J.~H. 2000, \apj, 533, 710

\bibitem[{{Ichikawa} {et~al.}(2019){Ichikawa}, {Ricci}, {Ueda}, {Bauer}, {Kawamuro}, {Koss}, {Oh}, {Rosario}, {Shimizu}, {Stalevski}, {Fuller}, {Packham}, \& {Trakhtenbrot}}]{2019ApJ...870...31I}
{Ichikawa}, K., {Ricci}, C., {Ueda}, Y., {et~al.} 2019, \apj, 870, 31

\bibitem[{{Isobe} {et~al.}(1990){Isobe}, {Feigelson}, {Akritas}, \& {Babu}}]{1990ApJ...364..104I}
{Isobe}, T., {Feigelson}, E.~D., {Akritas}, M.~G., \& {Babu}, G.~J. 1990, \apj, 364, 104

\bibitem[{{Jin} {et~al.}(2024){Jin}, {Lusso}, {Ward}, {Done}, \& {Middei}}]{2024MNRAS.527..356J}
{Jin}, C., {Lusso}, E., {Ward}, M., {Done}, C., \& {Middei}, R. 2024, \mnras, 527, 356

\bibitem[{{Jin} {et~al.}(2012){Jin}, {Ward}, \& {Done}}]{2012MNRAS.425..907J}
{Jin}, C., {Ward}, M., \& {Done}, C. 2012, \mnras, 425, 907

\bibitem[{{Just} {et~al.}(2007){Just}, {Brandt}, {Shemmer}, {Steffen}, {Schneider}, {Chartas}, \& {Garmire}}]{2007ApJ...665.1004J}
{Just}, D.~W., {Brandt}, W.~N., {Shemmer}, O., {et~al.} 2007, \apj, 665, 1004

\bibitem[{{Kawamuro} {et~al.}(2022){Kawamuro}, {Ricci}, {Imanishi}, {Mushotzky}, {Izumi}, {Ricci}, {Bauer}, {Koss}, {Trakhtenbrot}, {Ichikawa}, {Rojas}, {Smith}, {Shimizu}, {Oh}, {den Brok}, {Baba}, {Balokovi{\'c}}, {Chang}, {Kakkad}, {Pfeifle}, {Privon}, {Temple}, {Ueda}, {Harrison}, {Powell}, {Stern}, {Urry}, \& {Sanders}}]{2022ApJ...938...87K}
{Kawamuro}, T., {Ricci}, C., {Imanishi}, M., {et~al.} 2022, \apj, 938, 87

\bibitem[{{Kelly}(2007)}]{2007ApJ...665.1489K}
{Kelly}, B.~C. 2007, \apj, 665, 1489

\bibitem[{{Kinney} {et~al.}(1993){Kinney}, {Bohlin}, {Calzetti}, {Panagia}, \& {Wyse}}]{1993ApJS...86....5K}
{Kinney}, A.~L., {Bohlin}, R.~C., {Calzetti}, D., {Panagia}, N., \& {Wyse}, R. F.~G. 1993, \apjs, 86, 5

\bibitem[{{Koratkar} \& {Blaes}(1999)}]{1999PASP..111....1K}
{Koratkar}, A. \& {Blaes}, O. 1999, \pasp, 111, 1

\bibitem[{Kormendy \& Richstone(1995)}]{doi:10.1146/annurev.aa.33.090195.003053}
Kormendy, J. \& Richstone, D. 1995, Annual Review of Astronomy and Astrophysics, 33, 581

\bibitem[{{Koss} {et~al.}(2011){Koss}, {Mushotzky}, {Veilleux}, {Winter}, {Baumgartner}, {Tueller}, {Gehrels}, \& {Valencic}}]{2011ApJ...739...57K}
{Koss}, M., {Mushotzky}, R., {Veilleux}, S., {et~al.} 2011, \apj, 739, 57

\bibitem[{{Koss} {et~al.}(2017){Koss}, {Trakhtenbrot}, {Ricci}, {Lamperti}, {Oh}, {Berney}, {Schawinski}, {Balokovi{\'c}}, {Baronchelli}, {Crenshaw}, {Fischer}, {Gehrels}, {Harrison}, {Hashimoto}, {Hogg}, {Ichikawa}, {Masetti}, {Mushotzky}, {Sartori}, {Stern}, {Treister}, {Ueda}, {Veilleux}, \& {Winter}}]{2017ApJ...850...74K}
{Koss}, M., {Trakhtenbrot}, B., {Ricci}, C., {et~al.} 2017, \apj, 850, 74

\bibitem[{{Koss} {et~al.}(2022{\natexlab{a}}){Koss}, {Ricci}, {Trakhtenbrot}, {Oh}, {den Brok}, {Mej{\'\i}a-Restrepo}, {Stern}, {Privon}, {Treister}, {Powell}, {Mushotzky}, {Bauer}, {Ananna}, {Balokovi{\'c}}, {B{\"a}r}, {Becker}, {Bessiere}, {Burtscher}, {Caglar}, {Congiu}, {Evans}, {Harrison}, {Heida}, {Ichikawa}, {Kamraj}, {Lamperti}, {Pacucci}, {Ricci}, {Riffel}, {Rojas}, {Schawinski}, {Temple}, {Urry}, {Veilleux}, \& {Williams}}]{2022ApJS..261....2K}
{Koss}, M.~J., {Ricci}, C., {Trakhtenbrot}, B., {et~al.} 2022{\natexlab{a}}, \apjs, 261, 2

\bibitem[{{Koss} {et~al.}(2022{\natexlab{b}}){Koss}, {Trakhtenbrot}, {Ricci}, {Bauer}, {Treister}, {Mushotzky}, {Urry}, {Ananna}, {Balokovi{\'c}}, {den Brok}, {Cenko}, {Harrison}, {Ichikawa}, {Lamperti}, {Lein}, {Mej{\'\i}a-Restrepo}, {Oh}, {Pacucci}, {Pfeifle}, {Powell}, {Privon}, {Ricci}, {Salvato}, {Schawinski}, {Shimizu}, {Smith}, \& {Stern}}]{2022ApJS..261....1K}
{Koss}, M.~J., {Trakhtenbrot}, B., {Ricci}, C., {et~al.} 2022{\natexlab{b}}, \apjs, 261, 1

\bibitem[{{Koss} {et~al.}(2022{\natexlab{c}}){Koss}, {Trakhtenbrot}, {Ricci}, {Oh}, {Bauer}, {Stern}, {Caglar}, {den Brok}, {Mushotzky}, {Ricci}, {Mej{\'\i}a-Restrepo}, {Lamperti}, {Treister}, {B{\"a}r}, {Harrison}, {Powell}, {Privon}, {Riffel}, {Rojas}, {Schawinski}, \& {Urry}}]{2022ApJS..261....6K}
{Koss}, M.~J., {Trakhtenbrot}, B., {Ricci}, C., {et~al.} 2022{\natexlab{c}}, \apjs, 261, 6

\bibitem[{{Krawczyk} {et~al.}(2013){Krawczyk}, {Richards}, {Mehta}, {Vogeley}, {Gallagher}, {Leighly}, {Ross}, \& {Schneider}}]{2013ApJS..206....4K}
{Krawczyk}, C.~M., {Richards}, G.~T., {Mehta}, S.~S., {et~al.} 2013, \apjs, 206, 4

\bibitem[{{Krimm} {et~al.}(2013){Krimm}, {Holland}, {Corbet}, {Pearlman}, {Romano}, {Kennea}, {Bloom}, {Barthelmy}, {Baumgartner}, {Cummings}, {Gehrels}, {Lien}, {Markwardt}, {Palmer}, {Sakamoto}, {Stamatikos}, \& {Ukwatta}}]{2013ApJS..209...14K}
{Krimm}, H.~A., {Holland}, S.~T., {Corbet}, R.~H.~D., {et~al.} 2013, \apjs, 209, 14

\bibitem[{{Kuraszkiewicz} {et~al.}(2003){Kuraszkiewicz}, {Wilkes}, {Hooper}, {McLeod}, {Wood}, {Bjorkman}, {Delain}, {Hughes}, {Elvis}, {Impey}, {Lonsdale}, {Malkan}, {McDowell}, \& {Whitney}}]{2003ApJ...590..128K}
{Kuraszkiewicz}, J.~K., {Wilkes}, B.~J., {Hooper}, E.~J., {et~al.} 2003, \apj, 590, 128

\bibitem[{{Lamperti} {et~al.}(2017){Lamperti}, {Koss}, {Trakhtenbrot}, {Schawinski}, {Ricci}, {Oh}, {Land t}, {Riffel}, {Rodr{\'\i}guez-Ardila}, {Gehrels}, {Harrison}, {Masetti}, {Mushotzky}, {Treister}, {Ueda}, \& {Veilleux}}]{2017MNRAS.467..540L}
{Lamperti}, I., {Koss}, M., {Trakhtenbrot}, B., {et~al.} 2017, \mnras, 467, 540

\bibitem[{{Lawrence}(2012)}]{2012MNRAS.423..451L}
{Lawrence}, A. 2012, \mnras, 423, 451

\bibitem[{{Lusso} {et~al.}(2012){Lusso}, {Comastri}, {Simmons}, {Mignoli}, {Zamorani}, {Vignali}, {Brusa}, {Shankar}, {Lutz}, {Trump}, {Maiolino}, {Gilli}, {Bolzonella}, {Puccetti}, {Salvato}, {Impey}, {Civano}, {Elvis}, {Mainieri}, {Silverman}, {Koekemoer}, {Bongiorno}, {Merloni}, {Berta}, {Le Floc'h}, {Magnelli}, {Pozzi}, \& {Riguccini}}]{2012MNRAS.425..623L}
{Lusso}, E., {Comastri}, A., {Simmons}, B.~D., {et~al.} 2012, \mnras, 425, 623

\bibitem[{{Lusso} {et~al.}(2010){Lusso}, {Comastri}, {Vignali}, {Zamorani}, {Brusa}, {Gilli}, {Iwasawa}, {Salvato}, {Civano}, {Elvis}, {Merloni}, {Bongiorno}, {Trump}, {Koekemoer}, {Schinnerer}, {Le Floc'h}, {Cappelluti}, {Jahnke}, {Sargent}, {Silverman}, {Mainieri}, {Fiore}, {Bolzonella}, {Le F{\`e}vre}, {Garilli}, {Iovino}, {Kneib}, {Lamareille}, {Lilly}, {Mignoli}, {Scodeggio}, \& {Vergani}}]{2010A&A...512A..34L}
{Lusso}, E., {Comastri}, A., {Vignali}, C., {et~al.} 2010, \aap, 512, A34

\bibitem[{{Lusso} \& {Risaliti}(2016)}]{2016ApJ...819..154L}
{Lusso}, E. \& {Risaliti}, G. 2016, \apj, 819, 154

\bibitem[{{Lusso} \& {Risaliti}(2017)}]{2017A&A...602A..79L}
{Lusso}, E. \& {Risaliti}, G. 2017, \aap, 602, A79

\bibitem[{{Lusso} {et~al.}(2020){Lusso}, {Risaliti}, {Nardini}, {Bargiacchi}, {Benetti}, {Bisogni}, {Capozziello}, {Civano}, {Eggleston}, {Elvis}, {Fabbiano}, {Gilli}, {Marconi}, {Paolillo}, {Piedipalumbo}, {Salvestrini}, {Signorini}, \& {Vignali}}]{2020A&A...642A.150L}
{Lusso}, E., {Risaliti}, G., {Nardini}, E., {et~al.} 2020, \aap, 642, A150

\bibitem[{{Magdziarz} \& {Zdziarski}(1995)}]{1995MNRAS.273..837M}
{Magdziarz}, P. \& {Zdziarski}, A.~A. 1995, \mnras, 273, 837

\bibitem[{{Makishima} {et~al.}(1986){Makishima}, {Maejima}, {Mitsuda}, {Bradt}, {Remillard}, {Tuohy}, {Hoshi}, \& {Nakagawa}}]{1986ApJ...308..635M}
{Makishima}, K., {Maejima}, Y., {Mitsuda}, K., {et~al.} 1986, \apj, 308, 635

\bibitem[{{Malkan}(1983)}]{1983ApJ...268..582M}
{Malkan}, M.~A. 1983, \apj, 268, 582

\bibitem[{{Malkan} \& {Sargent}(1982)}]{1982ApJ...254...22M}
{Malkan}, M.~A. \& {Sargent}, W.~L.~W. 1982, \apj, 254, 22

\bibitem[{{Marchese} {et~al.}(2012){Marchese}, {Della Ceca}, {Caccianiga}, {Severgnini}, {Corral}, \& {Fanali}}]{2012A&A...539A..48M}
{Marchese}, E., {Della Ceca}, R., {Caccianiga}, A., {et~al.} 2012, \aap, 539, A48

\bibitem[{{Marconi} {et~al.}(2004){Marconi}, {Risaliti}, {Gilli}, {Hunt}, {Maiolino}, \& {Salvati}}]{2004IAUS..222...49M}
{Marconi}, A., {Risaliti}, G., {Gilli}, R., {et~al.} 2004, in IAU Symposium, Vol. 222, The Interplay Among Black Holes, Stars and ISM in Galactic Nuclei, ed. T.~{Storchi-Bergmann}, L.~C. {Ho}, \& H.~R. {Schmitt}, 49--52

\bibitem[{{Markowitz} {et~al.}(2014){Markowitz}, {Krumpe}, \& {Nikutta}}]{2014MNRAS.439.1403M}
{Markowitz}, A.~G., {Krumpe}, M., \& {Nikutta}, R. 2014, \mnras, 439, 1403

\bibitem[{{Marshall} {et~al.}(2022){Marshall}, {Auger-Williams}, {Banerji}, {Maiolino}, \& {Bowler}}]{2022MNRAS.515.5617M}
{Marshall}, A., {Auger-Williams}, M.~W., {Banerji}, M., {Maiolino}, R., \& {Bowler}, R. 2022, \mnras, 515, 5617

\bibitem[{{Mart{\'\i}nez-Ram{\'\i}rez} {et~al.}(2024){Mart{\'\i}nez-Ram{\'\i}rez}, {Rivera}, {Lusso}, {Bauer}, {Nardini}, {Buchner}, {Brown}, {Pineda}, {Temple}, {Banerji}, {Stalevski}, \& {Hennawi}}]{2024A&A...688A..46M}
{Mart{\'\i}nez-Ram{\'\i}rez}, L.~N., {Rivera}, G.~C., {Lusso}, E., {et~al.} 2024, \aap, 688, A46

\bibitem[{{Massaro} {et~al.}(2009){Massaro}, {Giommi}, {Leto}, {Marchegiani}, {Maselli}, {Perri}, {Piranomonte}, \& {Sclavi}}]{2009A&A...495..691M}
{Massaro}, E., {Giommi}, P., {Leto}, C., {et~al.} 2009, \aap, 495, 691

\bibitem[{{McKaig} {et~al.}(2023){McKaig}, {Ricci}, {Paltani}, {Gupta}, {Abel}, \& {Ueda}}]{2023MNRAS.526.5072M}
{McKaig}, J., {Ricci}, C., {Paltani}, S., {et~al.} 2023, \mnras, 526, 5072

\bibitem[{{Mej{\'\i}a-Restrepo} {et~al.}(2022){Mej{\'\i}a-Restrepo}, {Trakhtenbrot}, {Koss}, {Oh}, {den Brok}, {Stern}, {Powell}, {Ricci}, {Caglar}, {Ricci}, {Bauer}, {Treister}, {Harrison}, {Urry}, {Ananna}, {Asmus}, {Assef}, {B{\"a}r}, {Bessiere}, {Burtscher}, {Ichikawa}, {Kakkad}, {Kamraj}, {Mushotzky}, {Privon}, {Rojas}, {Sani}, {Schawinski}, \& {Veilleux}}]{2022ApJS..261....5M}
{Mej{\'\i}a-Restrepo}, J.~E., {Trakhtenbrot}, B., {Koss}, M.~J., {et~al.} 2022, \apjs, 261, 5

\bibitem[{{Miller} {et~al.}(2023){Miller}, {Cackett}, {Goad}, {Horne}, {Barth}, {Romero-Colmenero}, {Fausnaugh}, {Gelbord}, {Korista}, {Landt}, {Treu}, \& {Winkler}}]{2023ApJ...953..137M}
{Miller}, J.~A., {Cackett}, E.~M., {Goad}, M.~R., {et~al.} 2023, \apj, 953, 137

\bibitem[{{Mitchell} {et~al.}(2023){Mitchell}, {Done}, {Ward}, {Kynoch}, {Hagen}, {Lusso}, \& {Landt}}]{2023MNRAS.524.1796M}
{Mitchell}, J. A.~J., {Done}, C., {Ward}, M.~J., {et~al.} 2023, \mnras, 524, 1796

\bibitem[{{Mitsuda} {et~al.}(1984){Mitsuda}, {Inoue}, {Koyama}, {Makishima}, {Matsuoka}, {Ogawara}, {Shibazaki}, {Suzuki}, {Tanaka}, \& {Hirano}}]{1984PASJ...36..741M}
{Mitsuda}, K., {Inoue}, H., {Koyama}, K., {et~al.} 1984, \pasj, 36, 741

\bibitem[{{Narayan}(2005)}]{2005Ap&SS.300..177N}
{Narayan}, R. 2005, \apss, 300, 177

\bibitem[{{Netzer}(2013)}]{2013peag.book.....N}
{Netzer}, H. 2013, {The Physics and Evolution of Active Galactic Nuclei}

\bibitem[{{Noll} {et~al.}(2009){Noll}, {Burgarella}, {Giovannoli}, {Buat}, {Marcillac}, \& {Mu{\~n}oz-Mateos}}]{2009A&A...507.1793N}
{Noll}, S., {Burgarella}, D., {Giovannoli}, E., {et~al.} 2009, \aap, 507, 1793

\bibitem[{{Novikov} \& {Thorne}(1973)}]{1973blho.conf..343N}
{Novikov}, I.~D. \& {Thorne}, K.~S. 1973, in Black Holes (Les Astres Occlus), 343--450

\bibitem[{{Oh} {et~al.}(2022){Oh}, {Koss}, {Ueda}, {Stern}, {Ricci}, {Trakhtenbrot}, {Powell}, {den Brok}, {Lamperti}, {Mushotzky}, {Ricci}, {B{\"a}r}, {Rojas}, {Ichikawa}, {Riffel}, {Treister}, {Harrison}, {Urry}, {Bauer}, \& {Schawinski}}]{2022ApJS..261....4O}
{Oh}, K., {Koss}, M.~J., {Ueda}, Y., {et~al.} 2022, \apjs, 261, 4

\bibitem[{{Padovani} {et~al.}(2017){Padovani}, {Alexander}, {Assef}, {De Marco}, {Giommi}, {Hickox}, {Richards}, {Smol{\v{c}}i{\'c}}, {Hatziminaoglou}, {Mainieri}, \& {Salvato}}]{2017A&ARv..25....2P}
{Padovani}, P., {Alexander}, D.~M., {Assef}, R.~J., {et~al.} 2017, \aapr, 25, 2

\bibitem[{{Paliya} {et~al.}(2019){Paliya}, {Koss}, {Trakhtenbrot}, {Ricci}, {Oh}, {Ajello}, {Stern}, {Powell}, {Urry}, {Harrison}, {Lamperti}, {Mushotzky}, {Marcotulli}, {Mej{\'\i}a-Restrepo}, \& {Hartmann}}]{2019ApJ...881..154P}
{Paliya}, V.~S., {Koss}, M., {Trakhtenbrot}, B., {et~al.} 2019, \apj, 881, 154

\bibitem[{{Pei}(1992)}]{1992ApJ...395..130P}
{Pei}, Y.~C. 1992, \apj, 395, 130

\bibitem[{{Peng} {et~al.}(2002){Peng}, {Ho}, {Impey}, \& {Rix}}]{2002AJ....124..266P}
{Peng}, C.~Y., {Ho}, L.~C., {Impey}, C.~D., \& {Rix}, H.-W. 2002, \aj, 124, 266

\bibitem[{{Peng} {et~al.}(2010){Peng}, {Ho}, {Impey}, \& {Rix}}]{2010AJ....139.2097P}
{Peng}, C.~Y., {Ho}, L.~C., {Impey}, C.~D., \& {Rix}, H.-W. 2010, \aj, 139, 2097

\bibitem[{{Poole} {et~al.}(2008){Poole}, {Breeveld}, {Page}, {Land sman}, {Holland}, {Roming}, {Kuin}, {Brown}, {Gronwall}, {Hunsberger}, {Koch}, {Mason}, {Schady}, {vanden Berk}, {Blustin}, {Boyd}, {Broos}, {Carter}, {Chester}, {Cucchiara}, {Hancock}, {Huckle}, {Immler}, {Ivanushkina}, {Kennedy}, {Marshall}, {Morgan}, {Pandey}, {de Pasquale}, {Smith}, \& {Still}}]{2008MNRAS.383..627P}
{Poole}, T.~S., {Breeveld}, A.~A., {Page}, M.~J., {et~al.} 2008, \mnras, 383, 627

\bibitem[{{Ramos Almeida} \& {Ricci}(2017)}]{2017NatAs...1..679R}
{Ramos Almeida}, C. \& {Ricci}, C. 2017, Nature Astronomy, 1, 679

\bibitem[{{Rankine} {et~al.}(2024){Rankine}, {Aird}, {Ruiz}, \& {Georgakakis}}]{2024MNRAS.527.9004R}
{Rankine}, A.~L., {Aird}, J., {Ruiz}, A., \& {Georgakakis}, A. 2024, \mnras, 527, 9004

\bibitem[{{Reeves} {et~al.}(2008){Reeves}, {Done}, {Pounds}, {Terashima}, {Hayashida}, {Anabuki}, {Uchino}, \& {Turner}}]{2008MNRAS.385L.108R}
{Reeves}, J., {Done}, C., {Pounds}, K., {et~al.} 2008, \mnras, 385, L108

\bibitem[{{Ricci} {et~al.}(2023){Ricci}, {Chang}, {Kawamuro}, {Privon}, {Mushotzky}, {Trakhtenbrot}, {Laor}, {Koss}, {Smith}, {Gupta}, {Dimopoulos}, {Aalto}, \& {Ros}}]{2023ApJ...952L..28R}
{Ricci}, C., {Chang}, C.-S., {Kawamuro}, T., {et~al.} 2023, \apjl, 952, L28

\bibitem[{{Ricci} {et~al.}(2018){Ricci}, {Ho}, {Fabian}, {Trakhtenbrot}, {Koss}, {Ueda}, {Lohfink}, {Shimizu}, {Bauer}, {Mushotzky}, {Schawinski}, {Paltani}, {Lamperti}, {Treister}, \& {Oh}}]{2018MNRAS.480.1819R}
{Ricci}, C., {Ho}, L.~C., {Fabian}, A.~C., {et~al.} 2018, \mnras, 480, 1819

\bibitem[{{Ricci} \& {Trakhtenbrot}(2023)}]{2023NatAs...7.1282R}
{Ricci}, C. \& {Trakhtenbrot}, B. 2023, Nature Astronomy, 7, 1282

\bibitem[{{Ricci} {et~al.}(2017){Ricci}, {Trakhtenbrot}, {Koss}, {Ueda}, {Del Vecchio}, {Treister}, {Schawinski}, {Paltani}, {Oh}, {Lamperti}, {Berney}, {Gand hi}, {Ichikawa}, {Bauer}, {Ho}, {Asmus}, {Beckmann}, {Soldi}, {Balokovi{\'c}}, {Gehrels}, \& {Markwardt}}]{2017ApJS..233...17R}
{Ricci}, C., {Trakhtenbrot}, B., {Koss}, M.~J., {et~al.} 2017, \apjs, 233, 17

\bibitem[{Ricci {et~al.}(2015)Ricci, Ueda, Koss, Trakhtenbrot, Bauer, \& Gandhi}]{Ricci_2015}
Ricci, C., Ueda, Y., Koss, M.~J., {et~al.} 2015, The Astrophysical Journal, 815, L13

\bibitem[{{Ricci} {et~al.}(2022){Ricci}, {Treister}, {Bauer}, {Mej{\'\i}a-Restrepo}, {Koss}, {den Brok}, {Balokovi{\'c}}, {B{\"a}r}, {Bessiere}, {Caglar}, {Harrison}, {Ichikawa}, {Kakkad}, {Lamperti}, {Mushotzky}, {Oh}, {Powell}, {Privon}, {Ricci}, {Riffel}, {Rojas}, {Sani}, {Smith}, {Stern}, {Trakhtenbrot}, {Urry}, \& {Veilleux}}]{2022ApJS..261....8R}
{Ricci}, F., {Treister}, E., {Bauer}, F.~E., {et~al.} 2022, \apjs, 261, 8

\bibitem[{{Richards} {et~al.}(2006){Richards}, {Lacy}, {Storrie-Lombardi}, {Hall}, {Gallagher}, {Hines}, {Fan}, {Papovich}, {Vanden Berk}, {Trammell}, {Schneider}, {Vestergaard}, {York}, {Jester}, {Anderson}, {Budav{\'a}ri}, \& {Szalay}}]{2006ApJS..166..470R}
{Richards}, G.~T., {Lacy}, M., {Storrie-Lombardi}, L.~J., {et~al.} 2006, \apjs, 166, 470

\bibitem[{{Risaliti} \& {Elvis}(2004)}]{2004ASSL..308..187R}
{Risaliti}, G. \& {Elvis}, M. 2004, Astrophysics and Space Science Library, Vol. 308, {A Panchromatic View of AGN}, ed. A.~J. {Barger}, 187

\bibitem[{{Risaliti} {et~al.}(2002){Risaliti}, {Elvis}, \& {Nicastro}}]{2002ApJ...571..234R}
{Risaliti}, G., {Elvis}, M., \& {Nicastro}, F. 2002, \apj, 571, 234

\bibitem[{{Risaliti} \& {Lusso}(2015)}]{2015ApJ...815...33R}
{Risaliti}, G. \& {Lusso}, E. 2015, \apj, 815, 33

\bibitem[{{Rosario} {et~al.}(2018){Rosario}, {Burtscher}, {Davies}, {Koss}, {Ricci}, {Lutz}, {Riffel}, {Alexander}, {Genzel}, {Hicks}, {Lin}, {Maciejewski}, {M{\"u}ller-S{\'a}nchez}, {Orban de Xivry}, {Riffel}, {Schartmann}, {Schawinski}, {Schnorr-M{\"u}ller}, {Saintonge}, {Shimizu}, {Sternberg}, {Storchi-Bergmann}, {Sturm}, {Tacconi}, {Treister}, \& {Veilleux}}]{2018MNRAS.473.5658R}
{Rosario}, D.~J., {Burtscher}, L., {Davies}, R.~I., {et~al.} 2018, \mnras, 473, 5658

\bibitem[{{Runnoe} {et~al.}(2012{\natexlab{a}}){Runnoe}, {Brotherton}, \& {Shang}}]{2012MNRAS.422..478R}
{Runnoe}, J.~C., {Brotherton}, M.~S., \& {Shang}, Z. 2012{\natexlab{a}}, \mnras, 422, 478

\bibitem[{{Runnoe} {et~al.}(2012{\natexlab{b}}){Runnoe}, {Brotherton}, \& {Shang}}]{2012MNRAS.426.2677R}
{Runnoe}, J.~C., {Brotherton}, M.~S., \& {Shang}, Z. 2012{\natexlab{b}}, \mnras, 426, 2677

\bibitem[{{Saccheo} {et~al.}(2023){Saccheo}, {Bongiorno}, {Piconcelli}, {Testa}, {Bischetti}, {Bisogni}, {Bruni}, {Cresci}, {Feruglio}, {Fiore}, {Grazian}, {Luminari}, {Lusso}, {Mainieri}, {Maiolino}, {Marconi}, {Ricci}, {Tombesi}, {Travascio}, {Vietri}, {Vignali}, {Zappacosta}, \& {La Franca}}]{2023A&A...671A..34S}
{Saccheo}, I., {Bongiorno}, A., {Piconcelli}, E., {et~al.} 2023, \aap, 671, A34

\bibitem[{{Schlegel} {et~al.}(1998){Schlegel}, {Finkbeiner}, \& {Davis}}]{1998ApJ...500..525S}
{Schlegel}, D.~J., {Finkbeiner}, D.~P., \& {Davis}, M. 1998, \apj, 500, 525

\bibitem[{{Schmidt}(1968)}]{1968ApJ...151..393S}
{Schmidt}, M. 1968, \apj, 151, 393

\bibitem[{{Sergeev} {et~al.}(2005){Sergeev}, {Doroshenko}, {Golubinskiy}, {Merkulova}, \& {Sergeeva}}]{2005ApJ...622..129S}
{Sergeev}, S.~G., {Doroshenko}, V.~T., {Golubinskiy}, Y.~V., {Merkulova}, N.~I., \& {Sergeeva}, E.~A. 2005, \apj, 622, 129

\bibitem[{{Setoguchi} {et~al.}(2023){Setoguchi}, {Ueda}, {Toba}, {Li}, {Silverman}, \& {Uematsu}}]{2023arXiv231203552S}
{Setoguchi}, K., {Ueda}, Y., {Toba}, Y., {et~al.} 2023, arXiv e-prints, arXiv:2312.03552

\bibitem[{{Shakura} \& {Sunyaev}(1973)}]{1973A&A....24..337S}
{Shakura}, N.~I. \& {Sunyaev}, R.~A. 1973, \aap, 500, 33

\bibitem[{{Shang} {et~al.}(2011){Shang}, {Brotherton}, {Wills}, {Wills}, {Cales}, {Dale}, {Green}, {Runnoe}, {Nemmen}, {Gallagher}, {Ganguly}, {Hines}, {Kelly}, {Kriss}, {Li}, {Tang}, \& {Xie}}]{2011ApJS..196....2S}
{Shang}, Z., {Brotherton}, M.~S., {Wills}, B.~J., {et~al.} 2011, \apjs, 196, 2

\bibitem[{{Shen} {et~al.}(2020){Shen}, {Hopkins}, {Faucher-Gigu{\`e}re}, {Alexander}, {Richards}, {Ross}, \& {Hickox}}]{2020MNRAS.495.3252S}
{Shen}, X., {Hopkins}, P.~F., {Faucher-Gigu{\`e}re}, C.-A., {et~al.} 2020, \mnras, 495, 3252

\bibitem[{{Shields}(1978)}]{1978Natur.272..706S}
{Shields}, G.~A. 1978, \nat, 272, 706

\bibitem[{{Shimizu} {et~al.}(2017){Shimizu}, {Mushotzky}, {Mel{\'e}ndez}, {Koss}, {Barger}, \& {Cowie}}]{2017MNRAS.466.3161S}
{Shimizu}, T.~T., {Mushotzky}, R.~F., {Mel{\'e}ndez}, M., {et~al.} 2017, \mnras, 466, 3161

\bibitem[{{Signorini} {et~al.}(2023){Signorini}, {Risaliti}, {Lusso}, {Nardini}, {Bargiacchi}, {Sacchi}, \& {Trefoloni}}]{2023A&A...676A.143S}
{Signorini}, M., {Risaliti}, G., {Lusso}, E., {et~al.} 2023, \aap, 676, A143

\bibitem[{{Slone} \& {Netzer}(2012)}]{2012MNRAS.426..656S}
{Slone}, O. \& {Netzer}, H. 2012, \mnras, 426, 656

\bibitem[{{Stalin} {et~al.}(2010){Stalin}, {Petitjean}, {Srianand}, {Fox}, {Coppolani}, \& {Schwope}}]{2010MNRAS.401..294S}
{Stalin}, C.~S., {Petitjean}, P., {Srianand}, R., {et~al.} 2010, \mnras, 401, 294

\bibitem[{{Steffen} {et~al.}(2006){Steffen}, {Strateva}, {Brandt}, {Alexand er}, {Koekemoer}, {Lehmer}, {Schneider}, \& {Vignali}}]{2006AJ....131.2826S}
{Steffen}, A.~T., {Strateva}, I., {Brandt}, W.~N., {et~al.} 2006, \aj, 131, 2826

\bibitem[{{Strateva} {et~al.}(2005){Strateva}, {Brandt}, {Schneider}, {Vanden Berk}, \& {Vignali}}]{2005AJ....130..387S}
{Strateva}, I.~V., {Brandt}, W.~N., {Schneider}, D.~P., {Vanden Berk}, D.~G., \& {Vignali}, C. 2005, \aj, 130, 387

\bibitem[{{Svoboda} {et~al.}(2015){Svoboda}, {Beuchert}, {Guainazzi}, {Longinotti}, {Piconcelli}, \& {Wilms}}]{2015A&A...578A..96S}
{Svoboda}, J., {Beuchert}, T., {Guainazzi}, M., {et~al.} 2015, \aap, 578, A96

\bibitem[{{Tananbaum} {et~al.}(1979){Tananbaum}, {Avni}, {Branduardi}, {Elvis}, {Fabbiano}, {Feigelson}, {Giacconi}, {Henry}, {Pye}, {Soltan}, \& {Zamorani}}]{1979ApJ...234L...9T}
{Tananbaum}, H., {Avni}, Y., {Branduardi}, G., {et~al.} 1979, \apjl, 234, L9

\bibitem[{{Tang} {et~al.}(2007){Tang}, {Zhang}, \& {Hopkins}}]{2007MNRAS.377.1113T}
{Tang}, S.~M., {Zhang}, S.~N., \& {Hopkins}, P.~F. 2007, \mnras, 377, 1113

\bibitem[{{Temple} {et~al.}(2021){Temple}, {Hewett}, \& {Banerji}}]{2021MNRAS.508..737T}
{Temple}, M.~J., {Hewett}, P.~C., \& {Banerji}, M. 2021, \mnras, 508, 737

\bibitem[{{Temple} {et~al.}(2023){Temple}, {Matthews}, {Hewett}, {Rankine}, {Richards}, {Banerji}, {Ferland}, {Knigge}, \& {Stepney}}]{2023MNRAS.523..646T}
{Temple}, M.~J., {Matthews}, J.~H., {Hewett}, P.~C., {et~al.} 2023, \mnras, 523, 646

\bibitem[{{Thorne}(1974)}]{1974ApJ...191..507T}
{Thorne}, K.~S. 1974, \apj, 191, 507

\bibitem[{{Timlin} {et~al.}(2021){Timlin}, {Brandt}, \& {Laor}}]{2021MNRAS.504.5556T}
{Timlin}, John~D., I., {Brandt}, W.~N., \& {Laor}, A. 2021, \mnras, 504, 5556

\bibitem[{{Timlin} {et~al.}(2020){Timlin}, {Brandt}, {Ni}, {Luo}, {Pu}, {Schneider}, {Vivek}, \& {Yi}}]{2020MNRAS.492..719T}
{Timlin}, J.~D., {Brandt}, W.~N., {Ni}, Q., {et~al.} 2020, \mnras, 492, 719

\bibitem[{{Ueda} {et~al.}(2003){Ueda}, {Akiyama}, {Ohta}, \& {Miyaji}}]{2003ApJ...598..886U}
{Ueda}, Y., {Akiyama}, M., {Ohta}, K., \& {Miyaji}, T. 2003, \apj, 598, 886

\bibitem[{{Ulrich} {et~al.}(1997){Ulrich}, {Maraschi}, \& {Urry}}]{1997ARA&A..35..445U}
{Ulrich}, M.-H., {Maraschi}, L., \& {Urry}, C.~M. 1997, \araa, 35, 445

\bibitem[{{Urry} \& {Padovani}(1995)}]{1995PASP..107..803U}
{Urry}, C.~M. \& {Padovani}, P. 1995, \pasp, 107, 803

\bibitem[{{Vasudevan} \& {Fabian}(2009)}]{2009MNRAS.392.1124V}
{Vasudevan}, R.~V. \& {Fabian}, A.~C. 2009, \mnras, 392, 1124

\bibitem[{{Vasudevan} {et~al.}(2010){Vasudevan}, {Fabian}, {Gandhi}, {Winter}, \& {Mushotzky}}]{2010MNRAS.402.1081V}
{Vasudevan}, R.~V., {Fabian}, A.~C., {Gandhi}, P., {Winter}, L.~M., \& {Mushotzky}, R.~F. 2010, \mnras, 402, 1081

\bibitem[{{Vasudevan} {et~al.}(2009){Vasudevan}, {Mushotzky}, {Winter}, \& {Fabian}}]{2009MNRAS.399.1553V}
{Vasudevan}, R.~V., {Mushotzky}, R.~F., {Winter}, L.~M., \& {Fabian}, A.~C. 2009, \mnras, 399, 1553

\bibitem[{Virtanen {et~al.}(2020)Virtanen, Gommers, Oliphant, Haberland, Reddy, Cournapeau, Burovski, Peterson, Weckesser, Bright, {van der Walt}, Brett, Wilson, Millman, Mayorov, Nelson, Jones, Kern, Larson, Carey, Polat, Feng, Moore, {VanderPlas}, Laxalde, Perktold, Cimrman, Henriksen, Quintero, Harris, Archibald, Ribeiro, Pedregosa, {van Mulbregt}, \& {SciPy 1.0 Contributors}}]{2020SciPy-NMeth}
Virtanen, P., Gommers, R., Oliphant, T.~E., {et~al.} 2020, Nature Methods, 17, 261

\bibitem[{{Ward} {et~al.}(1987){Ward}, {Elvis}, {Fabbiano}, {Carleton}, {Willner}, \& {Lawrence}}]{1987ApJ...315...74W}
{Ward}, M., {Elvis}, M., {Fabbiano}, G., {et~al.} 1987, \apj, 315, 74

\bibitem[{{Wilkes} {et~al.}(1994){Wilkes}, {Tananbaum}, {Worrall}, {Avni}, {Oey}, \& {Flanagan}}]{1994ApJS...92...53W}
{Wilkes}, B.~J., {Tananbaum}, H., {Worrall}, D.~M., {et~al.} 1994, \apjs, 92, 53

\bibitem[{{Yang} {et~al.}(2022){Yang}, {Boquien}, {Brandt}, {Buat}, {Burgarella}, {Ciesla}, {Lehmer}, {Ma{\l}ek}, {Mountrichas}, {Papovich}, {Pons}, {Stalevski}, {Theul{\'e}}, \& {Zhu}}]{2022ApJ...927..192Y}
{Yang}, G., {Boquien}, M., {Brandt}, W.~N., {et~al.} 2022, \apj, 927, 192

\bibitem[{{Yang} {et~al.}(2020){Yang}, {Boquien}, {Buat}, {Burgarella}, {Ciesla}, {Duras}, {Stalevski}, {Brandt}, \& {Papovich}}]{2020MNRAS.491..740Y}
{Yang}, G., {Boquien}, M., {Buat}, V., {et~al.} 2020, \mnras, 491, 740

\bibitem[{{Young} {et~al.}(2010){Young}, {Elvis}, \& {Risaliti}}]{2010ApJ...708.1388Y}
{Young}, M., {Elvis}, M., \& {Risaliti}, G. 2010, \apj, 708, 1388

\bibitem[{{Yuan} \& {Narayan}(2014)}]{2014ARA&A..52..529Y}
{Yuan}, F. \& {Narayan}, R. 2014, \araa, 52, 529

\bibitem[{{Yuan} {et~al.}(1998){Yuan}, {Siebert}, \& {Brinkmann}}]{1998A&A...334..498Y}
{Yuan}, W., {Siebert}, J., \& {Brinkmann}, W. 1998, \aap, 334, 498

\bibitem[{{Zdziarski} {et~al.}(1996){Zdziarski}, {Johnson}, \& {Magdziarz}}]{1996MNRAS.283..193Z}
{Zdziarski}, A.~A., {Johnson}, W.~N., \& {Magdziarz}, P. 1996, \mnras, 283, 193

\bibitem[{{{\.Z}ycki} {et~al.}(1999){{\.Z}ycki}, {Done}, \& {Smith}}]{1999MNRAS.309..561Z}
{{\.Z}ycki}, P.~T., {Done}, C., \& {Smith}, D.~A. 1999, \mnras, 309, 561

\end{thebibliography}


\begin{appendix}


\section{Additional material}\label{sect:appendixc}

\begin{table*}
\setlength{\tabcolsep}{0.7\tabcolsep}
\centering
\caption{Source magnitudes as calculated by the \textit{Swift}/UVOT command \texttt{uvotsource} via 5$"$ aperture photometry.}
\begin{tabular}{cccccccc}
\hline
\hline
\specialrule{0.1em}{0em}{0.5em}
\vspace{1mm}
BAT ID & Swift ID & $M_{\rm V}$ & $M_{\rm B}$ & $M_{\rm U}$ & $M_{\rm W1}$ & $M_{\rm M2}$ & $M_{\rm W2}$\\
\hline\\
\vspace{1mm}
2 & SWIFTJ0001.6-7701 & $-$ & $-$ & $-$ & $17.16\pm0.03$ & $17.37\pm0.03$ & $17.30\pm0.03$\\
\vspace{1mm}
3 & SWIFTJ0002.5+0323 & $15.01\pm0.05$ & $15.61\pm0.04$ & $16.16\pm0.04$ & $16.60\pm0.03$ & $16.83\pm0.03$ & $16.91\pm0.03$\\
\vspace{1mm}
6 & SWIFTJ0006.2+2012 & $14.24\pm0.02$ & $14.42\pm0.02$ & $14.39\pm0.02$ & $14.72\pm0.02$ & $14.93\pm0.02$ & $14.96\pm0.02$\\
\vspace{1mm}
10 & SWIFTJ0021.2-1909 & $-$ & $-$ & $19.21\pm0.06$ & $21.20\pm0.08$ & $-$ & $-$\\
\vspace{1mm}
14 & SWIFTJ0026.5-5308 & $-$ & $-$ & $-$ & $17.80\pm0.08$ & $18.30\pm0.03$ & $-$\\
\vspace{1mm}
16 & SWIFTJ0029.2+1319 & $15.40\pm0.04$ & $15.62\pm0.03$ & $15.68\pm0.02$ & $16.17\pm0.03$ & $16.43\pm0.03$ & $16.32\pm0.03$\\
\vspace{1mm}
19 & SWIFTJ0034.5-7904 & $14.76\pm0.04$ & $15.00\pm0.03$ & $14.95\pm0.03$ & $15.34\pm0.02$ & $15.52\pm0.03$ & $15.50\pm0.02$\\
 \vspace{1mm}
34 & SWIFTJ0051.6+2928 & $15.09\pm0.04$ & $15.81\pm0.03$ & $16.81\pm0.04$ & $17.57\pm0.05$ & $18.03\pm0.05$ & $18.08\pm0.02$\\
 \vspace{1mm}
36 & SWIFTJ0051.9+1724 & $15.44\pm0.05$ & $15.63\pm0.04$ & $15.45\pm0.02$ & $15.88\pm0.03$ & $16.19\pm0.03$ & $16.09\pm0.02$\\
 \vspace{1mm}
39 & SWIFTJ0054.9+2524 & $15.61\pm0.05$ & $15.69\pm0.03$ & $15.61\pm0.03$ & $16.07\pm0.03$ & $16.31\pm0.02$ & $16.04\pm0.03$\\
\hline
\end{tabular}
\tablefoot{These magnitude values include potential contribution from the host galaxy to the AGN (within 5$"$ region; see Sect. \ref{sect:uvot}). The table in its entirety is available at the CDS.}
\label{tab:UVOT_mag}
\end{table*}


\begin{table*}
\setlength{\tabcolsep}{0.001\tabcolsep}
\centering
\caption{AGN magnitudes (corrected) as estimated by \textsc{GALFIT} image decomposition (Sect. \ref{sect:galfit}).}
\begin{tabular}{ccccccccc}
\hline
\hline
\specialrule{0.1em}{0em}{0.5em}
\vspace{1mm}
BAT ID & Swift ID & $M_{\rm V}$ & $M_{\rm B}$ & $M_{\rm U}$ & $M_{\rm W1}$ & $M_{\rm M2}$ & $M_{\rm W2}$ & Flag\\
\hline\\
\vspace{1mm}
3 & SWIFTJ0002.5+0323 & $16.74\pm0.09$ & $17.12\pm0.05$ & $16.91\pm0.03$ & $17.61\pm0.03$ & $17.52\pm0.02$ & $17.72\pm0.02$ & -\\
\vspace{1mm}
6 & SWIFTJ0006.2+2012 & \,\,\,$14.51\pm0.01$\tablefootmark{o} & \,\,\,$14.84\pm0.00$\tablefootmark{o} & $14.82\pm0.01$ & \,\,\,$14.88\pm0.00$\tablefootmark{o} & \,\,\,$14.99\pm0.00$\tablefootmark{o} & \,\,\,$15.12\pm0.00$\tablefootmark{o} & -\\
\vspace{1mm}
16 & SWIFTJ0029.2+1319 & \,\,\,$15.17\pm0.04$\tablefootmark{o} & \,\,\,\,\,\,$15.69\pm0.004$\tablefootmark{o} & \,\,\,$15.97\pm0.01$\tablefootmark{o} & \,\,\,$16.39\pm0.01$\tablefootmark{o} & \,\,\,\,\,\,$16.33\pm0.005$\tablefootmark{o} & \,\,\,$16.66\pm0.04$\tablefootmark{o} & c\\
\vspace{1mm}
19 & SWIFTJ0034.5-7904 & $16.54\pm0.08$ & $16.81\pm0.03$ & $16.40\pm0.05$ & \,\,\,$15.58\pm0.02$\tablefootmark{o} & \,\,\,$15.50\pm0.01$\tablefootmark{o} & \,\,\,$15.87\pm0.04$\tablefootmark{o} & c\\
\vspace{1mm}
34 & SWIFTJ0051.6+2928 & $16.25\pm0.04$ & $16.79\pm0.03$ & $17.25\pm0.03$ & $17.97\pm0.03$ & $18.36\pm0.03$ & \,\,\,$18.37\pm0.01$\tablefootmark{o} & -\\
\vspace{1mm}
36 & SWIFTJ0051.9+1724 & \,\,\,$15.08\pm0.06$\tablefootmark{o} & \,\,\,\,\,\,$15.71\pm0.002$\tablefootmark{o} & \,\,\,$15.77\pm0.01$\tablefootmark{o} & \,\,\,$16.01\pm0.01$\tablefootmark{o} & \,\,\,$16.16\pm0.01$\tablefootmark{o} & \,\,\,$16.60\pm0.05$\tablefootmark{o} & c\\
\vspace{1mm}
39 & SWIFTJ0054.9+2524 & $15.86\pm0.02$ & \,\,\,\,\,\,\,\,\,$15.79\pm0.0003$\tablefootmark{o} & \,\,\,$15.72\pm0.01$\tablefootmark{o} & \,\,\,$16.28\pm0.01$\tablefootmark{o} & \,\,\,\,\,\,$16.21\pm0.004$\tablefootmark{o} & \,\,\,$16.39\pm0.04$\tablefootmark{o} & c\\
 \vspace{1mm}
43 & SWIFTJ0059.4+3150 & $16.25\pm0.05$ & $17.75\pm0.18$ & \,\,\,$17.45\pm0.12$\tablefootmark{o} & \,\,\,$16.61\pm0.02$\tablefootmark{p} & \,\,\,$16.96\pm0.01$\tablefootmark{o} & $17.22\pm0.00$ & -\\
 \vspace{1mm}
45 & SWIFTJ0101.5-0308 & $17.43\pm0.11$ & $19.01\pm0.31$ & \,\,\,$18.51\pm0.09$\tablefootmark{o} & \,\,\,$19.44\pm0.12$\tablefootmark{o} & \,\,\,$19.78\pm0.09$\tablefootmark{o} & \,\,\,$19.82\pm0.09$\tablefootmark{o} & -\\
 \vspace{1mm}
51 & SWIFTJ0105.7-1414 & $15.71\pm0.07$ & $15.91\pm0.07$ & $16.03\pm0.00$ & $16.15\pm0.01$ & $15.70\pm0.18$ & $16.23\pm0.01$ & c\\
\hline
\end{tabular}
\tablefoot{The last column shows in the form of flags (c) if the GALFIT-estimated magnitudes were corrected using corrections calculated in Sect. \ref{sect:galfit_corr} to create the final SEDs. We report an uncertainty of 0.00 when it is very low ($\lesssim 10^{-6}$ mag). The table in its entirety is available at the CDS.\\
\tablefoottext{o}{A sky $+$ PSF model was used to get the best \textsc{GALFIT} residual as \textsc{GALFIT} was unable to fit a S\'ersic profile and hence, the corresponding correction was employed (see Fig. \ref{fig:galfit_corr}), if and when needed.}\\
\tablefoottext{p}{UVOT pipeline-estimated magnitude was used in the final SED.}
}
\label{tab:galfit_mag}
\end{table*}
\FloatBarrier


\section{GALFIT simulations}\label{sect:appendixa}

The AGN flux can be contaminated by the host galaxy's light. To extract reliable AGN flux measurements, we used \textsc{GALFIT} to fit the 2D surface brightness of our sources. However, considering the wide variety of sources in our sample, varying over redshift, morphology and total magnitude, we carried out some simulations to check the reliability of the flux values estimated by \textsc{GALFIT}. We simulated a population of fake type\,I AGN (similar to a procedure on optical images of \citealp{2011ApJ...739...57K}) covering the same redshift range as our original sample of type\,1 AGN ($0.001 < z < 0.3$). We used ten Seyfert 2 galaxies, classified so by the optical classification provided by \cite{2017ApJ...850...74K}, with X-ray column densities above $10^{23}\,{\rm cm^{-2}}$ to eliminate the possibility of any significant optical/UV radiation coming from the SMBH at the center. We pre-processed, aspect corrected, and reduced the raw \textit{Swift}/UVOT images of these ten sources in the three UV filters (UVW1, UVM2, UVW2), for start. The final sky images were then used to extract the total galaxy magnitudes of these sources to be used in \textsc{GALFIT} later. We then randomly selected a field star from our images and placed it at the center of the galaxy to replicate a type\,I AGN. The magnitude of the star was varied by changing its normalization to reciprocate different AGN light ratios (AGN magnitude to total magnitude calculated using 5$"$ aperture photometry), ranging from 10\% to 95\%. We calculated the star's magnitude using \textsc{GALFIT} before placing it on the galaxy to get estimates of the star flux that would later be compared to the PSF flux estimated by \textsc{GALFIT}. We included different possible morphologies in this sample of ten galaxies: edge-on galaxies, face-on galaxies, and indeterminate. We did not go with the traditional galaxy morphology classification of elliptical and spiral galaxies because on scales as small as the ones in our UVOT images, \textsc{GALFIT} cannot really distinguish between bulge-dominated and disk-dominated morphologies. Also, we are not interested in fitting the extended features of the host galaxy, such as spiral arms, starburst regions and clumps. Instead, we included Seyfert\,2s with different orientations to understand \textsc{GALFIT}'s response to circular (face-on) versus elliptical (edge-on) galaxy shapes in 2D. In total, we simulated 270 type\,I AGN across the three \textit{Swift}/UVOT UV filters with ten different AGN light ratios.


\begin{figure}
\centering
\includegraphics[width=0.57\textwidth]{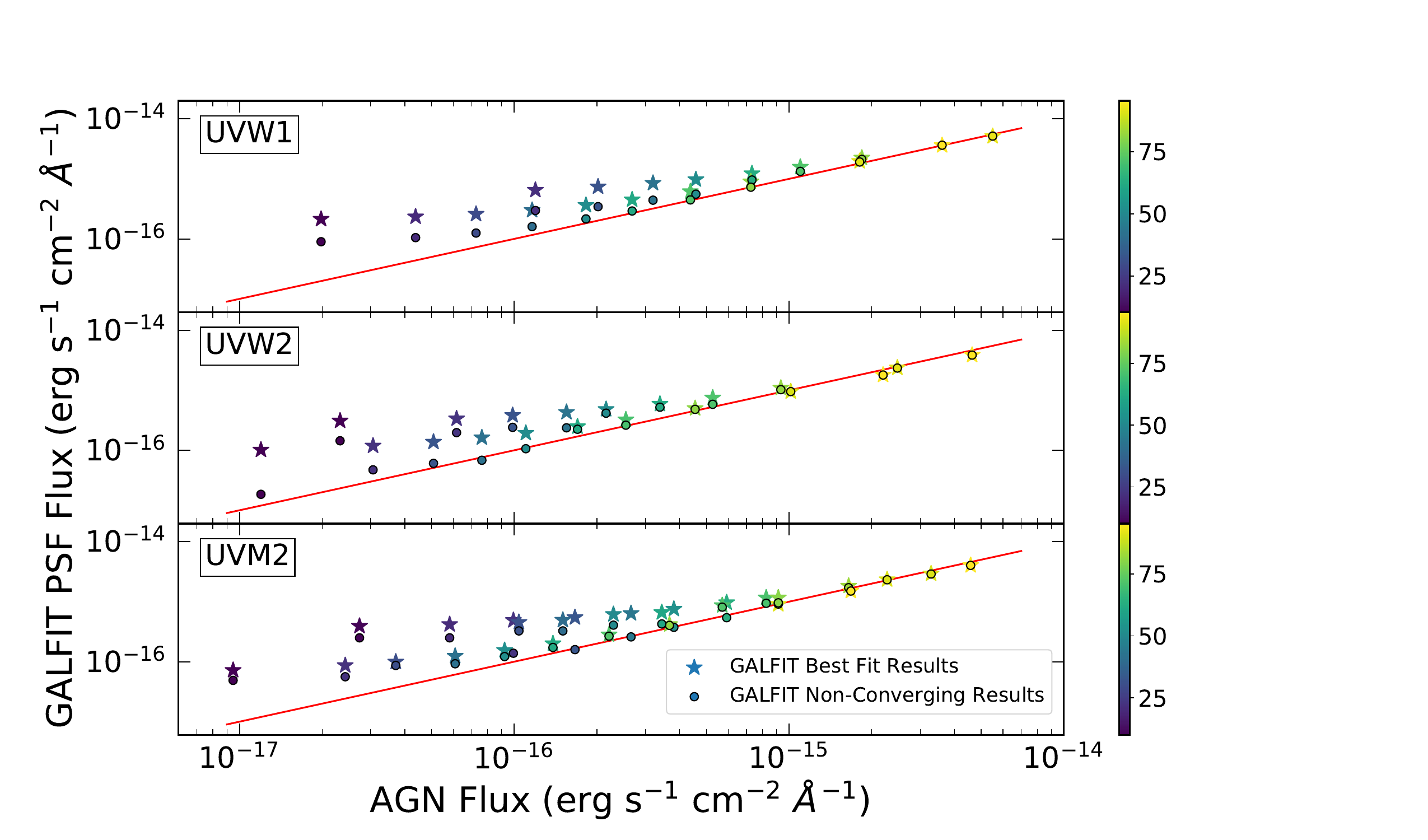} 
\caption{Comparing the simulated AGN (star) flux to the PSF flux estimated by \textsc{GALFIT} for the three UV filters. The star symbols show the best-fit results without a S\'ersic profile, while the circle symbols show the results when the fitting with a S\'ersic profile does not converge but we still get better constraints on the PSF profile compared to the fit without a S\'ersic profile. The $x = y$ line is shown in red. The color bar indicates the AGN light ratio in percent.}
\label{fig:galfit_sim-noconv}
\end{figure}


The \textsc{GALFIT} analysis of these 270 simulated AGN was done following the same procedure as the one followed for our original sample (described in Sect. \ref{sect:galfit}):
\begin{enumerate}
    \item A relatively empty region was selected from the image to fit the background using the sky component in \textsc{GALFIT}. In the case of crowded fields, a bad pixel mask was incorporated into the \textsc{GALFIT} parameter file to mask the background sources.
    \item A region around the source with the same dimension as the one used in the previous step was selected and fit using the sky and PSF component in \textsc{GALFIT}. Like before, a bad pixel mask was used to mask the light from all the extra sources around our galaxy of interest. The sky estimated in the previous step was used as the input sky value in this step. However, we left the sky value to vary in general and fixed it only in cases when it was way off from the one estimated in the previous step. To fit the PSF component, \textsc{GALFIT} requires a PSF image that it, in turn, fits to the image. The PSF image was created using the \texttt{photutils} package in Python. This package uses a defined set of stars from the field of the source and reiterates over these star profiles to produce a PSF image. We used three to six bright background stars to create a PSF image for each of our ten sources in each UV filter. The input value used for the PSF magnitude was the one calculated by \textsc{GALFIT} previously for the star in the absence of any host galaxy. All the residuals were visually examined to check for incredibly useless fits.
    \item Finally, the same source region as before was fit using the sky, PSF, and S\'ersic profile component in \textsc{GALFIT}. The sky and PSF input parameters used were the ones we obtained from the previous step. The input S\'ersic magnitude was defined as the galaxy magnitude obtained from the UVOT pipeline. The input value for effective S\'ersic radius was visually determined from the source images. The input values for the coordinates of the S\'ersic profile were put as the ones for the PSF profile from the results of the previous step. We started with an input S\'ersic index value of 2.5 and moved to 2.0, 1.5, 1.0, 0.5, 3.0, 3.5, and 4.0 in cases when \textsc{GALFIT} was not able to converge. All the other parameters of the S\'ersic profile were left to their default values. In multiple cases, when \textsc{GALFIT} was unable to converge with these input parameters, we fixed the controversial parameters (generally, axis ratio or effective radius, sometimes the coordinates of the S\'ersic profile) to specific acceptable values so as to get \textsc{GALFIT} to converge and produce reliable fits, estimates, and residuals. All the residuals were visually inspected to confirm the reliability of the fits. In some cases, even after multiple repetitions, when \textsc{GALFIT} was not able to fit all the three components to the source, we quote results from step 2; that is, just the sky $+$ PSF fits.
\end{enumerate}

\begin{figure*}
  \begin{subfigure}[t]{0.56\textwidth}
    \centering
    \caption{}
    \vspace{-7mm}
    \includegraphics[width=\textwidth]{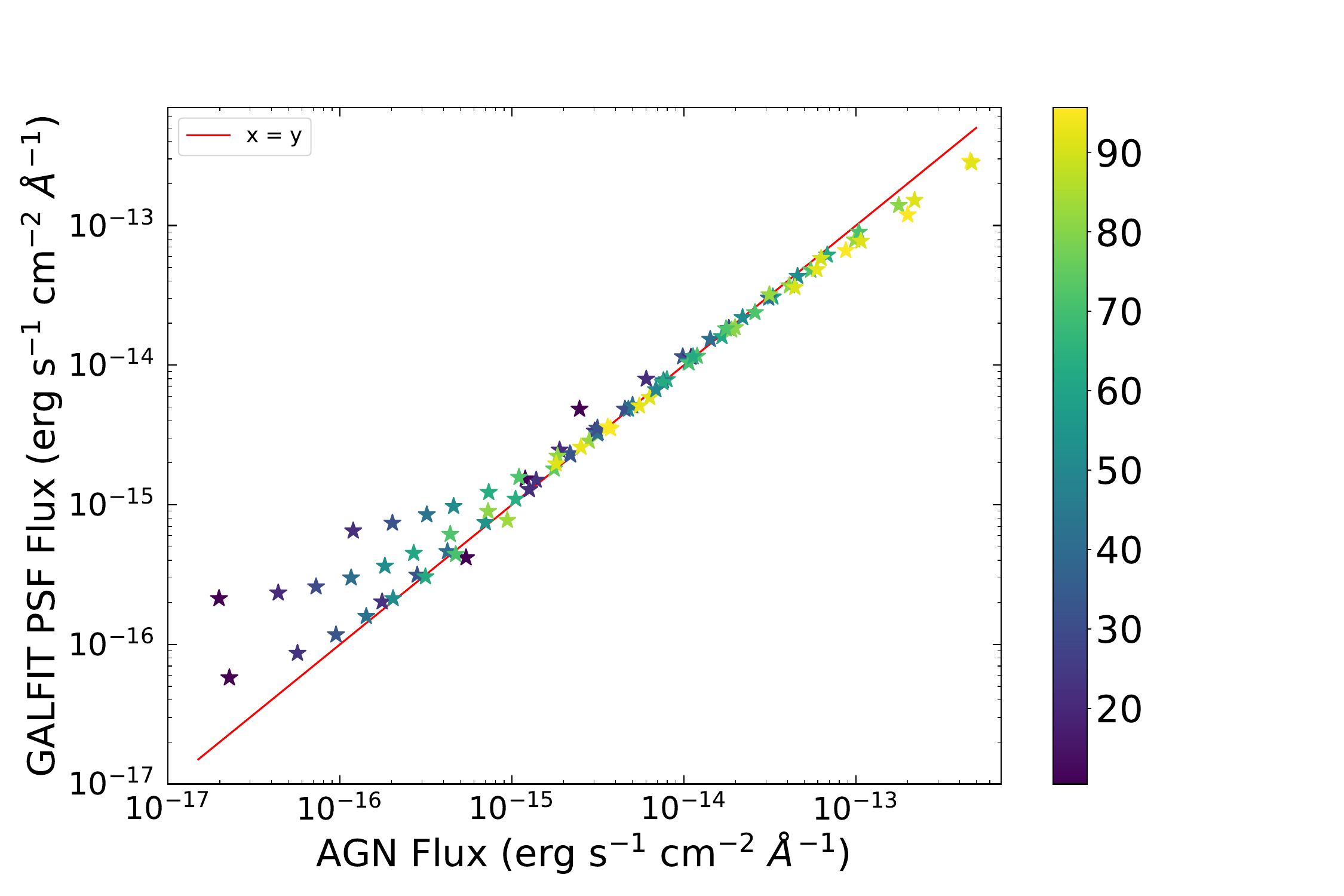}
    \label{fig:galfit_sim_W1-all}
  \end{subfigure}
  \hspace{-1.2cm}
  \begin{subfigure}[t]{0.52\textwidth}
    \centering
    \caption{}
    \vspace{-7mm}
    \includegraphics[width=\textwidth]{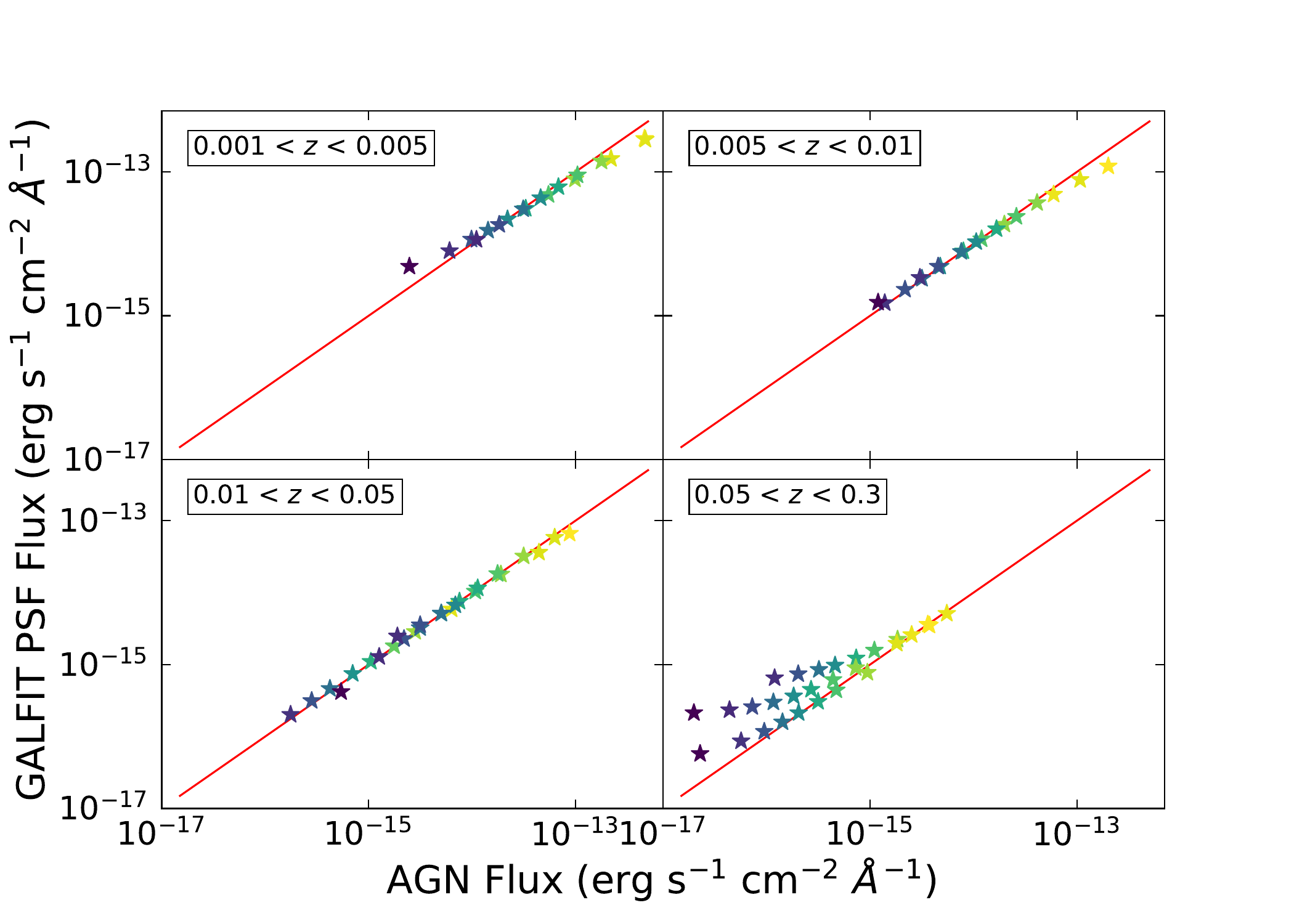}
    \label{fig:galfit_sim_W1-z1}
  \end{subfigure}
\vspace{-8mm}
\caption{Comparison of the simulated AGN (star) flux to the PSF flux estimated by \textsc{GALFIT} in UVOT filter UVW1, for (a) all simulated AGN and (b) sources divided into different redshift bins. The $x = y$ line is shown in red. The color bar indicates the AGN light ratio in percent.}
\label{fig:galfit_sim_W1}
\end{figure*}

\begin{figure*}
  \begin{subfigure}[t]{0.56\textwidth}
    \centering
    \caption{} 
    \vspace{-7mm}
    \includegraphics[width=\textwidth]{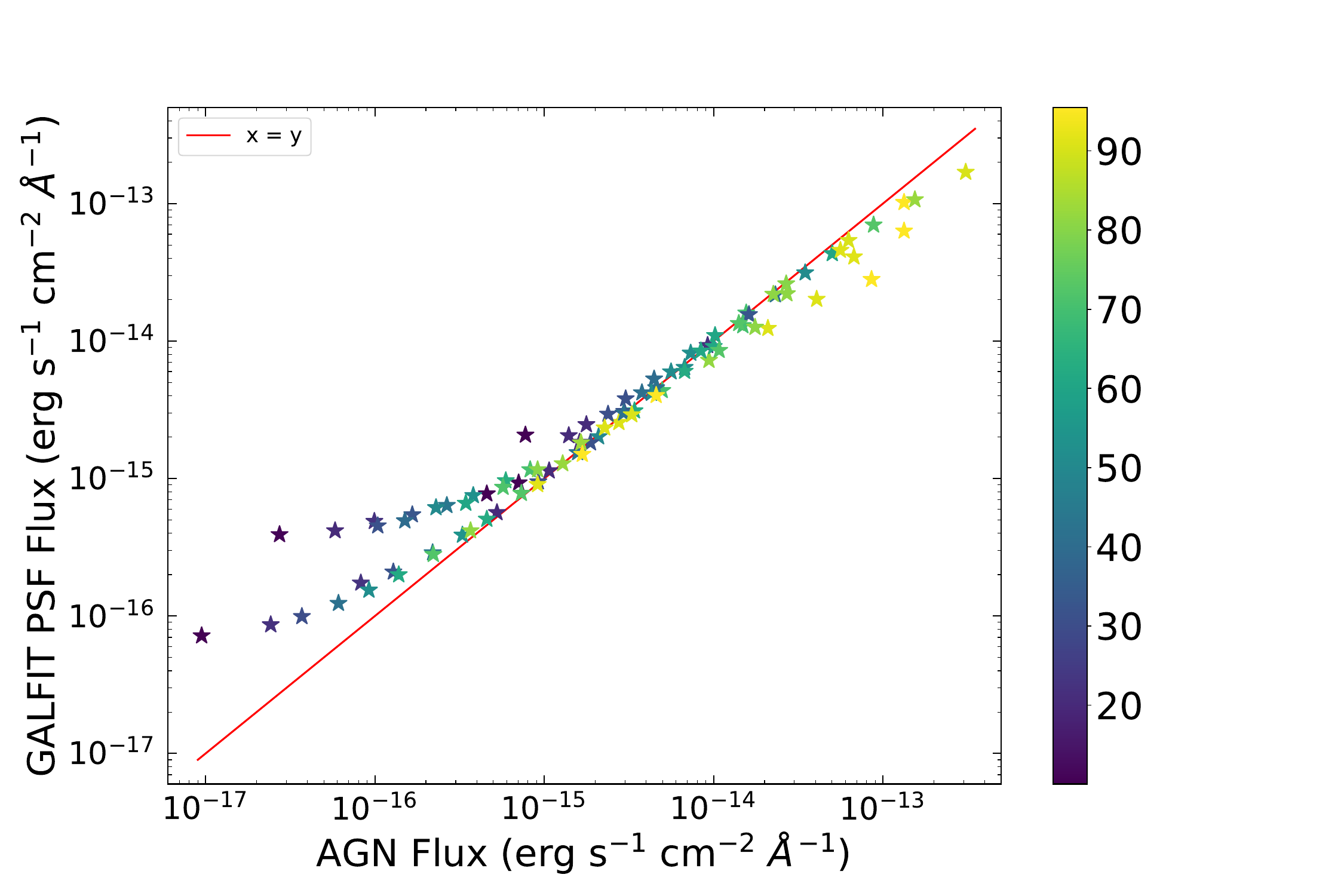}
    \label{fig:galfit_sim_M2-all}
  \end{subfigure}
  \hspace{-1.2cm}
  \begin{subfigure}[t]{0.52\textwidth}
    \centering
    \caption{} 
    \vspace{-7mm}
    \includegraphics[width=\textwidth]{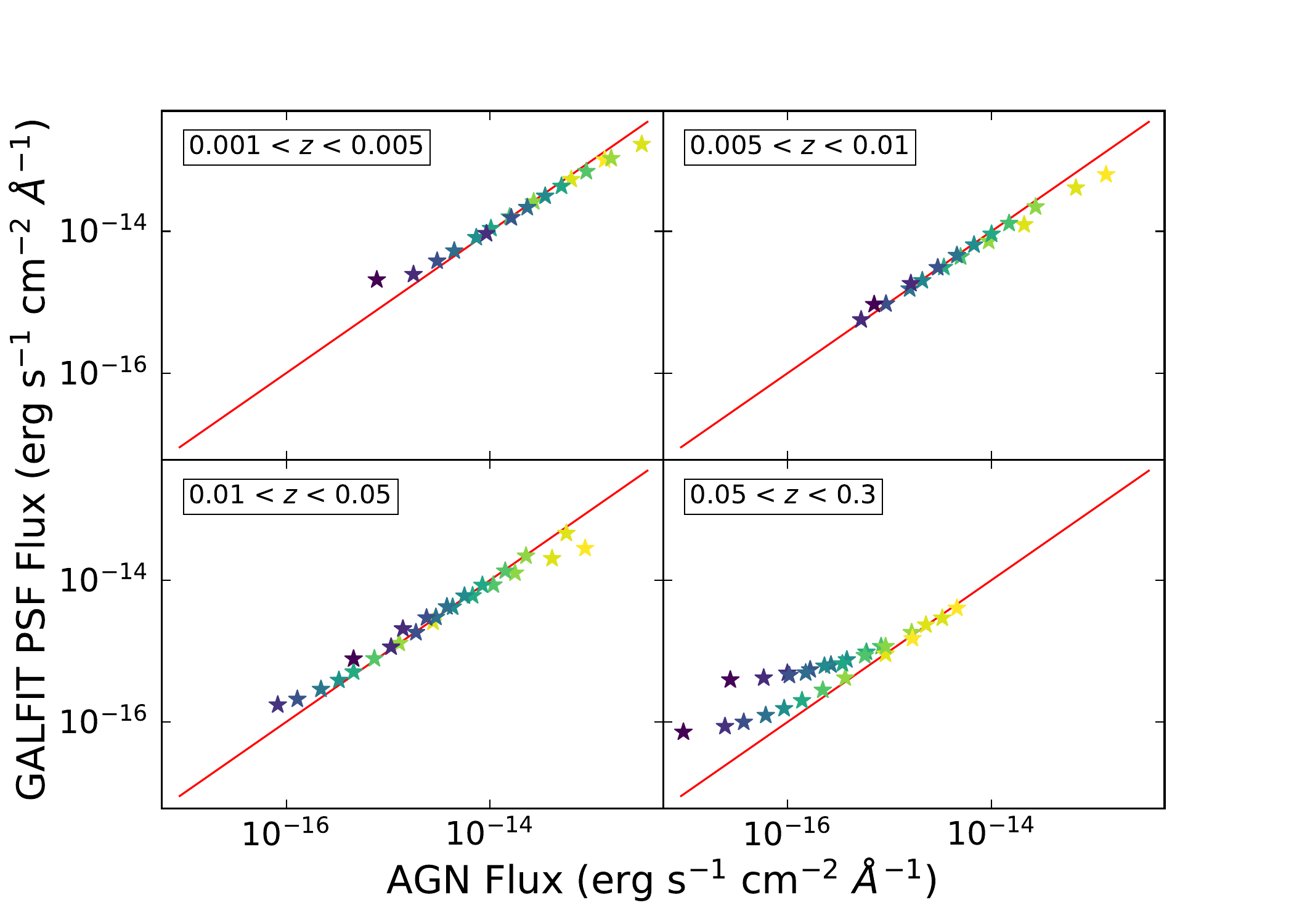}
    \label{fig:galfit_sim_M2-z1}
  \end{subfigure}
\vspace{-8mm}
\caption{Comparison of the simulated AGN (star) flux to the PSF flux estimated by \textsc{GALFIT} in UVOT filter UVM2, for (a) all simulated AGN and (b) sources divided into different redshift bins. The $x = y$ line is shown in red. The color bar indicates the AGN light ratio in percent.}
\label{fig:galfit_sim_M2}
\end{figure*}

\begin{figure*}
  \begin{subfigure}[t]{0.56\textwidth}
    \centering
    \caption{} 
    \vspace{-7mm}
    \includegraphics[width=\textwidth]{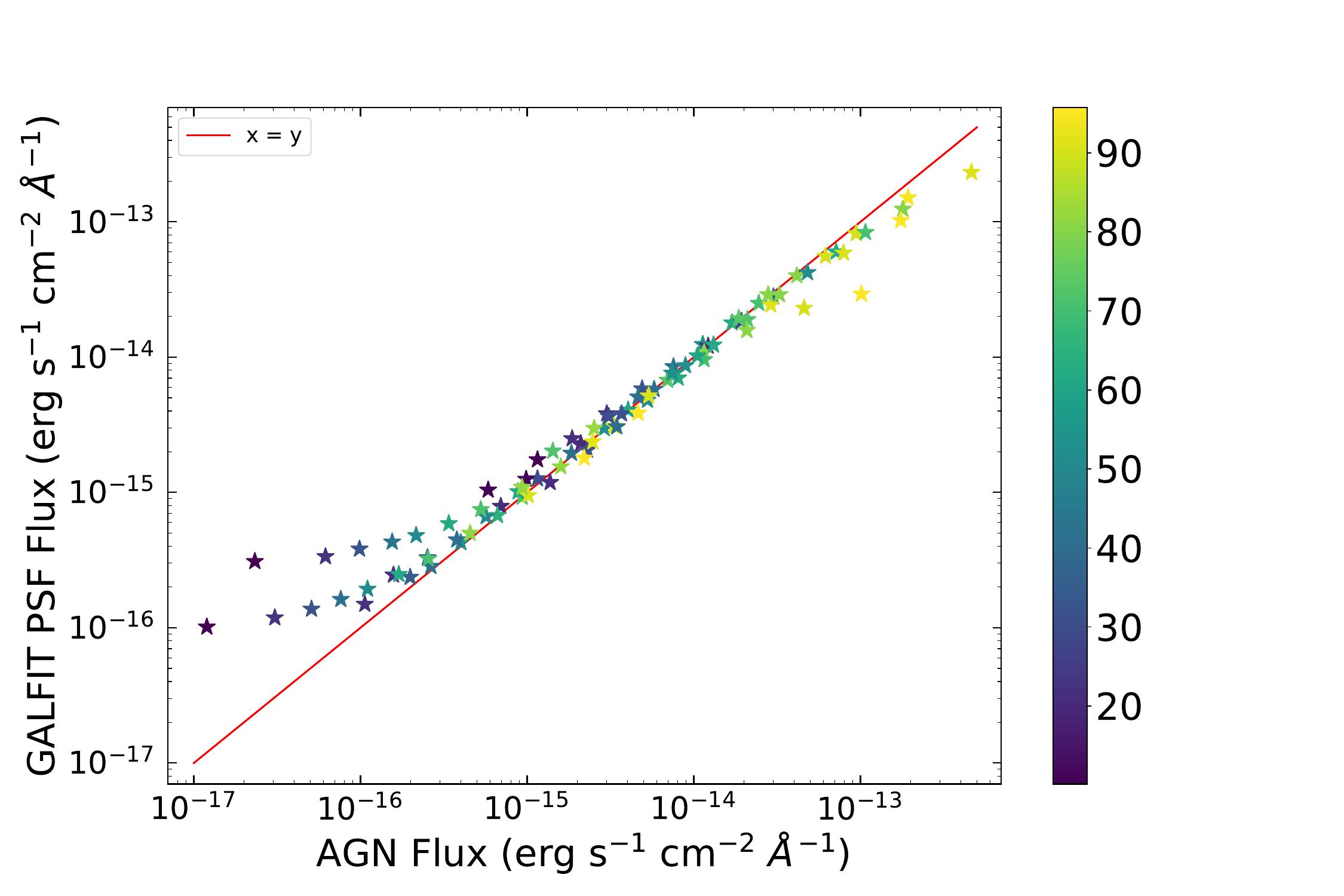} 
    \label{fig:galfit_sim_W2-all}
  \end{subfigure}
  \hspace{-1.2cm}
  \begin{subfigure}[t]{0.52\textwidth}
    \centering
    \caption{} 
    \vspace{-7mm}
    \includegraphics[width=\textwidth]{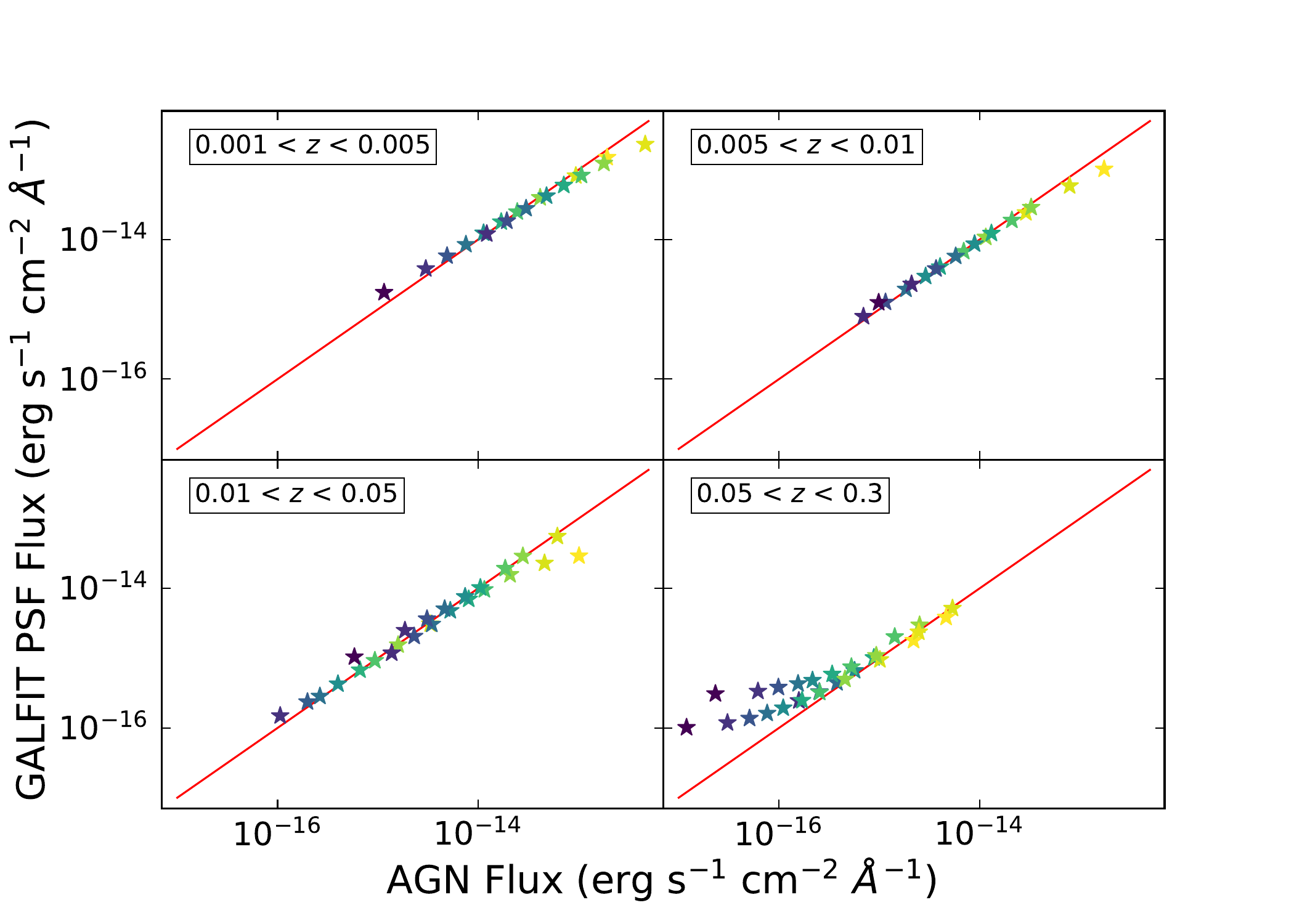}
    \label{fig:galfit_sim_W2-z1}
  \end{subfigure}
\vspace{-8mm}
\caption{Comparison of the simulated AGN (star) flux to the PSF flux estimated by \textsc{GALFIT} in UVOT filter UVW2, for (a) all simulated AGN and (b) sources divided into different redshift bins. The $x = y$ line is shown in red. The color bar indicates the AGN light ratio in percent.}
\label{fig:galfit_sim_W2}
\end{figure*}

In Figs. \ref{fig:galfit_sim_W1}, \ref{fig:galfit_sim_M2}, and \ref{fig:galfit_sim_W2}, we plot our results from the \textsc{GALFIT} analysis of these 270 simulated type\,I AGN for the UVW1, UVM2, and UVW2 filters, respectively. In each case, the plot on the left shows the AGN (PSF) flux estimated by \textsc{GALFIT} with respect to the actual flux of the star used as an AGN replica in these fake type\,I galaxies, while the plots on the right show the same information but segregated into different redshift bins. The symbols are color-coded based on the AGN to total light ratio. In general, the figures show a broad agreement between the AGN fluxes estimated by \textsc{GALFIT} and the actual star flux. However, significant deviations (up to 1 dex) appear at fluxes $<10^{-15}\,{\rm erg\,\,cm^{-2}\,s^{-1}\,\AA^{-1}}$ and specifically for AGN light ratios below 80\% in that flux regime. Deviations from the $x=y$ line (up to 0.3 dex) are also seen at the highest fluxes ($>10^{-13}\,{\rm erg\,\,cm^{-2}\,s^{-1}\,\AA^{-1}}$) and AGN light ratios around and above 90\%. The differences at higher fluxes are more prominent for filters UVW2 and UVM2 as compared to the UVW1 filter. However, looking at the redshift-specific plots, it is evident that the large differences in the \textsc{GALFIT} estimates when compared to the star flux are restricted to the highest redshift range ($0.05 < z < 0.3$). On closer inspection, we found that these are, without exception, the sources for which \textsc{GALFIT} was not able to fit a S\'ersic profile to the source, and the flux estimates are thereby from the sky $+$ PSF fit only. This could be because of the very faint detection of the host galaxy at such large distances, the whole galaxy is pretty much undetected. The significant difference in the fluxes is therefore quite understandable, considering the fact that \textsc{GALFIT} tried to fit the host as well the AGN light at the center with only the PSF component, and hence gives much larger flux values for the AGN compared to the expected star's flux at the center. In the case of small deviations seen at the highest flux and AGN light ratios above 90\%, in filters UVW2 and UVM2 for redshift range $0.01 < z < 0.05$, these cases again correspond to scenarios where \textsc{GALFIT} was not able to get reliable S\'ersic fits for the source images. This could be because the AGN, in these cases, is responsible for most of the light coming from the galaxy, and hence the AGN flux is extremely high when compared to the host galaxy flux. As a result, \textsc{GALFIT} is not able to fit the host galaxy in the background of such a strong central AGN light source that dominates the entire galaxy. Therefore, we get \textsc{GALFIT} flux estimates that are lower than the actual AGN flux.


\FloatBarrier


\begin{figure}[h]
  \begin{subfigure}[t]{0.54\textwidth}
    \centering
    \includegraphics[width=\textwidth]{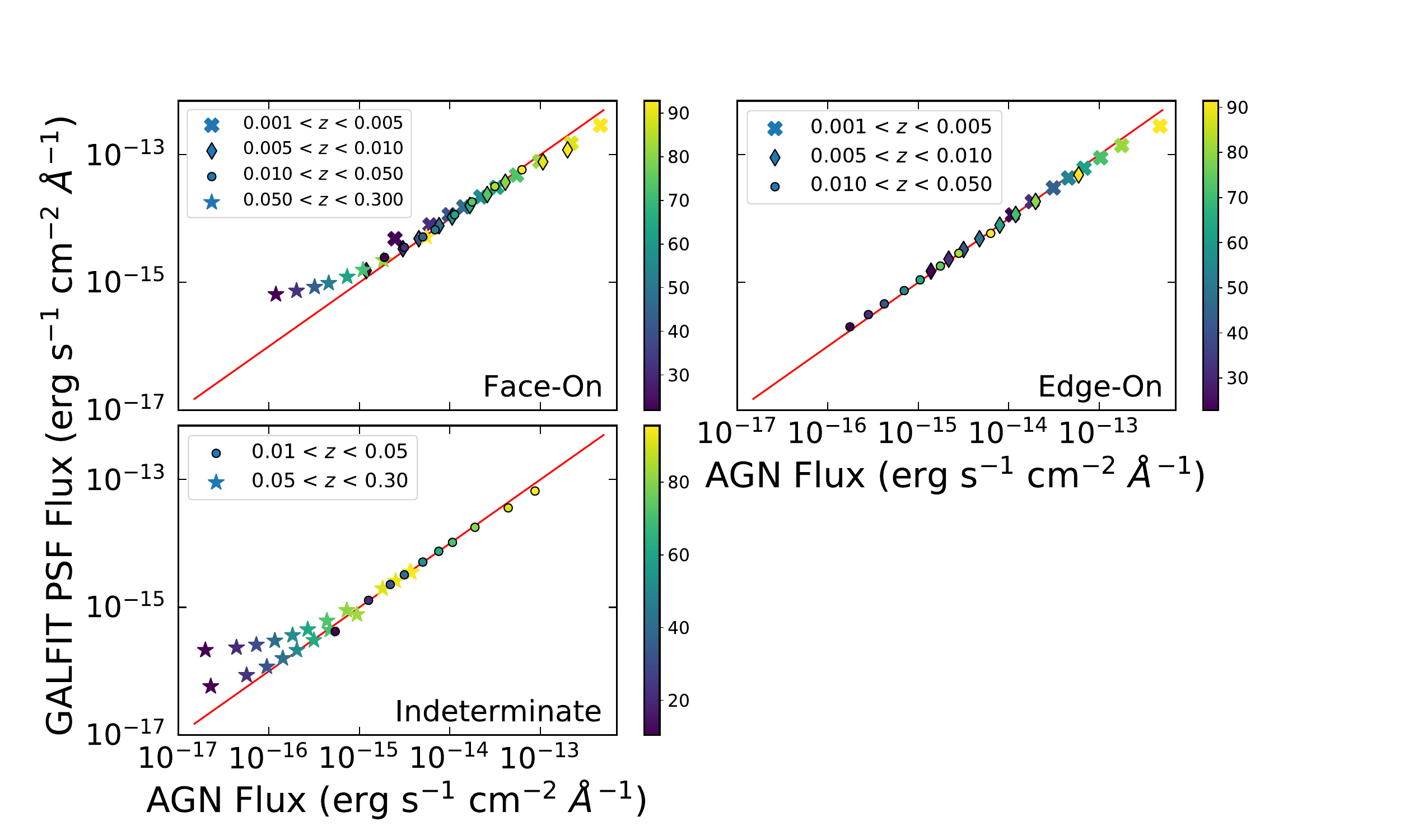} 
    \caption{UVW1}        
    \label{fig:galfit_sim_W1-morph-z}
  \end{subfigure}
  \begin{subfigure}[t]{0.54\textwidth}
    \centering
    \includegraphics[width=\textwidth]{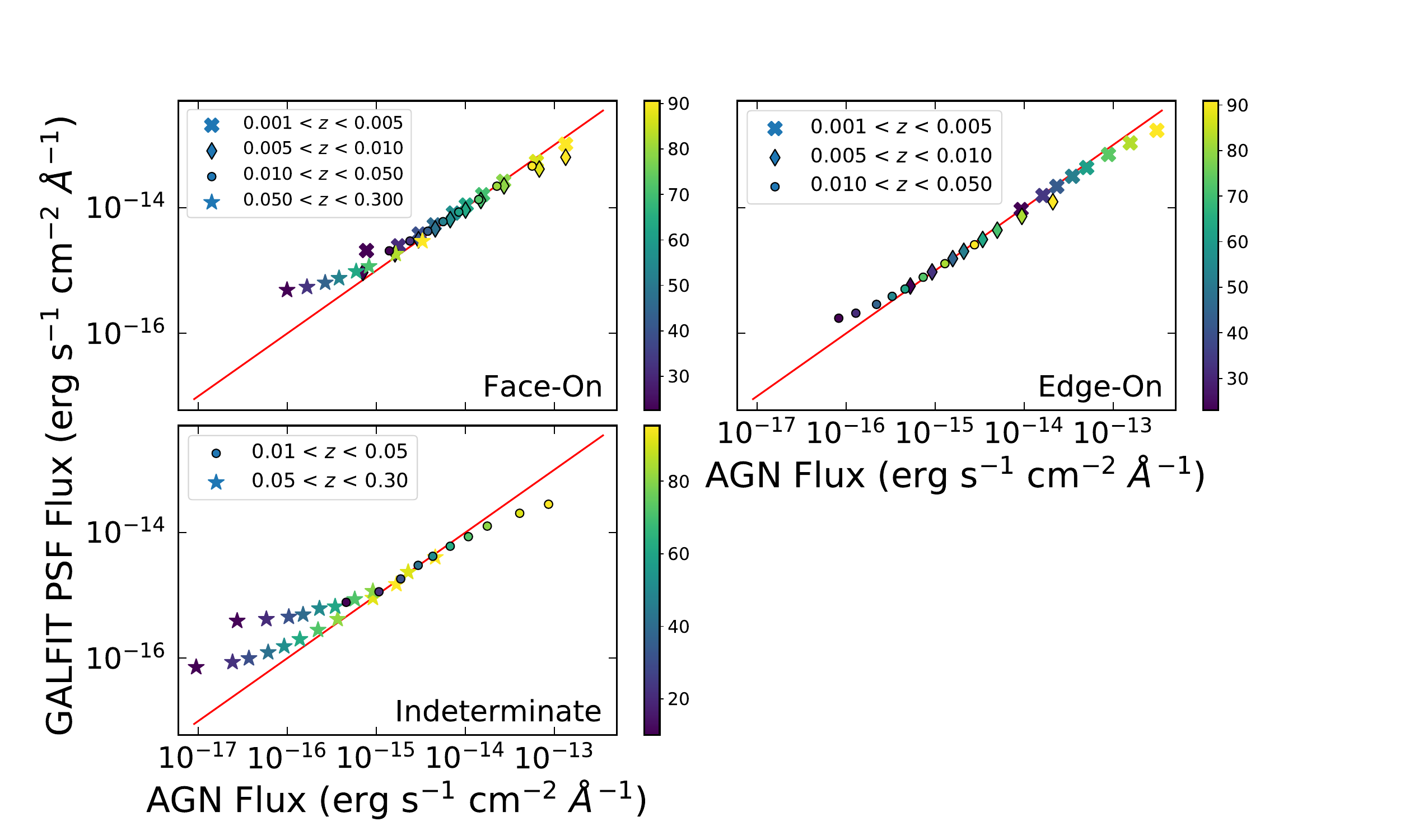}
    \caption{UVM2}
    \label{fig:galfit_sim_M2-morph-z}
  \end{subfigure}
  \newpage
  \begin{subfigure}[t]{0.54\textwidth}
    \centering
    \includegraphics[width=\textwidth]{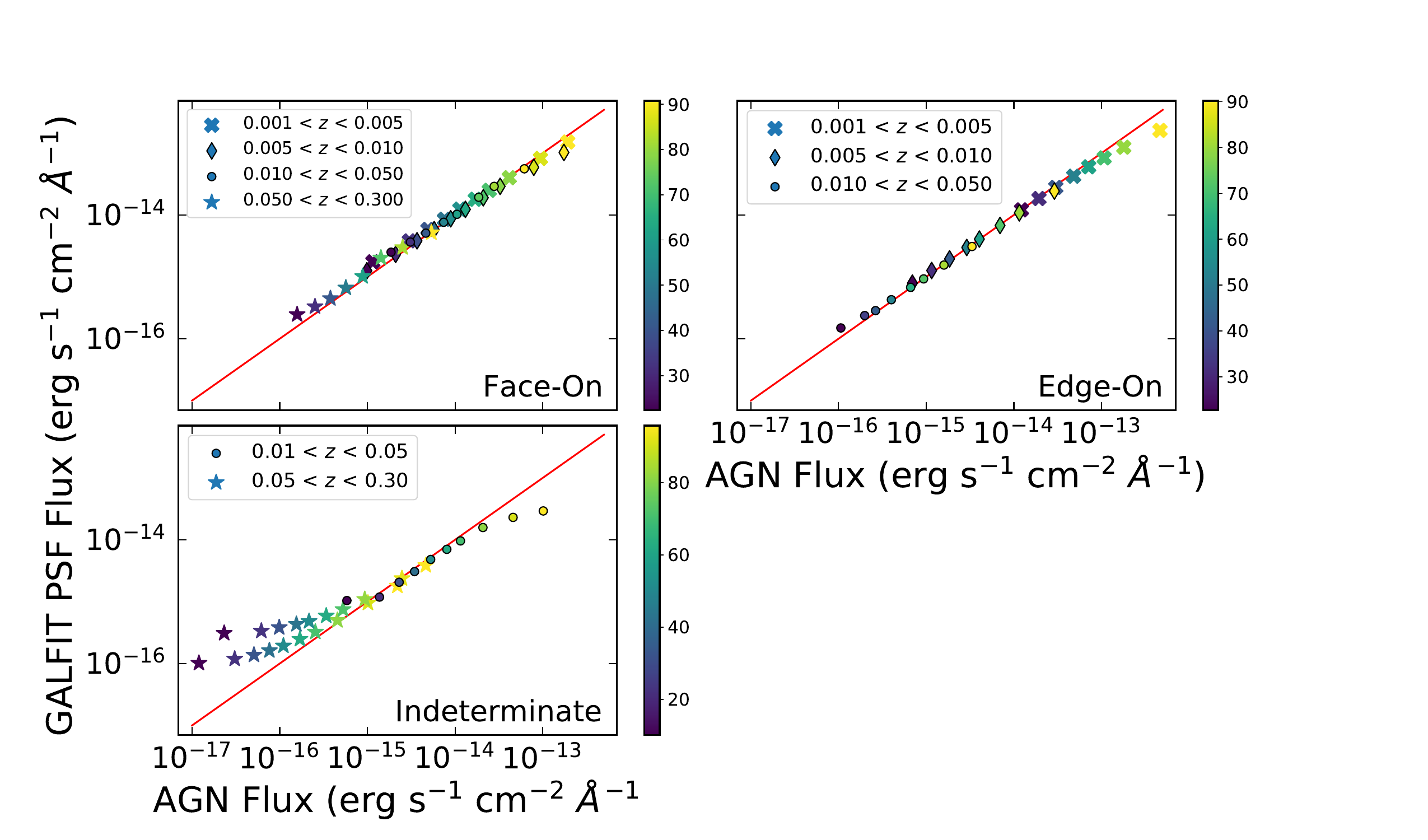}
    \caption{UVW2}
    \label{fig:galfit_sim_W2-morph-z}
  \end{subfigure}
\caption{Comparison of the simulated AGN (star) flux to the PSF flux estimated by \textsc{GALFIT} for filters (a) UVW1, (b)UVM2, and (c) UVW2. The three subplots in each figure show the three different morphologies considered and the different symbols correspond to the redshift range. The $x = y$ line is shown in red. The color bar indicates the AGN light ratio in percent.}
\label{fig:galfit_sim-morph-z}
\end{figure}


In Fig. \ref{fig:galfit_sim-morph-z}, we plot the simulated AGN flux versus the \textsc{GALFIT} estimated PSF flux for the three UV filters in different morphology classes. We have also included redshift information in these plots in the form of different symbols. At first look, based on these plots, it might seem like \textsc{GALFIT} works much better for edge-on galaxies as compared to the indeterminates and face-on galaxies in filters UVW1 and UVM2, and for filter UVW2, it works quite nicely for both edge-ons and face-ons. However, if we take a look at the redshift distribution in all these plots, it is abundantly clear that the differences in the flux estimates are strongly dependent on the redshift. The simulated edge-on galaxies of our test sample are all lying in the lower redshift bins and hence give good fit results. Whereas, the indeterminates belong to the highest redshift range. The few sources that show significant deviations from the $x = y$ line with face-on orientation are also part of the highest redshift bin.

\begin{figure}
\centering
\includegraphics[width=0.52\textwidth]{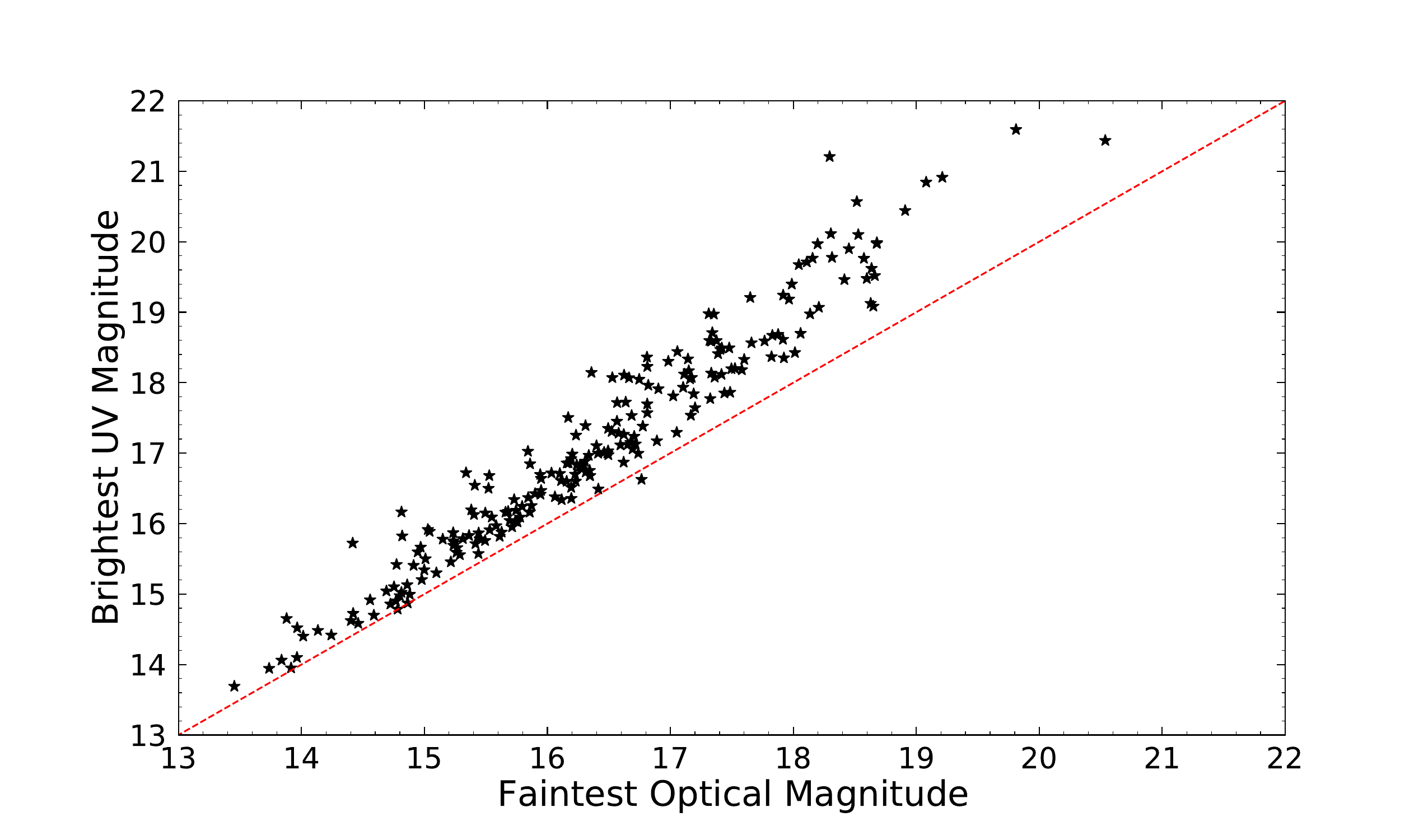} 
\caption{Source magnitudes in the faintest optical band against the brightest UV band. It is evident from this figure that the AGN in our sample are brighter in the optical when compared to the UV and even the faintest optical band is brighter than the brightest UV band. Therefore, based on the \textsc{GALFIT} simulations carried out in the UV band to obtain the correction factors for \textsc{GALFIT}-estimated PSF magnitudes and fluxes, only the faintest and farthest sources need to be corrected in the optical as well. The $x = y$ line is shown in red.}
\label{fig:uv_op_comp}
\end{figure}

Based on the results we obtained from these simulations in the UV filters, we repeat a similar analysis for the three \textit{Swift}/UVOT optical filters but only in the highest redshift bin. This is a reasonable approach, considering that the sources in our sample are brighter in the optical compared to the UV (see Fig. \ref{fig:uv_op_comp}). And as the UV simulations show that it is only the faintest sources that the \textsc{GALFIT}-estimated magnitudes are not reliable, we calculate the optical corrections for the same set of sources ($z>0.05$).


\section{Comparison with HST}\label{sect:appendixb}

Thanks to ongoing follow-up observations of type\,I \textit{Swift}/BAT AGN with optical HST imaging (as part of the optical campaign of BASS), we have high-quality HST images in filter F225W (similar to UVM2) for several sources in our sample ($\sim$ 20). We select four sources (SWIFTJ0630.7+6342, SWIFTJ1535.9+5751, SWIFTJ1708.6+2155, and SWIFTJ1925.0+5041) that have HST observations in the F225W filter, contemporaneous with our \textit{Swift}/UVOT observations. We fit the HST images with \textsc{GALFIT} following the procedure mentioned in Sect. \ref{sect:galfiting} and extract their AGN flux to compare with those obtained from the \textit{Swift}/UVOT images in the UV filter M2. Considering the high resolution of HST images with detailed and extended galaxy profiles, we focus on fitting just a small region around the nucleus to estimate the AGN flux. For all four sources, the magnitudes estimated from the HST images are within 0.1 mag of the ones from the UVOT images. This is acceptable, since the central wavelengths of the two filters are slightly different (F225W: 2371 {\AA} and UVM2: 2246 {\AA}). Therefore, we can conclude that the \textsc{GALFIT}-estimated AGN magnitudes for our \textit{Swift}/UVOT images are consistent with those from the HST images. An effort to study AGN SEDs constructed using HST wide band optical imaging for type\,I \textit{Swift}/BAT AGN ($\sim$ 140) is ongoing.


\end{appendix}

\end{document}